\documentclass[a4paper, oneside, appendixprefix]{scrbook}

\usepackage{amsmath}
\usepackage{amssymb}
\usepackage{hyperref}
\usepackage{graphicx}
\usepackage[numbers, sort&compress]{natbib}
\setcapindent{0pt}
\setcapwidth[c]{\textwidth}
\addtokomafont{caption}{\footnotesize}
\begin{document}

\begin{titlepage}
\centering
\vspace*{1cm}
\renewcommand{\baselinestretch}{1.1}
\textsc{\huge Dissertation} \\
\vskip 1cm
\textsc{\large submitted to the\\
Combined Faculties for the Natural Sciences and for Mathematics \\
of the Ruperto-Carola University of Heidelberg, Germany\\
for the degree of \\
Doctor of Natural Sciences\\
}

\vskip 13cm 
\renewcommand{\baselinestretch}{1.5}
presented by\\
{\bf \large Stefan Fl\"orchinger}  \\
born in Friedrichshafen, Germany\\
Oral examination: 24. June 2009

\end{titlepage}
\pagenumbering{roman}

\thispagestyle{empty}

\begin{titlepage}
\mbox{}
\end{titlepage}

\begin{titlepage}
\centering
\vspace*{1cm}
\textsc{\huge Functional renormalization}\\
\vspace*{0.2cm}
\textsc{\huge and} \\
\vspace*{0.2cm}
\textsc{\huge ultracold quantum gases}
\vfill
\large 
\begin{tabular}{lcl}
Referees: &&\textbf{Prof. Dr. Christof Wetterich} \\
          &&\textbf{Prof. Dr.  Holger Gies } \\
\end{tabular}

\end{titlepage}

\begin{center}
  {\bf \Large Funktionale Renormierung und ultrakalte Quantengase}\\[.2cm]
  {\itshape Zusammenfassung}
\end{center}

{\small
Die Funktionale Renormierungsgruppen-Methode wird zur theoretischen Untersuchung ultrakalter Quantengase angewandt. Flussgleichungen werden abgeleitet f\"{u}r Bosonen mit n\"{a}h\-er\-ungs\-weise punktf\"{o}rmiger Wechselwirkung, f\"{u}r Fermionen mit zwei (Hyperfein-) Komponenten im Crossover vom Bardeen-Cooper-Schrieffer (BCS) Zustand zu einem Bose-Einstein Kondensat (BEC) sowie f\"{u}r ein Gas von Fermionen mit drei Komponenten. Die L\"{o}sungen der Flussgleichungen be\-stim\-men die Eigenschaften dieser Systeme sowohl f\"{u}r wenige Teilchen als auch im thermischen Gleichgewicht.

Im Fall der Bosonen werden die Eigenschaften sowohl f\"{u}r drei als auch f\"{u}r zwei r\"{a}umliche Dimensionen diskutiert, insbesondere das Quantenphasendiagramm, der kondensierte und superfluide Anteil, die kritische Temperatur, die Korrelationsl\"{a}nge, die spezifische W\"{a}rme und die Schallausbreitung. Die Diskussion des Fermionen-Gases im BCS-BEC-Crossover konzentriert sich auf den Effekt von Teilchen-Loch Fluktuationen, betrifft aber das gesamte Phasendiagramm. F\"{u}r drei verschiedene Fermionen zeigen die Flussgleichungen einen Grenzzyklus sowie das Efimov-Spektrum von Dreiteilchen-Bindungszust\"{a}nden. Angewandt auf Lithium kann ein k\"{u}rzlich beobachteter Dreiteilchen-Verlust erkl\"{a}rt werden. Mit Hilfe eines Kontinuit\"{a}ts-Ar\-gu\-men\-tes findet sich f\"{u}r drei Fermionen-Komponenten in der N\"{a}he einer gemeinsamen Resonanz eine neue Trionen-Phase, welche die BCS- und die BEC-Phase trennt.

Etwas formaler ist die Herleitung einer neuen exakten Flussgleichung f\"{u}r skalenabh\"{a}ngige zusammengesetzte Operatoren. Diese erm\"{o}glicht etwa eine verbesserte Behandlung gebundener Zust\"{a}nde.

}

\vspace{1.0cm}

\begin{center}
  {\bf \Large Functional renormalization and ultracold quantum gases}\\[.2cm]
  {\itshape  Abstract}
\end{center}

{\small
The method of functional renormalization is applied to the theoretical investigation of ultracold quantum gases. Flow equations are derived for a Bose gas with approximately pointlike interaction, for a Fermi gas with two (hyperfine) spin components in the Bardeen-Cooper-Schrieffer (BCS) to Bose-Einstein condensation (BEC) crossover and for a Fermi gas with three components. The solution of the flow equations determine the properties of these systems both in the few-body regime and in thermal equilibrium.

For the Bose gas this covers the quantum phase diagram, the condensate and superfluid fraction, the critical temperature, the correlation length, the specific heat or sound propagation. The properties are discussed both for three and two spatial dimensions. The discussion of the Fermi gas in the BCS-BEC crossover concentrates on the effect of particle-hole fluctuations but addresses the complete phase diagram. For the three component fermions, the flow equations in the few-body regime show a limit-cycle scaling and the Efimov tower of three-body bound states. Applied to the case of Lithium they explain recently observed three-body loss features. Extending the calculations by continuity to nonzero density, it is found that a new trion phase separates a BCS and a BEC phase for three component fermions close to a common resonance.

More formal is the derivation of a new exact flow equation for scale dependent composite operators. This equation allows for example a better treatment of bound states.

}
\vfill

\tableofcontents

\chapter{Introduction}
\label{ch:Introduction}
\pagenumbering{arabic}
Functional renormalization in its modern formulation contributes a central part and a valuable tool to our understanding of theoretical physics. It describes how different theories, each of them valid on some momentum scale, are connected to each other. In our modern understanding most theories of physics are ``effective'' theories. They describe phenomena connected with some typical momentum scale $k$ to a good approximation -- often with very high precision. On the other side, they neglect phenomena that are not relevant at this momentum scale. Even the Standard Model of elementary particle physics is of this kind, since it neglects e.\ g.\ gravity. Often, the relevant degrees of freedom change with the scale. For example, in Quantum Chromodynamics (QCD) the high energy theory is described in terms of quarks and gluons, while the low energy limit is governed by mesons and baryons. Functional renormalization describes this transition between different descriptions -- from one (effective) theory to another. 

The functional renormalization group method is also useful for a different task -- the statistical description of complex systems with many particles. In atomic systems, the physics at the momentum scale given by the inverse Bohr radius $k=\frac{\hbar}{a_0}$ is well known. The solution of Schr\"odinger's equation determines the stationary wave functions for electrons, the orbitals. From the structure of the orbitals, the electrostatic, magnetic and spin-related properties, one can calculate predictions for scattering experiments, binding energies and so on. However, if we increase the number of particles, the complexity of the problem rapidly increases as well. It is impossible to find exact solutions to the Schr\"odinger equation for several -- say ten -- atoms including all their electrons. To make progress, sensible approximations are needed. Finding a way to describe complex systems in terms of simple, but nevertheless accurate effective theories is not easy. A lot of physical intuition and insight is needed to make the right approximations. Functional renormalization helps us in this important task. An exact renormalization group flow equation connects field theories on different scales. Its close investigation often shows which terms become relevant or irrelevant if the characteristic momentum scale is changed. 

The renormalization group was first developed for the description of critical phenomena close to phase transitions in statistical systems \cite{PhysRevB.4.3174, PhysRevA.8.401, Wegner1976}. Subsequently, it was realized that the idea of a renormalization group flow as a continuous version of Kadanoff's block-spin transformation \cite{Kadanoff1966} is of great value also for quantum field theory, see e.\ g.\ \cite{WilsonKogut1974}. The development of the renormalization group allowed for a deeper understanding of the formalism including the ``mysterious divergences'' in perturbation theory and the ``renormalization of coupling constants'' by introducing counterterms. 

The modern formulation of functional renormalization uses the concept of the average action (or flowing action) as a modification of the quantum effective action \cite{Wetterich1991, Wetterich:1991be, WetterichZPhys601993}. The effective action is the generating functional of the one-particle irreducible correlation functions, see e.\ g.\ \cite{Weinberg}. A simple, intuitive, but nevertheless exact renormalization group flow equation describes the evolution from microscopic to macroscopic scales \cite{Wetterich1993b}. The approach has proven to be successful in many applications ranging from Quantum Chromodynamics (QCD) to critical phenomena, for reviews see \cite{Berges2000ew, Bagnuls200191, Aoki2000, WetterichIntJModPhys2001, PTP.105.1, PTPS.160.58, Pawlowski2007a}. However, the method is not yet developed completely. Open issues concern for example the description on non-equilibrium dynamics or the improvement of various approximation schemes. A conceptual point addressed in this thesis is a flow equation for scale-dependent composite operators which allows for a flow from ultraviolet to infrared degrees of freedom \cite{Gies:2001nw, Gies:2002kd}.

Ultracold quantum gases provide an ideal testing ground for the flow equation method. The ultraviolet physics in the form of atomic physics is well known. In the few-particle sector, exact results are available from quantum mechanical treatments. Many parameters, such as interaction strengths or more obvious temperature and density, can be tuned experimentally over a wide range. Using tightly confining trap potentials, it is even possible to realize different dimensionalities of space. The experimental methods were developed rapidly and improved steadily in the last years and allow for the preparation of very clean samples, nearly without any impurities. It can be expected that future developments lead to more and more accurate measurement techniques which would allow for precision tests of theoretical predictions.

At the current stage, many experiments are rather well described by perturbative theories for small coupling constants. Examples are the free theory where interaction effects are neglected completely, or Gaussian approximations such as Bogoliubov theory, Hartree-Fock, or various variants of Mean-Field theories. In principle, the flow equation method can reproduce the results of all these approaches in the regime where the corresponding approximations are valid. Moreover, it can give corrections and also describe (non-perturbative) features that are not captured in a Gaussian treatment, for example critical phenomena. Experimentally, these corrections should become relevant for strongly interacting systems such as fermions in the BCS-BEC crossover or for lower dimensional systems where fluctuation effects are more important. Also the regions around phase transitions -- either quantum or classical -- are interesting in this respect. Flow equations have the potential to constitute a systematic extension of perturbative treatments. In contrast to Monte-Carlo methods, the numerical effort is very small. Physical insight is easier to gain from inspecting flow equations then from complex numerical simulations. Even exact statements can sometimes be made from considering the flow equations in an interesting limit where they can be solved exactly.

A large part of the original work presented in this thesis has been published in different articles. For the Bose gas these are {\itshape ``Functional renormalization for Bose-Einstein condensation''} \cite{floerchinger:053603}, {\itshape ``Superfluid Bose gas in two dimensions''} \cite{floerchinger:013601} and {\itshape ``Nonperturbative thermodynamics of an interacting Bose gas''} \cite{FloerchingerNpB2009} and for the BSC-BEC crossover {\itshape ``Particle-hole fluctuations in BCS-BEC crossover''} \cite{Floerchinger2008} as well as {\itshape ``Functional renormalization group approach for the BCS-BEC crossover''} \cite{Diehl:2008}. The work on three-component fermions is published in {\itshape ``Functional renormalization for trion formation in ultracold fermion gases''} \cite{floerchinger:013603}, {\itshape ``Efimov effect from functional renormalization''} \cite{Moroz2008} and {\itshape ``Three-body loss in lithium from functional renormalization''} \cite{Schmidt2009}. Finally, the new exact flow equation for composite operators is published in {\itshape ``Exact flow equation for composite operators''} \cite{FloerchingerWetterich2009exact}. 

In this thesis, the first chapters are devoted to more conceptual issues while concrete applications to ultracold quantum gases are discussed in later chapters. In the remainder of the present chapter, we explain some mathematical ideas underlying the flow equation method at the example of a one-dimensional integral. The functional integral formulation of quantum field theory including the Matsubara formalism is briefly reviewed thereafter. Chapter \ref{ch:TheWetterichequation} discusses the flow equation first obtained by C.\ Wetterich. We re-derive it starting from the functional integral representation of the partition function. A somewhat generalized form is derived in chapter \ref{ch:Generalizedflowequation}. In chapter \ref{ch:Truncations} we discuss the idea and use of truncations as an approximate method to solve the flow equation. Chapter \ref{ch:Cutoffchoices} is devoted to the choice of the appropriate cutoff function with an emphasis on the particular problems occurring in nonrelativistic field theories. 

Contact with concrete physics is first made in chapter \ref{ch:Variousphysicalsystems} where we introduce the microscopic models investigated in this thesis. Besides the repulsive Bose gas in three and two dimensions, this includes two component fermions in the BCS-BEC crossover and fermions with three hyperfine species (``BCS-Trion-BEC transition''). Chapters \ref{ch:Symmetries} and \ref{ch:Truncationsandflowequations} are again a bit more formal and discuss the different symmetries of our models and the used approximation schemes. Results concerning the few-body physics are presented in chapter \ref{ch:Few-bodyphysics}. However, the results of this thesis concern mainly the many-body regime and are presented in chapter \ref{ch:Many-bodyphysics}. We discuss the phase diagram and thermodynamic observables for bosons in three spatial dimensions, the superfluidity of an interacting Bose gas in two dimensions, the phase diagram of a two-component Fermi gas in the BCS-BEC crossover with emphasis on the particle-hole fluctuations and the BCS-Trion-BEC transition expected for a gas of three fermion species close to a common Feshbach resonance. Finally, we draw some conclusions in chapter \ref{ch:Conclusions}. 

In appendix \ref{ch:Foundationsofquantumtheory:someideas} we present some ideas concerning the connection between the functional integral and probability in the foundations of quantum theory. More specific, we discuss a reformulation of the functional integral representation in terms of (quasi-) probabilities. More technical additions such as concrete flow equations for the effective potential and the proof of a theorem concerning the flow equations in vacuum, are given in appendix \ref{ch:Technicaladditions}.

\section{Flow equations to solve an integral}
\label{ch:Flowequationstosolveanintegral}
In this introductory section we develop a method to calculate simple one-dimensional integrals. This may not seem very useful since many integration techniques are known and the method we devise here is not particularly simple. However, it has some advantages, the most important of which is that it can be generalized easily to higher-dimensional or even infinite-dimensional (functional) integrals. No attempt is made to present the following discussion in mathematical rigor. It should be seen as an introductory warm-up to come into contact with some tools and ideas used in later chapters. In particular we will assume that all involved functions have nice enough properties concerning smoothness and convergence. 

Our goal is to calculate an integral of the form
\begin{equation}
Z=\int_{-\infty}^\infty dx \,f(x).
\label{eq:2.1}
\end{equation}	
For simplicity we take the function $f$ to be positive semi-definite, $f(x)\geq0$. One may think of $f$ as describing a probability distribution of the variable $x$. (For $Z=1$ this probability distribution would be normalized.) For convenience we introduce the function $S(x)$ defined by
\begin{equation}
f(x)=e^{-S(x)}.
\end{equation}
The special case where $S$ is quadratic in $x$
\begin{equation}
S(x)=\frac{1}{2}m^2 x^2
\end{equation}
with $m^2>0$ can be treated exactly. In that case the integral in Eq.\ \eqref{eq:2.1} is of Gaussian form and we obtain
\begin{equation}
Z=\int_{-\infty}^\infty dx\, e^{-\frac{1}{2}m^2 x^2} = \sqrt{2\pi}/m.
\end{equation}

It is useful to generalize $Z$ somewhat by introducing the ``source'' $j$. We define
\begin{equation}
Z(j) = \int_{-\infty}^{\infty} dx\,e^{-S(x)+jx}.
\label{eq:2partitionfunctionnocutoff}
\end{equation}
From $Z(j)$ we can easily derive expectation values, for example
\begin{equation}
\langle x\rangle=\frac{1}{Z(j)} \int_{-\infty}^\infty dx \,x \,e^{-S(x)+jx} = \frac{1}{Z(j)}\frac{\partial}{\partial j} Z(j).
\end{equation}
Higher momenta of the probability distribution are obtained as
\begin{equation}
\langle x^n\rangle = \frac{1}{Z(j)}\frac{\partial^n}{\partial j^n}Z(j).
\end{equation}
From the ``Schwinger function'' $W(j)=\ln Z(j)$ we can directly obtain the cumulants of the probability distribution. For example, the variance is given by
\begin{equation}
\sigma = \langle x^2\rangle -\langle x \rangle^2 = \frac{\partial^2}{\partial j^2} W(j).
\end{equation}
More general, the $n$-th cumulant is given by
\begin{equation}
\kappa_n = \frac{\partial^n}{\partial j^n} W(j).
\end{equation}
For this reason the functions $Z(j)$ and $W(j)$ are also known as the moment-generating function and cumulant generating function, respectively.

We also introduce the ``effective action'' as the Legendre transform of the Schwinger function 
\begin{equation}
\Gamma(\varphi) = j \varphi - W(j).
\label{eq:2Legendretransform}
\end{equation}
On the right hand side of Eq.\ \eqref{eq:2Legendretransform} the source $j$ has to be taken such that is fulfills the implicit equation
\begin{equation}
\varphi = \frac{\partial}{\partial j} W(j).
\label{eq2:impliciteqj}
\end{equation}
In other words $\varphi$ is the expectation value, $\varphi=\langle x \rangle$. It is straightforward to derive some properties of the effective action
\begin{eqnarray}
\nonumber
\frac{\partial}{\partial\varphi}\Gamma(\varphi) &=& j,\\
\frac{\partial^2}{\partial\varphi^2}\Gamma = \frac{\partial j}{\partial \varphi} &=& \left(\frac{\partial \varphi}{\partial j}\right)^{-1} = \left(\frac{\partial^2 }{\partial j^2}W\right)^{-1}.
\end{eqnarray}
We note that the Legendre transform of the effective action $\Gamma(\varphi)$ is again the Schwinger function $W(j)$. This shows that if  both functions are well defined they carry the same information. 

After this excursion to probability theory let us now come back to the issue of calculating the integral in Eq.\ \eqref{eq:2.1}. We show in the following that solving the integral in Eq.\ \eqref{eq:2.1} is equivalent to solving a partial differential equation for the flowing action or average action $\Gamma_k(\varphi)$, a generalization of the effective action $\Gamma(\varphi)$. We start by generalizing Eq.\ \eqref{eq:2partitionfunctionnocutoff}
\begin{equation}
Z_k(j) = e^{W_k(j)} = \int_{-\infty}^\infty dx \, e^{-S(x)-\frac{1}{2}k^2 x^2+jx}
\end{equation}
where we introduced a ``cutoff'' term $e^{-\frac{1}{2}k^2 x^2}$. For large $k^2$ this factor suppresses the contribution from large values of $x$ to the integral. The $k$-dependence of $W_k(j)$ is obtained as
\begin{eqnarray}
\nonumber
\partial_k W_k(j) &=& -\frac{1}{2}(\partial_k k^2) \langle x^2\rangle\\
&=& -\frac{1}{2}(\partial_k k^2) \left(\frac{\partial^2 W_k}{\partial j^2}+\langle x\rangle^2\right).
\label{eq2:scaledepofSchwingerf}
\end{eqnarray}
The flowing action is defined by subtracting from the Legendre transform
\begin{equation}
\tilde \Gamma_k(\varphi) = j\varphi-W_k(j),
\label{eq2:tildeflowingaction}
\end{equation}
the cutoff term
\begin{equation}
\Gamma_k(\varphi) = \tilde \Gamma_k(\varphi) -\frac{1}{2} k^2 \varphi^2.
\end{equation}
As for the effective action the argument of the flowing action is  given by $\varphi=\frac{\partial}{\partial j}W_k$. We obtain the ``field equation'' for the average action as
\begin{equation}
\frac{\partial \Gamma_k}{\partial \varphi} = j - k^2 \varphi.
\end{equation}
Similarly as for the effective action one has
\begin{equation}
\frac{\partial^2 \tilde \Gamma_k}{\partial \varphi^2}\frac{\partial^2 W_k}{\partial j^2} = 1\quad \text{or} \quad \frac{\partial^2 W_k}{\partial j^2} = \left(\frac{\partial^2\Gamma_k}{\partial\varphi^2}+k^2\right)^{-1}.
\end{equation}
In order to derive a flow equation for $\Gamma_k$ we use Eq.\ \eqref{eq2:tildeflowingaction}
\begin{eqnarray}
\nonumber
\partial_k \tilde \Gamma_k(\varphi){\big |}_\varphi &=& -\partial_k W_k(j){\big |}_j +\left(\varphi-\frac{\partial W_k}{\partial j}\right)\partial_k j{\big |}_\varphi\\
&=& -\partial_k W_k(j){\big |}_j.
\end{eqnarray}
Together with Eq.\ \eqref{eq2:scaledepofSchwingerf} this leads us to
\begin{equation}
\partial_k \Gamma_k = \frac{1}{2}(\partial_k k^2)\left(\frac{\partial^2\Gamma_k}{\partial \varphi^2}+k^2\right)^{-1}.
\label{eq2:flowequationzerodimension}
\end{equation}
This is the simplest form of the more general flow equation first obtained by C. Wetterich \cite{Wetterich1993b}.

For the case of the one-dimensional integral as in Eq.\ \eqref{eq:2.1} corresponding to a zero-dimensional field theory (zero time- and zero space-dimensions), this is just a partial differential equation for $\Gamma_k(\varphi)$ as a function of $k$ and $\varphi$.

The practical consequences of Eq.\ \eqref{eq2:flowequationzerodimension} are as follows. 
Suppose that we know the form of $\Gamma_k$ as a function of $\varphi$ for some value of $k$ (say for large $k$).
From solving the flow equation \eqref{eq2:flowequationzerodimension} we can then obtain $\Gamma_k(\varphi)$ for all values of $k$ (at least in principle). The limit $k\to 0$ is especially interesting since it follows directly  from the definitions that the flowing action approaches the effective action
\begin{equation}
\lim_{k\to0} \Gamma_k(\varphi) = \Gamma(\varphi).
\end{equation}
Not only can we infer from $\Gamma(\varphi)$ the value of our integral $Z$ in Eq.\ \eqref{eq:2.1}, but also all correlation functions or cumulants. To obtain $Z$ we use
\begin{equation}
Z=e^{W_{k=0}(j=0)}
\end{equation}
and $\Gamma=-W$ for $j=0$. In other words, we have
\begin{equation}
Z=e^{-\Gamma(\varphi_\text{eq})},
\end{equation}
where $\varphi_\text{eq}$ is determined such that 
\begin{equation}
\frac{\partial}{\partial\varphi}\Gamma(\varphi){\big |}_{\varphi=\varphi_\text{eq}}=0.
\end{equation}
The cumulants can be obtained from $W(j)$, i.~e. the Legendre transform of $\Gamma(\varphi)$.

This discussion shows already that the flow equation \eqref{eq2:flowequationzerodimension} is quite powerful and useful if we are interested in the normalization of the probability distribution $S(x)$ as well as in its properties such as cumulants or probabilistic moments. 

It remains to be shown that the flowing action has a simple form in the limit of large $k^2$ which can serve as an initial condition for the flow equation \eqref{eq2:flowequationzerodimension}. To see that this is indeed the case we consider the integral representation that is easily derived from the definitions
\begin{eqnarray}
e^{-\Gamma_k(\varphi)} = \int_{-\infty}^\infty dx \, e^{-S(\varphi+x)-\frac{1}{2}k^2x^2+\frac{\partial \Gamma_k}{\partial \varphi}x}.
\end{eqnarray}
For large $k^2$ we can use
\begin{equation}
\lim_{k\to\infty} \frac{k}{\sqrt{2\pi}} e^{-\frac{1}{2}k^2 x^2} = \delta(x)
\end{equation}
in order to obtain
\begin{equation}
\lim_{k\to\infty}\Gamma_k(\varphi)=S(\varphi)+\ln \frac{k}{\sqrt{2\pi}}.
\end{equation}
With this we have a simple initial condition for the function $\Gamma_k$ and thus for the flow equation \eqref{eq2:flowequationzerodimension}.

The reader may wonder what we have won so far. Solving a partial differential equation is usually at least as complicated as solving an integral. We emphasize again that the big advantage of the flow equation method is that it can be generalized to integrals with many dimensions. In fact we have used besides general arguments only the usual result for Gaussian integration. Higher dimensional Gaussian integrals can be treated very similar. Although it is usually not possible to find exact solutions to the flow equation, it is very valuable as a starting point for different sorts of approximations.

\section{Functional integral representation of quantum field theory}
\label{ch:Functionalintegralrepresentationofquantumfieldtheory}
In the functional integral formulation of quantum field theory one calculates expectation values and correlation functions on a very large configuration space. For example, for a single complex scalar field one allows the value $\varphi(x)\in \mathbb{C}$ to be different for every space-time point $x$.  Different field configurations are weighted by a factor that is determined by the microscopic action $S[\varphi]$. For (stationary) quantum fields in thermal equilibrium this weighting factor
\begin{equation}
e^{-S[\varphi]}
\end{equation}
is real and positive. It has the meaning of a probability for the microscopic field configuration $\varphi$ (``functional probability''). In the more general case of dynamical (time-dependent) quantum fields, the weighting factor is a complex number. It looses its direct interpretation as a probability. However, on a formal level there is still a close relationship between quantum field theory and statistical theories. For many purposes one can use analytic continuation from real to imaginary time variables to map dynamical quantum field theories with complex weighting factors to statistical theories where the weighting factor is real. Since we are here mainly interested in the properties of thermal and chemical equilibrium, we concentrate on the imaginary time formulation in the following subsections. In Appendix \ref{ch:Foundationsofquantumtheory:someideas} we present some ideas how it might be possible to reformulate the functional integral description of time-dependent quantum fields in such a way that it deals with real (and positive) probabilities. This shows the probabilistic character of quantum field theory explicitly and might be useful for a more detailed understanding of quantum field theory as well as its philosophical consequences.

\subsubsection{From the lattice to field theory}

One way to approach the infinitely many degrees of freedom of a continuous field theory comes from a discrete lattice of space-time points. Consider a lattice of points
\begin{equation}
\vec x_{ijk} = a \begin{pmatrix}i  \\ j \\ k\end{pmatrix}, \quad i,j,k\in \mathbb{Z}
\end{equation}
at times
\begin{equation}
t_n = b\, n, \quad n\in \mathbb{Z}.
\end{equation}
For every set of indices $(ijk,n)$ the field $\varphi(x_{ijk},t_n)$ has some value, e.~g. $\varphi(x_{ijk},t_n)\in \mathbb{C}$ for a complex scalar. The partition sum, i.~e. the sum over all possible configurations weighted by the corresponding functional probability is given by
\begin{equation}
Z = \left( \prod_{(ijk,n)\in \mathbb{Z}^4} \int d \varphi(\vec x_{ijk},t_n)\right)\, e^{-S(\varphi(\vec x_{ijk},t_n))}.
\label{eq:PartitionfunctionSTlattice}
\end{equation}
The action $S$ depends on the values of $\varphi(\vec x_{ijk},t_n)$ for the different lattice points. Eq.\ \eqref{eq:PartitionfunctionSTlattice} describes a theory on a discrete space-time lattice. From the probabilities $e^{-S}$ we can calculate all sorts of expectation values, correlation functions and so on. 

Our theory becomes a continuum field theory in the limit where $a\to 0$ and $b\to 0$. The partition function reads then 
\begin{equation}
Z = \lim_{a,b\to0} \left(\prod_{(ijk,n)\in \mathbb{Z}^4}\int d\varphi(\vec x_{ijk},t_n)\right) e^{-S(\varphi(\vec x_{ijk},t_n))}.
\end{equation}
This can also be written as
\begin{equation}
Z = \int D\varphi \,e^{-S[\varphi]}.
\end{equation}
The functional integral $\int D\varphi$ might be defined by the limiting procedure above. The microscopic action $S$ is now a functional of the field configuration $\varphi(\vec x, t)$, where space and time are now continuous, $(\vec x, t)\in \mathbb{R}^4$.

\subsubsection{Expectation values, correlation functions}

With our formalism we aim for a statistical description of fields. Important concepts are expectation values of operators and correlation functions. For simplicity, we denote the field degrees of freedom by $\tilde \Phi_\alpha$. The collective index $\alpha$ labels both continuous degrees of freedom such as position or momentum and discrete variables such as spin, flavor or simply ``particle species''. The field $\tilde \Phi_\alpha$ might consist of both bosonic and fermionic parts. The fermionic components are described by Grassmann numbers while the bosonic components correspond to ordinary ($\mathbb{C}$) numbers. As an example we consider a theory with a complex scalar field $\varphi$ and a fermionic complex two-component spinor $\psi=(\psi_1,\psi_2)$. It is useful to decompose the complex scalar into real and imaginary parts
\begin{equation}
\varphi = \frac{1}{\sqrt{2}}(\varphi_1+i \varphi_2).
\end{equation}
In momentum space the Nambu spinor of fields reads
\begin{equation}
\Phi(q) = \left( \varphi_1(q), \varphi_2(q), \psi_1(q), \psi_2(q), \psi_1^*(-q), \psi_2^*(-q)\right)
\end{equation}
The index $\alpha$ stands in this case for the momentum $q$ and the position in the Nambu-spinor, e.q. $\Phi_\alpha = \psi_1(q)$ for $\alpha=(q,3)$.

The field expectation value is given by
\begin{equation}
\Phi_\alpha = \langle \tilde \Phi_\alpha\rangle= \frac{1}{Z}\int D\tilde \Phi \,\tilde\Phi_\alpha \, e^{-S[\tilde\Phi]},
\label{eq3:fieldexpectationvalue}
\end{equation}
with
\begin{equation}
Z = \int D\tilde \Phi\, e^{-S[\tilde \Phi]}.
\end{equation}
Quite similar one defines correlation functions as
\begin{equation}
c_{\alpha\beta\gamma\dots} = \langle\tilde\Phi_\alpha\tilde \Phi_\beta\tilde\Phi_\gamma\dots\,\rangle = \frac{1}{Z}\int D\tilde\Phi \,\tilde\Phi_\alpha \tilde\Phi_\beta\tilde\Phi_\gamma\dots e^{-S[\tilde\Phi]}.
\end{equation}
As an example let us consider the two-point function. It is sensible to decompose it into a connected and a disconnected part like
\begin{equation}
\langle\tilde\Phi_\alpha\tilde\Phi_\beta\rangle = \langle\tilde\Phi_\alpha\tilde\Phi_\beta\rangle_c + \langle\tilde\Phi_\alpha\rangle \langle\tilde\Phi_\beta\rangle.
\end{equation}
The connected part is the (full) propagator 
\begin{equation}
G_{\alpha\beta} = \langle\tilde\Phi_\alpha\tilde\Phi_\beta\rangle_c.
\end{equation}
Although we discussed here the statistical formulation of the theory (``imaginary time'') the concepts of expectation values and correlation functions are also useful for the real-time formalism. Formally, the main difference is that the weighting factor $e^{-S[\tilde\Phi]}$ becomes complex after analytic continuation
\begin{equation}
e^{-S[\tilde\Phi]}\to e^{iS_t[\tilde\Phi]},
\end{equation}  
where $S_t[\tilde\Phi]$ is now the real-time action.

\subsubsection{Functional derivatives, generating functionals}

To calculate expectation values and correlation functions it is useful to work with sources, functional derivatives and generating functionals. We first explain what a functional derivative is. In some sense  it is a natural generalization of the usual derivative to functionals, i.\ e.\ to objects that depend on an argument which is itself a function on some space. The basic axiom for the functional derivative is
\begin{equation}
\frac{\delta}{\delta f(x)} f(y) = \delta(x-y) \quad \text{or} \quad \frac{\delta}{\delta f(x)}\int_y f(y)g(y) = g(x).
\label{eq:axiomfunctionalderivative}
\end{equation} 
Here we use a notation where the precise meaning of $\delta(x-y)$ and $\int_x$ depends on the situation. For example when we consider a space with $3+1$ dimensions we have
\begin{equation}
\delta(x-y) = \delta^{(4)}(x-y) = \delta(x_0-y_0) \delta^{(3)}(\vec x-\vec y)
\end{equation}
and
\begin{equation}
\int_x = \int dx_0 \int d^3x.
\end{equation}
It should always be clear from the context what is meant. Eq.\ \eqref{eq:axiomfunctionalderivative} is the natural extension of the corresponding rule for vectors $x,y\in \mathbb{R}^n$
\begin{equation}
\frac{\partial}{\partial x_i} x_j = \delta_{ij}\quad \text{or} \quad \frac{\partial}{\partial x_i}\sum_j x_j y_j = y_i.
\end{equation}
In addition to Eq.\ \eqref{eq:axiomfunctionalderivative} the functional derivative should obey the usual derivative rules such as product rule, chain rule etc. Using the abstract index notation introduced before Eq.\ \eqref{eq3:fieldexpectationvalue} we write the axiom in Eq.\ \eqref{eq:axiomfunctionalderivative} as
\begin{equation}
\frac{\delta}{\delta f_\alpha} f_\beta = \delta_{\alpha\beta} \quad \text{or}\quad \frac{\delta}{\delta f_\alpha} \sum_\beta f_\beta g_\beta = g_\alpha. 
\end{equation}

With this formalism at hand we can now come back to our task of calculating expectation values and correlation functions. We introduce the source-dependent partition function by the definition
\begin{equation}
Z[J]=\int D\tilde\Phi \, e^{-S[\tilde \Phi]+J_\alpha \tilde\Phi_\alpha}.
\label{eq3:sourcedeppartfunction}
\end{equation}
Expectation values are obtained as functional derivatives
\begin{equation}
\Phi_\alpha = \langle\tilde\Phi_\alpha\rangle = \frac{1}{Z} \frac{\delta}{\delta J_\alpha} Z[J],
\label{eq1:expectvalue}
\end{equation}
and similarly correlation functions
\begin{equation}
c_{\alpha\beta\gamma\dots} = \frac{1}{Z} \frac{\delta}{\delta J_\alpha} \frac{\delta}{\delta J_\beta} \frac{\delta}{\delta J_\gamma}\dots Z[J].
\end{equation}
The connected part of the correlation functions can be obtained more direct from the Schwinger functional
\begin{equation}
W[J] = \ln Z[J].
\label{eq3:Schwingerfunctaslog}
\end{equation}
For example the propagator $G$, the connected two-point function, is given by
\begin{equation}
G_{\alpha\beta} = \frac{\delta}{\delta J_\alpha} \frac{\delta}{\delta J_\beta} W[J].
\end{equation}
Due to these properties one calls $Z[J]$ ($W[J]$) the generating functional for the (connected) correlation functions. 

\subsubsection{Microscopic actions in real time and analytic continuation}

In this subsection we discuss the relation between the real-time and the imaginary-time action as well as the analytic continuation in more detail. For concreteness we consider a nonrelativistic repulsive Bose gas in three-dimensional homogeneous space and in vacuum ($\mu=0$). It is straightforward to transfer the discussion also to other cases. 

In real time the microscopic action is given by
\begin{equation}
S_t = -\int dt \int d^3 x \left\{\varphi^*(-i\partial_t-\Delta-i\epsilon)\varphi+\frac{1}{2}\lambda(\varphi^*\varphi)^2\right\}.
\label{eq3:microscopicactionrealtime}
\end{equation}
The overall minus sign is to match the standard convention. After Fourier transformation the term quadratic in $\varphi$ that determines the propagator reads
\begin{equation}
S_{t,2} = \int \frac{d \omega}{2\pi} \int \frac{d^3p}{(2\pi)^3} \, \varphi^*\left(\omega-\vec p^2+i\epsilon\right)\varphi.
\label{eq3:quadraticpartrealtimeforier}
\end{equation}
In the basis with the complex fields $\varphi$, $\varphi^*$ the inverse microscopic propagator reads
\begin{equation}
G(p)^{-1}\,\delta(p-p^\prime) = \frac{\delta}{\delta \varphi^*(p)} \frac{\delta}{\delta \varphi(p^\prime)} S_{t,2} = \left(\omega-\vec p^2+i\epsilon\right)\, \delta (p-p^\prime).
\label{eq3:propagatorrealtime}
\end{equation}
From $\det G^{-1}(p)=0$ we obtain for $\epsilon\to0$ the dispersion relation $\omega=\vec p^2$. 

For the action in Eq.\ \eqref{eq3:microscopicactionrealtime} one can determine the field theoretic expectation values and correlation functions using the formalism described in the previous subsection with the complex weighting function
\begin{equation}
e^{iS_t[\varphi]}.
\label{eq3:weightingfactorrealtime}
\end{equation}
In Eqs. \eqref{eq3:microscopicactionrealtime}, \eqref{eq3:quadraticpartrealtimeforier} and \eqref{eq3:propagatorrealtime} the small imaginary term $i\epsilon$ is introduced to enforce the correct frequency integration contour (Feynman prescription). In Eq.\ \eqref{eq3:weightingfactorrealtime} it leads to a Gaussian suppression for large values of $\varphi^*\varphi$,
\begin{equation}
e^{iS_t[\varphi]} = e^{i \text{Re} S_t[\varphi]} e^{-\epsilon \int_x \varphi^*\varphi},
\end{equation}
which makes the functional integral convergent. Let us now consider the analytic continuation to imaginary time
\begin{equation}
t\to e^{-i\alpha}\tau,\quad 0\leq \alpha <\pi/2.
\end{equation}
For $\alpha\to \pi/2$ we have $t\to - i \tau$ and 
\begin{equation}
(-i\frac{\partial}{\partial t}-i\epsilon)\to \frac{\partial}{\partial \tau}, \quad \int dt\to -i\int d\tau.
\end{equation}
The weighting factor in Eq.\ \eqref{eq3:weightingfactorrealtime} becomes
\begin{equation}
e^{-S[\varphi]}
\end{equation}
with
\begin{equation}
S[\varphi] = \int d\tau \int d^3 x \left\{\varphi^*(\partial_\tau-\Delta)\varphi+\frac{1}{2}\lambda (\varphi^*\varphi)^2\right\}.
\label{eq3:imaginarytimemicroscopicaction}
\end{equation}

\subsubsection{Matsubara formalism}

In statistical physics one is often interested in properties of the thermal (and chemical) equilibrium. For quantum field theories the thermal equilibrium is conveniently described using the Matsubara formalism. In this section we give a short account of the formalism and refer for a more detailed discussion to the literature \cite{Mahan1981}. 

The grand canonical partition function is defined as
\begin{equation}
Z=\text{Tr}\, e^{-\beta(H-\mu N)}.
\label{eq3:grandcanparttrace}
\end{equation}
Here we use $\beta =\frac{1}{T}$ and recall our units for temperature with $k_B=1$. The trace operation in Eq.\ \eqref{eq3:grandcanparttrace} sums over all possible states of the system, including varying particle number. The operator $H$ is the Hamiltonian and $N$ the particle number operator. The factor
\begin{equation}
e^{-\beta(H-\mu N)}
\label{eq3:imaginarytimeevoloperator}
\end{equation}
is quite similar to an unitary time evolution operator $e^{i\Delta t H}$ evolving the system over some time interval $\Delta t = t_2-t_1$. Indeed, we can define $\tilde H = H-\mu N$ and evolve the system from time $t_1=0$ to the imaginary time $t_2=-i\beta$ with the operator in Eq.\ \eqref{eq3:imaginarytimeevoloperator}. If we take a (generalized) torus with circumference $\beta$ in the imaginary time direction as our space-time manifold we can use the functional integral formulation of quantum field theory to write Eq.\ \eqref{eq3:grandcanparttrace} as
\begin{equation}
Z = \int D \tilde\varphi e^{-S[\tilde\varphi]}
\end{equation}
where $S$ is an action with imaginary and periodic time. From the imaginary time action described in the last subsection it is obtained by replacing also the Hamiltonian $H$ by $H-\mu N$. For our Bose gas example this results in
\begin{equation}
S[\varphi] = \int_0^\beta d\tau \int d^3 x \left\{\varphi^*(\partial_\tau-\Delta-\mu)\varphi+\frac{1}{2}\lambda (\varphi^*\varphi)^2\right\}.
\label{eq3:matsubaraaction}
\end{equation}
Since time is now periodic, the Fourier transform leads to discrete frequencies. The quadratic part of $S$ in Eq.\ \eqref{eq3:matsubaraaction} reads in momentum space
\begin{equation}
S_2[\varphi] = T\sum_{n=-\infty}^\infty \int \frac{d^3 p}{(2\pi)^3} \varphi^*(i\omega_n+\vec p^2-\mu)\varphi
\label{eq3:matsubactionquadraticpart}
\end{equation}
with the Matsubara frequency $\omega_n=2\pi T n$. In the limit $T\to 0$ the summation over Matsubara frequencies becomes again an integration
\begin{equation}
T \sum_n \to \int \frac{d\omega}{2\pi}.
\end{equation}
For the Fourier decomposition in Eq.\ \eqref{eq3:matsubactionquadraticpart} we used the boundary condition
\begin{eqnarray}
\nonumber
\varphi(\tau=\beta, \vec x) &=& \varphi(\tau=0,\vec x),\\
\varphi^*(\tau=\beta, \vec x) &=& \varphi^*(\tau=0,\vec x),
\end{eqnarray}
as appropriate for bosonic fields. For fermionic or Grassmann-valued fields $\psi$ a careful analysis (see e.g. \cite{WegnerGrassmannVariable}) leads to the boundary conditions
\begin{eqnarray}
\nonumber
\psi(\tau=\beta,\vec x) &=& -\psi (\tau=0,x)\\
\psi^*(\tau=\beta,\vec x) &=& -\psi^*(\tau=0,\vec x).
\end{eqnarray}
In this case the Matsubara frequencies appearing in Eq.\ \eqref{eq3:matsubactionquadraticpart} are of the form
\begin{equation}
\omega_n = 2\pi T \left(n+\frac{1}{2}\right), \quad n\in \mathbb{Z}.
\end{equation}

\chapter{The Wetterich equation}
\label{ch:TheWetterichequation}
In this section we review the properties and derivation of the flow equation first published by Christof Wetterich in 1993 \cite{Wetterich1993b}. We will derive this equation from the functional integral representation of quantum field theory. The relation of the flow equation to other methods such as perturbation theory for small interaction strength is then particular clear. In principle one might also consider the flow equation as an own formulation of quantum field theory from which other formulations such as the functional integral representation or the operator formalism can be derived. For practical purposes all this different formulations have their advantages and disadvantages. While some problems are best solved with the flow equation formalism, different methods might be more suitable for other problems. Our physical insight and intuition grows if we look at physics from different perspectives. It is therefore a sensible ambition to further develop the flow equation method and learn how to apply it to various problems in modern physics.

\section{Scale dependent Schwinger functional}

We start  with the Schwinger functional in Eqs. \eqref{eq3:sourcedeppartfunction}, \eqref{eq3:Schwingerfunctaslog}, which we modify by introducing an cutoff term $\Delta S_k$
\begin{equation}
e^{W_k[J]} = \int D\tilde \Phi\, e^{-S[\tilde\Phi]-\Delta S_k[\tilde\Phi]+J_\alpha\tilde\Phi_\alpha}
\label{eq4:scaledepSchwingerfunc}
\end{equation}
with
\begin{equation}
\Delta S_k[\tilde \Phi] = \frac{1}{2}\tilde \Phi_\alpha (R_k)_{\alpha\beta}\tilde\Phi_\alpha.
\end{equation}
Again we work with an abstract index notation where e.~g. $\alpha$ stands for both continuous and discrete variables. For simplicity we will sometimes drop this index when the meaning is clear and write for example
\begin{equation}
\Delta S_k[\tilde \Phi] = \frac{1}{2}\tilde \Phi R_k  \tilde\Phi.
\end{equation}
In praxis one chooses $R_k$ to be an infrared cutoff which is diagonal in momentum space. For example, for a single complex scalar field $\tilde \varphi$ we use
\begin{equation}
\Delta S_k = \int_q \tilde \varphi^*(q) R_k(q) \tilde \varphi(q)
\end{equation}
with
\begin{equation}
R_k(q)= A_\varphi (k^2-\vec q^2)\theta(k^2-\vec q^2).
\end{equation}
More general the function $R_k(q)$ should have the properties
\begin{eqnarray}
\nonumber
R_k(q) & \to \infty \quad & (k\to\infty),\\
\nonumber
R_k(q) & \to \,\,0\, \quad & (k\to0),\\
R_k(q) & \approx \,\,k^2 \quad & (q\to0).
\label{eq4:cutoffreq}
\end{eqnarray}
The cutoff term $R_k$ plays a similar role for the functional integral as the parameter $k^2$ in chapter \ref{ch:Flowequationstosolveanintegral} where we developed a flow equation to solve one-dimensional integrals. For large $R_k$ the term $\Delta S_k[\tilde \Phi]$ in Eq.\ \eqref{eq4:scaledepSchwingerfunc} suppresses the contribution of large values of $\tilde \Phi$ to the functional integral in Eq.\ \eqref{eq4:scaledepSchwingerfunc}. These fluctuations are included as the cutoff $R_k$ is lowered. An additional feature to the one-dimensional case in chapter \ref{ch:Flowequationstosolveanintegral} is the $q$-dependence of $R_k$. One can choose the form of $R_k(q)$ such that it vanishes for momenta with $q^2\gg k^2$. For a given $k$ only the contribution of modes with $q^2\ll k^2$ in Eq.\ \eqref{eq4:scaledepSchwingerfunc} is then suppressed while the contribution of modes with $q^2\gg k^2$ is not modified. We will discuss the choice of an appropriate form of the cutoff function in chapter \ref{ch:Cutoffchoices}. 

Since the only explictly scale-dependence of the Schwinger functional $W_k[J]$ comes from the cutoff term $\Delta S_k$ we can easily calculate its scale-derivative
\begin{eqnarray}
\nonumber
\partial_k W_k{\big |}_J &=& -\frac{1}{2} \langle \tilde \Phi_\alpha (\partial_k R_k)_{\alpha\beta} \tilde \Phi_\beta\rangle\\
&=& -\frac{1}{2}(\partial_k R_k)_{\alpha\beta} \left(\langle\tilde \Phi_\alpha\tilde\Phi_\beta\rangle_c+\Phi_\alpha\Phi_\beta\right).
\label{eq4:scalederW1}
\end{eqnarray}
We can use for the connected part $\langle\tilde \Phi_\alpha \tilde\Phi_\beta\rangle_c$ the functional derivative
\begin{equation}
\langle\tilde \Phi_\alpha\tilde\Phi_\beta\rangle_c = {\big (}W_k^{(2)}{\big )}_{\alpha\beta} = \frac{\delta}{\delta J_\alpha}\frac{\delta}{\delta J_\beta} W_k = \frac{\delta}{\delta J_\alpha} \Phi_\beta
\label{eq4:W2}
\end{equation}
in order to write Eq.\ \eqref{eq4:scalederW1} as 
\begin{equation}
\partial_k W_k{\big |}_J = -\frac{1}{2} \text{STr}\left\{ (\partial_k R_k) W_k^{(2)} \right\} -\frac{1}{2} \Phi (\partial_k R_k)\Phi.
\label{eq4:scalederW2}
\end{equation}
The STr-operation sums over equal indices and includes an extra minus sign for fermionic degrees of freedom. This comes from the fact that $\langle\tilde \Phi_\alpha \tilde\Phi_\beta\rangle_c=-\langle\tilde \Phi_\beta \tilde\Phi_\alpha\rangle_c$ if $\tilde\Phi_\alpha$ and $\tilde\Phi_\beta$ are fermionic Grassmann-valued fields.

\section{The average action and its flow equation}

From the scale dependent Schwinger functional we can now go to the average action or flowing action. It is defined by subtracting from the Legendre transform
\begin{equation}
\tilde\Gamma_k[\Phi] = J_\alpha \Phi_\alpha - W_k[J]\quad \text{with} \quad \Phi_\alpha=\frac{\delta W_k}{\delta J_\alpha}
\label{eq4:deftildegamma}
\end{equation}
the cutoff term
\begin{equation}
\Gamma_k[\Phi] = \tilde \Gamma_k[\Phi] -\frac{1}{2} \Phi R_k \Phi.
\label{eq4:defgamma}
\end{equation}
From the definition it is immediately clear that the average action equals the quantum effective action
\begin{equation}
\Gamma[\Phi] = J_\alpha \Phi_\alpha - W[J]\quad \text{with} \quad \Phi_\alpha=\frac{\delta W}{\delta J_\alpha}
\end{equation}
for $k\to 0$. The quantum effective action is the generating functional of the one-particle irreducible correlation functions. 
It is straightforward to show a number of properties of the average action. The field equation follows by taking the functional derivative
\begin{equation}
\frac{\delta}{\delta \Phi_\alpha} \tilde \Gamma_k = \pm J_\alpha.
\label{eq4:FieldequationtildeGamma}
\end{equation}
The upper (lower) sign is for bosonic (fermionic) field components $\Phi_\alpha$. The functional derivative in Eq.\ \eqref{eq4:FieldequationtildeGamma} is a left derivative for Grassmann valued $\Phi_\alpha$. For a right-handed derivative we obtain
\begin{equation}
\tilde \Gamma_k \frac{\overset{\leftharpoonup}{\delta}}{\delta \Phi_\alpha} = J_\alpha.
\end{equation}
The second functional derivative is
\begin{equation}
{\big (}\tilde\Gamma_k^{(2)}{\big )}_{\alpha\beta} = \frac{\delta}{\delta \Phi_\alpha} {\bigg (} \Gamma_k \frac{\overset{\leftharpoonup}{\delta}}{\delta \Phi_\beta}{\bigg )} = \frac{\delta}{\delta \Phi_\alpha}  J_\beta.
\end{equation}
Comparing this to Eq.\ \eqref{eq4:W2} we find the useful relation
\begin{equation}
{\big (}\tilde\Gamma_k^{(2)}{\big )}_{\alpha\beta} {\big (}W_k^{(2)}{\big )}_{\beta\gamma} = \delta_{\alpha\gamma}.
\end{equation}
or
\begin{equation}
W_k^{(2)} = {\big (}\Gamma_k^{(2)}+R_k{\big )}^{-1}.
\label{eq4:w2asinverseG2}
\end{equation}
To calculate the flow equation for $\Gamma_k$ we use
\begin{equation}
\partial_k \tilde \Gamma_k{\big |}_\Phi = -\partial_k W_k{\big |}_J + \partial_k J_\alpha \left(\Phi_\alpha -\frac{\delta W_k}{\delta J_\alpha}\right) = -\partial_k W_k{\big |}_J
\end{equation}
and
\begin{equation}
\partial_k \Gamma_k[\Phi] = \partial_k \tilde \Gamma_k[\Phi] -\frac{1}{2}\Phi (\partial_k R_k) \Phi.
\end{equation}
Together with Eqs. \eqref{eq4:scalederW2} and \eqref{eq4:w2asinverseG2} we obtain then the central result of this chapter, the Wetterich equation \cite{Wetterich1993b}
\begin{equation}
\partial_k \Gamma_k = \frac{1}{2} \text{STr}\left\{ {\big (}\partial_k R_k{\big )}{\big (}\Gamma_k^{(2)}+R_k{\big )}^{-1}\right\}.
\label{eq4:Wettericheqn}
\end{equation}

\section{Functional integral representation and initial condition}

From the definition of the average action in Eqs. \eqref{eq4:deftildegamma}, \eqref{eq4:defgamma} and the scale dependent Schwinger functional in Eq.\ \eqref{eq4:scaledepSchwingerfunc} we obtain the functional integral representation
\begin{eqnarray}
\nonumber
e^{-\Gamma_k[\Phi]} &=& \int D\tilde \Phi \, e^{-S[\tilde \Phi]-\frac{1}{2}\tilde \Phi R_k \tilde\Phi+J\tilde\Phi-J\Phi+\frac{1}{2}\Phi R_k \Phi}\\
&=& \int D \tilde \Phi \, e^{-S[\Phi+\tilde \Phi]-\frac{1}{2}\tilde \Phi R_k \tilde \Phi+\frac{1}{2}\left[(\delta \Gamma_k/\delta \Phi)\tilde \Phi+\tilde\Phi\frac{\delta \Gamma_k}{\delta \Phi}\right]}.
\label{eq4:functintrepaverageaction}
\end{eqnarray} 
In the last line we performed a change of the integration measure
\begin{equation}
\tilde\Phi\to \tilde\Phi+\Phi
\end{equation}
and used 
\begin{eqnarray}
\frac{1}{2}\left[(\delta \Gamma_k/\delta \Phi)\tilde \Phi+\tilde\Phi\frac{\delta \Gamma_k}{\delta \Phi}\right] &=& \frac{1}{2}{\bigg [}\Gamma_k \frac{\overset{\leftharpoonup}{\delta}}{\delta \Phi} \tilde\Phi + \tilde \Phi \frac{\overset{\rightharpoonup}{\delta}}{\delta \Phi}\Gamma_k{\bigg ]}\\
&=& J\tilde\Phi -\frac{1}{2}\Phi R_k \tilde\Phi-\frac{1}{2}\tilde\Phi R_k \Phi.
\end{eqnarray}
If the cutoff is chosen such that for $k\to\infty$
\begin{eqnarray}
\frac{1}{2}\tilde\Phi R_k \tilde\Phi \to \frac{1}{2} \sum_\alpha r_{k,\alpha} \Phi^*_\alpha \Phi_\alpha \quad \text{with}\quad r_{k,\alpha}\to \infty,
\end{eqnarray}
we find from Eq.\ \eqref{eq4:functintrepaverageaction}
\begin{equation}
\lim_{k\to\infty} \Gamma_k[\Phi] =S[\Phi]+\text{const}.
\label{eq4:limitGammaklargek}
\end{equation}
This is a remarkable and very useful result. In the limit of large $k$ the average action $\Gamma_k$ approaches the microscopic action $S$. Eq.\ \eqref{eq4:limitGammaklargek} serves as an initial condition for the flow equation \eqref{eq4:Wettericheqn}.

\chapter{Generalized flow equation}
\label{ch:Generalizedflowequation}
In this section we derive a generalization of the Wetterich equation discussed in the last chapter. This exact flow equation for composite operators is published in \cite{FloerchingerWetterich2009exact}. In contrast to Eq. \eqref{eq4:Wettericheqn} we introduce scale-dependent composite operators which describe for example bound states.

\section{Scale-dependent Bosonization}

Let us consider a scale-dependent Schwinger functional for a theory formulated in terms of the field $\tilde \psi$
\begin{equation}
e^{W_k[\eta]}=\int D\tilde\psi\, e^{-S_\psi[\tilde\psi]-\frac{1}{2}\tilde\psi_\alpha(R_k^\psi)_{\alpha\beta}\tilde\psi_\beta+\eta_\alpha\tilde\psi_\alpha}.
\label{eq:scaledepSF}
\end{equation}
Again we use the abstract index notation where e.g. $\alpha$ stands for both continuous variables such as position or momentum and internal degrees of freedom. We now multiply the right hand side of Eq.\ \eqref{eq:scaledepSF} by a term that is for $R_k^\varphi=0$ only a field independent constant. It has the form of the functional integral over the field $\tilde\varphi$ with a Gaussian weighting factor
\begin{equation}
\int D \tilde\varphi \, e^{-S_\text{pb}-\frac{1}{2}\tilde \varphi_\epsilon(R_k^\varphi)_{\epsilon\sigma}\tilde\varphi_\sigma+j_\epsilon\tilde\varphi_\epsilon},
\end{equation}
where
\begin{eqnarray}
\nonumber
S_\text{pb} &=& \frac{1}{2}\left(\tilde\varphi_\epsilon-\chi_\tau Q^{-1}_{\tau\epsilon}\right)Q_{\epsilon\sigma}(\tilde\varphi_\sigma-Q^{-1}_{\sigma\rho}\chi_\rho).\\
&=& \frac{1}{2}\left(\tilde\varphi-\chi Q^{-1}\right)Q(\tilde\varphi-Q^{-1}\chi),
\label{eq:Spb}
\end{eqnarray}
and $\chi$ depends on the ``fundamental field'' $\tilde \psi$. We will often suppress the abstract index as in the last line of Eq.\ \eqref{eq:Spb}. We assume that the field $\tilde\varphi$ and the operator $\chi$ are bosonic. Without further loss of generality we can then also assume that $Q$ and $R_k^\varphi$ are $k$-dependent symmetric matrices.

As an example, we consider an operator $\chi$ which is quadratic in the original field $\tilde\psi$,
\begin{equation}
\chi_\epsilon = H_{\epsilon\alpha\beta}\tilde\psi_\alpha\tilde\psi_\beta.
\end{equation}
The Schwinger functional reads now
\begin{equation}
e^{W_k[\eta,j]} = \int D\tilde \psi\,D\tilde\varphi \, e^{-S_k[\tilde \psi, \tilde\varphi]+\eta\tilde\psi+j\tilde\varphi}
\label{eq:SFwithbosonfi}
\end{equation}
with
\begin{eqnarray}
\nonumber
S_k[\tilde\psi,\tilde\varphi] &=& S_\psi[\tilde\psi]+\frac{1}{2}\tilde\psi R_k^\psi\tilde \psi + \frac{1}{2}\tilde\varphi (Q+R_k^\varphi)\tilde\varphi\\
&&+ \frac{1}{2}\chi Q^{-1}\chi -\tilde\varphi \chi. 
\label{eq:actionferbos}
\end{eqnarray}
In the integration over $\tilde\varphi$, we can easily shift the variables to obtain
\begin{eqnarray}
\nonumber
e^{W_k[\eta,j]} &=& \int D \tilde\psi \, e^{-S_\psi[\tilde \psi]-\frac{1}{2}\tilde\psi R_k^\psi\tilde\psi+\eta \tilde\psi}\\
\nonumber
&& \times e^{\frac{1}{2}(j+\chi)(Q+R_k^\varphi)^{-1}(j+\chi)-\frac{1}{2}\chi Q^{-1}\chi}\\
&&\times \int D\tilde\varphi\, e^{-\frac{1}{2}\tilde\varphi(Q+R_k^\varphi)\tilde\varphi}.
\label{eq:Schingerfas}
\end{eqnarray}
The remaining integral over $\tilde\varphi$ gives only a ($k$-dependent) constant. For $R_k^\varphi=0$ and $j=0$ we note that $W_k[\eta,j]$ coincides with $W_k[\eta]$ in Eq.\ \eqref{eq:scaledepSF}.

We next derive identities for correlation functions of composite operators which follow from the equivalence of the equations\eqref{eq:SFwithbosonfi} and \eqref{eq:Schingerfas}. Taking the derivative with respect to $j$ we can calculate the expectation value for $\tilde \varphi$
\begin{eqnarray}
\nonumber
\varphi_\epsilon &=& \langle\tilde\varphi_\epsilon\rangle = \frac{\delta}{\delta j_\epsilon}  W_k[\eta,j]\\
&=& (Q+R_k^\varphi)^{-1}_{\epsilon\sigma}\,\left(j_\sigma+H_{\sigma\alpha\beta}\langle\tilde\psi_\alpha\tilde\psi_\beta\rangle\right).
\label{eq:varphiintermsofpsi}
\end{eqnarray}
This can also be written as
\begin{equation}
\langle\chi\rangle = Q\varphi - l
\label{eq:expvchi}
\end{equation}
with the modified source $l$
\begin{equation}
l_\epsilon = j_\epsilon-(R_k^\varphi)_{\epsilon\sigma}\varphi_\sigma.
\end{equation}
For the connected two-point function
\begin{equation}
(\delta_j\delta_j W_k)_{\epsilon\sigma}= \frac{\delta^2}{\delta j_\epsilon \delta j_\sigma} W_k =\langle\tilde\varphi_\epsilon\tilde\varphi_\sigma\rangle_c
\end{equation}
we obtain from Eq.\ \eqref{eq:Schingerfas} 
\begin{eqnarray}
\nonumber
&&(Q+R_k)(\delta_j\delta_j W_k)(Q+R_k) \\
\nonumber
&&= \langle(j+\chi)(j+\chi)\rangle -\langle(j+\chi)\rangle\langle(j+\chi)\rangle+(Q+R_k^\varphi)\\
&&= \langle\chi\chi\rangle-\langle\chi\rangle\langle\chi\rangle+(Q+R_k^\varphi)
\end{eqnarray}
or
\begin{eqnarray}
\nonumber
\langle\chi_\epsilon\chi_\sigma\rangle &=& \left[(Q+R_k^\varphi) (\delta_j\delta_j W_k)(Q+R_k^\varphi)\right]_{\epsilon\sigma} \\
&&+ (Q\varphi-l)_\epsilon(Q\varphi-l)_\sigma -(Q+R_k^\varphi)_{\epsilon\sigma}.
\label{eq:idk12}
\end{eqnarray}
Similarly, the derivative of Eq.\ \eqref{eq:expvchi} with respect to $j$ yields
\begin{eqnarray}
\langle\tilde\varphi_\epsilon\chi_\sigma\rangle = \langle\tilde\varphi_\epsilon\tilde\varphi_\tau\rangle (Q+R_k^\varphi)_{\tau\sigma}-\varphi_\epsilon j_\sigma -\delta_{\epsilon\sigma}\\
\nonumber
= \varphi_\epsilon (Q\varphi)_\sigma + \left[(\delta_j\delta_j W_k)(Q+R_k^\varphi)\right]_{\epsilon\sigma}-\varphi_\epsilon l_\sigma-\delta_{\epsilon\sigma}.
\label{eq:idk13}
\end{eqnarray}

We now turn to the scale-dependence of $W_k[\eta,j]$. In addition to $R_k^\psi$ and $R_k^\varphi$ also $Q$ and $H$ are $k$-dependent. For $H$ we assume
\begin{equation}
\partial_k H_{\epsilon\alpha\beta} = (\partial_k F_{\epsilon\rho}) H_{\rho\alpha\beta}
\end{equation}
where we take the dimensionless matrix $F$ to be symmetric for simplicity. For the operator $\chi$ this gives
\begin{equation}
\partial_k \chi_\epsilon = \partial_k H_{\epsilon\alpha\beta}\tilde\psi_\alpha\tilde\psi_\beta = \partial_k F_{\epsilon\rho} \chi_\rho.
\end{equation}
From Eqs. \eqref{eq:SFwithbosonfi} and \eqref{eq:actionferbos} we can derive (for fixed $\eta$, $j$)
\begin{eqnarray}
\nonumber
\partial_k W_k &=& -\frac{1}{2}\langle\tilde\psi(\partial_k R_k^\psi)\tilde\psi\rangle - \frac{1}{2}\langle\tilde\varphi(\partial_k R_k^\varphi+\partial_k Q)\tilde\varphi\rangle\\
\nonumber
&&-\frac{1}{2}\langle\chi\left(\partial_k Q^{-1}+Q^{-1}(\partial_k F)+(\partial_k F)Q^{-1}\right)\chi\rangle\\
&&+\langle\tilde\varphi(\partial_k F)\chi\rangle.
\end{eqnarray}
Now we insert Eqs. \eqref{eq:idk12} and \eqref{eq:idk13}
\begin{eqnarray}
\nonumber
\partial_k W_k &=& -\frac{1}{2}\psi (\partial_k R_k^\psi)\psi - \frac{1}{2}\varphi (\partial_k R_k^\varphi) \varphi\\
\nonumber
&&-\frac{1}{2} \text{STr}\,\{  (\delta_\eta\delta_\eta W_k)(\partial_k R_k^\psi) \}\\
\nonumber
&&- \frac{1}{2}\text{Tr}{\big \{} (\delta_j\delta_j W_k)(\partial_k R_k^\varphi){\big \}}\\
\nonumber
&&-\frac{1}{2}\text{Tr} {\big \{}{\big [}Q(\partial_k Q^{-1})R_k^\varphi+R_k^\varphi(\partial_k Q^{-1})Q\\
\nonumber
&&\,\,\,+R_k^\varphi(\partial_k Q^{-1})R_k^\varphi+ R_k^\varphi Q^{-1}(\partial_k F)(Q+R_k)\\
\nonumber
&&\,\,\,+(Q+R_k^\varphi)(\partial_kF)Q^{-1}R_k^\varphi{\big ]}(\delta_j\delta_j W_k){\big \}}\\
\nonumber
&&+\frac{1}{2}l\left[(\partial_k Q^{-1})Q+Q^{-1}(\partial_k F)Q\right]\varphi\\
\nonumber
&&+\frac{1}{2}\varphi \left[Q(\partial_k Q^{-1})+Q(\partial_k F)Q^{-1}\right]l\\
\nonumber
&&-\frac{1}{2}l \left[\partial_k Q^{-1}+Q^{-1}(\partial_k F)+(\partial_kF)Q^{-1}\right]l\\
\nonumber
&&+\frac{1}{2} \text{Tr} \left\{\left[\partial_k Q^{-1}+Q^{-1}(\partial_k F)+(\partial_k F)Q^{-1}\right]R_k^\varphi\right\}\\
&&+\frac{1}{2}\text{Tr} \left\{Q\partial_k Q^{-1}\right\}.
\label{eq:longflowW}
\end{eqnarray}
The supertrace $\text{STr}$ contains the appropriate minus sign in the case that $\psi_\alpha$ are fermionic Grassmann variables.

Equation \eqref{eq:longflowW} can be simplified substantially when we restrict the $k$-dependence of $F$ and $Q$ such that
\begin{equation}
\partial_k F = -Q(\partial_k Q^{-1}) = -(\partial_k Q^{-1})Q.
\label{eq:restrSQ}
\end{equation}
In fact, one can show that the freedom to choose $F$ and $Q$ independent from each other that is lost by this restriction, is equivalent to the freedom to make a linear change in the source $j$, or at a later stage of the flow equation in the expectation value $\varphi$. With the choice in Eq.\ \eqref{eq:restrSQ} we obtain
\begin{eqnarray}
\nonumber
\partial_k W_k &=& -\frac{1}{2}\psi (\partial_k R_k^\psi)\psi -\frac{1}{2}\varphi(\partial_k R_k^\varphi)\varphi\\
\nonumber
&&-\frac{1}{2}\text{STr}{\big \{}(\partial_k R_k^\psi)(\delta_\eta\delta_\eta W_k){\big \}}\\
\nonumber
&&-\frac{1}{2}\text{Tr}{\big \{} \left[\partial_k R_k^\varphi-R_k^\varphi(\partial_k Q^{-1})R_k^\varphi\right](\delta_j\delta_j W_k){\big \}}\\
\nonumber
&&+\frac{1}{2}l(\partial_k Q^{-1})l+\frac{1}{2}\text{Tr}{\{}\partial_kQ^{-1}(Q-R_k^\varphi){\}}.
\label{eq:shortflowW}
\end{eqnarray}
The last term is independent of the sources $\eta$ and $j$ and is therefore irrelevant for many purposes.

\section{Flowing action}

The average action or flowing action is defined by subtracting from the Legendre transform
\begin{equation}
\tilde\Gamma_k[\psi,\varphi] = \eta \psi + j \varphi - W_k[\eta,j]
\end{equation}
the cutoff terms
\begin{equation}
\Gamma_k[\psi,\varphi]=\tilde\Gamma_k[\psi,\varphi]-\frac{1}{2}\psi R_k^\psi\psi -\frac{1}{2}\varphi R_k^\varphi\varphi.
\label{eq:defflowingaction}
\end{equation}
As usual, the arguments of the effective action are given by
\begin{equation}
\psi_\alpha=\frac{\delta}{\delta \eta_\alpha}W_k \quad \text{and} \quad \varphi_\epsilon=\frac{\delta}{\delta j_\epsilon} W_k.
\end{equation}
By taking the derivative of Eq.\ \eqref{eq:defflowingaction} it follows
\begin{equation}
\frac{\delta}{\delta \psi_\alpha}\Gamma_k = \pm \eta_\alpha - (R_k^\psi)_{\alpha\beta} \psi_\beta,
\end{equation}
where the upper (lower) sign is for a bosonic (fermionic) field $\psi$. Similarly,
\begin{equation}
\frac{\delta}{\delta \varphi_\epsilon}\Gamma_k = j_\epsilon - (R_k^\varphi)_{\epsilon\sigma} \varphi_\sigma=l_\epsilon.
\end{equation}
In the matrix notation
\begin{eqnarray}
\nonumber
W_k^{(2)} &=& \begin{pmatrix}\delta_\eta\delta_\eta W_k, && \delta_\eta\delta_j W_k \\ \delta_j\delta_\eta W_k, && \delta_j\delta_j W_k\end{pmatrix},\\
\nonumber
\Gamma_k^{(2)} &=& \begin{pmatrix}\delta_\psi\delta_\psi \Gamma_k, && \delta_\psi\delta_\varphi \Gamma_k \\\delta_\varphi\delta_\psi \Gamma_k, && \delta_\varphi\delta_\varphi \Gamma_k\end{pmatrix},\\
R_k&=&\begin{pmatrix} R_k^\psi, &&  0 \\ 0,&& R_k^\varphi \end{pmatrix},
\end{eqnarray}
it is straight forward to establish
\begin{equation}
W_k^{(2)} \,\tilde\Gamma_k^{(2)} =1,\quad\quad W_k^{(2)} = (\Gamma_k^{(2)}+R_k)^{-1}.
\end{equation}

In order to derive the exact flow equation for the average action we use the identity
\begin{equation}
\partial_k \tilde \Gamma_k{\big |}_{\psi,\varphi} = -\partial_k W_k{\big |}_{\eta,j}.
\end{equation}
This yields the central result of this chapter
\begin{eqnarray}
\nonumber
\partial_k \Gamma_k &=& \frac{1}{2}\text{STr}\, \left\{(\Gamma_k^{(2)}+R_k)^{-1}\left(\partial_k R_k-R_k(\partial_k Q^{-1})R_k\right)\right\}\\
&&-\frac{1}{2}\Gamma_k^{(1)} \left(\partial_k Q^{-1}\right)\Gamma_k^{(1)}+\gamma_k
\label{eq:flowequationGamma}
\end{eqnarray}
with
\begin{equation}
\gamma_k=-\frac{1}{2} \text{Tr}\left\{ (\partial_k Q^{-1})(Q-R_k)\right\}.
\end{equation}
As it should be, this reduces to the standard flow equation for a framework with fixed partial bosonization in the limit $\partial_k Q^{-1}=0$. The additional term is quadratic in the first derivative of $\Gamma_k$ with respect to $\varphi$ -- we recall that $\partial_k Q^{-1}$ has non-zero entries only in the $\varphi$-$\varphi$ block. Furthermore there is a field independent term $\gamma_k$ that can be neglected for many purposes. At this point a few remarks are in order.

(i) For $k\to 0$ the cutoffs $R_k^\psi$, $R_k^\varphi$ should vanish. This ensures that the correlation functions of the partially bosonized theory are simply related to the original correlation functions generated by $W_0[\eta]$, Eq.\ \eqref{eq:scaledepSF}, namely
\begin{eqnarray}
\nonumber
W_0[\eta, j] &=& \ln \left(\int D\tilde \psi \, e^{-S_\psi[\tilde \psi]+\eta\tilde\psi+jQ^{-1}\chi}\right)+\frac{1}{2}j\, Q^{-1} \,j+\text{const.},\\
W_0[\eta,j=0] &=& W_0[\eta] +\text{const.}
\end{eqnarray}
Knowledge of the dependence on $j$ permits the straightforward computation of correlation functions for composite operators $\chi$.
\newline

(ii) For solutions of the flow equation one needs a well known ``initial value'' which describes the microscopic physics. This can be achieved by letting the cutoffs $R_k^\psi$, $R_k^\varphi$ diverge for $k\to\Lambda$ (or $k\to\infty$). In this limit the functional integral in Eqs. \eqref{eq:SFwithbosonfi}, \eqref{eq:actionferbos} can be solved exactly and one finds
\begin{equation}
\Gamma_\Lambda[\psi,\varphi] = S_\psi[\psi] +\frac{1}{2}\varphi Q_\Lambda \varphi +\frac{1}{2}\chi[\psi]Q_\Lambda^{-1} \chi[\psi]-\varphi \chi[\psi].
\label{eq:averageactionmicrosc}
\end{equation}
This equals the ``classical action'' obtained from a Hubbard-Stratonovich transformation, with $\chi$ expressed in terms of $\psi$. 

(iii) In our derivation we did not use that $\chi$ is quadratic in $\psi$. We may therefore take for $\chi$ an arbitrary bosonic functional of $\psi$. It is straightforward to adapt our formalism such that also fermionic composite operators can be considered.

The flow equation \eqref{eq:flowequationGamma} has a simple structure of a one loop expression with a cutoff insertion -- $\text{STr}$ contains the appropriate integration over the loop momentum -- supplemented by a ``tree-contribution'' $\sim \left(\Gamma_k^{(1)}\right)^2$. Nevertheless, it is an exact equation, containing all orders of perturbation theory as well as non-perturbative effects. The simple form of the tree contributions will allow for easy implementations of a scale dependent partial bosonization. The details of this can be found in \cite{FloerchingerWetterich2009exact}.

\section{General coordinate transformations}

It is sometimes useful to perform a change of coordinates in the space of fields during the renormalization flow. In this section we discuss the transformation behavior of the Wetterich equation \eqref{eq4:Wettericheqn} under such a change of the basis for the fields. We follow here the calculation in \cite{Wetterich1996, Gies:2001nw} For simplicity we restrict the discussion to bosonic fields. It is straightforward to transfer this to fermions as well \cite{Wetterich1996, Gies:2001nw}. Similarly, one might also consider a general coordinate transformation for the generalized flow equation \eqref{eq:flowequationGamma}. 

Let us consider a transformation of the form
\begin{equation}
\Phi \to \Psi[\Phi].
\label{eq2:Fieldmap}
\end{equation}
Here we denote by $\Phi$ the original fields. The functional $\Psi[\Phi]$ is a $k$-dependent map of the old coordinates to the new ones. We assume that the map in Eq.\ \eqref{eq2:Fieldmap} is invertible and write the inverse
\begin{equation}
\Phi[\Psi].
\end{equation}
In terms of the fields $\Psi$ the definition of the flowing action reads
\begin{equation}
\Gamma_k[\Psi] = J_\alpha \Phi_\alpha [\Psi]-W_k[J]-\frac{1}{2} \Phi[\Psi] R_k \Phi[\Psi]
\end{equation}
where $J$ is determined by
\begin{equation}
\Phi_\alpha[\Psi]=\frac{\delta W_k}{\delta J_\alpha}.
\end{equation}
In the limit $k\to0$ the flowing action is a Legendre transform with respect to the old fields $\Phi$ but not with respect to the new fields $\Psi$. This implies for example that $\Gamma[\Psi]$ is not necessarily convex with respect to the fields $\Psi$. In addition, only the fields $\Phi$ are expectation values of fields as in Eq.\ \eqref{eq1:expectvalue}. The relation of the fields $\Psi$ to the microscopic fields $\tilde \Phi$ is more complicated. The field equation reads in terms of the new fields
\begin{equation}
\frac{\delta \Gamma_k}{\delta \Psi_\alpha} = J_\beta \frac{\delta \Phi_\beta}{\delta \Psi_\alpha} - \frac{\delta}{\delta \Psi_\alpha} \Delta S_k.
\end{equation}
Note that the cutoff term
\begin{equation}
\Delta S_k = \frac{1}{2}\Phi_\alpha[\Psi](R_k)_{\alpha\beta} \Phi_\beta[\psi]
\end{equation}
is not necessarily quadratic in the fields $\Psi$. The matrix $R_k$ obtains from
\begin{eqnarray}
\nonumber
(R_k)_{\alpha\beta} &=& \frac{\delta}{\delta \Phi_\alpha}\frac{\delta }{\delta \Phi_\beta} \Delta S_k\\
&=& \frac{\delta \Psi_\mu}{\delta \Phi_\alpha} \frac{\delta \Psi_\mu}{\delta \Phi_\beta}\left(\frac{\delta}{\delta \Psi_\mu}\frac{\delta }{\delta \Psi_\nu} \Delta S_k\right) + \frac{\delta^2 \Psi_\nu}{\delta \Phi_\alpha \delta \Phi_\beta}\left(\frac{\delta}{\delta \Psi_\nu}\Delta S_k\right).
\label{eq3:defRk}
\end{eqnarray}
Similarly we obtain for the matrix $\Gamma_k^{(2)}$
\begin{eqnarray}
\nonumber
(\Gamma_k^{(2)})_{\alpha\beta} &=& \frac{\delta}{\delta \Phi_\alpha}\frac{\delta }{\delta \Phi_\beta} \Gamma_k\\
&=& \frac{\delta \Psi_\mu}{\delta \Phi_\alpha} \frac{\delta \Psi_\mu}{\delta \Phi_\beta}\left(\frac{\delta}{\delta \Psi_\mu}\frac{\delta }{\delta \Psi_\nu} \Gamma_k\right) + \frac{\delta^2 \Psi_\nu}{\delta \Phi_\alpha \delta \Phi_\beta}\left(\frac{\delta}{\delta \Psi_\nu}\Gamma_k\right).
\label{eq3:defGamma2}
\end{eqnarray}
We now come to the scale dependence of $\Gamma_k$. It is given by
\begin{equation}
\partial_k \Gamma_k[\Psi]=\partial_k \Gamma_k[\Psi]{\big |}_\Phi - \frac{\delta \Gamma_k}{\delta \Psi_\alpha} \partial_k \Psi_\alpha{\big |}_\Phi.
\label{eq2:scaledepGammaphipsi}
\end{equation}
For the first term on the right hand side of Eq.\ \eqref{eq2:scaledepGammaphipsi} we can use the Wetterich equation\eqref{eq4:Wettericheqn} and obtain
\begin{equation}
\partial_k \Gamma_k[\Psi] = \frac{1}{2} \text{Tr} (\Gamma_k^{(2)}+R_k)^{-1} \partial_k R_k-\frac{\delta \Gamma_k}{\delta \Psi_\alpha} \partial_k \Psi_\alpha{\big |}_\Phi.
\label{eq3:Wettericheqwithcoordtransf}
\end{equation}
We emphasize that $\Gamma_k^{(2)}$ and $R_k$ are now somewhat more complicated objects then usually. They are defined by Eqs. \eqref{eq3:defRk} and \eqref{eq3:defGamma2}. One might also define the transformed matrices
\begin{eqnarray}
\nonumber
(\widehat \Gamma_k^{(2)})_{\mu\nu} &=& \frac{\delta}{\delta \Psi_\mu} \frac{\delta}{\delta \Psi_\nu} \Gamma_k + \frac{\delta \Phi_\alpha}{\delta \Psi_\mu} \frac{\delta \Phi_\beta}{\delta \Psi_\nu} \frac{\delta^2 \Psi_\nu}{\delta \Phi_\alpha \delta \Phi_\beta}\left(\frac{\delta}{\delta \Psi_\nu} \Gamma_k\right),\\
(\widehat R_k)_{\mu\nu} &=& \frac{\delta}{\delta \Psi_\mu} \frac{\delta}{\delta \Psi_\nu} \Delta S_k + \frac{\delta \Phi_\alpha}{\delta \Psi_\mu} \frac{\delta \Phi_\beta}{\delta \Psi_\nu} \frac{\delta^2 \Psi_\nu}{\delta \Phi_\alpha \delta \Phi_\beta}\left(\frac{\delta}{\delta \Psi_\nu} \Delta S_k\right),
\label{eq3:hatteddef}
\end{eqnarray}
and similar
\begin{eqnarray}
(\widehat{\partial_k R_k})_{\mu\nu} &=& \frac{\delta \Phi_\alpha}{\delta \Psi_\mu} \frac{\delta \Phi_\beta}{\delta \Psi_\nu}  (\partial_k R_k)_{\alpha\beta}.
\label{eq3:hatted2}
\end{eqnarray}
In Eq.\ \eqref{eq3:hatteddef} the second functional derivatives are supplemented by connection terms as appropriate for general (non-linear) coordinate systems. With Eqs.\ \eqref{eq3:hatteddef} \eqref{eq3:hatted2} the flow equation for $\Gamma_k$ reads
\begin{equation}
\partial_k \Gamma_k[\Psi] = \frac{1}{2} \text{Tr} (\widehat \Gamma_k+\widehat R_k)^{-1} \widehat{\partial_k R_k} - \frac{\delta \Gamma_k}{\delta \Psi_\alpha} \partial_k \Psi_\alpha{\big |}_\Phi.
\end{equation}
Unfortunately this equation has lost its one-loop structure due to the connection terms in Eq.\ \eqref{eq3:hatteddef}. An important exception is a linear coordinate transformation
\begin{equation}
\Psi_\alpha[\Phi] = \Xi_\alpha + M_{\alpha\beta} \Phi_\beta.
\end{equation}
In that case the terms $\frac{\delta^2 \Psi}{\delta \Phi\delta \Phi}$ vanish and the one-loop structure is preserved.

\chapter{Truncations}
\label{ch:Truncations}
The exact flow equations derived in the previous sections are powerful and elegant but also complicated. They are functional differential equations i.~e. differential equations for the object $\Gamma_k[\Phi]$ which depends on the scale parameter $k$ and is a functional of the field configuration $\Phi$. The mathematics of these kind of differential equations is hardly developed and it is in most cases not possible to find exact solutions in a closed form. However, it is possible to find approximate solutions using truncations in the space of possible functionals. The idea is to take an ansatz for the flowing action of the form
\begin{equation}
\Gamma_k[\Phi] = \int_x \sum_{i=1}^N g_i\, {\mathcal O}_i[\Phi]
\label{eq6:truncationexpansion}
\end{equation}
where ${\mathcal O}_i[\Phi]$ are some operators and the $g_i$ are (generalized) coupling constants. In the ideal case the set of operators ${\mathcal O}_i$ builds a complete set for $N\to \infty$. Some physical insight into the problem at hand is necessary in order to choose a convenient form of the truncation. In any case it must be possible to write the microscopic action $\Gamma_\Lambda = S$ in the form \eqref{eq6:truncationexpansion}. Plugging the ansatz in Eq.\ \eqref{eq6:truncationexpansion} into the flow equation \eqref{eq4:Wettericheqn} or \eqref{eq:flowequationGamma} for the average action one obtains a set of coupled ordinary differential equations
\begin{equation}
\partial_k g_i = \beta_i (g_1,\dots,g_N;k)
\end{equation}
These equations can now be solved either analytically or numerically, depending on the complexity of the problem. For increasing $N$ the approximate solutions found by such a procedure should become better and better. They were even exact if the flow of the couplings $g_i$ with $i>N$ would vanish
\begin{equation}
\beta_i = 0\quad \text{for}\quad i>N.
\end{equation}
These couplings $g_i$ would then vanish on all scales and the expansion in Eq.\ \eqref{eq6:truncationexpansion} would give an exact solution to the flow equation. For the example of a free or Gaussian theory this is indeed the case. Since all interaction couplings are zero, most of the functions $\beta_i$ also vanish. 

Exact solutions can also be obtained in another scenario. Suppose that we have a hierarchy in the flow equations of the $g_i$, for example
\begin{eqnarray}
\nonumber
\partial_k g_1 &=& \beta_1(g_1;k)\\
\nonumber
\partial_k g_2 &=& \beta_2(g_1,g_2;k)\\
\nonumber
\partial_k g_3 &=& \beta_3(g_1,g_2,g_3;k)\\
\ldots. &&
\label{eq6:floweqhierachy}
\end{eqnarray}
The flow equations for every coupling $g_i$ depends only on the couplings $g_i$ with $i\leq j$. Eq.\ \eqref{eq6:floweqhierachy} can then be solved step by step. First we solve the differential equation for $g_1(k)$ (the first line in Eq.\ \eqref{eq6:floweqhierachy}), plugg this solution into the second equation for $g_2$, solve it, and so on. In praxis one has to stop at some $i=l$, of course. The only error in this solution comes from the fact that it may be necessary to solve the (ordinary) differential equations for the $g_i$ numerically. What we obtain by this process is not an exact solution for the complete functional $\Gamma_k[\Phi]$ but for the coefficients $g_i$ with $i\leq l$. However, all observables that depend on these couplings only, can be determined exactly. As we will see later, an hierarchy of flow equations similar to Eq.\ \eqref{eq6:floweqhierachy} is indeed found for the nonrelativistic few-body problem. From a flow equation point of view this hierarchy is the reason why no ``renormalization of coupling constants'' is needed in quantum mechanics. 

\subsubsection{Symmetries as a guiding principles}

How should one choose a truncation? The choice of the appropriate ansatz for the flowing action is certainly one of the most important points for someone who wants to work with the flow equation method in praxis. Besides the necessary physical insight there is one major guiding principle: symmetries. As will be discussed in chapter \ref{ch:Symmetries} the flowing action $\Gamma_k$ respects the same symmetries as the microscopic action $S$ if no anomalies of the functional integral measure are present and if the cutoff term $\Delta S_k$ is also invariant. In the notation of Eq.\ \eqref{eq6:truncationexpansion} this implies that the coefficient $g_i$ of an operator ${\mathcal O}_i[\Phi]$ that is not invariant under all symmetries will not be generated by the flow equation such that $g_i=0$ for all $k$. As an example we consider the microscopic action of a Bose gas
\begin{equation}
S=\int \varphi^* (\partial_\tau-\Delta-\mu)\varphi +\frac{1}{2}\lambda_\varphi (\varphi^*\varphi)^2.
\label{eq6:actionBosegas}
\end{equation}
It is invariant under the global U(1) symmetry
\begin{eqnarray}
\nonumber
\varphi &\to& e^{i\alpha} \varphi \\
\varphi^* &\to& e^{-i\alpha} \varphi^*. 
\end{eqnarray}
This implies that only operators that are invariant under this transformation may appear in the flowing action $\Gamma_k[\Phi]$. For example, the part that describes homogeneous fields, the effective potential is of the form
\begin{equation}
\Gamma_k = \ldots + \int_x U(\varphi^*\varphi)
\end{equation}
where $U(\rho)$ is a function of the U(1)-invariant combination $\rho=\varphi^*\varphi$, only. The action in Eq.\ \eqref{eq6:actionBosegas} has more symmetries such as translation, rotation or, at zero temperature, Galilean invariance. 

A useful strategy to find a sensible truncation is to start from the microscopic action $S$ or an effective action $\Gamma$ calculated in some (perturbative) approximation scheme such as for example mean-field theory. One now renders the appearing coefficients to become $k$-dependent ``running couplings'' and adds also additional terms after checking that they are allowed by the symmetries of the microscopic action. 
 
\subsubsection{Separation of scales}

Some symmetries are realized only in some range of the renormalization group flow. For example, Galilean symmetry is broken explicitly by the thermal heat bath for $T>0$. Nevertheless, for $k^2\gg T$ the flowing action $\Gamma_k$ (or its real-time version obtained from analytic continuation) will still be invariant under Galilean boost transformations. This comes since the scale parameter $k$ sets the infrared scale on the right hand side of the flow equation. As long as $k^2/T$ is large, the flow equations are essentially the same as for $T\to0$. In other words, the flow only ``feels'' the temperature once the scale $k$ is of the same order of magnitude $k^2\approx T$. On the other side, for $k^2\ll T$ the flow equations may simplify again. Now they are equivalent to those obtained in the large temperature limit $T\to \infty$. Different symmetries may apply to the action in this limit. This separation of scales is often very useful for practical purposes. In different regimes of the flow different terms are important, while others might be neglected. For example, the universal critical properties such as the critical exponents or amplitude ratios can be calculated in the framework of the classical theory, i.~e. in the large temperature limit $T\to\infty$. The flow equations in this limit are much simpler then the ones obtained for arbitrary temperature $T$. 

The scale-separation is also useful for the fixing of the initial coupling constants at the initial scale $k=\Lambda$. If this scale is much larger then the temperature $\Lambda^2\gg T$ and the relevant momentum scale for the density, the inverse interparticle distance $\Lambda\gg n^{1/3}$, the initial flow is the same as in vacuum where $T=n=0$. One can then also use the same initial values for most of the couplings and only change the temperature and the chemical potential appropriately to describe points in the phase diagram that correspond to $T>0$ and $n>0$.  

\subsubsection{Derivative expansion}

A central part of a truncation is the form of the propagator. It follows from the second functional derivative of the flowing action. For the example of a Bose gas one has in the normal phase
\begin{equation}
G_k^{-1}(p) \,\delta(p-p^\prime) = \frac{\delta}{\delta \varphi^*(p)} \frac{\delta}{\delta \varphi(p^\prime)} \Gamma_k[\Phi].
\label{eqn6:derexpansion}
\end{equation}
The inverse propagator $G_k^{-1}$ may be a quite complicated function of the momentum $p$ which consists of the spatial momentum and the (Matsubara-) frequency, $p=(p_0, \vec p)$. From rotational invariance it follows that $G_k^{-1}$ depends on the spatial momentum only in the invariant combination $\vec p^2$. At zero temperature it follows from Galilean invariance that $G_k^{-1}$ is a (analytic) function of the combination $ip_0+\vec p^2$, provided that Galilean invariance is not broken by the cutoff. 

Using a derivative expansion, one truncates the flowing action in the form
\begin{equation}
\Gamma_k = \int_x \varphi^* (Z\partial_\tau-A\vec \nabla^2 - V \partial_\tau^2+\ldots)\varphi + U(\varphi^*\varphi).
\end{equation}
The ``kinetic coefficients'' $Z$, $A$, $V$ etc.\ depend on the scale parameter $k$ and for more advanced approximations also on the U(1) invariant combination $\rho=\varphi^* \varphi$. One can improve the expansion in Eq.\ \eqref{eqn6:derexpansion} by promoting the coefficients $Z$, $A$, $V$ etc.\ to functions of $p_0$ and $\vec p^2$. 

In praxis one usually neglects terms higher then quadratic in the momenta. Nevertheless, derivative expansion often leads to quite good results. The reason is the following. On the right hand side of the flow equation the cutoff insertion $R_k$ in the propagator $(\Gamma^{(2)}+R_k)^{-1}$ suppresses the contribution of the modes with small momenta. On the other side, the cutoff derivative $\partial_k R_k$ suppresses the contribution of very large momenta provided that $R_k(q)$ falls of sufficiently fast for large $q$. Effectively mainly modes with momenta of the order $k^2$ contribute. It would therefore be sensible to use on the right hand side of the flow equations the coefficients
\begin{equation}
Z(p_0 = k^2,\vec p^2=k^2), A(p_0=k^2,\vec p^2=k^2), \ldots.
\end{equation}
One main effect of the external frequencies and momenta in $Z(p_0,\vec p^2)$ etc.\ is to provide an infrared cutoff scale of order $\text{Max}(p_0,\vec p^2)$. Such an infrared cutoff scale is of course also provided by $R_k$ itself and one might therefore also work with the $k$-dependent couplings
\begin{equation}
Z(p_0=0,\vec p^2=0), A(p_0=0,\vec p^2=0),\ldots.
\end{equation}
We emphasize that it is important that the cutoff $R_k(q)$ falls off sufficiently fast for large $q$. If this is not the case, the derivative expansion might lead to erroneous results since the kinetic coefficients as appropriate for small momenta and frequencies are then also used for large momenta and frequencies. Only when the scale derivative $\partial_k R_k$ provides for a sufficient ultraviolet cutoff does the derivative expansion work properly.
		
\chapter{Cutoff choices}
\label{ch:Cutoffchoices}
In this section we discuss the choice of the cutoff function $R_k(q)$. In principle, this function can be chosen quite arbitrary as long as the general requirements in Eq.\ \eqref{eq4:cutoffreq} are fulfilled. In praxis, the choice of $R_k$ is more involved than it seems on first sight. There is a number of points that have to be taken into account for nonrelativistic systems.

\paragraph{(i) Hierarchy of flow equations.} In nonrelativistic few-body physics there exists an interesting and important hierarchy: The $n$-body problem partly decouples from the $n+1$-body problem. More precisely, one can solve the $n$-body (scattering)-problem without any information about the additional couplings of the $n+1$-body problem. In other words, there are no quantum corrections to a $n$-point function involving couplings that describe interactions between more then $n$ particles. We will describe this in more detail in chapter \ref{ch:Few-bodyphysics}. Formally, this hierarchy is closely connected with the frequency pole structure of the correlation functions. For example, the microscopic propagator for a Bose gas
\begin{equation}
\frac{1}{i q_0 + \vec q^2-\mu}
\end{equation}
has a pole for $q_0 = i(\vec q^2-\mu)$. In vacuum, i.~e. for $\mu\leq0$ this pole is always in the upper half of the complex plane. For loop expressions where all particle-number arrows point is the same direction (therefore constituting a closed tour of particles) this is similar: All frequency poles are in the same half-plane. If it is possible to close the frequency integration contour in the other half-plane this expressions vanish. 

In choosing a regulator function $R_k(q)$ (with $q=(q_0,\vec q)$) one has to be careful to maintain this feature. The cutoff may shift the frequency pole or give rise to additional poles, but (at least for $\mu\leq 0$) these poles must remain in the original half plane. If the cutoff depends on the frequency in such a way that
\begin{equation}
\frac{1}{i q_0 +\vec q^2-\mu+R_k(q_0,\vec q)}
\end{equation}
is a non-analytic function in $q_0$ (in the sense of complex analysis), one has to check directly that the hierarchy feature is not violated. This is very important for practical purposes. 

\paragraph{(ii) Matsubara summation.} At nonzero temperature the Matsubara frequency is discrete, $q_0=2\pi T n$ (for bosons). Loop expressions involve a summation over the integer number $n$ from $-\infty$ to $\infty$. In praxis one has to find a way to perform these summations -- either analytically or numerically. An analytic treatment has the advantage that it is also well controlled in the limit of zero temperature $T\to \infty$ or in the high temperature limit $T\to\infty$. In addition, it usually allows for simpler expressions and lower numerical effort. Methods to perform the Matsubara summation analytically are explained in the literature, e.~g. \cite{Mahan1981}. Usually they require that the integrand in the loop expressions depend on $q_0$ in an analytic way (again in the sense of complex analysis). It is then important to choose the cutoff $R_k$ appropriately. 

\paragraph{(iii) Ultraviolet regulator property.} The cutoff function $R_k(q)$ does not only provide an infrared cutoff but may also provide for an ultraviolet regulator for large momenta $q$ within loop expressions. When $R_k(q)$ falls off sufficiently fast for large $q$, the cutoff derivative $\partial_k R_k$ on the right hand side of the flow equation suppresses the contribution of large loop-momenta $q$. This feature is especially important if a derivative expansion is used to treat the momentum dependence of the propagator (and higher correlation functions). As discussed at the end of chapter \ref{ch:Truncations} derivative expansion might lead to fake results if loop expressions are not sufficiently regularized in the ultraviolet. 

\paragraph{(iv) Symmetry.} As emphasized already in chapter \ref{ch:Truncations}, the symmetries of a problem play a very important role for the flow equation method. If no anomalies of the functional integral measure are present, the effective action $\Gamma=\Gamma_{k=0}$ shows the same invariances as the microscopic action $S$. This is also the case for the flowing action $\Gamma_k$ for intermediate scales $0<k<\Lambda$ provided that the cutoff term $\Delta S_k$ is also invariant under the corresponding symmetry. This is a very useful feature since it strongly restricts the space of possible functionals $\Gamma_k$ and helps in finding an effective truncation. Under any circumstances should one choose $\Delta S_k$ to be invariant under global ``internal'' symmetries such as U(1) or SU(N). A cutoff that is invariant under Galilean symmetry (at zero temperature and after analytic continuation to real time) would also be of advantage. 

\paragraph{(v) Optimization ideas.} In a given truncation one observes that different choices of the cutoff lead to different results for physical observables. The range of observed outcomes should become smaller as the truncation becomes better. It may therefore be taken as a rough error estimation. One might ask which cutoff (or which class of cutoffs) leads to the ``best'' result. It would be sensible to work with such an optimized cutoff function. These ideas are discussed in more detail for systems with relativistic invariance in the literature \cite{Litim2000b, Litim2001a, Litim2001b, Pawlowski2007a} and apply in slightly modified form also to the nonrelativistic case.

\paragraph{(vi) Simplicity.} The cutoff function should be chosen such that the loop expressions become as simple as possible. Not only does this simplify life but often also makes analytic investigations possible that would be very cumbersome otherwise. If it is possible to find analytic expressions for the flow equations (no numerical summation or integration left), the numerical effort to solve them is strongly reduced. Often, this makes larger truncations possible. This in turn should lead to more precise results (which again makes life simpler).

\paragraph{(vii) Fermi surface.} For fermionic systems at nonvanishing density one has to deal with additional complications from the fact that the actual pole or infrared singularity of the (not regularized) propagator at zero frequency does not occur at $\vec q^2=0$ but at the Fermi surface $\vec q^2=f(\mu)$. (In the simplest approximation and for homogeneous systems one has $f(\mu)=\mu$.) The cutoff function $R_k$ has to be constructed such that it regularizes this singularity at the Fermi surface. 

\paragraph{}
After this discussion of the general requirements we now turn to the regulator functions used in this thesis. To cope with points (i) (hierarchy), (ii) (Matsubara summation) and (vi) (simplicity) of the above listing the cutoff is chosen to be independent of the frequency $q_0$, i.~e. 
\begin{equation}
R_k = R_k(\vec q).
\end{equation}
This is certainly not optimal with respect to the points (iii) (ultraviolet regulator property) and (iv) (symmetry). More explicit, we use for the Bose gas
\begin{eqnarray}
\nonumber
\Delta S_k &=& \int_q \bar \varphi^* \bar A (k^2-\vec q^2-m^2)\theta(k^2-\vec q^2-m^2) \bar \varphi\\
&=& \int_q \varphi^* (k^2-\vec q^2-m^2)\theta(k^2-\vec q^2-m^2)\varphi,
\label{eq7:DeltaSkbosons}
\end{eqnarray}
with $m^2=U^\prime(\rho){\big |}_{\rho_0}$ and $m^2=0$ in the regime with spontaneous U(1) symmetry breaking. The cutoff in Eq.\ \eqref{eq7:DeltaSkbosons} is similar to the optimized cutoff proposed by Litim for relativistic systems \cite{Litim2001a}. It respects U(1)-symmetry but breaks Galilean invariance. The loop expressions obtained with the cutoff \eqref{eq7:DeltaSkbosons} are very simple since all Matsubara summations and momentum integrations can be performed analytically for the truncations investigated in this thesis. A drawback is that $\partial_k R_k$ does not serve as an ultraviolet cutoff for the Matsubara summation. In the flow equation for the pressure this leads to convergence problems for a truncation with linear and quadratic frequency terms, indeed. 

For the BCS-BEC crossover we use a slide modification of Eq.\ \eqref{eq7:DeltaSkbosons},
\begin{eqnarray}
\nonumber
\Delta S_k &=& \int_q \psi^\dagger \left(\text{sign}(\vec q^2-\mu)k^2-(\vec q^2-\mu)\right)\theta(k^2-|\vec q^2-\mu|)\psi \\
&&+ \int_q \varphi^*\left(k^2-\vec q^2/2\right)\theta(k^2-\vec q^2/2) \varphi.
\label{eq7:DeltaSkfermions}
\end{eqnarray}
Again, this is an optimized choice in the sense of \cite{Litim2001a, Pawlowski2007a} but now properly regularizes around the Fermi-surface for the fermions $\psi$. In ``pure'' diagrams involving only fermionic or only bosonic lines, it is still possible to perform frequency summations and momentum integrations in closed form. For mixed diagrams, one has to perform the momentum integration numerically. 

The question arises whether one can find a cutoff which fulfills all the requirements (i) - (vii). Already the combination of points (i), (ii) and (iii) restrict the space of possible functions $R_k(q)$ quite a bit. A cutoff that respects Galilean invariance must be a function of the combination $i q_0+\vec q^2$. A possible choice for bosons would be of the form
\begin{equation}
\Delta S_k = \int_q \varphi^* A\, k^2 r_k(z)\varphi
\end{equation} 
with $z=(i q_0+ \xi_q)/k^2$, $\xi_q=\vec q^2+m^2$ and the dimensionless function
\begin{equation}
r_k(z) = \frac{1}{1+c_1 z+c_2 z^2+\ldots +c_n z^n}.
\label{eq7:Algebraiccutoff}
\end{equation}
The coefficients $c_i$ might be chosen for convenience, a simple choice is $c_i=1$. The ultraviolet properties become better if $n$ is large. On the other hand, expressions are expected to be simpler for small $n$. The cutoff in Eq.\ \eqref{eq7:Algebraiccutoff} is analytic in the sense of complex analysis and fulfills therefore criterion (ii). It respects the symmetries of the microscopic propagator (iv) and at least for small $n$ it also fulfills the requirement of simplicity (vi). For large $n$, real and positive argument $z$ and $c_i=1$ we find that the regulator in Eq.\ \eqref{eq7:Algebraiccutoff} approaches the optimized regulator in Eq.\ \eqref{eq7:DeltaSkbosons}. This can be seen from summing the geometric series in the denominator of Eq.\ \eqref{eq7:Algebraiccutoff}. In order to fulfill point (i) (hierarchy of flow equations) one has to choose the coefficients $c_i$ in Eq.\ \eqref{eq7:Algebraiccutoff} conveniently.

\chapter{Investigated models}
\label{ch:Variousphysicalsystems}
In this chapter we explain the physical systems investigated in this thesis. We introduce the microscopic models in the form of Lagrange densities, discuss their scope and comment on experimental realizations. After a discussion of the symmetries of the microscopic models in the next chapter and a presentation of the approximation scheme thereafter, we discuss our results for few-body physics in chapter \ref{ch:Few-bodyphysics} and for many-body physics in chapter \ref{ch:Many-bodyphysics}.

\section{Bose gas in three dimensions}
\label{sec:Bosegasinthreedimensions}
A gas of non-relativistic bosons with a repulsive pointlike interaction is one of the simplest interacting statistical systems. Since the first experimental realization \cite{Andersonetal1995, PhysRevLett.75.1687, PhysRevLett.75.3969} of Bose-Einstein condensation (BEC) \cite{Einstein1924, Einstein1925, Bose1924} with ultracold gases of bosonic atoms, important experimental advances  have been achieved, for reviews see \cite{RevModPhys.71.463, RevModPhys.73.307, morsch:179, bloch:885, PethickSmith2002, PitaevsikiiStringari2003}. Thermodynamic observables like the specific heat \cite{1367-2630-8-9-189} or properties of the phase transition like the critical exponent $\nu$ \cite{Donneretal2007} have been measured in harmonic traps. Still, the theoretical description of these apparently simple systems is far from being complete.

For ultracold dilute non-relativistic bosons in three dimensions, Bogoliubov theory gives a successful description of most quantities of interest \cite{Bogoliubov}. This approximation breaks down, however, near the critical temperature for the phase transition, as well as for the low temperature phase in lower dimensional systems, due to the importance of fluctuations. One would therefore like to have a systematic extension beyond the Bogoliubov theory, which includes the fluctuation effects beyond the lowest order in a perturbative expansion in the scattering length. Such extensions have encountered obstacles in the form of infrared divergences in various expansions \cite{Beliaev1958, Gavoret1964, Nepomnyashchii1975}. Only recently, a satisfactory framework has been found to cure these problems \cite{PhysRevLett.78.1612, PhysRevB.69.024513, Wetterich:2007ba}.
In this thesis, we extend this formalism to a nonvanishing temperature. We present a quantitative rather accurate picture of Bose-Einstein condensation in three dimensions and find that the Bogoliubov approximation is indeed valid for many quantities. The same method is also applied for two spatial dimensions (see section \ref{sec:Bosegasintwodimensions}) and can also be applied for one dimension.

For dilute non-relativistic bosons in three dimensions with repulsive interaction we find an upper bound on the scattering length $a$. This is similar to the "triviality bound" for the Higgs scalar in the standard model of elementary particle physics. As a consequence, the scattering length is at most of the order of the inverse effective ultraviolet cutoff $\Lambda^{-1}$, which indicates the breakdown of the pointlike approximation for the interaction at short distances. Typically, $\Lambda^{-1}$ is of the order of the range of the Van der Waals interaction. For dilute gases, where the interparticle distance $n^{-1/3}$ is much larger than $\Lambda^{-1}$, we therefore always find a small concentration $c=a n^{1/3}$. This provides for a small dimensionless parameter, and perturbation theory in $c$ becomes rather accurate for most quantities. For typical experiments with ultracold bosonic alkali atoms one has $\Lambda^{-1}\approx 10^{-7} \,\text{cm}$, $n^{1/3}\approx 10^4 \,\text{cm}^{-1}$, such that $c\lesssim 10^{-3}$ is really quite small.

Bosons with pointlike interactions can also be employed for an effective description of many quantum phase transitions at zero temperature, or phase transitions at low temperature $T$. In this case, they correspond to quasi-particles, and their dispersion relation may differ from the one of non-relativistic bosons, $\omega=\frac{\vec{p}^2}{2M}$. We describe the quantum phase transitions for a general microscopic dispersion relation, where the inverse classical propagator in momentum and frequency space takes the form $G_0^{-1}=-S\omega-V\omega^2+\vec{p}^2$ (in units where the particle mass $M$ is set to $1/2$). We present the quantum phase diagram at $T=0$ in dependence on the scattering length $a$ and a dimensionless parameter $\tilde{v}\sim V/S^2$, which measures the relative strength of the term quadratic in $\omega$ in $G_0^{-1}$. In the limit $S\rightarrow 0$ ($\tilde{v}\rightarrow\infty$) our model describes relativistic bosons.

\subsubsection{Lagrangian}

Our microscopic action describes nonrelativistic bosons, with an effective interaction between two particles given by a contact potential. It is assumed to be valid on length scales where the microscopic details of the interaction are irrelevant and the scattering length  is sufficient to characterize the interaction. The microscopic action reads
\begin{equation}
S[\varphi]=\int_x \,{\Big \{}\varphi^*\,(S\partial_\tau-V\partial_\tau^2-\Delta-\mu)\,\varphi\,+\,\frac{1}{2}\lambda(\varphi^*\varphi)^2{\Big \}},
\label{microscopicaction}
\end{equation}
with 
\begin{equation}
x=(\tau,\vec{x}), \,\,\int_x=\int_0^{\frac{1}{T}}d\tau\int d^3x.
\end{equation}
The integration goes over the whole space as well as over the imaginary time $\tau$, which at finite temperature is integrated on a circle of circumference $\beta=1/T$ according to the Matsubara formalism. We use natural units $\hbar=k_B=1$. We also scale time and energy units with appropriate powers of $2M$, with $M$ the particle mass. In other words, our time units are set such that effectively $2M=1$. In these units time has the dimension of length squared. For standard non-relativistic bosons one has $V=0$ and $S=1$, but we also consider quasiparticles with a more general dispersion relation described by nonzero $V$.

After Fourier transformation, the kinetic term reads
\begin{equation}
\int_q \varphi^*(q)(i S q_0+V q_0^2+\vec{q}^2)\varphi(q),
\label{eqMicroscopicFourier}
\end{equation}
with
\begin{eqnarray}
q=(q_0,\vec{q}),\quad \int_q=\int_{q_0}\int_{\vec{q}},\quad \int_{\vec{q}}=\frac{1}{(2\pi)^3}\int d^3q.
\end{eqnarray}
At nonzero temperature, the frequency $q_0=2\pi T n$ is discrete, with
\begin{equation}
\int_{q_0}=T \sum_{n=-\infty}^\infty,
\end{equation}
while at zero temperature this becomes
\begin{equation}
\int_{q_0}=\frac{1}{2\pi}\int_{-\infty}^\infty dq_0.
\end{equation}
The dispersion relation encoded in Eq.\ \eqref{eqMicroscopicFourier} obtains by analytic continuation
\begin{equation}
S\omega+V\omega^2=\vec{q}^2/2M.
\end{equation}

In this thesis, we consider homogeneous situations, i.e. an infinitely large volume without a trapping potential. Many of our results can be translated to the inhomogeneous case in the framework of the local density approximation. One assumes that the length scale relevant for the quantum and statistical fluctuations is much smaller than the characteristic length scale of the trap. In this case, our results can be transferred by taking the chemical potential position dependent in the form $\mu\left(\vec{x})=2M(\mu-V_t(\vec{x}\right))$, where $V_t(\vec{x})$ is the trapping potential.

The microscopic action \eqref{microscopicaction} is invariant under the global $U(1)$ symmetry which is associated to the conserved particle number,
\begin{equation}
\varphi\rightarrow e^{i\alpha}\varphi.
\end{equation}
On the classical level, this symmetry is broken spontaneously when the chemical potential $\mu$ is positive. In this case, the minimum of $-\mu\varphi^*\varphi+\frac{1}{2}\lambda(\varphi^*\varphi)^2$ is situated at $\varphi^*\varphi=\frac{\mu}{\lambda}$. The ground state of the system is then characterized by a macroscopic field $\varphi_0$, with $\varphi_0^*\varphi_0=\rho_0=\frac{\mu}{\lambda}$. It singles out a direction in the complex plane and thus breaks the $U(1)$ symmetry. Nevertheless, the action itself and all modifications due to quantum and statistical fluctuations respect the symmetry. For $V=0$ and $S=1$, the situation is similar for Galilean invariance. At zero temperature, we can perform an analytic continuation to real time and the microscopic action \eqref{microscopicaction} is then invariant under transformations that correspond to a change of the reference frame in the sense of a Galilean boost. It is easy to see that in the phase with spontaneous $U(1)$ symmetry breaking also the Galilean symmetry is broken spontaneously: A condensate wave function, that is homogeneous in space and time, would be represented in momentum space by
\begin{equation} 
\varphi(\omega,\vec{p})=\varphi_0 \,(2\pi)^4\, \delta^{(3)}(\vec{p})\delta(\omega).
\end{equation}
Under a Galilean boost transformation with a boost velocity $2\vec{q}$, this would transform according to
\begin{eqnarray}
\nonumber
\varphi(\omega,\vec{p})\rightarrow&&\varphi(\omega-\vec{q}^2,\vec{p}-\vec{q})\\
&&=\varphi_0\,(2\pi)^4\,\delta^{(3)}(\vec{p}-\vec{q})\delta(\omega-\vec{q}^2).
\end{eqnarray} 
This shows that the ground state is not invariant under such a change of reference frame. This situation is in contrast to the case of a relativistic Bose-Einstein condensate, like the Higgs boson field after electroweak symmetry breaking. A relativistic scalar transforms under Lorentz boost transformations according to
\begin{equation}
\varphi(p^\mu)\rightarrow\varphi((\Lambda^{-1})^\mu_{\,\,\nu}\,p^\nu),
\end{equation}
such that a condensate wave function
\begin{eqnarray}
\nonumber
\varphi_0\,(2\pi)^4 \,\delta^{(4)}(p^\mu)\rightarrow&&\varphi_0\,(2\pi)^4\,\delta^{(4)}((\Lambda^{-1})^\mu_{\,\,\nu}\,p^\nu)\\
&&=\varphi_0\,(2\pi)^4\, \delta^{(4)}(p^\mu)
\end{eqnarray}
transforms into itself. We will investigate the implications of Galilean symmetry for the form of the effective action in chapter \ref{ch:Symmetries}. An analysis of general coordinate invariance in nonrelativistic field theory can be found in \cite{Son2006}.


\section{Bose gas in two dimensions}
\label{sec:Bosegasintwodimensions}

Bose-Einstein condensation and superfluidity for cold nonrelativistic atoms can be experimentally investigated in systems of various dimensions \cite{morsch:179, bloch:885, PethickSmith2002}. Two dimensional systems can be achieved by building asymmetric traps, resulting in different characteristic sizes for one ``transverse extension'' $l_T$ and two ``longitudinal extensions'' $l$  of the atom cloud \cite{PhysRevLett.87.130402, PhysRevLett.92.173003, 0953-4075-38-3-007, 0295-5075-57-1-001, Koehl2005, C.Orzel03232001, spielman:080404, PhysRevLett.93.180403, Hazibabic2006}. For $l \gg l_T$ the system behaves effectively two-dimensional for all modes with momenta $\vec{q}^2\lesssim l_T^{-2}$. From the two-dimensional point of view, $l_T$ sets the length scale for microphysics -- it may be as small as a characteristic molecular scale. On the other hand, the effective size of the probe $l$ sets the scale for macrophysics, in particular for the thermodynamic observables.

Two-dimensional superfluidity shows particular features. In the vacuum, the interaction strength $\lambda$ is dimensionless such that the scale dependence of $\lambda$ is logarithmic \cite{lapidus:459}. The Bogoliubov theory with a fixed small $\lambda$ predicts at zero temperature a divergence of the occupation numbers for small $q=|\vec{q}|$, $n(\vec{q})\sim n_C\, \delta^{(2)}(\vec{q})$ \cite{Bogoliubov}. In the infinite volume limit, a nonvanishing condensate $n_c=\bar{\rho}_0$ is allowed only for $T=0$, while it must vanish for $T>0$ due to the Mermin-Wagner theorem \cite{PhysRevLett.17.1133, PhysRev.158.383}.  On the other hand, one expects a critical temperature $T_c$ where the superfluid density $\rho_0$ jumps by a finite amount according to the behavior for a Kosterlitz-Thouless phase transition \cite{Berezinskii1971, Berezinskii1972, 0022-3719-6-7-010, PhysRevLett.39.1201}. We will see that $T_c/n$ (with $n$ the atom-density) vanishes in the infinite volume limit $l\to \infty$. Experimentally, however, a Bose-Einstein condensate can be observed for temperatures below a nonvanishing critical temperature $T_c$ -- at first sight in contradiction to the theoretical predictions for the infinite volume limit.

A resolution of these puzzles is related to the simple observation that for all practical purposes the macroscopic size $l$ remains finite. Typically, there will be a dependence of the characteristic dimensionless quantities as $\bar{\rho}_0/n$, $T_c/n$ or $\lambda$ on the scale $l$. This dependence is only logarithmic. While $\lambda(n=T=0, l\to \infty)=0$, $(\bar{\rho}_0/n)(T\neq0, l\to \infty)=0$, $(T_c/n)(l\to0)=0$, in accordance with general theorems, even a large finite $l$ still leads to nonzero values of these quantities, as observed in experiment. 

The description within a two-dimensional renormalization group context starts with a given microphysical or classical action at the ultraviolet momentum scale $\Lambda_\text{UV}\sim l_T^{-1}$. When the scale parameter $k$ reaches the scale $k_\text{ph}\sim l^{-1}$, all fluctuations are included since no larger wavelength are present in a finite size system. The experimentally relevant quantities and the dependence on $l$ can be obtained from $\Gamma_{k_\text{ph}}$. For a system with finite size $l$ we are interested in $\Gamma_{k_\text{ph}}$, $k_\text{ph}=l^{-1}$. If statistical quantities for finite size systems depend only weakly on $l$, they  can be evaluated from $\Gamma_{k_\text{ph}}$ in the same way as their thermodynamic infinite volume limit follows from $\Gamma$. Details of the geometry etc. essentially concern the appropriate factor between $k_\text{ph}$ and $l^{-1}$. 

The microscopic model we use for the two-dimensional Bose gas is basically the one for the three-dimensional case in Eq.\ \eqref{microscopicaction}. The difference is that now $\vec x$ and the space-integral are two-dimensional
\begin{equation}
x=(\tau,\vec{x}), \,\,\int_x=\int_0^{\frac{1}{T}}d\tau\int d^2x,
\end{equation}
and similarly in momentum space. The dimensionless interaction parameter $\lambda$ in \eqref{microscopicaction} describes now a reduced two-dimensional interaction strength and is directly related to the scattering length in units of the transverse extension $a/l_T$. The few-body physics and the logarithmic scale-dependence of $\lambda$ is discussed in section \ref{sec:Repulsiveinteractingbosons}. 


\section{BCS-BEC Crossover}
\label{sec:BCS-BECCrossover}

Besides the bosons we also investigate systems with ultracold fermions. A qualitative new feature for fermions in comparison to bosons is the antisymmetry of the wavefunction and the tightly connected ``Pauli blocking''. Due to the antisymmetry of the wavefunction it is not possible to have two identical fermions in the same state. This feature has many interesting consequences. For example, a $s$-wave interaction between two identical fermions is not possible. This in turn implies that a gas of fermions in the same spin- (and hyperfine-) state has many properties of a free Fermi gas provided the $p$-wave and higher interactions are suppressed. The situation changes for a Fermi gas with two spin or hyperfine states. $S$-wave interactions and pairing are now possible. In the simplest case the densities of the two components are equal. Depending on the microscopic interaction the system has different properties. For a repulsive interaction one expects Landau Fermi liquid behavior (for not too small temperature) where many qualitative properties are as for the free Fermi gas \cite{Landau1957}. For weak attractive interaction the theory of Baarden, Cooper and Schrieffer (BCS) \cite{Bardeen:1957kj, Bardeen:1957mv} is valid. Cooper-pairs are expected to form at small temperatures and the system is then superfluid. On the other hand, for strong attractive interaction one expects the formation of bound states of two fermions. These bound states are then bosons and undergo Bose-Einstein condensation (BEC) at small temperatures. Again, the system shows superfluidity. As first pointed out by Eagles \cite{PhysRev.186.456} and Leggett \cite{Leggett1980} there is a smooth and continuous crossover (BCS-BEC crossover) between the two limits described above. 

\begin{figure}
\centering
\includegraphics[width=0.35\textwidth]{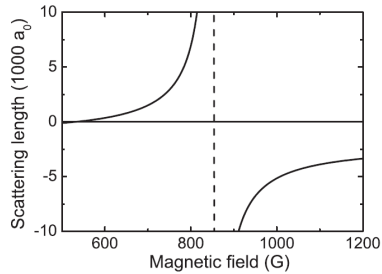}
\caption{Scattering length $a$ in units of the Bohr radius $a_0$ as a function of the magnetic field $B$ for the lowest hyperfine states of $^6$Li \cite{PhysRevA.66.041401, PhysRevLett.94.103201}.}
\label{fig:Feshbach}
\end{figure}

Experimental realizations of this crossover can be realized using Feshbach resonances. The detailed mechanism how these resonances work can be found in the literature, e.~g. \cite{PitaevsikiiStringari2003, PethickSmith2002}. It is important that the scattering length $a$ which serves as a measure for the $s$-wave interaction can be tuned to arbitrary values. As an example we consider the case of $^6$Li where the resonance was investigated in Refs. \cite{PhysRevA.66.041401, PhysRevLett.94.103201} and is shown in Fig.\ \ref{fig:Feshbach}. For magnetic fields in the range around $B=1200\, \text{G}$ the scattering length is relatively small and negative. In this regime the many-body ground state is of the BCS-type. Fermions with different spin and with momenta on opposite points on the Fermi surface form pairs. These Cooper pairs are (hyperfine-) spin singlets and have small or vanishing momentum. They are condensed in a Bose-Einstein condensate (BEC). The system is superfluid and the U(1) symmetry connected with particle number conservation is spontaneously broken. The macroscopic wavefunction of the BEC can be seen as an order parameter which is quadratic in the fermion field $\varphi_0\sim \langle\psi_1\psi_2\rangle$. Increasing the temperature, the system will at some point undergo a second order phase transition to a normal state where the order parameter vanishes, $\varphi_0=0$.

In the magnetic field range around $B=600 \,\text{G}$ in Fig. \ref{fig:Feshbach} the scattering length $a$ is small and positive. There is now a bound state of two fermions in the spectrum and the ground state of the many-body system is BEC-like. Pairs of fermions with different spin constitute bound states (dimers) which are pairs in position space. The interaction between these dimers is repulsive and proportional to the scattering length between fermions. When this repulsive interaction is weak the dimers are completely condensed in a BEC at zero temperature (no quantum depletion of the condensate). Again the order parameter is the macroscopic wavefunction of this condensate which is quadratic in the fermion fields $\varphi_0\sim\langle\psi_1\psi_2\rangle$. The phase transition between the superfluid state at small temperatures and the normal state is of second order, again. 

Now we come to the magnetic field in the intermediate crossover regime $700\, \text{G} \lesssim B \lesssim 1000 \,\text{G}$. The scattering length is now large and positive or large and negative with a divergence at $B\approx 834\, \text{G}$ \cite{PhysRevLett.94.103201}. Since the two-body scattering properties are solely governed by the requirement of unitarity of the scattering matrix for $a\to\pm \infty$, the point $B=834\, \text{G}$ is also called the ``unitarity point''. Due to the divergent scattering length one speaks of strongly interacting fermions. Perturbative methods for small coupling constants fail in the crossover regime. Non-perturbative methods show that the ground state is superfluid and governed by a order parameter $\varphi_0\sim \langle \psi_1\psi_2\rangle$ as before. 

The crossover from the BCS- to the BEC-like ground state is conveniently parameterized by the inverse scattering length in units of the Fermi momentum $c^{-1}=(a k_F)^{-1}$ where the Fermi momentum is determined by the density $n=\frac{1}{3\pi^2}k_F^3$ (in units with $\hbar=k_B=2M=1$). The dimensionless parameter $c^{-1}$ varies from large negative values on the BCS side to large positive values on the BEC side of the crossover. It crosses zero at the unitarity point. We will also use the Fermi energy which equals the Fermi temperature in our units $E_F=T_F=k_F^2$.

The quantitatively precise understanding of BCS-BEC crossover physics  is a  challenge for theory. Experimental breakthroughs as the realization of molecule condensates and the subsequent crossover to a BCS-like state of weakly attractively interacting fermions have been achieved \cite{PhysRevLett.92.040403, PhysRevLett.92.120403, PhysRevLett.92.150402, PhysRevLett.93.050401, C.Chin08202004, PhysRevLett.95.020404}. Future experimental precision measurements could provide a testing ground for non-perturbative methods. An attempt in this direction are the recently published measurements of the critical temperature \cite{Luo2007} and collective dynamics \cite{altmeyer:040401, wright:150403}.

A wide range of qualitative features of the BCS-BEC crossover is already well described by extended mean-field theories which account for the contribution of both fermionic and bosonic degrees of freedom \cite{Nozieres1985, PhysRevLett.71.3202}. In the limit of narrow Feshbach resonances mean-field theory becomes exact \cite{Diehl:2005an, Gurarie2007}. Around this limit perturbative methods for small Yukawa couplings \cite{Diehl:2005an} can be applied. Using $\epsilon$-expansion \cite{nussinov:053622, nishida:050403, nishida:063617, nishida:063618, arnold:043605, chen:043620} or $1/N$-expansion \cite{Sachdev06} techniques one can go beyond the case of small Yukawa couplings.

Quantitative understanding of the crossover at and near the resonance has been developed through numerical calculations using various quantum Monte-Carlo (QMC) methods \cite{PhysRevLett.91.050401, PhysRevLett.93.200404, bulgac:090404, bulgac:023625, burovski:160402, akkineni:165116}. Computations of the complete phase diagram have been performed from functional field-theoretical techniques, in particular from $t$-matrix approaches \cite{Haussmann1993, PhysRevLett.85.2801, PhysRevB.61.15370, PhysRevLett.92.220404, PhysRevB.70.094508}, Dyson-Schwinger equations \cite{Diehl:2005an,Diehl:2005ae}, 2-partice irreducible (2-PI) methods \cite{haussmann:023610}, and renormalization-group flow equations \cite{Birse2005,Diehl:2007th,Diehl:2007ri,Gubbels:2008zz}. These unified pictures of the whole phase diagram \cite{Sachdev06, Haussmann1993, PhysRevLett.85.2801, PhysRevB.61.15370, PhysRevLett.92.220404, PhysRevB.70.094508, Diehl:2005an, Diehl:2005ae, haussmann:023610, Diehl:2007th, Diehl:2007ri, Gubbels:2008zz}, however,  do not yet reach a similar quantitative precision as the QMC calculations.

In this thesis we discuss mainly the limit of broad Fesh\-bach resonances for which all thermodynamic quantities can be expressed in terms of two dimensionless parameters, namely the temperature in units of the Fermi temperature $T/T_F$ and the concentration $c=ak_F$. In the broad resonance regime, macroscopic observables are to a large extent independent of the concrete microscopic physical realization, a property referred to as universality \cite{Diehl:2005an, Sachdev06, Diehl:2007th}. This universality includes the unitarity regime where the scattering length diverges, $a^{-1}=0$ \cite{PhysRevLett.92.090402}, however it is not restricted to that region. Macroscopic quantities are independent of the microscopic details and can be expressed in terms of only a few parameters. In our case this is the two-body scattering length $a$ or, at finite density, the concentration $c=ak_F$. At nonzero temperature, an additional parameter is given by $T/T_F$. 

For small and negative scattering length $c^{-1}<0, |c|\ll 1$ (BCS side), the system can be treated with perturbative methods. However, there is a significant decrease in the critical temperature as compared to the original BCS result. This was first recognized by Gorkov and Melik-Barkhudarov \cite{Gorkov}. The reason for this correction is a screening effect of particle-hole fluctuations in the medium \cite{Heiselberg}. There has been no systematic analysis of this effect in approaches encompassing the full BCS-BEC crossover so far.

In section \ref{sec:Particle-holefluctuationsandtheBCS-BECCrossover}, we present an approach using the flow equation described in chapter \ref{ch:TheWetterichequation}. We include the effect of particle-hole fluctuations and recover the Gorkov correction on the BCS side. We calculate the critical temperature for the second-order phase transition between the normal and the superfluid phase throughout the whole crossover.

We also calculate the critical temperature at the point $a^{-1}=0$ for different resonance widths $\Delta B$. As a function of the microscopic Yukawa coupling $h_\Lambda$, we find a smooth crossover between the exact narrow resonance limit and the broad resonance result. The resonance width is connected to the Yukawa coupling via $\Delta B=h_\Lambda^2/(8\pi\mu_M a_b)$ where $\mu_M$ is the magnetic moment of the bosonic bound state and $a_b$ is the background scattering length.

\subsubsection{Lagrangian}

We start with a microscopic action including a two-component Grassmann field $\psi=(\psi_1,\psi_2)$, describing fermions in two hyperfine states. Additionally, we introduce a complex scalar field $\varphi$ as the bosonic degrees of freedom. In different regimes of the crossover, it can be seen as a field describing molecules, Cooper pairs or simply an auxiliary field. Using the resulting two-channel model we can describe both narrow and broad Feshbach resonances in a unified setting. Explicitly, the microscopic action at the ultraviolet scale $\Lambda$ reads

\begin{eqnarray}
\nonumber
S[\psi, \varphi] & = & \int_0^{1/T} d\tau \int d^3x{\Big \{}\psi^\dagger(\partial_\tau-\Delta-\mu)\psi\\
\nonumber
& & +\varphi^*(\partial_\tau-\frac{1}{2}\Delta-2\mu+ \nu_\Lambda)\varphi\\
& & - h_\Lambda(\varphi^*\psi_1\psi_2+h.c.){\Big \}}\,,
\label{eqMicroscopicAction}
\end{eqnarray}
where we choose nonrelativistic natural units with $\hbar=k_B=2M=1$, with $M$ the mass of the atoms.
The system is assumed to be in thermal equilibrium, which we describe using the Matsubara formalism. In addition to the position variable $\vec{x}$, the fields depend on the imaginary time variable $\tau$ which parameterizes a torus with circumference $1/T$. The variable $\mu$ is the chemical potential. The Yukawa coupling $h$ couples the fermionic and bosonic fields. It is directly related to the width of the Feshbach resonance. The parameter $\nu$ depends on the magnetic field and determines the detuning from the Feshbach resonance. Both $h$ and $\nu$ get renormalized by fluctuations, and the microscopic values $h_\Lambda$, and $\nu_\Lambda$ have to be determined by the properties of two body scattering in vacuum. For details, we refer to \cite{Diehl:2007th} and section \ref{sec:Twofermionspecies:Dimerformation}. 

More formally, the bosonic field $\varphi$ appears quadratically in the microscopic action in Eq.\ \eqref{eqMicroscopicAction}. The functional integral over $\varphi$ can be carried out. This shows that our model is equivalent to a purely fermionic theory with an interaction term
\begin{eqnarray}
\nonumber
S_{\text{int}} & = & \int_{p_1,p_2,p_1^\prime,p_2^\prime}  \left\{-\frac{h^2}{P_\varphi(p_1+p_2)}\right\}\psi_1^{\ast}(p_1^\prime){\psi_1}(p_1)\\
&& \times \psi_2^{\ast}(p_2^\prime){\psi_2}(p_2)\,\delta(p_1+p_2-p_1^\prime-p_2^\prime),
\label{lambdapsieff}
\end{eqnarray}
where $p = (p_0,\vec p)$ and the classical inverse boson propagator is given by
\begin{equation}
	P_{\varphi}(q)= i q_0 + \frac{\vec{q}^2}{2}+ \nu_\Lambda-2\mu\,.
\label{eq:Bosonpropagator}
\end{equation}

On the microscopic level the interaction between the fermions is described by the tree level expression
\begin{equation}
\lambda_{\psi,\text{eff}}=-\frac{h^2}{-\omega+\frac{1}{2}\vec q^2+\nu_\Lambda-2\mu}.
\end{equation}
Here, $\omega$ is the real-time frequency of the exchanged boson $\varphi$. It is connected to the Matsubara frequency $q_0$ via analytic continuation $\omega=-iq_0$. Similarly, $\vec q=\vec p_1+\vec p_2$ is the center of mass momentum of the scattering fermions $\psi_1$ and $\psi_2$ with momenta $\vec p_1$ and $\vec p_2$, respectively.

The limit of broad Feshbach resonances, which is realized in current experiments, e.g. with $\mathrm{^6Li}$ and $\mathrm{^{40}K}$ corresponds to the limit $h\to\infty$, for which the microscopic interaction becomes pointlike, with strength $-h^2/\nu_\Lambda$. 


\section{BCS-Trion-BEC Transition}
\label{sec:BCS-Trion-BECTransitionshort}

In the last section we discussed the interesting BCS-BEC crossover that is realized in a system consisting of two fermion species. We restricted ourselves to the case where the density for both components is equal. Interesting physics is also found if this constraint is released. The phase diagram of the imbalanced Fermi gas shows also first order phase transitions and phase separation, see \cite{KetterleZwierlein2007, Chevy2007} and references therein. 

Another interesting generalization is to take a third fermion species into account. A very rich phase diagram can be expected for the general case where the total density is arbitrarily distributed to the different components. Even the simpler case where the densities for all three components are equal is far less understood as the analogous two-component case. For simplicity we restrict much of the discussion to the case where all properties of the three components apart from the hyperfine-spin are the same. In particular, we assume that they have equal mass, chemical potential and scattering properties. We label the different hyperfine states by 1, 2 and 3. The $s$-wave scattering length $a_{12}$ for scattering between fermions of components 1 and 2 is the same as for scattering between fermions of species 2 and 3 or 3 and 1, $a_{12}=a_{23}=a_{31}=a$. 

Close to a common resonance where $a\to\pm\infty$ one expects the three-body problem to be dominated by the Efimov effect \cite{Efimov1970, Efimov1973}. This implies the formation of a three-body bound state (the ``trion''). Directly at the resonance an infinite tower of three-body bound states, the Efimov-trimers, exists. We refer to the Efimov trimer with the lowest lying energy as trion. The few-body physics is discussed in more detail in section \ref{sec:Threefermionspecies:ThomasandEfimoveffect}. 

The many-body phase diagram is far less understood. Not too close to the resonance one expects a superfluid ground state which is similar to the BCS ground state for $a<0$ or a BEC-like ground state for $a>0$. However, there are also some important differences. While in the two-component case the order parameter is a singlet under the corresponding SU(2) spin symmetry, the order parameter for the three component case with SU(3) spin symmetry is a (conjugate) triplet. In the superfluid phase the spin symmetry is therefore broken spontaneously. Due to some similarities with QCD this was called color superfluidity \cite{PhysRevB.70.094521, paananen:053606, paananen:023622, cherng:130406, zhai:031603, Bedaque2007}.

Between the extended BCS and BEC phase one can expect the ground state to be dominated by trions. Since trions are SU(3) singlets, the spin symmetry is unbroken in this regime such that there have to be true quantum phase transitions at the border to the BCS and BEC regimes. Such a trion phase has first been proposed for fermions in an optical lattice by Rapp, Zarand, Honerkamp and Hofstetter \cite{rapp:160405, rapp:144520}, see also \cite{Wilczek2007}. We will further discuss the many-body physics in section \ref{sec:BCS-Trion-BECTransitionlong}. To the knowledge of the author, there have been no experiments addressing the many-body issues so far. Only recently, experiments with $^6$Li probing the few-body physics found interesting phenomena \cite{ottenstein:203202, Huckans2008}. For the case of $^6$Li the assumption of equal scattering properties for the three different species are not fulfilled. We will present a more general model where SU(3) symmetry is broken explicitly and where the parameters can be chosen to describe $^6$Li in section \ref{sec:Threefermionspecies:ThomasandEfimoveffect}. We also discuss the experiments and show how their results can be explained in our framework. The remainder of this section is devoted to the discussion of the microscopic model in the SU(3) symmetric case. 

\subsubsection{Lagrangian}

As our microscopic model we use an action similar to the one for the BCS-BEC crossover in Eq.\ \eqref{eqMicroscopicAction}
\begin{eqnarray}
\nonumber
S&=&\int_x {\bigg \{} \psi^\dagger\partial_\tau-\Delta-\mu)\psi+\varphi^\dagger(\partial_\tau-\frac{1}{2}\Delta-2\mu+\nu_\varphi)\varphi\\
\nonumber
&&+\chi^*(\partial_\tau-\frac{1}{3}\Delta-3\mu+\nu_\chi)\chi\\
\nonumber
&&+\frac{1}{2} h\,\epsilon_{ijk}\,\left(\varphi_i^*\psi_j\psi_k-\varphi_i\psi_j^*\psi_k^*\right)\\
\nonumber
&&+g\left(\varphi_i^*\psi_i^*\chi-\varphi_i\psi_i\chi^*\right){\bigg \}}.
\label{eq8:microscopicactiontrionmodel}
\end{eqnarray}
The (Grassmann valued) fermion field has now three components $\psi=(\psi_1,\psi_2,\psi_3)$ and similar the boson field $\varphi=(\varphi_1,\varphi_2,\varphi_3)\hat{=} (\psi_1\psi_2,\psi_2\psi_3,\psi_3\psi_1)$. In addition we also include a single component fermion field $\chi$. This trion field represents the totally antisymmetric combination $\psi_1\psi_2\psi_3$. One choose the parameters such that $g=0$ and $\nu_\chi\to\infty$ at the microscopic scale. The trion field $\chi$ is then only an non-dynamical auxiliary field. 

We assumed in Eq.\ \eqref{eq8:microscopicactiontrionmodel} that the fermions $\psi_1$, $\psi_2$, and $\psi_3$ have equal mass $M$ and chemical potential $\mu$. We also assume that the interactions are independent of the spin (or hyperspin) so that our microscopic model is invariant under a global SU(3) symmetry transforming the fermion species into each other. While the fermion field $\psi=(\psi_1,\psi_2,\psi_3)$ transforms as a triplet ${\bf 3}$, the boson field $\varphi=(\varphi_1,\varphi_2,\varphi_3)$ transforms as a conjugate triplet $\bar {\bf 3}$. The trion field $\chi$ is a singlet under SU(3). In concrete experiments, for example with $^6\text{Li}$ \cite{ottenstein:203202}, the SU(3) symmetry may be broken explicitly since the Feshbach resonances of the different channels occur for different magnetic field values and have different widths. In addition to the SU(3) spin symmetry our model is also invariant under a global U(1) symmetry $\psi\to e^{i\alpha}\psi$, $\varphi\to e^{2i\alpha}\varphi$, and $\chi\to e^{3i\alpha}\chi$. The conserved charge related to this symmetry is the total particle number. Since we do not expect any anomalies the quantum effective action $\Gamma=\Gamma_{k=0}$ will also be invariant under $\text{SU(3)}\times \text{U(1)}$.

Apart from the terms quadratic in the fields that determine the propagators, Eq.\ \eqref{eq8:microscopicactiontrionmodel} contains the Yukawa-type interactions $\sim h$ and $\sim g$. The energy gap parameters $\nu_\varphi$ for the bosons and $\nu_\chi$ for the trions are sometimes written as $m_\varphi^2=\nu_\varphi-2\mu$, $m_\chi^2=\nu_\chi-3\mu$, absorbing an explicit dependence on the chemical potential $\mu$.

In Eq.\ \eqref{eq8:microscopicactiontrionmodel}, the fermion field $\chi$ can be ``integrated out'' by inserting the $(\psi,\varphi)$-dependent solution of its field equation into $\Gamma_k$. For $m_\chi^2\rightarrow \infty$ this results in a contribution to a local three-body interaction, $\lambda_{\varphi\psi}=-g^2/m_\chi^2$. Furthermore one may integrate out the boson field $\varphi$, such that (for large $m_\varphi^2$) one replaces the parts containing $\varphi$ and $\chi$ in $\Gamma_k$ by an effective pointlike fermionic interaction
\begin{equation}
\Gamma_{k,\text{int}}=\int_x\frac{1}{2}\lambda_\psi(\psi^\dagger\psi)^2+\frac{1}{3!}\lambda_3 \left(\psi^\dagger \psi\right)^3,
\end{equation}
with
\begin{equation}
\lambda_\psi= -\frac{h^2}{m_\varphi^2}, \quad \quad \lambda_3=-\frac{h^2 g^2}{m_\varphi^4 m_\chi^2}.
\label{eq:subst}
\end{equation}
We note that the contribution of trion exchange to $\lambda_{\varphi\psi}$ or $\lambda_3$ depends only on the combination $g^2/m_\chi^2$. The sign of $g$ can be changed by $\chi\rightarrow - \chi$, and the sign of $g^2$ can be reversed by a sign flip of the term quadratic in $\chi$. Keeping the possible reinterpretation by this mapping in mind, we will formally also admit negative $g^2$ (imaginary $g$).

\chapter{Symmetries}
\label{ch:Symmetries}
The symmetries of the microscopic action play an important role for the flow equation method. Provided the measure of the functional integral shows no anomalies, the effective action $\Gamma$ is invariant under the same symmetry transformations as the microscopic action $S$. This also holds for the flowing action $\Gamma_k$ if the cutoff term $\Delta S_k$ is invariant. If this is not the case, the constraints from symmetries on $\Gamma_k$ (Ward identities) are modified \cite{Ellwanger1994}. In devising truncations, the symmetries are a very useful guiding principle as was already emphasized in chapter \ref{ch:Truncations}. In the following section we will discuss the implication of Ward identities on the form of the effective action $\Gamma$ (and the flowing action $\Gamma_k$) in more detail. In section \ref{sec:Noetherstheorem} we discuss the second important implication of symmetries, the conservation of Noether currents. We concentrate the discussion to the Bose gas but it applies with minor modifications also for the systems with fermions. A discussion of the symmetries for the BCS-BEC crossover model \eqref{eqMicroscopicAction} can be found in \cite{Diehl:2008}.

\section{Derivative expansion and ward identities}
\label{sec:Derivativeexpansionandwardidentities}

Let us consider the microscopic model in Eq.\ \eqref{microscopicaction} using the flow equation \eqref{eq4:Wettericheqn}. We use a derivative expansion for the truncation of the effective average action with derivative operators up to four momentum dimensions 
\begin{eqnarray}
\nonumber
\Gamma_k &=& \int_x{\Bigg \{}U(\rho,\mu)\\
\nonumber
&&+\frac{1}{2}Z_1(\rho,\mu) \left(\varphi^*\partial_\tau\varphi-\varphi\partial_\tau\varphi^*\right)\\
\nonumber
&&+\frac{1}{2}Z_2(\rho,\mu) \left(\varphi^*(-\Delta)\varphi+\varphi(-\Delta)\varphi^*\right)\\
\nonumber
&&+\frac{1}{2}V_1(\rho,\mu) \left(\varphi^*(-\partial_\tau^2)\varphi+\varphi(-\partial_\tau^2)\varphi^*\right)\\
\nonumber
&&+V_2(\rho,\mu) \left(\varphi^*(\partial_\tau\Delta)\varphi-\varphi(\partial_\tau\Delta)\varphi^*\right)\\
&&+\frac{1}{2}V_3(\rho,\mu) \left(\varphi^*(-\Delta^2)\varphi+\varphi(-\Delta^2)\varphi^*\right){\Bigg \}}
\label{derivativeexpansion}
\end{eqnarray}
Here, we employ the renormalized fields 
\begin{eqnarray}
\nonumber
\varphi &=& \bar{A}^{1/2}\bar{\varphi},\\
\rho &=& \varphi^*\varphi=\bar{A}\bar{\rho}=\bar{A}\bar{\varphi}^*\bar{\varphi}
\end{eqnarray}
and coupling functions $U$, $Z_i$, $V_i$. We fix the wave function renormalization factor $\bar{A}$ such that $Z_1(\rho_0,\mu_0)=1$. 
Terms of the form $\rho(-\Delta)\rho$ or $\rho(-\partial_\tau^2)\rho$ are not included here, since they are expected to play a sub-leading role. For a systematic derivative expansion they have to be added - the terms with up to two derivatives can be found in \cite{Wetterich:2007ba}. In terms of dimensions, the operator $\partial_\tau$ counts as two space derivatives for the nonrelativistic model with $V=0$, while it counts as one space dimension for the relativistic model with $S=0$. We expand the $k$-dependent functions $U(\rho,\mu)$, $Z_1(\rho,\mu)$, $Z_2(\rho,\mu)$, $V_1(\rho,\mu)$,$V_2(\rho,\mu)$ and $V_3(\rho,\mu)$ around the $k$-dependent minimum $\rho_0(k)$ of the effective potential and the $k$-independent value of the chemical potential $\mu_0$ that corresponds to the physical particle number density $n$. For example, with $Z_1=Z_1(\rho_0,\mu_0)$, one has
\begin{eqnarray}
\nonumber
Z_1(\rho,\mu)&=&Z_1+Z_1^\prime(\rho_0,\mu_0)(\rho-\rho_0)\\
&&+Z_1^{(\mu)}(\rho_0,\mu_0)(\mu-\mu_0)+\ldots.
\end{eqnarray}

Let us concentrate on the non-relativistic model where $S=1$, $V=0$ in the microscopic action. At zero temperature, we can perform an analytic continuation to real time $\tau=it$. The microscopic action \eqref{microscopicaction} is then
\begin{eqnarray}
\nonumber
S[\varphi]&=&-\int_{-\infty}^{\infty} dt\int d^3x \\
&&{\Big \{}\varphi^*\,(-i\partial_t-\mu-\Delta)\,\varphi\,+\,\frac{1}{2}\lambda(\varphi^*\varphi){\Big \}}.
\label{realtimemicroscopicaction}
\end{eqnarray}
In addition to the global U(1) symmetry $\varphi\to e^{i\alpha}\varphi$, $\varphi^*\to e^{-i\alpha}\varphi^*$, space translations, rotations, time translations and the discrete symmetries parity and time reflection, two further symmetries constrain the form of the effective action $\Gamma$. In order to derive these constraints, we extend Eq. \eqref{realtimemicroscopicaction} to a $t$-dependent source $\mu(t)$. First, there is a semi-local $U(1)$ symmetry of the form
\begin{eqnarray}
\nonumber
\varphi(t,\vec{x})& \rightarrow & e^{i\alpha(t)}\varphi(t,\vec{x})\\
\nonumber
\varphi^*(t,\vec{x}) & \rightarrow & e^{-i\alpha(t)}\varphi^*(t,\vec{x})\\
\mu & \rightarrow  & \mu+\partial_t\alpha.
\end{eqnarray}
This holds since the combination $(-i\partial_t-\mu)$ acts as a covariant derivative. In addition, we have the invariance under Galilean boost transformations of the fields
\begin{eqnarray}
\nonumber
\varphi(t,\vec{x}) & \rightarrow & \varphi'(t,\vec{x})=e^{-i(\vec{q}^2t-\vec{q}\vec{x})}\varphi(t,\vec{x}-2\vec{q}t)\\
\varphi^*(t,\vec{x}) & \rightarrow & \varphi^{*\prime}(t,\vec{x})=e^{i(\vec{q}^2t-\vec{q}\vec{x})}\varphi^*(t,\vec{x}-2\vec{q}t).\label{galileanboost}
\end{eqnarray}
While the invariance of the interaction term under this symmetry is obvious, its realization for the kinetic term is more involved. Performing the transformation explicitly, one finds
\begin{eqnarray}
\nonumber
\varphi^*\Delta\varphi & \rightarrow & \varphi^*\Delta\varphi-\vec{q}^2\varphi^*\varphi+2 i\vec{q}\varphi^*\vec{\nabla}\varphi\\
\varphi^*i\partial_t\varphi & \rightarrow & \varphi^*i\partial_t\varphi+\vec{q}^2\varphi^*\varphi-2 i\vec{q}\varphi^*\vec{\nabla}\varphi,
\end{eqnarray}
such that indeed the combination
\begin{equation}
i\partial_t+\Delta \label{schroedingeroperator}
\end{equation}
leads to this invariance. On the other hand, the validity of the Galilean symmetry for an effective action guarantees that only the combination \eqref{schroedingeroperator} or powers of this operator act on $\varphi$. An operator of the form $(i\partial_t+\gamma\Delta)$ with $\gamma\neq 1$ would break the symmetry. (Note, that $\Delta\rho$ is also invariant.)

Both the semi-local $U(1)$ symmetry and the Galilean symmetry are helpful only at zero temperature. At nonzero temperature, the analytic continuation to real time is no longer useful. An analog version of the semi-local $U(1)$ transformation for Euclidean time $\tau$ would involve the imaginary part of the chemical potential $\mu$, which has no physical meaning. The dependence of physical quantities on $\mu+\mu^*$ is not restricted. In addition, the Galilean symmetry is broken explicitly by the thermal heat bath.
 
Combining semi-local $U(1)$ symmetry and Galilean symmetry at $T=0$, we find that the derivative operators $i\partial_t$, $\Delta$ and the chemical potential term $(\mu-\mu_0)$ are combined to powers of the operator
\begin{equation}
D=(-i\partial_t-(\mu-\mu_0)-\Delta).
\end{equation}
In addition to powers of that operator acting on $\varphi$, only spatial derivatives of terms, that are invariant under $U(1)$ transformations, like $\rho\Delta\rho$, may appear. Since the symmetry transformations act linearly on the fields, the full effective action $\Gamma[\varphi]$ is also invariant. This also holds for the average action $\Gamma_k[\varphi]$, provided that the cutoff term $\Delta S_k[\varphi]$ is invariant. We can write the effective action as an expansion in the operator $D$
\begin{eqnarray}
\nonumber
\Gamma[\varphi]&=&\int_x \bigg{\{}U_0(\rho)\\
\nonumber
&&+\frac{1}{2}\tilde{Z}(\rho)\left(\varphi^*(-i\partial_t-(\mu-\mu_0)-\Delta)\varphi+c.c\right)\\
\nonumber
&&+\frac{1}{2}\tilde{V}(\rho)\left(\varphi^*(-i\partial_t-(\mu-\mu_0)-\Delta)^2\varphi+c.c\right)\\
&&+\dots\bigg{\}}
\label{eqEffectiveactionGalilean}
\end{eqnarray}
Performing the Wick rotation back to Euclidean time, we can compare this to Eq.\ \eqref{derivativeexpansion}, and find for $T=0$ the relations
\begin{eqnarray}
\nonumber
Z_1(\rho,\mu_0)&=&Z_2(\rho,\mu_0)=\tilde{Z}(\rho),\\
\nonumber
V_1(\rho,\mu_0)&=&V_2(\rho,\mu_0)=V_3(\rho,\mu_0)=\tilde{V}(\rho),\\
\nonumber
Z_1^{(\mu)}(\rho_0,\mu_0)&=&2\left(\tilde{V}(\rho_0)+\rho_0\tilde{V}^\prime(\rho_0)\right),\\
Z_2^{(\mu)}(\rho_0,\mu_0)&=&2\tilde{V}(\rho_0),
\end{eqnarray}
and therefore
\begin{eqnarray}
\nonumber
\alpha&=&-\left(\tilde{Z}(\rho_0)+\rho_0\tilde{Z}^\prime(\rho_0)\right),\\
\nonumber
n_k&=&\tilde{Z}(\rho_0)\rho_0,\\
\beta &=&-\left(2\tilde{Z}^\prime(\rho_0)+\rho_0\tilde{Z}^{\prime\prime}(\rho_0)\right).
\end{eqnarray}

We next compute the inverse propagator in a constant background field by expanding $\Gamma_k$ to second order in the fluctuations around this background. For this purpose, it is convenient to decompose 
\begin{equation}
\varphi(\tau,\vec{x})=\varphi_0+\frac{1}{\sqrt{2}}\left(\varphi_1(\tau, \vec{x})+i\varphi_2(\tau,\vec{x})\right).
\end{equation}
The constant condensate field $\varphi_0$ can be chosen to be real without loss of generality. The fluctuating real fields are the radial mode $\varphi_1$ and the Goldstone mode $\varphi_2$, and $\rho=\rho_0+\sqrt{2}\varphi_0\varphi_1+\frac{1}{2}\varphi_1^2+\frac{1}{2}\varphi_2^2$. The truncation of the effective average action \eqref{derivativeexpansion} reads in that basis
\begin{eqnarray}
\nonumber
\Gamma_k[\varphi] = \int_x{\Bigg \{}U(\rho,\mu)
&+& \frac{1}{2}Z_1(\rho,\mu) \left(i\sqrt{2}\varphi_0\partial_\tau\varphi_2+i\varphi_1\partial_\tau \varphi_2-i\varphi_2\partial_\tau\varphi_1\right)\\
\nonumber
&+&\frac{1}{2}Z_2(\rho,\mu) \left(\sqrt{2}\varphi_0(-\Delta)\varphi_1+\varphi_1(-\Delta) \varphi_1+\varphi_2(-\Delta)\varphi_2\right)\\
\nonumber
&+&\frac{1}{2}V_1(\rho,\mu) \left(\sqrt{2}\varphi_0(-\partial_\tau^2)\varphi_1+\varphi_1(-\partial_\tau^2) \varphi_1+\varphi_2(-\partial_\tau^2)\varphi_2\right)\\
\nonumber
&+&V_2(\rho,\mu) \left(i\sqrt{2}\varphi_0(\partial_\tau\Delta)\varphi_2+i\varphi_1(\partial_\tau\Delta) \varphi_2-i\varphi_2(\partial_\tau\Delta)\varphi_1\right)\\
\nonumber
&+&\frac{1}{2}V_3(\rho,\mu) \left(\sqrt{2}\varphi_0(-\Delta^2)\varphi_1+\varphi_1(-\Delta^2) \varphi_1+\varphi_2(-\Delta^2)\varphi_2\right){\Bigg \}}.\\
\end{eqnarray}
The inverse propagator $\Gamma_k^{(2)}$ can be inferred from an expansion to second order in $\varphi_1$ and $\varphi_2$. We keep the linear order in $\mu-\mu_0$, which will be needed for the flow equation for the density. This yields
\begin{eqnarray}
\nonumber
\Gamma_k[\varphi] &=&\int_x{\Bigg \{}U(\rho_0,\mu_0)+U^{(\mu)}\,(\mu-\mu_0)+\frac{1}{2}(U^\prime+2\rho_0 U^{\prime\prime})\varphi_1^2+\frac{1}{2}U^\prime \varphi_2^2\\
\nonumber
&+&\frac{1}{2}\left(Z_1+Z_1^\prime \rho_0+Z_1^{(\mu)}(\mu-\mu_0)\right) \left(i\varphi_1\partial_\tau \varphi_2-i\varphi_2\partial_\tau\varphi_1\right)\\
\nonumber
&+&\frac{1}{2}\left(1+2 Z_2^\prime \rho_0+Z_2^{(\mu)}(\mu-\mu_0)\right) \left(\varphi_1(-\Delta) \varphi_1\right)\\
\nonumber
&+&\frac{1}{2}\left(1+Z_2^{(\mu)}(\mu-\mu_0)\right) \left(\varphi_2(-\Delta) \varphi_2\right)\\
\nonumber
&+&\frac{1}{2}\left(V_1+2 V_1^\prime\rho_0+V_1^{(\mu)}(\mu-\mu_0)\right) \left(\varphi_1(-\partial_\tau^2) \varphi_1\right)\\
\nonumber
&+&\frac{1}{2}\left(V_1+V_1^{(\mu)}(\mu-\mu_0)\right) \left(\varphi_2(-\partial_\tau^2) \varphi_2\right)\\
\nonumber
&+&\left(V_2+V_2^\prime\rho_0+V_2^{(\mu)}(\mu-\mu_0)\right) \left(i\varphi_1(\partial_\tau\Delta) \varphi_2-i\varphi_2(\partial_\tau\Delta)\varphi_1\right)\\
\nonumber
&+&\frac{1}{2}\left(V_3+2V_3^\prime \rho_0+V_3^{(\mu)}(\mu-\mu_0)\right) \left(\varphi_1(-\Delta^2) \varphi_1\right)\\
&+&\frac{1}{2}\left(V_3+V_3^{(\mu)}(\mu-\mu_0)\right)\left(\varphi_2(-\Delta^2) \varphi_2\right){\Bigg \}},
\end{eqnarray}
where we dropped the argument $(\rho_0,\mu_0)$ at several places and used the implicit rescaling condition $Z_2(\rho_0,\mu_0)=1$.
In a simple truncation, we take at $\mu=\mu_0$ only 
\begin{eqnarray}
\nonumber
S&=&Z_1(\rho_0,\mu_0)+Z_1^\prime(\rho_0,\mu_0) \,\rho_0,\\
V&=&V_1(\rho_0,\mu_0)
\end{eqnarray} 
into account. We neglect the contribution of the other couplings, i.e. set $Z_2^\prime=V_2=V_3=V_1^\prime=V_2^\prime=V_3^\prime=0$. As shown above, it follows from the symmetry requirements at zero temperature, that  $V_1=V_2=V_3=\tilde{V}$, $Z_1^{(\mu)}=2(\tilde{V}+\tilde{V}^\prime\rho_0)$ and $Z_2^{(\mu)}=2\tilde{V}$. The truncation $V_2=V_3=0$ therefore violates the Galilean symmetry, as does our choice of the cutoff term $\sim R_k$. Within our approximation, it is consistent to set $Z_1^{(\mu)}=Z_2^{(\mu)}=2V$ at zero temperature. Also the deviations from this relation at finite temperature are neglected for simplicity in this thesis. This yields the truncation used in order to obtain the numerical results which will be discussed in chapter \ref{ch:Many-bodyphysics}.

\subsubsection{Propagator and dispersion}

The inverse propagator is given by the second functional derivative of the effective action
\begin{eqnarray}
\nonumber
\Gamma^{(2)}&=&\begin{pmatrix} \overset{\rightharpoonup}{\delta}_{\varphi_1(-q)} \\ \overset{\rightharpoonup}{\delta}_{\varphi_2(-q)} \end{pmatrix} \Gamma_k \begin{pmatrix} \overset{\leftharpoonup}{\delta}_{\varphi_1(p)}, & \overset{\leftharpoonup}{\delta}_{\varphi_2(p)} \end{pmatrix} \\
&=& G^{-1} \delta(p-q),
\end{eqnarray}
and we find from the truncation \eqref{derivativeexpansion}
\begin{eqnarray}
G^{-1}&=&\begin{pmatrix} H+2J+(V_1+2\rho V_1^\prime)q_0^2 &\hspace{-0.2cm},& -q_0 \sqrt{2K} \\ q_0 \sqrt{2K}&\hspace{-0.2cm},& H+V_1 q_0^2 \end{pmatrix}.\quad
\end{eqnarray}
Here we use the abbreviations
\begin{eqnarray}
\nonumber
H&=&Z_2\vec{p}^2-V_3\vec{p}^4+U^\prime\\
\nonumber
J&=&\rho Z_2^\prime\vec{p}^2-\rho V_3^\prime \vec{p}^4+\rho U^{\prime\prime}\\
2K&=&\left( Z_1+\rho Z_1^\prime-2(V_2+\rho V_2^\prime)\vec{p}^2\right)^2.
\end{eqnarray}
At zero temperature, we can analytically continue to real time $q_0\rightarrow i\omega$, and find
\begin{eqnarray}
G^{-1}&=&\begin{pmatrix} H+2J-(V_1+2\rho V_1^\prime)\omega^2 &\hspace{-0.2cm},& -i\omega \sqrt{2K} \\ i\omega \sqrt{2K} &\hspace{-0.2cm},& H-V_1\omega^2 \end{pmatrix}.\quad
\end{eqnarray}

The dispersion relation is found from the on shell condition
\begin{equation}
\text{det}\,G^{-1}=0
\end{equation}
which yields
\begin{eqnarray}
\nonumber
&& H^2+2HJ-2\left(H(V_1+\rho V_1^\prime)+J V_1+K)\right)\,\omega^2\\
&& +\,V_1(V_1+2\rho V_1^\prime)\,\omega^4=0.
\end{eqnarray}
The solutions for $\omega$ define the dispersion relation. We find two branches, according to
\begin{eqnarray}
\nonumber
(\omega_\pm^2) &=& \frac{1}{V_1(V_1+2\rho V_1^\prime)}\Bigg{(}H(V_1+\rho V_1^\prime)+J V_1+K\\
&& \pm\bigg{(}(K+J V_1)^2+2H\left(K(V_1+\rho V_1^\prime)-J V \rho V_1^\prime\right)+H^2(\rho V_1^\prime)^2\bigg{)}^{1/2}\Bigg{)}.\,\,
\end{eqnarray}
In the phase with spontaneous symmetry breaking, the $(+)$ branch of this solution is an "optical mode", while the $(-)$ branch is a sound mode. The microscopic sound velocity is $c_S=\frac{\partial \omega}{\partial p}\big{|}_{p=0}$. Using $\rho=\rho_0$, $U^\prime=0$, $U^{\prime\prime}=\lambda$ and $Z_2=1$, we find
\begin{equation}
c_S^2=\frac{1}{\frac{(Z_1+\rho_0 Z_1^\prime)^2}{2\lambda\rho_0}+V_1}=\frac{2\lambda \rho_0}{S^2+2\lambda \rho_0 V}.
\end{equation}
The "optical mode" has at vanishing spatial momentum the frequency
\begin{equation}
\omega_{+}^2(\vec{q}^2=0)=\frac{2\lambda \rho_0}{V_1+2\rho_0V_1^\prime}+\frac{(Z_1+\rho_0Z_1^\prime)}{V_1(V_1+2\rho_0V_1^\prime)}
\end{equation}
which diverges $\omega_{+}^2\rightarrow\infty$ in the limit $V_1\rightarrow0$.

\section{Noethers theorem}
\label{sec:Noetherstheorem}

In the following we further discuss the role of continuous symmetries of the microscopic action $S[\varphi]$. Since all these symmetries are linear in the fields, the full effective action $\Gamma[\varphi]$ is also symmetric. From Noether's theorem it follows that there exists a conserved current $j^\mu=(j^0,\vec{j})$ connected with every such symmetry. If the action is formulated as an integral over the imaginary time $\tau$ the conservation equation implies for the current
\begin{equation}
\partial_\tau j^{(\tau)}+\vec{\nabla}\vec{j}=0. \label{euklidconservedcurrent}
\end{equation}
At zero temperature, we can perform a Wick rotation to real time, $\tau\rightarrow it$, and Eq.\  \eqref{euklidconservedcurrent} takes the usual form
\begin{equation}
\partial_t j^{(t)}+\vec{\nabla}\vec{j}=0. \label{realtimeconservedcurrent}
\end{equation}
The Noether charge $C=\int d^3 x j^{(t)}$ is conserved in time, i.e. $\frac{d}{dt}C=0$. This holds if $\vec{j}$ falls off sufficiently fast at spatial infinity. At finite temperature however, the situation is different. A simple analytic continuation to real time is no longer possible, since the configuration space is now a torus with periodicity $1/T$ in the $\tau$-direction. Instead, we can integrate Eq.\  \eqref{euklidconservedcurrent} over complex time $\tau$, giving
\begin{equation}
\vec{\nabla}\vec{J}=\vec{\nabla}\int_0^{1/T}d\tau \vec{j}=j^{(\tau)}(0)-j^{(\tau)}(1/T)=0.
\end{equation}
From the symmetry, it now follows that there exists a solenoidal vector field or three component current $\vec{J}=\int_{\tau}\vec{j}$.

A global symmetry of an action $\Gamma[\varphi]$ (where $\Gamma$ could be replaced by $S$ or $\Gamma_k$ if appropriate) can be formulated in its infinitesimal form as
\begin{equation}
\Gamma[\varphi+\epsilon s\varphi]=\Gamma[\varphi],
\end{equation}
with $\epsilon$ independent of $x$. Here $s$ is the infinitesimal generator of the symmetry transformation. For a local transformation, where $\epsilon$ depends on $x$, $\epsilon=\epsilon(x)$, we can expand
\begin{equation}
\Gamma [\varphi+\epsilon s \varphi]=\Gamma[\varphi]+\int_x \left\{ (\partial_\mu\epsilon){\cal J}^\mu+(\partial_\mu\partial_\nu\epsilon){\cal K}^{\mu\nu}+\dots\right\}.
\label{Noetherexpansion}
\end{equation}
The global symmetry implies that the expansion on the r.h.s. of Eq.\ \eqref{Noetherexpansion} starts with $\partial_\mu\epsilon$. Here and in the following it is implied that $\epsilon$ as well as its derivatives are infinitesimal, i.e. we keep only terms that are linear in $\epsilon$. The index $\mu$ goes over $(0,1,2,3)$, representing $(t,x^1,x^2,x^3)$ in the real time case and $(\tau,x^1,x^2,x^3)$ for imaginary time. Eq.\ \eqref{Noetherexpansion} implies for arbitrary $\varphi(x)$
\begin{equation}
\int_x\left\{\frac{\delta\Gamma[\varphi]}{\delta\varphi}\epsilon s\varphi-(\partial_\mu\epsilon){\cal J}^\mu-(\partial_\mu\partial_\nu\epsilon){\cal K}^{\mu\nu}+\dots\right\}=0.
\label{eqNoetherIntegral}
\end{equation}
Our notation is for real fields and implies a summation over components, if appropriate. In a complex basis one replaces $\frac{\delta \Gamma}{\delta \varphi}\epsilon s\varphi$ by $\frac{\delta \Gamma}{\delta \varphi}\epsilon s\varphi+\frac{\delta \Gamma}{\delta \varphi^*}\epsilon s\varphi^*$.

Eq.\ \eqref{eqNoetherIntegral} is valid for all field configurations $\varphi$ and not only for those that fulfill the field equation $\frac{\delta\Gamma[\varphi]}{\delta \varphi}=0$. In consequence, the integrand is a total derivative
\begin{eqnarray}
\nonumber
& \frac{\delta\Gamma[\varphi]}{\delta\varphi}\epsilon s\varphi-(\partial_\mu\epsilon){\cal J}^\mu-(\partial_\mu\partial_\nu\epsilon){\cal K}^{\mu\nu}+\ldots &\\
& =-\partial_\mu\left(j^\mu \epsilon+\kappa^{\mu\nu}\partial_\nu\epsilon+\ldots\right). &
\end{eqnarray}
We can now specialize to $\partial_\mu\epsilon=\partial_{\mu}\partial_\nu\epsilon=\ldots=0$ and find
\begin{equation}
\frac{\delta\Gamma[\varphi]}{\delta\varphi} s\varphi(x)=-\partial_\mu j^\mu.
\end{equation}
This defines the Noether current $j^\mu$. For solutions of the field equation, $\frac{\delta\Gamma[\varphi]}{\delta\varphi}=0$, the current $j^\mu$ is conserved, $\partial_\mu j^\mu=0$.

For a given $x$ we can also specialize to
\begin{equation}
\epsilon(x)=0, \,\,\,\partial_\mu\epsilon(x)\neq0, \,\,\,\partial_\mu\partial_\nu\epsilon(x)=0,\,\,\,\ldots,
\end{equation}
which leads to 
\begin{equation}
j^\mu={\cal J}^\mu-\partial_\nu \kappa^{\nu\mu}.
\label{currentcorrection}
\end{equation}
This process can be continued, leading us to a whole tower of identities for the conserved current $j^\mu$. 

If the action $\Gamma[\varphi]$ includes derivatives only up to a finite order $n$, i.e. can be written in the form
\begin{equation}
\Gamma[\varphi]=\int_x{\cal L}(\varphi,\partial\varphi, \partial\partial\varphi, \ldots, \partial^{(n)}\varphi),
\end{equation}
the expansion on the right hand side of \eqref{Noetherexpansion} only contains terms up to order $\partial^{(n)}\epsilon$ such that the tower of equations for $j^\mu$ can be solved.
Moreover, for homogeneous situations where $\frac{\delta\Gamma[\varphi]}{\delta\varphi}$ is solved by a constant $\varphi$, the second term on the r.h.s. of Eq.\ \eqref{currentcorrection} vanishes since it includes a derivative. We have then $j^\mu={\cal J}^\mu$.

A convenient way to find the local currents employs parameters $\epsilon(x)$ that decay sufficiently fast at infinity such that we can partially integrate Eq.\ \eqref{eqNoetherIntegral}
\begin{equation}
\int_x\epsilon(x)\left\{\frac{\delta\Gamma[\varphi]}{\delta\varphi} s\varphi+\partial_\mu{\cal J}^\mu-\partial_\mu\partial_\nu{\cal K}^{\mu\nu}+\dots\right\}=0.
\end{equation}
This yields the local identity
\begin{equation}
\frac{\delta \Gamma[\varphi]}{\delta \varphi}s\varphi=-\partial_\mu{\cal J}^\mu+\partial_\mu\partial_\nu {\cal K}^{\mu\nu}-\dots.
\end{equation}
An expansion of the l.h.s. in derivatives often yields substantial information on ${\cal J}^\mu$ etc. by inspection. 

Our construction yields a unique conserved local current $j^\mu$ for every generator of a continuous symmetry. We note, however, that $\alpha j^\mu+b^\mu$ is also a conserved local current if $\alpha$ and $b^\mu$ are independent of $x$. This remark is important if we want to associate $j^\mu$ with the current for a physical quantity. A rotation invariant setting implies $b^i=0$, but $b^0$ and $\alpha$ may differ from zero. 

After these general considerations we now specialize to nonrelativistic real time actions of the form
\begin{equation}
\Gamma[\varphi]=\int_{-\infty}^{\infty}dt \int d^3x{\cal L}(\varphi, (i\partial_t+\Delta)\varphi,(i\partial_t+\Delta)^2\varphi,\dots).
\end{equation}
We assume, that $\Gamma$ invariant under the same symmetries as the action \eqref{realtimemicroscopicaction}. From the symmetry under time translations
\begin{eqnarray}
\nonumber
\varphi & \rightarrow & \varphi+\epsilon (s_t)\varphi=\varphi+\epsilon \partial_t \varphi\\
{\cal L}& \rightarrow & {\cal L}+\epsilon \partial_t{\cal L}={\cal L}+\epsilon\partial_\mu(\delta^\mu_0 {\cal L}),
\end{eqnarray}
we find a conserved current $(j_E)^\mu$. Up to a possible additive constant its $t$-component is the energy density, while the spatial components describe energy flux density. The multiplicative constant $\alpha$ gets fixed if we choose the units to measure energy. The choice $\hbar=1$ corresponds to $\alpha=1$. Similarly, the invariance under spatial translations
\begin{eqnarray}
\nonumber
\varphi & \rightarrow  & \varphi+\epsilon^i(s_M)_i \varphi=\varphi-\epsilon^i\partial_i\varphi\\
{\cal L} & \rightarrow & {\cal L}-\epsilon^i\partial_i{\cal L}={\cal L}-\epsilon^i\partial_\mu(\delta_i^\mu{\cal L})
\end{eqnarray}
implies a conserved current $(j_M)_i^\mu$ for each spatial direction $i=1,2,3$. Up to an additive constant $(b_M)^0_i$ the $t$-component is the conserved momentum density, $p_i=(j_{M})^0_i+(b_M)^0_i$, while the spatial components can be interpreted as a momentum flux density, with the diagonal components $(j_{M})^i_i$ describing pressure. 

From the global $U(1)$ symmetry
\begin{eqnarray}
\nonumber
\varphi & \rightarrow & \varphi+\epsilon (s_C)\varphi=\varphi-i \epsilon \varphi\\
\nonumber
\varphi^* & \rightarrow & \varphi^*+\epsilon (s_C) \varphi^*=\varphi^*+i\epsilon \varphi^*\\
{\cal L} & \rightarrow & {\cal L}
\end{eqnarray}
we can infer the conservation of the current $(j_C)^\mu$
associated to the conserved particle number.  In order to identify the total particle number with the charge of this current, $\int d^3x (j_C)^0$, we need to fix a possible multiplicative constant $\alpha$. For this purpose, we use the Galilean boost invariance, described already in Eq.\ \eqref{galileanboost}. It reads in its infinitesimal form
\begin{eqnarray}
\nonumber
\varphi & \rightarrow & \varphi+\epsilon^i(s_G)_i\varphi=\varphi+2\epsilon^it\partial_i\varphi-i\epsilon^i x_i\varphi\\
\nonumber
\varphi^* & \rightarrow & \varphi^*+\epsilon^i(s_G)_i \varphi^* = \varphi^*+2\epsilon^i t\partial_i\varphi^*+i\epsilon^i x_i\varphi^*\\
{\cal L} & \rightarrow & {\cal L}+\epsilon^i\partial_\mu(2\,\delta^\mu_i t{\cal L}),
\end{eqnarray}
and the conserved charge of $(s_G)$ is the center of mass, again up to an additive constant. The generator $(s_G)$ can be decomposed as
\begin{equation}
(s_G)_i=x_i(s_C)-2t(s_M)_i. 
\end{equation}
This implies for the current
\begin{equation}
(j_G)^\mu_i=x_i\,(j_C)^\mu-2t\,(j_M)^\mu_i.
\end{equation}
Specializing to the $t$-component, identifying the momentum density $p_i=(j_M)^0_i+(b_M)^0_i$ and reintroducing the particle mass $2M=1$ we find
\begin{equation}
(j_G)^0_i=x_i\,(j_C)^0-t\frac{p_i-(b_M)^0_i}{M}. 
\end{equation}
From this we can conclude that up to an additive constant $(j_C)^0$ is the particle density $n=(j_C)^0+(b_C)^0$. 

For the effective action \eqref{eqEffectiveactionGalilean} we find for $\mu=\mu_0$ and constant $\varphi(x)=\sqrt{\rho_0}$ the current
\begin{equation}
(j_C)^0=\tilde{Z}(\rho_0)\rho_0.
\end{equation}
Using the normalization condition $\tilde{Z}(\rho_0)=1$, this gives $(j_C)^0=\rho_0$. At zero temperature, this is the particle density and the additive constant $(b_C)^0$ vanishes. At nonzero temperature we can compare to Eq.\ \eqref{eqDensityatT} and find $(b_C)^0=n_T$.

For completeness we also mention the symmetry under spatial rotations
\begin{eqnarray}
\nonumber
\varphi(t,\vec{x})\rightarrow\varphi(t,R^{-1}\vec{x})\\
{\cal L}(t,\vec{x})\rightarrow{\cal L}(t,R^{-1}\vec{x}),
\end{eqnarray}
with orthogonal matrix $R^i_{\,j}=(e^{i\vec{\eta}\vec{J}})^i_{\,j}$, generators $(J_i)^j_{\,k}=i\varepsilon_{ijk}$, and $\varepsilon_{ijk}$ the antisymmetric tensor in three dimensions.
The infinitesimal transformation reads
\begin{eqnarray}
\nonumber
\varphi(t,\vec{x})\rightarrow\varphi(t,\vec{x})+\eta^i\varepsilon_{ijk}x^k\partial_j\varphi(t,\vec{x})\\
{\cal L}(t,\vec{x})\rightarrow{\cal L}(t,\vec{x})+\eta^i\partial_l(\varepsilon_{ijk}x^k\delta^l_j{\cal L}).
\end{eqnarray}
The time component of the conserved current $(j_R)_i^\mu$ is, of course, the angular momentum density.

\chapter{Truncated flow equations}
\label{ch:Truncationsandflowequations}
In this chapter we present the concrete truncations used to investigate the different models described in chapter \ref{ch:Variousphysicalsystems}.

\section{Bose gas}
\label{sec:TruncationBosegas}
We start with the nonrelativistic Bose gas. The microscopic model for this system is shown in Eq.\ \eqref{microscopicaction}. The infrared cutoff function we use is shown in Eq.\ \eqref{eq7:DeltaSkbosons}. 
For the approximate solution of the flow equation \eqref{eq4:Wettericheqn} we use a truncation with up to two derivatives
\begin{eqnarray}
\nonumber
\Gamma_k&=&\int_x\bigg{\{} \bar{\varphi}^*\left(\bar{S}\partial_\tau-\bar{A}\Delta-\bar{V}\partial_\tau^2\right)\bar{\varphi}\\
&&+2\bar{V}(\mu-\mu_0)\,\bar{\varphi}^*\left(\partial_\tau-\Delta\right)\bar{\varphi}+\bar{U}(\bar{\rho},\mu)\bigg{\}},
\end{eqnarray}
with $\bar{\rho}=\bar{\varphi}^*\bar{\varphi}$. This particular form is motivated by a more systematic derivative expansion and an analysis of symmetry constraints (Ward identities) in section \ref{sec:Derivativeexpansionandwardidentities}. We introduce the renormalized fields $\varphi=\bar{A}^{1/2}\bar{\varphi}$, $\rho=\bar{A}\bar{\rho}$, the renormalized kinetic coefficients $S=\frac{\bar{S}}{\bar{A}}$, $V=\frac{\bar{V}}{\bar{A}}$ and we express the effective potential in terms of the renormalized invariant $\rho$, with
\begin{equation}
U(\rho,\mu)=\bar{U}(\bar{\rho},\mu).
\end{equation}
This yields
\begin{eqnarray}
\nonumber
\Gamma_k &=& \int_x\bigg{\{} \varphi^*\left(S\partial_\tau-\Delta-V\partial_\tau^2\right)\varphi\\
&&+2V(\mu-\mu_0)\, \varphi^*\left(\partial_\tau-\Delta\right)\varphi+U(\rho,\mu)\bigg{\}}.
\label{eqSimpleTruncation}
\end{eqnarray}

For the effective potential, we use an expansion around the $k$-dependent minimum $\rho_0(k)$ of the effective potential and the $k$-independent value of the chemical potential $\mu_0$ that corresponds to the physical particle number density $n$. We determine $\rho_0(k)$ and $\mu_0$ by the requirements
\begin{eqnarray}
\nonumber
(\partial_\rho U)(\rho_0(k),\mu_0)&=0\quad&\text{for all}\,k\\
-(\partial_\mu U)(\rho_0,\mu_0)&=n\quad&\text{at}\,k=0.
\end{eqnarray}
More explicitly we take a truncation for $U(\rho,\mu)$ of the form
\begin{eqnarray}
\nonumber
U(\rho,\mu)&=&U(\rho_0,\mu_0)-n_k(\mu-\mu_0)\\
\nonumber
&&+\left(m^2+\alpha(\mu-\mu_0)\right)(\rho-\rho_0)\\
&&+\frac{1}{2}\left(\lambda+\beta(\mu-\mu_0)\right)(\rho-\rho_0)^2.
\label{eq10:truncationU}
\end{eqnarray}
In the symmetric phase we have $\rho_0=0$, while in the phase with spontaneous symmetry breaking, we have $m^2=0$.
In summary, the flow of $\Gamma_k$ for fixed $\mu=\mu_0$ is described by four running renormalized couplings $\rho_0$, $\lambda$, $S$ and $V$. In addition, we need the anomalous dimension $\eta=-k\,\partial_k \text{ln}\bar{A}$. A computation of $n$ requires a flow equation of $n_k$, which involves the couplings linear in $\mu-\mu_0$, namely $\alpha$ and $\beta$. The pressure is calculated by following the $k$-dependence of the height of the minimum $p_k=-U(\rho_0,\mu_0)$. All couplings $\rho_0$, $\lambda$, $S$, $V$, $\bar{A}$, $n_k$, $p_k$, $\alpha$, $\beta$ depend on $k$ and $T$. The physical renormalized couplings obtain for $k\rightarrow 0$. They specify the thermodynamic potential $U(\rho_0,\mu_0)$ as well as suitable derivatives of the potential and the correlation function. 

The "initial values" at the scale $k=\Lambda$ are determined by the requirement
\begin{equation}
\Gamma_\Lambda[\varphi]=S[\varphi],
\end{equation}
using the microscopic action $S[\varphi]$ in Eq. \eqref{microscopicaction}. This implies the initial values
\begin{eqnarray}
\nonumber
&\rho_{0,\Lambda}=n_\Lambda=\theta(\mu_0)\mu_0/\lambda_\Lambda,\quad m_\Lambda^2=-\theta(-\mu_0)\mu_0,&\\
&\bar{A}_\Lambda=1,\quad \alpha_\Lambda=-1, \quad \beta_\Lambda=0.&
\end{eqnarray}
We remain with the free microscopic couplings $\lambda_\Lambda$, $S_\Lambda=\bar{S}_\Lambda$, $V_\Lambda=\bar{V}_\Lambda$. The coupling $\lambda_\Lambda$ will be replaced by the scattering length $a$ in section \ref{sec:Repulsiveinteractingbosons}. We further choose units for $\tau$ where $S_\Lambda=1.$ Then our second free coupling is
\begin{equation}
\tilde{v}=\frac{V_\Lambda\Lambda^2}{S_\Lambda^2}=V_\Lambda \Lambda^2.
\end{equation}
In consequence, besides the thermodynamic variables $T$ and $\mu_0$ our model is characterized by two free parameters, $a$ and $\tilde{v}$. Often, we will concentrate on "standard" non-relativistic bosons with a linear $\tau$ derivative, such that $\tilde{v}=0$. The scattering length $a$ remains then the only free parameter. In the vacuum, where $T=n=0$, this sets the relevant unit of length.

It is convenient to work with real fields $\varphi_{1,2}(x)$, $\varphi(x)=\frac{1}{\sqrt{2}}(\varphi_1(x)+i\varphi_2(x))$, with Fourier components \begin{equation}
\varphi_j(\tau,\vec{x})=\int_q e^{iqx} \varphi_j(q)=\int_{q_0}\int_{\vec{q}} e^{i(q_0\tau+\vec{q}\vec{x})}\varphi_j(q_0,\vec{q}).
\end{equation}
The inverse propagator for the fields $\bar{\varphi}$ becomes a $2\times2$ matrix in the space of $\bar{\varphi}_1$ and $\bar{\varphi}_2$, given by the second functional derivative of $\Gamma_k$. For a real constant background field $\bar{\varphi}_1(x)=\sqrt{2\bar{\rho}}$, $\bar{\varphi}_2(x)=0$ the latter becomes diagonal in momentum space
\begin{equation}
\Gamma_k^{(2)}(q,q^\prime)=G_k^{-1}(q)\delta(q-q^\prime).
\end{equation}
For our truncation one has at $\mu=\mu_0$
\begin{equation}
G^{-1}=\bar{A}\begin{pmatrix} \vec{q}^2+V q_0^2+U^\prime+2\rho U^{\prime\prime} & , & -S q_0 \\ S q_0 & , & \vec{q}^2+V q_0^2+U^\prime \end{pmatrix}.
\label{eqprop2}
\end{equation}
Here, primes denote derivatives with respect to $\rho$ (not $\bar{\rho}$). In the phase with spontaneous symmetry breaking, the infrared cutoff in Eq.\ \eqref{eq7:DeltaSkbosons} adds to the diagonal term in Eq.\ \eqref{eqprop2} a piece
\begin{equation}
\bar A\,  r_k(\vec q) = \bar{A}(k^2-\vec{q}^2)\theta(k^2-\vec{q}^2).
\end{equation}
This effectively replaces $\vec{q}^2\rightarrow k^2$ in Eq.\ \eqref{eq4:Wettericheqn} whenever $\vec{q}^2<k^2$, thus providing for an efficient infrared regularization.

\subsubsection{Flow equations for the effective potential}

\begin{figure}
\centering
\includegraphics[width=0.5\textwidth]{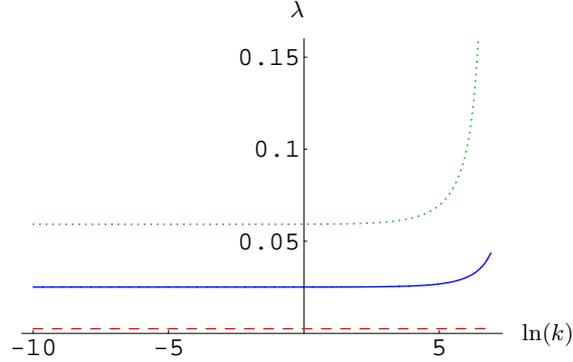}
\caption{Flow of the interaction strength $\lambda$ with the scale parameter $k$ for different start values, corresponding to $\lambda_\Lambda=10^3$ (dotted), $\lambda_\Lambda=0.044$ (solid) and $\lambda_\Lambda=0.0026$ (dashed). The first case is plotted for zero density ($n=0$) only, while the last two cases are plotted also for unit density ($n=1$). The curves $n=0$ and $n=1$ are identical within the plot resolution. The solid and the dashed curve correspond to $a=10^{-3}$ and $a=10^{-4}$, respectively.}
\label{figflowoflambdad3}
\end{figure}

\begin{figure}
\centering
\includegraphics[width=0.5\textwidth]{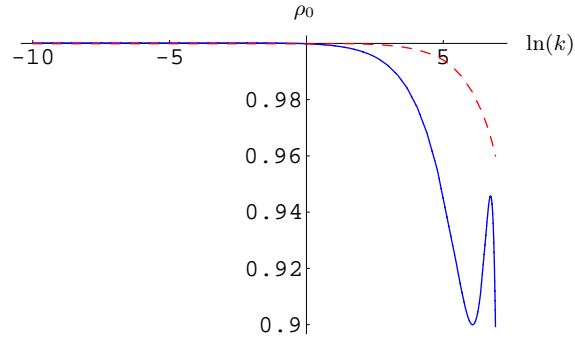}
\caption{Flow of the minimum of the effective potential for $n=1$. The parameters for the solid and the dashed curves are the same as in figure \ref{figflowoflambdad3}.}
\label{figflowofrhod3}
\end{figure}

We project the flow equation of the effective average action onto equations for the coupling constants by using appropriate background fields and taking functional derivatives. The flow equation for the effective potential obtains by using a space- and time-independent background field in Eq.\ \eqref{eq4:Wettericheqn}, with $t=\text{ln}(k/\Lambda)$
\begin{equation}
\partial_t U\big{|}_{\bar{\rho}}=k\,\partial_k U\big{|}_{\bar{\rho}}=\zeta=\frac{1}{2}\int_q \text{tr} \,G_k \,\partial_t(\bar{A}\,r_k).
\label{eqFlowpotentialMatrix}
\end{equation}
The propagator $G_k$ is here determined from
\begin{equation}
G_k^{-1}=G^{-1}+\bar{A}\,\begin{pmatrix} r_k & 0 \\ 0 & r_k \end{pmatrix}=\bar{A}\,\begin{pmatrix} \tilde{P}_{11}, & \tilde{P}_{12} \\ \tilde{P}_{21}, & \tilde{P}_{22} \end{pmatrix},
\end{equation}
with
\begin{eqnarray}
\nonumber
\tilde{P}_{11}&=&k^2+V q_0^2+U^\prime+2\rho U^{\prime\prime}+2V(\mu-\mu_0)\vec{q}^2,\\
\nonumber
\tilde{P}_{21}&=&-\tilde{P}_{12}=S q_0+2V(\mu-\mu_0)q_0,\\
\tilde{P}_{22}&=&k^2+V q_0^2+U^\prime+2V(\mu-\mu_0)\vec{q}^2.
\end{eqnarray}
Again primes denote a differentiation with respect to $\rho$. We switch to renormalized fields by making a change of variables in the differential equation \eqref{eqFlowpotentialMatrix}
\begin{equation}
\partial_t U\big{|}_\rho=\zeta+\eta\rho U^\prime.
\label{eqFlowpotentialrenorm}
\end{equation}
We can now derive the flow equations for the couplings $\rho_0(k)$ and $\lambda(k)$ by appropriate differentiation of \eqref{eqFlowpotentialrenorm} with respect to $\rho$. The flow equation for $U$ is given more explicitly in appendix \ref{sec:FlowoftheeffectivepotentialforBosegas}. Differentiation with respect to $\mu$ yields the flow of $n_k$, $\alpha$, $\beta$. We use in detail
\begin{eqnarray}
\nonumber
\frac{d}{dt}\lambda &=& \frac{d}{dt}(\partial_\rho^2U)(\rho_0,\mu_0)\\
\nonumber
&=& (\partial_\rho^2\partial_tU)(\rho_0,\mu_0)+(\partial_\rho^3U)(\rho_0,\mu_0) \,\partial_t\rho_0,\\
&=& \partial_\rho^2\zeta \big{|}_{\rho_0,\mu_0}+2\eta \lambda,
\end{eqnarray}
where we recall, that $\partial_\rho^3 U=0$ in our truncation. To determine the flow equation of $\rho_0$, we use the condition that $U^\prime(\rho_0)=0$ for all $k$, and therefore
\begin{eqnarray}
\nonumber
&\frac{d}{dt}(\partial_\rho U)(\rho_0,\mu_0)=0,&\\
\nonumber
&(\partial_\rho\partial_tU)(\rho_0,\mu_0)+(\partial_\rho^2 U)(\rho_0,\mu_0)\,\partial_t\rho_0=0,&\\
&\partial_t\rho_0=-\frac{1}{\lambda}(\partial_\rho\partial_t U)(\rho_0,\mu_0)=-\eta\rho_0-\frac{1}{\lambda}\partial_\rho \zeta\big{|}_{\rho_0,\mu_0}.&
\end{eqnarray}
We show the flow of $\lambda$ and $\rho_0$ in figs. \ref{figflowoflambdad3}, \ref{figflowofrhod3} for $n=1$, $T=0$ and different values of $\lambda_\Lambda$ (with $\tilde{v}=0$). The change in $\rho_0$ is rather modest. This will be different for nonzero temperature.

The flow of $n_k$ is given by
\begin{eqnarray}
\nonumber
\frac{d}{dt}n_k&=&\frac{d}{dt}(-\partial_\mu U)(\rho_0,\mu_0)\\
\nonumber
&=& -(\partial_\mu\partial_t U)(\rho_0,\mu_0)-(\partial_\rho\partial_\mu U)(\rho_0,\mu_0)\,\partial_t\rho_0\\
&=& -\partial_\mu\zeta\big{|}_{\rho_0,\mu_0}-\alpha \,\partial_t\rho_0,
\label{eqFlowprescriptionnk}
\end{eqnarray}
and similar for the flow of $\alpha$ and $\beta$,
\begin{eqnarray}
\nonumber
\frac{d}{dt}\alpha&=&\frac{d}{dt}(\partial_\rho\partial_\mu U)(\rho_0,\mu_0)\\
\nonumber
&=& (\partial_\rho\partial_\mu\partial_t U)(\rho_0,\mu_0)+(\partial_\rho^2\partial_\mu U)(\rho_0,\mu_0)\,\partial_t\rho_0\\
\nonumber
&=& \partial_\rho\partial_\mu\zeta\big{|}_{\rho_0,\mu_0}+\eta \alpha+\beta (\eta \rho_0+\partial_t\rho_0),\\
\nonumber
\frac{d}{dt}\beta &=&\frac{d}{dt}(\partial_\rho^2\partial_\mu U)(\rho_0,\mu_0)\\
\nonumber
&=& (\partial_\rho^2\partial_\mu\partial_t U)(\rho_0,\mu_0)+(\partial_\rho^3\partial_\mu U)(\rho_0,\mu_0)\partial_t \rho_0,\\
&=& \partial_\rho^2\partial_\mu \zeta\big{|}_{\rho_0,\mu_0}+2\eta \beta,
\label{eqflowalphabeta}
\end{eqnarray}
where the last equation holds since $\partial_\rho^3\partial_\mu U=0$ in our truncation.

\subsubsection{Kinetic coefficients}
\begin{figure}
\centering
\includegraphics[width=0.5\textwidth]{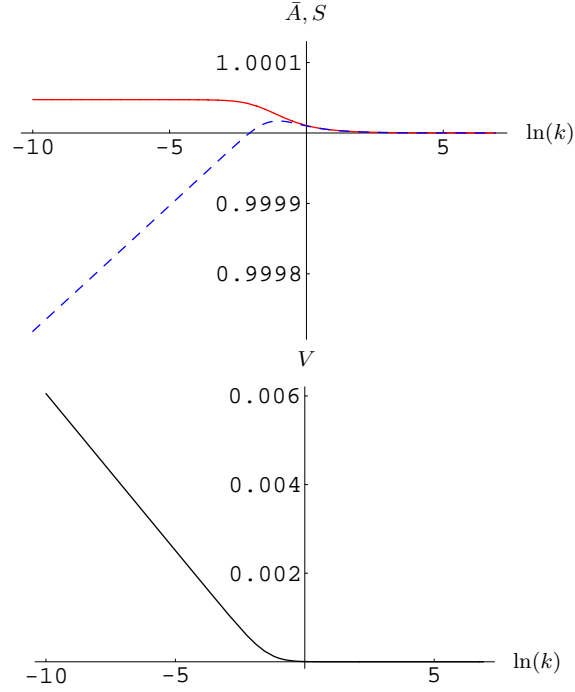}
\caption{Flow of the kinetic coefficients $\bar{A}$ (solid), $S$ (dashed) and $V$ for a scattering length $a=10^{-3}$, temperature $T=0$ and density $n=1$.}
\label{figFlowKineticd3}
\end{figure}

For a derivation of $\eta=-(\partial_t  \bar{A})/\bar{A}$ and the flow equations for $S$ and $V$, we have to evaluate the flow equation \eqref{eq4:Wettericheqn} for a background field depending on $q_0$ and $\vec{q}$. We use an analytic continuation $q_0=i\omega$ and obtain the flow equation for $S$ from
\begin{equation}
\partial_t(S\bar{A})=-i\Omega^{-1}\frac{\partial}{\partial \omega}\frac{\delta}{\delta \bar{\varphi}_2(-\omega,0)}\frac{\delta}{\delta \bar{\varphi}_1(\omega,0)}\partial_t \Gamma_k\bigg{|}_{\omega=0},
\label{eqflowprescriptionS}
\end{equation}
with four-volume $\Omega=\frac{1}{T}\int_{\vec{x}}$. The projection prescription for $V$ is
\begin{equation}
\partial_t (V\bar{A})=-\Omega^{-1}\frac{\partial}{\partial \omega^2}\frac{\delta}{\delta \bar{\varphi}_2(-\omega,0)}\frac{\delta}{\delta \bar{\varphi}_2(\omega,0)}\partial_t \Gamma_k\bigg{|}_{\omega=0},
\label{eqflowprescriptionV}
\end{equation}
and similar for $\bar{A}$
\begin{equation}
\partial_t\bar{A}=\Omega^{-1}\frac{\partial}{\partial \vec{p}^2}\frac{\delta}{\delta \bar{\varphi}_2(0,-\vec{p})}\frac{\delta}{\delta \bar{\varphi}_2(0,\vec{p})}\partial_t \Gamma_k\bigg{|}_{\vec{p}^2=0}.
\label{eqflowprescriptionA}
\end{equation}
After the functional differentiation, we evaluate the expressions in Eqs. \eqref{eqflowprescriptionS}, \eqref{eqflowprescriptionV}, and \eqref{eqflowprescriptionA} at homogeneous background fields. These calculations are a little intricate, but standard and straightforward in principle. A more detailed description of the calculation can be found in \cite{Wetterich:2007ba}. More explicit flow equations are given below. Eventually, it is always possible to perform the Matsubara sums and also the spatial momentum integration analytically. In figure \ref{figFlowKineticd3} we show the flow of $\bar{A}$, $S$ and $V$ at zero temperature and for density $n=1$. 
The kinetic coefficient $\bar{A}$ starts on the large scale with $\bar{A}=1$, increases a little around $k=n^{1/3}$ and saturates to a constant. In contrast, the coefficient $S$ starts to decrease after a short period of increase with $\bar{A}$. For very tiny scales $k$, $S$ would finally go to zero. The frequency dependence of the propagator is then governed by the quadratic frequency coefficient $V$. In three spatial dimensions, however, this decrease of $S$ is so slow that it is not relevant on the length scales of experiments. This is one of the reasons why Bogoliubov theory, which neglects the appearance of $V$, describes experiments with ultracold bosonic quantum gases in three dimensions with so much success. The coefficient $V$ is always generated in the phase with spontaneous symmetry breaking \cite{Wetterich:2007ba}. Its $k$-dependence is also shown in figure \ref{figFlowKineticd3}.

We show now our flow equation obtained for the kinetic coefficients $S$, $\bar{A}$ and $V$. We neglect all contributions from momentum dependent vertices. In other words, we use $\rho$-independent constants $S=Z_1+\rho_0Z_1^\prime$, $\bar{A}=\bar{Z}_2$ and $V=V_1$. In our truncation with $Z_2^\prime=V_2=V_3=0$, and with the cutoff \eqref{eq7:DeltaSkbosons}, we can perform all momentum integrations analytically, leading us to
\begin{eqnarray}
\nonumber
\partial_t V&=&\eta V-\left(1-\frac{\eta}{d+2}\right)\,T\,\sum_n \,32\,{v_d}\,k^{2 + d}{\lambda }^2{{\rho }_0}\\
\nonumber
&&\hspace{-8mm} \times{\bigg [} k^2\left( S^2 + k^2V \right)+\left( S^2 + 2k^2V \right) \lambda {{\rho }_0}-2V\left( S^2 + k^2V + V\lambda {{\rho }_0} \right) {{{\omega }_n}}^2 -3V^3{{{\omega }_n}}^4 {\bigg ]}\\
\nonumber
&&\hspace{-8mm} \times {\bigg [}d {\left( k^4 + 2k^2\lambda {{\rho }_0}+\left( S^2 + 2k^2V + 2V\lambda {{\rho }_0} \right) 
         {{{\omega }_n}}^2 + V^2{{{\omega }_n}}^4 \right) {\bigg ]}^{-3}},\\
         \nonumber
\partial_t S&=&\eta S-\left(1-\frac{\eta}{d+2}\right)\,T\,\sum_n\, \,32\,{v_d}\,k^{2 + d}S{\lambda }^2{{\rho }_0}\\
\nonumber
&&\hspace{-8mm} \times {\bigg [} k^4 - 2\lambda {{\rho }_0}\left( k^2 + \lambda {{\rho }_0} \right)+S^2{{{\omega }_n}}^2 + 2V\left( k^2 - \lambda {{\rho }_0} \right) {{{\omega }_n}}^2 + V^2{{{\omega }_n}}^4 {\bigg ]}\\
\nonumber
&&\hspace{-8mm} \times {\bigg [} d \left( k^4 + 2k^2\lambda {{\rho }_0}+\left( S^2 + 2k^2V + 2V\lambda {{\rho }_0} \right){{{\omega }_n}}^2 + V^2{{{\omega }_n}}^4 \right){\bigg ]}^{-3},\\
\nonumber
\eta &=& -\frac{\partial_t \bar{A}}{\bar{A}} = T\,\sum_n\, 16\,{v_d}\,k^{2 + d}{\lambda }^2{{\rho }_0}\\
&& \times
{\bigg [} d{\left( k^4 + 2k^2\lambda {{\rho }_0} + \left( S^2 + 2k^2V + 2V\lambda {{\rho }_0} \right) {{{\omega }_n}}^2 + V^2{{{\omega }_n}}^4 \right) }{\bigg ]}^{-2}.
\label{eqflowofVSEta}
\end{eqnarray}
Here, $d$ is the number of spatial dimensions, $v_d=(2^{d+1}\pi^{d/2}\Gamma(d/2))^{-1}$ and $\omega_n=2\pi T n$. The Matsubara sums over $n$ can be performed analytically by using Eq.\ \eqref{eqMatsubara} and derivatives thereof.
In the limit $T\rightarrow0$, the Matsubara frequencies are continuous $\omega_n\rightarrow q_0$ and the sum becomes an integral $T\sum_n\rightarrow \frac{1}{2\pi}\int_{-\infty}^\infty d q_0$.

\section{BCS-BEC Crossover}

Let us now come to the approximation used to investigate the BCS-BEC crossover model. Our truncation of the flowing action reads
\begin{eqnarray}
\nonumber
\Gamma_k[\chi] & =  & \int_0^{1/T}d\tau \int d^3x {\bigg \{} \psi^\dagger (\partial_\tau -\Delta -\mu) \psi\\
\nonumber
& + & \bar{\varphi}^*(\bar Z_\varphi \partial_\tau-\frac{1}{2}\bar A_\varphi \Delta)\bar\varphi + \bar U_k(\bar \rho,\mu) \\
& - & \bar h (\bar \varphi^* \psi_1\psi_2 + \bar \varphi \psi_2^\ast \psi_1^\ast) {\bigg \} }.
\label{eq:baretruncation}
\end{eqnarray}
Here the effective potential $\bar U(\bar \rho,\mu)$ contains no derivatives and is a function of $\bar{\rho}=\bar{\varphi}^*\bar{\varphi}$ and $\mu$. Besides the couplings parameterizing $\bar U$ (see below) our truncation contains three further $k$-dependent (``running'') couplings $\bar A_\varphi$, $\bar Z_\varphi$ and $\bar h$. The truncation in Eq.\ \eqref{eq:baretruncation} can be motivated by a systematic derivative expansion and analysis of symmetry constraints (Ward identities), see \cite{Diehl:2007th, Diehl:2008} and section \ref{sec:Derivativeexpansionandwardidentities}. The truncation in Eq.\ \eqref{eq:baretruncation} does not yet incorporate the effects of particle-hole fluctuations and we will come back to this issue in Sect.\ \ref{sec:Particle-holefluctuationsandtheBCS-BECCrossover}. In terms of renormalized fields $\varphi=\bar A_\varphi^{1/2}\bar\varphi$, $\rho=\bar A_\varphi \bar \rho$, renormalized couplings $Z_\varphi=\bar Z_\varphi / \bar A_\varphi$, $h=\bar h/\sqrt{\bar A_\varphi}$ and effective potential $U(\rho,\mu)=\bar U(\bar \rho, \mu)$, Eq.\ \eqref{eq:baretruncation} reads
\begin{eqnarray}
\nonumber
\Gamma_k[\chi] & = & \int_0^{1/T}d\tau \int d^3x {\bigg \{}  \psi^\dagger (\partial_\tau -\Delta -\mu) \psi\\
\nonumber
& + & \varphi^*( Z_\varphi \partial_\tau-\frac{1}{2} \Delta)\varphi +  U_k(\rho,\mu) \\
& - &  h \,(\varphi^* \psi_1\psi_2 + \varphi \psi_2^\ast \psi_1^\ast) {\bigg \} }.
\label{eq:truncation}
\end{eqnarray}
For the effective potential, we use an expansion around the $k$-dependent location of the minimum $\rho_0(k)$ and the $k$-independent value of the chemical potential $\mu_0$ that corresponds to the physical particle number density $n$. We determine $\rho_0(k)$ and $\mu_0$ by the requirements 
\begin{eqnarray}
\nonumber
(\partial_\rho U)(\rho_0(k),\mu_0)=0 &&\text{for all }k\\
-(\partial_\mu U) (\rho_0,\mu_0)=n && \text{at }k=0.  
\end{eqnarray}
More explicitly, we employ a truncation for $U(\rho,\mu)$ of the form
\begin{eqnarray}
\nonumber
U(\rho,\mu) &=& U(\rho_0,\mu_0)-n_k (\mu-\mu_0)\\
&& +(m^2+\alpha(\mu-\mu_0)) (\rho-\rho_0)\nonumber\\
&& +\frac{1}{2}\lambda (\rho-\rho_0)^2.
\end{eqnarray}
In the symmetric or normal gas phase, we have $\rho_0=0$, while in the regime with spontaneous symmetry breaking, we have $m^2=0$. The atom density $n=-\partial U/\partial \mu$ corresponds to $n_k$ in the limit $k\to 0$.

In total, we have the running couplings $m^2(k)$, $\lambda(k)$, $\alpha(k)$, $n_k$, $Z_\varphi(k)$ and $h(k)$. (In the phase with spontaneous symmetry breaking $m^2$ is replaced by $\rho_0$.)  In addition, we need the anomalous dimension $\eta=-k\partial_k \text{ln} \bar A_\varphi$. We project the flow of the average action $\Gamma_k$  on the flow of these couplings by taking appropriate (functional) derivatives on both sides of Eq.\ \eqref{eq4:Wettericheqn}.  We thereby obtain a set of coupled nonlinear differential equations which can be solved numerically. 

At the microscopic scale $k=\Lambda$ the initial values of our couplings are determined from $\Gamma_\Lambda=S$ with the microscopic action in Eq.\ \eqref{eqMicroscopicAction}. This gives $m^2(\Lambda)=\nu_\Lambda-2\mu_0$, $\rho_0=0$, $\lambda(\Lambda)=0$, $Z_\varphi(\Lambda)=1$, $h(\Lambda)=h_\Lambda$, $\alpha(\Lambda)=-2$ and $n_\Lambda=3\pi^2 \mu_0 \theta(\mu_0)$. The initial values $\nu_\Lambda$ and $h_\Lambda$ can be connected to the two particle scattering in vacuum close to a Feshbach resonance. For this purpose one follows the flow of $m^2(k)$ and $h(k)$ in vacuum, i.e. $\mu_0=T=n=0$ and extracts the renormalized parameters $m^2=m^2(k=0)$, $h=h(k=0)$. The scattering length $a$ obeys $a=-h^2/(8\pi m^2)$ and the renormalized Yukawa coupling $h$ determines the width of the resonance. Broad Feshbach resonances with large $h$ become independent of $h$.

\subsubsection{Flow of the effective potential}

For our choice of $R_k$ in Eq.\ \eqref{eq7:DeltaSkfermions} and with the approximation \eqref{eq:truncation}, we can perform
the momentum integration and the Matsubara sums explicitly 
\begin{eqnarray}\label{2J}
\nonumber
k\partial_kU_k &=& \eta_{A_\varphi}\,\rho\, U^\prime_k+\frac{\sqrt{2}k^5}{3\pi^2 Z_\varphi} \left(1-2\eta_{A_\varphi}/5\right) s_B^{(0)}\\
&& -\frac{k^5}{3\pi^2} \,l(\tilde \mu) \,s_F^{(0)},\\
\nonumber
l(\tilde \mu) &=& \left(\theta(\tilde\mu+1)(\tilde \mu+1)^{3/2}-\theta(\tilde \mu-1)(\tilde \mu-1)^{3/2}\right).
\end{eqnarray}
Here we use the dimensionless chemical potential $\tilde \mu=\mu/k^2$ and the anomalous dimension
$\eta_{\bar A_\varphi}=-\partial \text{ln} \bar A_\varphi /\partial \text{ln}
k$.  The \emph{threshold functions} $s_B$ and $s_F$ depend on
$w_1=U_k'/k^2$, $w_2=(U_k'+2\rho U_k'')/k^2$,
$w_3=h^2_\varphi\rho/ k^4$, as well as on and the dimensionless
temperature $\tilde T=T/k^2$. They describe the decoupling of modes if
the effective masses $w_j$ get large. They are normalized to unity for
vanishing arguments and $\tilde T\to 0$ and read
\begin{eqnarray}\label{12K}
\nonumber
s_{B}^{(0)}&=&\left[\sqrt{\frac{1+w_1}{1+w_2}}+\sqrt{\frac{1+w_2}{1+w_1}}\right]\\
\nonumber
&&\times \left[\frac{1}{2}+N_B(\sqrt{1+w_1}\sqrt{1+w_2}/S_\varphi)\right]\\
s_{\text{F}}^{(0)}&=&\frac{2}{\sqrt{1+w_3}}\left[\frac{1}{2}-N_F(\sqrt{1+w_3})\right].
\end{eqnarray}
(For $s_B^{(0)}$, only all its $\rho$ derivatives vanish for $w_1\sim w_2\to
\infty$.  
The remaining constant part is a shortcoming of the particular
choice of the cutoff acting only on spacelike momenta.)

In Eq.\ \eqref{12K} the temperature dependence arises through the Bose and Fermi functions
\begin{equation}
N_{B/F}(\epsilon)=\frac{1}{e^{\epsilon/\tilde T}\mp 1}.
\end{equation}
For $\tilde T \to 0$ the ```thermal parts'' $\sim N_{B,F}$ vanish, whereas for
large $\tilde T$ one has 
\begin{equation}
s_F^{(0)}\to \tilde T^{-1}, \quad s_B^{(0)}\to 2 \tilde T Z_\varphi (1+w_1)^{-1}(1+w_2)^{-1}.
\label{eq:Floweqhightemperature}
\end{equation}
In this high-temperature limit the fermionic fluctuations become unimportant. For the boson fluctuations only the $n=0$ Matsubara frequency contributes substantially. Inserting Eq.\ \eqref{eq:Floweqhightemperature} into Eq.\ \eqref{2J} yields the well known flow equations for the classical three-dimensional scalar theory with U(1) symmetry \cite{Wetterich1993b, Berges2000ew}. In appendix \ref{sec:FlowoftheeffectivepotentialforBCSBECcrossover} we derive the flow equations \eqref{2J} and discuss the threshold functions $s_B$ and $s_F$ more explicitly.

We recall that for $k\to 0$ $p_{k}=-U_k(\rho_0,\mu_0)$ is the pressure. The flow equations
for $p_k,m^2$ or $\rho_0(k)$, and $\lambda$ are given by
\begin{eqnarray}\label{2M}
\partial_k p_k &=& -\partial_kU_k{\big |}_{\rho_0}-U'_k{\big
  |}_{\rho_0}\partial_k\rho_0,  \nonumber\\ 
\partial_k m^2 &=& \partial_kU'_k{\big |}_{\rho=0}
\quad\quad\quad\quad\quad\,\,\,\, \text{for} \quad \rho_0=0,  \nonumber\\
\partial_k \rho_0 &=& -\big(U''_k{\big
  |}_{\rho_0}\big)^{-1}\partial_kU'_k{\big |}_{\rho_0}\quad \text{for} \quad
\rho_0>0, \nonumber\\
\partial_k \lambda &=& \partial_kU''_k{\big |}_{\rho_0}. 
\end{eqnarray}
Taking a derivative of Eq.\ \eqref{2J} with respect to $\rho$ one obtains for $\tilde T=0$
\begin{eqnarray}
\nonumber
k\partial_k U_k^\prime \!&=&\! \eta_{A_\varphi}(U_k^\prime-\rho U_k^{\prime\prime})+\frac{\sqrt{2}k}{3\pi^2 Z_\varphi}\left(1-\frac{2}{d+2}\eta_{A_\varphi}\right)\\
\nonumber
&&\!\!\times\! \left[ 2\rho (U_k^{\prime\prime})^2\left(s_{\text{B,Q}}^{(1,0)}
    +3 s_{\text{B,Q}}^{(0,1)}\right)+4\rho^2 U_k^{\prime\prime} U_k^{(3)}s_{\text{B,Q}}^{(0,1)}\right]\\
&&+\frac{k}{3\pi^2}h_\varphi^2\,l(\tilde \mu)\, s_{\text{F,Q}}^{(1)}.
\label{eq:Flowu1}
\end{eqnarray}
The threshold functions $s_{\text{B,Q}}^{(0,1)}$, $s_{\text{B,Q}}^{(1,0)}$, and
$s_{\text{F,Q}}^{(1)}$ are defined in App. \ref{sec:FlowoftheeffectivepotentialforBCSBECcrossover} and
describe again the decoupling of the heavy modes. They can be obtained
from $\rho$ derivatives of $s^{(0)}_{\text{B}}$ and $s^{(0)}_{\text{F}}$. Setting $\rho=0$
and $\tilde T\to0$, we can immediately infer from Eq.\ \eqref{eq:Flowu1} the
running of $m^2$ in the symmetric regime.
\begin{equation}
k\partial_k m^2=k\partial_k U_k^\prime =\eta_{A_\varphi}m^2 
+\frac{k}{3\pi^2} h^2\,l(\tilde \mu) \,s^{(1)}_{\text{F,Q}}(w_3=0).
\label{eq:flowmphi}
\end{equation}
One can see from Eq.\ \eqref{eq:flowmphi} that fermionic fluctuations lead to a
strong renormalization of the bosonic ``mass term'' $m^2$. In the course
of the renormalization group flow from large scale parameters $k$
(ultraviolet) to small $k$ (infrared) the parameter $m^2$
decreases strongly. When it becomes zero at some scale $k>0$ the flow
enters the regime where the minimum of the effective potential $U_k$ is at
some nonzero value $\rho_0$. This is directly related to spontaneous
breaking of the $U(1)$ symmetry and to local order. If $\rho_0\neq0$ persists
for $k\to0$ this indicates superfluidity.

For given $\bar A_\varphi,Z_\varphi,h_\varphi$, Eq.\ \eqref{2J} is a nonlinear
differential equation for $U_k$, which depends on two variables $k$
and $\rho$. It has to be supplemented by flow equations for
$\bar A_\varphi,Z_\varphi,h$. The flow equations for the wave
function renormalization $Z_\varphi$ and the gradient coefficient
$\bar A_\varphi$ cannot be extracted from the effective potential, but are
obtained from the following projection prescriptions,
\begin{eqnarray}
\nonumber
\partial_t \bar Z_\varphi &=& -\partial_t 
\frac{\partial}{\partial q_0} (\bar P_\varphi)_{12}(q_0,0){\big |}_{q_0=0} ,\\
\partial_t \bar A_\varphi &=&  \partial_t 2 \frac{\partial}{
  \partial \vec q\,^2} (\bar P_\varphi)_{22}(0,\vec q){\big |}_{\vec q=0},
\end{eqnarray}  
where the momentum dependent part of the propagator is defined by
\begin{equation}
\frac{\delta^2 \Gamma_k}{\delta\bar\varphi_a(q)
\delta\bar\varphi_b(q^\prime)}\Big|_{\varphi_1 =
\sqrt{2\rho_0}, \varphi_2=0} = (\bar P_\varphi)_{ab}(q)\delta(q+q^\prime).  
\end{equation}
The computation of the flow of the gradient coefficient is rather involved,
since the loop depends on terms of different type, $\sim (\vec q \cdot\vec
p)^2,~ \vec q\,^2$, where $\vec p$ is the loop momentum.  An outline of the calculation and explicit expressions can be found in \cite{Diehl:2008}.

\section{BCS-Trion-BEC Transition}

Now we turn to the truncation used to investigate the model with three fermion species in Eq.\ \eqref{eq8:microscopicactiontrionmodel}. For this model the focus will be on the few-body problem where the approximation scheme can be simpler in some respects. We use the following truncation for the average action
\begin{eqnarray}
\nonumber
\Gamma_k&=&\int_x {\bigg \{} \psi^\dagger\left(\partial_\tau-\Delta-\mu\right)\psi+\varphi^\dagger\left(\partial_\tau-\Delta/2+m_\varphi^2\right)\varphi\\
\nonumber
&&+h\,\epsilon_{ijk}\,\left(\varphi_i^*\psi_j\psi_k-\varphi_i\psi_j^*\psi_k^*\right)/2+\lambda_\varphi\left(\varphi^\dagger \varphi\right)^2/2\\
\nonumber
&&+\chi^*\left(\partial_\tau-\Delta/3+m_\chi^2\right)\chi+g\left(\varphi_i^*\psi_i^*\chi-\varphi_i\psi_i\chi^*\right)\\
&&+\lambda_{\varphi\psi}\left(\varphi_i^*\psi_i^*\varphi_j\psi_j\right){\bigg \}}.
\label{eq:triontruncation}
\end{eqnarray}
Here we use as always natural nonrelativistic units with $\hbar=k_B=2M=1$, where $M$ is the mass of the original fermions.
The integral in Eq.\ \eqref{eq:triontruncation} goes over homogeneous space and over imaginary time as appropriate for the Matsubara formalism $\int_x=\int d^3x \int_0^{1/T}d\tau$. On the level of the three-body sector, the symmetry of the problem would allow also for a term $\sim \psi^\dagger \psi\varphi^\dagger \varphi$ in Eq.\ \eqref{eq:triontruncation}. This term plays a similar role as for the case of two fermion species, where it was investigated in \cite{DKS}. The qualitative features of the three-body scattering are dominated by the term $\sim\lambda_{\varphi\psi}$ in Eq.\ \eqref{eq:triontruncation}. The quantitative influence of a term $\sim \psi^\dagger \psi \varphi^\dagger \varphi$ on the flow equations was also investigated in \cite{Moroz2008}.

At the microscopic scale $k=\Lambda$, we use the initial values of the couplings in Eq.\ \eqref{eq:triontruncation} $g=\lambda_\varphi=\lambda_{\varphi\psi}=0$ and $m^2_\chi\to\infty$. Then the fermionic field $\chi$ decouples from the other fields and is only an auxiliary field which is not propagating. However, depending on the parameters of our model we will find that $\chi$, which describes a composite bound state of three original fermions $\chi=\psi_1\psi_2\psi_3$, becomes a propagating degree of freedom in the infrared. The initial values of the boson energy gap $\nu_\varphi$ and the Yukawa coupling $h$ will determine the scattering length $a$ between fermions and the width of the resonance, see below. The pointlike limit (broad resonance) corresponds to $\nu_\varphi\to\infty$, $h^2\to\infty$ where the limits are taken such that the effective renormalized four fermion interaction remains fixed. In Eq.\ \eqref{eq:triontruncation} we use renormalized fields $\varphi=\bar A_\varphi^{1/2}(k)\, \bar{\varphi}$, $\psi=\bar A_\psi^{1/2}(k)\,\bar \psi$, $\chi = \bar A_\chi^{1/2}(k)\,\bar \chi$, with $\bar A_\varphi(\Lambda)=\bar A_\psi(\Lambda)=\bar A_\chi(\Lambda)=1$, and renormalized couplings $m_\varphi^2=\bar{m}_\varphi^2/\bar{A}_\varphi$, $h=\bar{h}/(\bar{A}_\varphi^{1/2}\bar A_\psi)$, $\lambda_\varphi=\bar \lambda_\varphi/\bar A_\varphi^2$, $m_\chi^2=\bar m_\chi^2/\bar A_\chi$, $g=\bar g /(\bar A_\chi^{1/2}\bar A_\varphi^{1/2}\bar A_\chi^{1/2})$, and $\lambda_{\varphi\psi}=\bar \lambda_{\varphi\psi}/(\bar A_\varphi \bar A_\psi)$.

To derive the flow equations for the couplings in Eq.\ \eqref{eq:triontruncation} we have to specify an infrared regulator function $R_k$. Here we use the particularly simple function
\begin{equation}
R_{k}=r(k^2-\vec p^2)\theta(k^2-\vec p^2),
\label{eq:cutofftrions}
\end{equation}
where $r=1$ for the fermions $\psi$, $r=1/2$ for the bosons $\varphi$, and $r=1/3$ for the composite fermionic field $\chi$. This choice has the advantage that we can derive analytic expressions for the flow equations and that it is optimized in the sense of \cite{Litim2000b}.

\chapter{Few-body physics}
\label{ch:Few-bodyphysics}
In this chapter we investigate the flow equation applied to nonrelativistic systems in the limit of vanishing temperature and density (``vacuum limit''). The effective action as the generating functional of the one-particle irreducible correlation functions contains then directly the information about the few-body physics such as as scattering properties, binding energies etc. Usually, the flow equations simplify substantially in the vacuum limit. Whenever they can be solved exactly, this leads to the same results as an quantum mechanical treatment. This is of course expected, since the flow equation itself is exact. Since the flow equation and the quantum mechanical treatment are equivalent, one might conjecture that many (all?) problems that can be solved in the quantum mechanical formalism find an equivalent solution in the flow equation context. (It is another question whether the flow equation solution is particular simple or gives any new insights.) This is especially interesting since the flow equation is more general. As briefly discussed at the beginning of chapter \ref{ch:TheWetterichequation}, one might consider functional renormalization as a formulation of quantum field theory. In the flow equation method both quantum mechanics and quantum field theory find a unified framework. 

An important feature of nonrelativistic field theory in the vacuum limit is a hierarchy between $n$-point correlation functions. This implies that the $n$-particle problem can be solved independent of the $n+1$-particle problem. More formally, the flow equation of the $n$-point function depends only on the correlation functions of lower order and the $n$-point function itself but not on correlation functions of higher order. To see this, let us consider a nonrelativistic field theory for the field $\varphi$. This may be a spinor containing both fermionic and bosonic degrees of freedom. We only require that all components have particle number one. In other words $\varphi$ must not contain any composite degrees of freedom such as bound states and no exchange particles such as photons or phonons. Technically, the requirement is that the microscopic model is invariant under the global $U(1)$ transformation
\begin{equation}
\varphi \to e^{i\alpha} \varphi
\end{equation}
where the charge $\alpha$ is the same for all components of $\varphi$. To fulfill the condition above it will sometimes be necessary to ``integrate out'' composite fields. 

The second premise is that the microscopic propagator for the field $\varphi$ is of the nonrelativistic form $i q_0+\vec q^2/(2M)+\nu-\mu$. Here we work with a imaginary-time (or Matsubara) frequency $q_0$. The mass $M$ and the gap parameter $\nu$ might be different for the different components of $\varphi$. One can then proof the following theorem: {\itshape The flow equation for the $n$-point function $G_n$ is independent of correlation functions $G_m$ with order $m>n$. Here we use a notation where $G_n = \Gamma_k^{(n)}$ for $n>2$ and $G_2=(\Gamma_k^{(2)}+R_k)^{-1}$ is the regularized propagator.} The proof will be given in appendix \ref{sec:Hierarchyofflowequationsinvacuum}. We now come to the discussion of the flow equations in the vacuum limit for the different models introduced in chapter \ref{ch:Variousphysicalsystems}.

	\section{Repulsive interacting bosons}
	\label{sec:Repulsiveinteractingbosons}
	In this section we will investigate the flow equations for the Bose gas in the limit of vanishing temperature and density. The correlation functions then directly contain the information about  few-body scattering. For a repulsive Bose gas with pointlike interaction in three or two dimensions, the main result is that the interaction strength is bounded from above. This is a vacuum screening effect and formally very similar to the triviality in the Higgs sector of the standard model of particle physics. In three dimensions the interaction strength has the canonical dimension of a length and the bound is of the order of the microscopical scale $\lambda\lesssim \Lambda^{-1}$. In two dimensions $\lambda$ is dimensionless and the running is only logarithmically. But let us first concentrate on the three-dimensional case and discuss the case of $d=2$ thereafter.

\subsubsection{Vacuum flow equations and their solution for $d=3$}

The vacuum is defined to have zero temperature $T=0$ and vanishing density $n=0$, which also implies $\rho_0=0$. The interaction strength $\lambda$ at the scale $k=0$ determines the four point vertex at zero momentum. It is directly related to the scattering length $a$ for the scattering of two particles in vacuum, which is experimentally observable. We therefore want to replace the microscopic coupling $\lambda_\Lambda$ by the renormalized coupling $a$. In our units ($2M=1$), one has the relation 
\begin{equation}
a=\frac{1}{8\pi}\lambda(k=0, T=0, n=0).
\end{equation}
The vacuum properties can be computed by taking for $T=0$ the limit $n\rightarrow0$. We may also perform an equivalent and technically simple computation in the symmetric phase by choosing $m^2(k=\Lambda)$ such, that $m^2(k\rightarrow0)=0$. This guarantees that the boson field $\varphi$ is a gap-less propagating degree of freedom. 

We first investigate the model with a linear $\tau$-derivative, $S_\Lambda=1$, $V_\Lambda=0$. Projecting the flow equation \eqref{eq4:Wettericheqn} to the truncation in Eqs. \eqref{eqSimpleTruncation}, \eqref{eq10:truncationU}, we find the following equations:
\begin{eqnarray}
\nonumber \partial_t m^2 & = & 0\\
\partial_t \lambda & = & \left(\frac{\lambda^2}{6}\right)\frac{{\left( k^2 - m^2 \right) }^{3/2}}{k^2\,
    {\pi }^2\,S}\,\Theta(k^2-m^2).
\label{eqflowvacuummlambda}
\end{eqnarray}
The propagator is not renormalized, $\partial_t S=\partial_tV=\partial_t \bar{A}=0$, $\eta=0$, $\partial_t\alpha=0$, and one finds $\partial_tn_k=0$. The coupling $\beta$ is running according to
\begin{equation}
\partial_t \beta = \left(\frac{1}{3}\alpha\lambda^2-\frac{1}{3}k^2\beta\lambda\right)\frac{\left(k^2 - m^2\right)^{3/2}}{k^4\,\pi^2 \,S}\Theta(k^2 - m^2).
\label{eqflowvacuumbeta}
\end{equation}
Since $\beta$ appears only in its own flow equation, it is of no further relevance in the vacuum. Also, no coupling $V$ is generated by the flow and we have therefore set $V=0$ on the r.h.s. of Eqs. \eqref{eqflowvacuummlambda} and \eqref{eqflowvacuumbeta}. 

\begin{figure}
\centering
\includegraphics[width=0.5\textwidth]{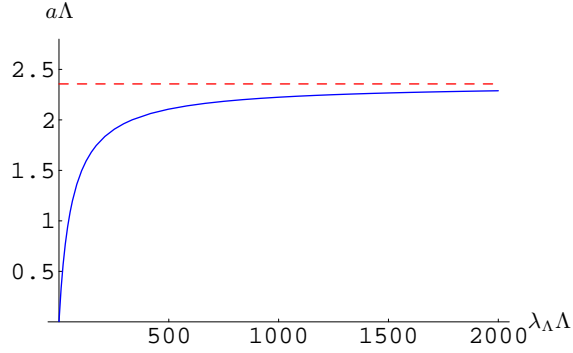}
\caption{Scatt\-er\-ing length $a$ in dependence on the microscopic interaction strength $\lambda_\Lambda$ (solid). The asymptotic maximum $a_{\text{max}}=\frac{3 \pi}{4\Lambda}$ is also shown (dashed).}
\label{figscatteringbound}
\end{figure}

Inserting in Eq. \eqref{eqflowvacuummlambda} the vacuum values $m^2=0$ and $S=1$, we find
\begin{equation}
\partial_t\lambda=\frac{k}{6\pi^2}\lambda^2.
\end{equation}
The solution 
\begin{equation}
\lambda(k)=\frac{1}{\frac{1}{\lambda_\Lambda}+\frac{1}{6\pi^2}(\Lambda-k)}
\end{equation}
tends to a constant for $k\rightarrow0$, $\lambda_0=\lambda(k=0)$. The dimensionless variable $\tilde{\lambda}=\frac{\lambda k}{S}$ goes to zero, when $k$ goes to zero. This shows the infrared freedom of the theory. For fixed ultraviolet cutoff, the scattering length
\begin{equation}
a=\frac{\lambda_0}{8\pi}=\frac{1}{\frac{8\pi}{\lambda_\Lambda}+\frac{4}{3\pi}\Lambda},
\end{equation}
as a function of the initial value $\lambda_\Lambda$, has an asymptotic maximum
\begin{equation}
a_{\text{max}}=\frac{3\pi}{4\Lambda}.
\label{eqscatteringbound}
\end{equation}
The relation between $a$ and $\lambda_\Lambda$ is shown in fig. \ref{figscatteringbound}.

As a consequence of Eq.\ \eqref{eqscatteringbound}, the nonrelativistic bosons in $d=3$ are a "trivial theory" in the sense that the bosons become noninteracting in the limit $\Lambda\rightarrow\infty$, where $a\rightarrow0$. The upper bound \eqref{eqscatteringbound} has important practical consequences. It tells us, that whenever the "macrophysical length scales" are substantially larger than the microscopic length $\Lambda^{-1}$, we deal with a weakly interacting theory. As an example, consider a boson gas with a typical inter-particle distance substantially larger than $\Lambda^{-1}$. (For atom gases $\Lambda^{-1}$may be associated with the range of the Van der Waals force.) We may set the units in terms of the particle density $n$, $n=1$. In these units $\Lambda$ is large, say $\Lambda=10^3$. This implies a very weak interaction, $a\lesssim 2.5\cdot10^{-3}$. In other words, the scattering length cannot be much larger than the microscopic length $\Lambda^{-1}$. For such systems, perturbation theory will be valid in many circumstances. We will find that the Bogoliubov theory indeed gives a reliable account of many properties. Even for an arbitrary large microphysical coupling $(\lambda_\Lambda\rightarrow\infty)$, the renormalized physical scattering length $a$ remains finite.

Let us mention, however, that the weak interaction strength does not guarantee the validity of perturbation theory in all circumstances. For example, near the critical temperature of the phase transition between the superfluid and the normal state, the running of $\lambda(k)$ will be different from the vacuum. As a consequence, the coupling will vanish proportional to the inverse correlation length $\xi^{-1}$ as $T$ approaches $T_c$, $\lambda \sim T^{-2}\xi^{-1}$. Indeed, the phase transition will be characterized by the non-perturbative critical exponents of the Wilson-Fisher fixed point. Also for lower dimensional systems, the upper bound \eqref{eqscatteringbound} for $\lambda_0$ is no longer valid - for example the running of $\lambda$ is logarithmic for $d=2$. For our models with $V_\Lambda\neq0$, the upper bound becomes dependent on $V_\Lambda$. It increases for $V_\Lambda>0$. In the limit $S_\Lambda\rightarrow0$, it is replaced by the well known "triviality bound" of the four dimensional relativistic model, which depends only logarithmically on $\Lambda$. Finally, for superfluid liquids, as $^4\text{He}$, one has $n\sim\Lambda^3$, such that for $a\sim \Lambda^{-1}$ one finds a large concentration $c$.

The situation for dilute bosons seems to contrast with ultracold fermion gases in the unitary limit of a Feshbach resonance, where $a$ diverges. One may also think about a Feshbach resonance for bosonic atoms, where one would expect a large scattering length for a tuning close to resonance. In this case, however, the effective action does not remain local. It is best described by the exchange of molecules. The scale of nonlocality is then given by the gap for the molecules, $m_M$. Only for momenta $\vec{q}^2<m_M^2$ the effective action becomes approximately local, such that $\Lambda=m_M$ for our approximation. Close to resonance, the effective cutoff is low and again in the vicinity of $a^{-1}$.

\subsubsection{Logarithmic running in two dimensions}
It is well known that in two dimensions the scattering properties cannot be determined by a scattering length as it is the case in three dimensions. In experiments where a tightly confining harmonic potential restricts the dynamics of a Bose gas to two dimensions, the interaction strength has a logarithmic energy dependence in the two-dimensional regime. For low energies and in the limit of vanishing momentum the scattering amplitude vanishes. In our formalism this is reflected by the logarithmic running of the interaction strength $\lambda(k)$ in the vacuum, where both temperature and density vanish. In general, the flow equations in vacuum describe the physics of few particles like for example the scattering properties or binding energies. Following the above calculation for the three dimensional case, we find the flow equations for the interaction strength ($t=\text{ln}(k/\Lambda)$)
\begin{equation}
\partial_t \lambda = \frac{\lambda^2\left( k^2 - m^2 \right)}{4k^2\,
    {\pi }\,S}\,\theta(k^2-m^2).
\label{eqflowvacuummlambdad2}
\end{equation}
Since in vacuum the propagator is not renormalized, $\partial_t S=\partial_t V=\partial_t \bar{A}=\partial_t m^2=0$, we set $S=1$ and $V=0$ on the right-hand side of Eq.\ \eqref{eqflowvacuummlambdad2}. The vacuum corresponds to $m^2=0$ and we obtain the flow equation
\begin{equation}
\partial_t \lambda=\frac{\lambda^2}{4\pi}.
\label{eqflowoflambdainvacuum}
\end{equation}
The vacuum flow is purely driven by quantum fluctuations. It will be modified by the thermal fluctuations for $T\neq 0$ and for nonzero density $n$.

The solution of Eq.\ \eqref{eqflowoflambdainvacuum}
\begin{equation}
\lambda(k)=\frac{1}{\frac{1}{\lambda_\Lambda}+\frac{1}{4\pi}\text{ln}(\Lambda/k)}
\label{eqlambdakofk}
\end{equation}
goes to zero logarithmically for $k\rightarrow0$, $\lambda(k=0)= 0$. In contrast to the three-dimensional system, the flow of the interaction strength $\lambda$ does not stop in two dimensions. To relate the microscopic parameter $\lambda_\Lambda$ to experiments exploring the scattering properties, we have to choose a momentum scale $q_\text{exp}$, where experiments are performed. To a good approximation the relevant interaction strength can be computed from Eq.\ \eqref{eqlambdakofk} by setting $k=q_\text{exp}$. If not specified otherwise, we will use a renormalized coupling $\lambda=\lambda(k_\text{ph})$.

For our calculation we also have to use a microscopic scale $\Lambda$ below which our approximation of an effectively two dimensional theory with pointlike interaction becomes valid. Our two-dimensional computation only includes the effect of fluctuations with momenta smaller then $\Lambda$. In experiments $\Lambda^{-1}$ is usually given either by the range of the van der Waals interaction or by the length scale of the potential that confines the system to two dimensions. We choose in the following
\begin{equation}
\Lambda=10,\quad k_\text{ph}=10^{-2}. 
\end{equation}
At this stage the momentum or length units are arbitrary, but we will later often choose the density to be $n=1$, so that we measure length effectively in units of the interparticle spacing $n^{-1/2}$. For typical experiments with ultracold bosonic alkali atoms one has $n^{-1/2}\approx 10^{-4}\text{cm}$. 
 
\begin{figure}
\centering
\includegraphics[width=0.5\textwidth]{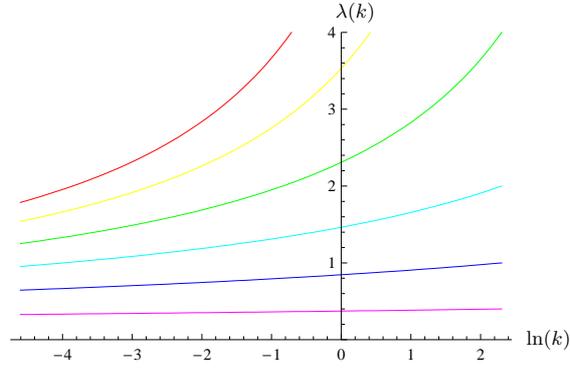}
\caption{Flow of the interaction strength $\lambda(k)$ at zero temperature and density for different initial values $\lambda_\Lambda=100$, $\lambda_\Lambda=10$, $\lambda_\Lambda=4$, $\lambda_\Lambda=2$, $\lambda_\Lambda=1$, and $\lambda_\Lambda=0.4$ (from top to bottom).}
\label{figflowoflambda}
\end{figure}
\begin{figure}
\centering
\includegraphics[width=0.5\textwidth]{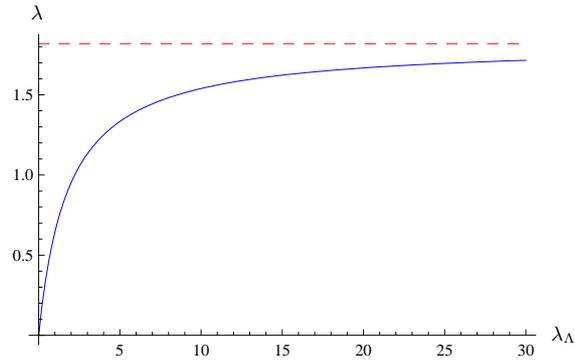}
\caption{Interaction strength $\lambda$ at the macroscopic scale $k_\text{ph}=10^{-2}$ in dependence on the microscopic interaction strength $\lambda_\Lambda$ at $\Lambda=10$ (solid). The upper bound $\lambda_{\text{max}}=\frac{4 \pi}{\text{ln}(\Lambda/k_\text{ph})}$ is also shown (dashed).}
\label{figinteraction}
\end{figure}

The flow of $\lambda(k)$ for different initial values $\lambda_\Lambda$ is shown in Fig. \ref{figflowoflambda}. 
Following the flow from $\Lambda$ to $k_\text{ph}$ yields the dependence of $\lambda=\lambda(k_\text{ph})$ on $\lambda_\Lambda$ as displayed in Fig. \ref{figinteraction}.
It follows from Eq.\ \eqref{eqlambdakofk} that for positive initial values $\lambda_\Lambda$ the interaction strength $\lambda$ is bounded by
\begin{equation}
\lambda<\frac{4\pi}{\text{ln}(\Lambda/k_\text{ph})}\approx1.82.
\end{equation}
The last equation holds for our choice of $\Lambda$ and $k_\text{ph}$. We emphasize that our bound holds only if the interactions are approximately pointlike for all momenta below $\Lambda$. Close to a Feshbach resonance this may not be true and our formalism would need to be extended by considering nonlocal interactions or introducing an additional field for the exchanged two-atom state in the ``closed channel''. In other words, close to a Feshbach resonance the effective cutoff $\Lambda$ for the validity of a two-dimensional model with pointlike interactions may be substantially lower, and the upper bound on $\lambda$ correspondingly higher. 

	\section{Two fermion species: Dimer formation}
	\label{sec:Twofermionspecies:Dimerformation}
	In this section, we discuss the few-body properties of the BCS-BEC crossover model in Eq.\ \eqref{eqMicroscopicAction}. The treatment here is condensed to the basics, a more extensive discussion can be found in \cite{Diehl:2008}. 

The vacuum problem can be structured in the following way: The two-body sector
describes the pointlike fermionic two-body interactions. It involves couplings up to
fourth order in the fermion field in a purely fermionic setting. In the language with a composite boson field $\varphi \sim
\psi\psi$ we have in addition terms quadratic in $\varphi$ or $\sim \psi\psi\varphi$. The two-body sector decouples from the sectors involving a higher number of particles, described by
higher-order interactions parameters such as the dimer-dimer scattering (which
corresponds to interactions of eighth order in $\psi$ in a fermionic
language). This decoupling reflects the situation in quantum mechanics, where a two-body calculation (in vacuum) never needs input from states with more than two particles.

For two fermion species and not too far from unitarity the solution of the two-body problem fixes the
independent renormalized couplings completely. The couplings in the sectors
with higher particle number then are derived quantities which can be computed
as functions of the parameters of the two-body problem. Here, we will study
the dimer-dimer or molecular scattering length $a_{\rm M}$ as an important
example.

On the technical side, the vacuum limit leads to a
massive simplification of the diagrammatic structure as compared to
the nonzero density and temperature system. This is discussed in \cite{Diehl:2008, Diehl:2007ri, DiehlPhD} With the aid of the residue theorem, one can prove the following statement \cite{Diehl:2008, Diehl:2007ri, DiehlPhD}: \emph{All diagrams with inner lines pointing into the same direction (thereby forming a closed tour of particle number) do not contribute to the flow in vacuum.} Here, the direction of a line is given by the $q_0$ flow of the propagator. 

For a proof, we first consider the form of the renormalized inverse fermion and boson propagators in the vacuum
limit,
\begin{eqnarray}
P_F(q) &=& i q_0 + \vec q^2 - \mu ,\nonumber\\
P_\varphi(q)  &=& i S_\varphi q_0 + \vec q^2 + m_\varphi^2 .
\end{eqnarray}
Lines pointing into the same direction represent integrals over products of $P_F^{-1},P_\varphi^{-1}$ with the \emph{same} sign of the frequency variable in the loop. Without loss of generality, we can choose it to be positive.

In the presence of a nonzero cutoff, the spacelike real part of the
regularized inverse propagators $P_{\text{F}} + R_k^{\text F}, P_\varphi +
R_k^\varphi$, including the mass terms, is always positive. Hence, the poles
all lie in the upper half of the complex plane. Closing the
integration contour in the lower half-plane, no residues are picked up, implying that
these integrals vanish. A derivative with respect to the cutoff $R_k$
increments the number of inner lines $P_{\text F}^{-1},P_\varphi^{-1}$ by one (it
changes the multiplicity of the poles), but does not affect the sign of the
real part of the propagator nor of $q_0$, such that our argument remains
valid for both the regularized loops and their $k$ derivative entering the
flow equation.

\subsubsection{Two-body problem}

The two-body problem is best solved in terms of the bare couplings. Their flow
equations read
\begin{eqnarray}
\nonumber
\partial_k \bar{m}_\varphi^2 &=& \frac{\bar{h}_\varphi^2}{6\pi^2
  k^3}\theta(k^2+\mu)(k^2+\mu)^{3/2}\\ 
\nonumber
\partial_k \bar Z_\varphi &=& -\frac{\bar{h}_\varphi^2}{6\pi^2
  k^5}\theta(k^2+\mu)(k^2+\mu)^{3/2}\\ 
\nonumber
\partial_k \bar A_\varphi &=& -\frac{\bar{h}_\varphi^2}{6\pi^2 k^5}\theta(k^2+\mu)(k^2+\mu)^{3/2}\\ 
\partial_k \bar{h}_\varphi &=&0.
\label{eq:vacuumflow}
\end{eqnarray}
The flow in the two-body sector is driven by fermionic diagrams only.  There
is no renormalization of the Yukawa coupling $\bar h$. The equations \eqref{eq:vacuumflow} are solved by direct
integration with the result
\begin{eqnarray}
\nonumber
\bar{m}_\varphi^2(k) &=&
\bar{m}_\varphi^2(\Lambda)\\
\nonumber
&&-\theta(\Lambda^2+\mu)\frac{\bar{h}_\varphi^2}{6\pi^2}{\bigg
  [}\sqrt{\Lambda^2+\mu}\,\left(1-\frac{\mu}{2\Lambda^2}\right)
-\frac{3}{2}\sqrt{-\mu}\,\,\text{arctan}
\left(\frac{\sqrt{\Lambda^2+\mu}}{\sqrt{-\mu}}\right){\bigg  ]} \\
\nonumber
&&+\theta(k^2+\mu)\frac{\bar{h}_\varphi^2}{6\pi^2}{\bigg
  [}\sqrt{k^2+\mu}\,\left(1-\frac{\mu}{2k^2}\right) 
-\frac{3}{2}\sqrt{-\mu}\,\,\text{arctan}
\left(\frac{\sqrt{k^2+\mu}}{\sqrt{-\mu}}\right){\bigg  ]}\\ 
\nonumber
\bar Z_\varphi(k) &=&
\bar Z_\varphi(\Lambda)\\
\nonumber
&& -\theta(\Lambda^2+\mu)
\frac{\bar{h}_\varphi^2}{48\pi^2}{\bigg [}\sqrt{\Lambda^2+\mu}
\,\frac{\left(5\Lambda^2+2\mu\right)}{\Lambda^4}
-\frac{3}{\sqrt{-\mu}}\,\,\text{arctan}
\left(\frac{\sqrt{\Lambda^2+\mu}}{\sqrt{-\mu}}\right){\bigg  ]}\\
\nonumber
&& +\theta(k^2+\mu)
\frac{\bar{h}_\varphi^2}{48\pi^2}{\bigg [}\sqrt{k^2+\mu}
\,\frac{\left(5k^2+2\mu\right)}{k^4}
-\frac{3}{\sqrt{-\mu}}\,\,\text{arctan}
\left(\frac{\sqrt{k^2+\mu}}{\sqrt{-\mu}}\right){\bigg  ]}\\
\nonumber 
\bar A_\varphi(k) &=& \bar A_\varphi(\Lambda)\\
\nonumber
&& -\theta(\Lambda^2+\mu)
\frac{\bar{h}_\varphi^2}{48\pi^2}{\bigg [}\sqrt{\Lambda^2+\mu}\,
\frac{\left(5\Lambda^2+2\mu\right)}{\Lambda^4}
-\frac{3}{\sqrt{-\mu}}\,\,\text{arctan}
\left(\frac{\sqrt{\Lambda^2+\mu}}{\sqrt{-\mu}}\right){\bigg ]}\\
\nonumber
&& +\theta(k^2+\mu)
\frac{\bar{h}_\varphi^2}{48\pi^2}{\bigg [}\sqrt{k^2+\mu}\,
\frac{\left(5k^2+2\mu\right)}{k^4}
-\frac{3}{\sqrt{-\mu}}\,\,\text{arctan}
\left(\frac{\sqrt{k^2+\mu}}{\sqrt{-\mu}}\right){\bigg ]}.\\
\label{eq:vacuumsolutions}
\end{eqnarray}
Here, $\Lambda$ is the initial ultraviolet scale. Let us discuss the initial value for the boson mass. It is given by
\begin{equation}
\bar{m}_\varphi^2(\Lambda)=\nu(B) -2\mu+\delta
\nu(\Lambda). \label{eq:new41} 
\end{equation}
The detuning $\nu(B)= \mu_{\text{M}} (B-B_0)$ describes the energy level of the microscopic state represented by the field $\varphi$ with respect to the fermionic state $\psi$. At a Feshbach resonance, this energy shift can be tuned by the magnetic field $B$, $\mu_{\text{M}}$ denotes the magnetic moment of the field $\varphi$, and $B_0$ is the resonance position. Physical observables such as the scattering length and the binding energy are obtained from the effective action and are therefore related to the coupling constants at the infrared scale $k=0$. The quantity $\delta\nu(\Lambda)$ denotes a renormalization counter term that has to be adjusted conveniently, see below.

\subsubsection{Renormalization}

We next show that close to a Feshbach resonance the microscopic parameters 
$\bar  m_{\varphi,\Lambda}\equiv\bar m_\varphi^2(k=\Lambda)$ and $\bar
h_{\varphi,\Lambda}^2\equiv\bar h_\varphi^2(k=\Lambda)$
are related to $B-B_0$ and $a$ by two simple relations
\begin{equation}
\bar m_{\varphi,\Lambda}^2=\mu_\text{M}(B-B_0)-2\mu+\frac{\bar h_{\varphi,\Lambda}^2}{6\pi^2}\Lambda
\label{eq:mvarphiLambda}
\end{equation}
and 
\begin{equation}
a=-\frac{\bar h_{\varphi, \Lambda}^2}{8\pi \mu_\text{M}(B-B_0)}.
\label{eq:hvarphiLambda}
\end{equation}
Away from the Feshbach resonance the Yukawa coupling may depend on $B$, $\bar
h_\varphi^2(B)=\bar h_\varphi^2+c_1(B-B_0)+\dots$. Also the microscopic
difference of energy levels between the open and closed channel may show
corrections to the linear $B$-dependence,
$\nu(B)=\mu_\text{M}(B-B_0)+c_2(B-B_0)^2+\dots$ or $\mu_\text{M}\to
\mu_\text{M}+c_2(B-B_0)+\dots$. Using $\bar h_\varphi^2(B)$ and
$\mu_\text{M}(B)$ our formalism can easily be adapted to a more general
experimental situation away from the Feshbach resonance. The relations in
Eqs. \eqref{eq:mvarphiLambda} and \eqref{eq:hvarphiLambda} hold for all
chemical potentials $\mu$ and temperatures $T$. For a different choice of the
cutoff function the coefficient $\delta\nu(\Lambda)$ being the term linear in $\Lambda$ in Eq.\ \eqref{eq:mvarphiLambda} might be modified.

We want to connect the bare parameters $\bar m_{\varphi,\Lambda}^2$ and $\bar
h_{\varphi,\Lambda}^2$ with the magnetic field $B$ and the scattering length $a$ for
fermionic atoms as renormalized parameters. In our units, $a$ is related to
the effective interaction $\lambda_{\psi,\text{eff}}$ by
\begin{equation}
a=\frac{\lambda_{\psi,\text{eff}}}{8\pi}.
\end{equation}
The fermion interaction
$\lambda_{\psi,\text{eff}}$ is determined by the molecule exchange process in
the limit of vanishing spatial momentum
\begin{equation}
\lambda_{\psi,\text{eff}}=-\frac{\bar
  h_{\varphi,\Lambda}^2}{\bar{P}_\varphi(\omega,\vec{p}^2=0,\mu)}. 
\label{eqlambdaeffmoleculeexchange}
\end{equation}
Even though \eqref{eqlambdaeffmoleculeexchange} is a tree-level process, it is not an approximation, since $\bar{P}_\varphi\equiv\bar P_{\varphi}|_{k\to0}$ denotes the full bosonic propagator which includes all fluctuation effects. The frequency in Eq.\ \eqref{eqlambdaeffmoleculeexchange} is the sum of the frequency of the incoming fermions which in turn is determined from the on-shell condition
\begin{equation}
\omega=2\omega_\psi=-2\mu.
\label{eqinitialvalueofmass}
\end{equation} 

On the BCS side we have
$\mu=0$ and find with 
\begin{equation}
\bar P_\varphi(\omega=0,\vec q=0)=\bar m_\varphi^2(k=0)\equiv \bar m^2_{\varphi,0}
\end{equation}
the relation
\begin{equation}
\lambda_{\psi,\text{eff}}=-\frac{\bar h_{\varphi,\Lambda}^2}{\bar m_{\varphi,0}^2},
\end{equation}
where $\bar m_{\varphi,0}^2=\bar m_\varphi^2(k=0)$. For the bosonic mass terms at $\mu=0$, we can read off from
Eqs. \eqref{eq:vacuumsolutions} and \eqref{eq:new41} that 
\begin{equation}
  \bar{m}_{\varphi,0}^2=\bar{m}_{\varphi,\Lambda}^2
  -\frac{\bar{h}_{\varphi,\Lambda}^2}{6\pi^2} \Lambda 
  = \mu_{\text{M}}(B-B_0)+\delta \nu(\Lambda)
  -\frac{\bar{h}_{\varphi,\Lambda}^2}{6\pi^2}\Lambda.
\end{equation}
To fulfill the resonance condition $a\to\pm\infty$ for $B=B_0$, $\mu=0$,
we choose
\begin{equation}
\delta \nu(\Lambda)=\frac{\bar{h}_{\varphi,\Lambda}^2}{6\pi^2}\Lambda.
\end{equation}
The shift $\delta\nu (\Lambda)$ provides for the additive UV renormalization
of $\bar{m}_\varphi^2$ as a relevant coupling.  It is exactly canceled by the
fluctuation contributions to the flow of the mass. This yields the general
relation \eqref{eq:mvarphiLambda} (valid for all $\mu$) between the bare mass
term $\bar m_{\varphi,\Lambda}^2$ and the magnetic field. On the BCS side we
find the simple vacuum relation
\begin{equation}
\bar m_{\varphi,0}^2=\mu_\text{M}(B-B_0).
\end{equation}
Furthermore, we obtain for the fermionic
scattering length
\begin{equation}
a=-\frac{\bar{h}_{\varphi,\Lambda}^2}{8\pi \mu_{\text{M}}(B-B_0)}.
\label{eqscatterinlengthandmageticfield}
\end{equation}
This equation establishes Eq.\ \eqref{eq:hvarphiLambda} and shows that
$\bar{h}_{\varphi,\Lambda}^2$ determines the width of the resonance. We have thereby
fixed all parameters of our model and can express $\bar m_{\varphi,\Lambda}^2$
and $\bar h_{\varphi,\Lambda}^2$ by $B-B_0$ and $a$. The relations
\eqref{eq:mvarphiLambda} and \eqref{eq:hvarphiLambda} remain valid also at
nonzero density and temperature. They fix the ``initial values'' of the flow
($\bar h_\varphi^2\to \bar h_{\varphi,\Lambda}^2$) at the microscopic scale
$\Lambda$ in terms of experimentally accessible quantities, namely $B-B_0$ and
$a$.

On the BEC side, we encounter $\mu<0$ and thus $\omega>0$. We therefore need
the bosonic propagator for $\omega\neq 0$. Even though we have computed
directly only quantities related to $\bar P_\varphi$ at $\omega=0$ and
derivatives with respect to $\omega$ ($Z_\varphi$), we can obtain information
about the boson propagator for nonvanishing frequency by using the semilocal
$U(1)$ invariance described in section \ref{sec:Derivativeexpansionandwardidentities}. In momentum space,
this symmetry transformation results in a shift of energy levels
\begin{eqnarray}
\nonumber
\psi(\omega, \vec{p}) &\to& \psi(\omega-\delta,\vec{p})\\
\nonumber
\varphi(\omega,\vec{p}) &\to& \varphi(\omega-2\delta,\vec{p})\\
\mu &\to& \mu+\delta.
\end{eqnarray}
Since the effective action is invariant under this symmetry, it follows for
the bosonic propagator that
\begin{equation}
\bar{P}_\varphi(\omega,\vec{p},\mu)
=\bar{P}_\varphi(\omega-2\delta,\vec{p},\mu+\delta).
\end{equation}
To obtain the propagator needed in Eq.\ \eqref{eqlambdaeffmoleculeexchange}, we
can use $\delta=-\mu$ and find as in Eq.\ \eqref{eqscatterinlengthandmageticfield}
\begin{equation}
\lambda_{\psi,\text{eff}}
=-\frac{\bar{h}_{\varphi,\Lambda}^2}{\bar{P}_\varphi(\omega=0,\vec{p}^2=0,\mu=0)}
=-\frac{\bar h_{\varphi,\Lambda}^2}{\mu_\text{M}(B-B_0)}.
\end{equation}
Thus the relations \eqref{eq:mvarphiLambda} and \eqref{eq:hvarphiLambda} for
the initial values $\bar m_{\varphi,\Lambda}$ and $\bar h_{\varphi,\Lambda}^2$
in terms of $B-B_0$ and $a$ hold for both the BEC and the BCS side of the
crossover.

\subsubsection{Binding energy}

We next establish the relation between the molecular binding energy
$\epsilon_\text{M}$, the scattering length $a$, and the Yukawa coupling $\bar
h_{\varphi,\Lambda}^2$. From Eq.\ \eqref{eq:vacuumsolutions}, we obtain for
$k=0$ and $\mu\leq 0$
\begin{eqnarray}\label{mphiFinal}
\nonumber
\bar{m}_{\varphi,0}^2 &=& \mu_{\text{M}} (B-B_0)-2\mu\\
\nonumber
&& +\frac{\bar h_{\varphi,\Lambda}^2}{6\pi^2} {\Bigg [}\Lambda-\sqrt{\Lambda^2+\mu}
\left(1-\frac{\mu}{2\Lambda^2}\right)\\
&& +\frac{3}{2}\sqrt{-\mu} \,\,\text{arctan}
\left(\frac{\sqrt{\Lambda^2+\mu}}{\sqrt{-\mu}}\right){\Bigg ]}.
\end{eqnarray}
In the limit $\Lambda/\sqrt{-\mu}\to\infty$ this yields
\begin{equation}
\bar m_{\varphi,0}^2 = \mu_{\text{M}} (B-B_0)-2\mu+\frac{\bar{h}_{\varphi,\Lambda}^2\sqrt{-\mu}}{8\pi}.
\end{equation}
Together with Eq.\ \eqref{eqscatterinlengthandmageticfield}, we can deduce
\begin{equation}
a=-\frac{\bar h_{\varphi,\Lambda}^2}{8\pi \left( \bar m_{\varphi,0}^2 +2\mu 
-\frac{\bar{h}_{\varphi,\Lambda}^2\sqrt{-\mu}}{8\pi}\right)},
\end{equation}
which holds in the vacuum for all $\mu$. On the BEC side where $\bar
m_{\varphi,0}^2=0$ this yields
\begin{equation}
  a=\frac{1}{\sqrt{-\mu}\left(1+\frac{16 \pi}{\bar h_{\varphi,\Lambda}^2}\sqrt{-\mu}\right)}.
\label{ScattLength}
\end{equation}
The binding energy of the bosons is given by the difference between the
energy for a boson ${\bar{m}_\varphi^2}/{\bar{Z}_\varphi}$ and the energy
for two fermions $-2\mu$. On the BEC side, we can use $\bar{m}_{\varphi,0}^2=0$ and
obtain
\begin{equation}
\epsilon_{\text{M}}=\frac{\bar{m}_\varphi^2}{\bar{Z}_\varphi}+2\mu\Big|_{k\to0}=2\mu.
\label{eq:bindingenergy}
\end{equation}
From Eqs.\ \eqref{ScattLength} and \eqref{eq:bindingenergy} we find a relation
between the scattering length $a$ and the binding energy $\epsilon_\text{M}$
\begin{equation}
\frac{1}{a^2}=\frac{-\epsilon_\text{M}}{2}
+(-\epsilon_\text{M})^{3/2} \frac{4\sqrt{2}\pi}{\bar h_{\varphi,\Lambda}^2}
+(-\epsilon_\text{M})^2\frac{(8\pi)^2}{\bar h_{\varphi,\Lambda}^4}.
\label{eq:scatteringlengthandbindingenergy}
\end{equation}
In the broad resonance limit $\bar{h}_{\varphi,\Lambda}^2\to\infty$, this is
just the well-known relation between the scattering length $a$ and the binding
energy $\epsilon_{\text{M}}$ of a dimer (see for example \cite{BraatenHammer})
\begin{equation}
\epsilon_{\text{M}}=-\frac{2}{a^2}=-\frac{1}{M a^2}.
\label{eq:scatteringlengthbroad}
\end{equation}
The last two terms in Eq.\ \eqref{eq:scatteringlengthandbindingenergy} give
corrections to Eq.\ \eqref{eq:scatteringlengthbroad} for more narrow
resonances.

The solution of the two-body problem turns out to be exact as expected. In our
formalism, this is reflected by the fact that the two-body sector decouples
from the flow equations of the higher-order vertices: no higher-order
couplings such as $\lambda_\varphi$ enter the set of equations
(\ref{eq:vacuumflow}). Extending the truncation to even higher order vertices or
by including a boson-fermion vertex $\psi^\dagger \psi \varphi^\ast\varphi$
does not change the situation.

\subsubsection{Dimer-Dimer Scattering}
\label{DimerDimer}

So far we have considered the sector of the theory up to order
$\varphi^\ast\psi\psi$, which is equivalent to the fermionic two-body problem
with pointlike interaction in the limit of broad resonances. Higher-order
couplings, in particular the four-boson coupling
$\lambda_\varphi(\varphi^\ast\varphi)^2$, do not couple to the two-body
sector. Nevertheless, a four-boson coupling emerges dynamically from the
renormalization group flow. 
In vacuum we have $\rho_0=0$ and $\lambda_\varphi$ is defined as $\lambda_\varphi=U^{\prime\prime}_k(0)$. The flow equation for $\lambda_\varphi$ can be found by taking the $\rho$-derivative of Eq.\ \eqref{eq:Flowu1}
\begin{eqnarray}
\nonumber
k \partial_k \lambda_\varphi &=& 2 \eta_{A_\varphi} U_k^{\prime\prime} - \frac{\sqrt{2} k^3}{3\pi^2 S_\varphi}\left(1-\frac{2}{d+2}\eta_{A_\varphi}\right)\\
\nonumber
&&\times 2 (U_k^{\prime\prime})^2 \left(s_{B,Q}^{(1,0)}+3 s_{B,Q}^{(0,1)}\right)+\frac{h_\varphi^4}{3\pi^2 k^3} s_{F,Q}^{(2)}\\
\nonumber
&=& 2\eta_{A_\varphi} \lambda_\varphi + \frac{\sqrt{2} k^5 \lambda_\varphi^2}{3\pi^2\, S_\varphi\, (m_\varphi^2+k^2)^2}(1-2\eta_{A_\varphi}/5)\\
&&-\frac{h_\varphi^4\,\theta(\mu+k^2)\,(\mu+k^2)^{3/2}}{4\pi^2k^6}.
\end{eqnarray} 
There are contributions from fermionic and bosonic vacuum fluctuations, but no contribution from higher $\rho$ derivatives of $U$. The fermionic diagram generates a four-boson coupling even for zero initial value. This coupling then feeds back into the flow equation via the bosonic diagram.

The scattering lengths are related to the corresponding couplings by the relation (cf. \cite{Diehl:2007th})
\begin{eqnarray}
  \frac{a_{\text{M}}}{a}=2\,
  \frac{\lambda_\varphi}{\lambda_{\psi,\text{eff}}},  
  \quad \lambda_{\psi,\text{eff}}= 8\pi a.
\end{eqnarray}
Omitting the bosonic fluctuations, a direct integration yields the mean field result $a_{\text{M}}/a=2$. This value is lowered when the bosonic fluctuations are taken into account. With our truncation and choice of cutoff one finds $a_{\text{M}}/a =0.718$. The calculation can be improved by extending the truncation to
include a boson-fermion vertex $\lambda_{\varphi\psi}$ which describes the
scattering of a dimer with a fermion \cite{DKS}. Inspection of the diagrammatic structure
shows that this vertex indeed couples into the flow equation for $\lambda_\varphi$.

The ratio $a_M/a$ has been computed by other methods. Diagrammatic approaches give $a_{\text{M}}/a =0.75(4)$
\cite{PhysRevB.61.15370}, whereas the solution of the 4-body Schr\"{o}dinger
equation yields $a_{\text{M}}/a =0.6$ \cite{PhysRevLett.93.090404}, confirmed in QMC
simulations \cite{PhysRevLett.93.200404} and with diagrammatic techniques \cite{Brodsky2005}.

	\section{Three fermion species: Efimov effect}
	\label{sec:Threefermionspecies:ThomasandEfimoveffect}
	In the following we discuss the few-body properties of the model for three fermion species in Eq.\ \eqref{eq8:microscopicactiontrionmodel}. We first concentrate on the SU(3) symmetric case where the mass and the scattering properties are equal for all three species. The three-body problem is governed by the Efimov effect. We later generalize the model in Eq.\ \eqref{eq8:microscopicactiontrionmodel} to cover also the case where SU(3) symmetry is broken explicitly. Subsequently we apply this to the case of $^6$Li and discuss the relation to some recent experiments. 

\subsection{SU(3) symmetric model}
\label{ssect:SU3symmetricmodel}

We use now the truncation in Eq.\ \eqref{eq:triontruncation} to derive flow equations for the model with global SU(3) symmetry in Eq.\ \eqref{eq8:microscopicactiontrionmodel} that describes three fermion species at a common two-body resonance. 

\subsubsection{Flow equations for two-body sector}

To obtain the flow equations that govern the two-body sector, we insert our ansatz equation \eqref{eq:triontruncation} into the Wetterich equation \eqref{eq4:Wettericheqn}. Functional derivatives with respect to the fields for zero temperature $T=0$ and density $n=0$ lead us to a system of ordinary coupled nonlinear differential equations for the couplings $\bar m_\varphi^2$, $\bar h$, $\bar A_\psi$, $\bar{A}_\varphi$, $\bar{A}_\chi$, $\bar m_\chi^2$, $\bar g$ and $\bar \lambda_{\varphi\psi}$. One finds that the propagator of the original fermions $\psi$ is not renormalized, $\bar A_\psi(k)=1$, $\bar h(k) = \bar h$. The flow equations for the couplings determining the two body sector, namely the boson gap parameter (with $t=\ln (k/\Lambda)$)
\begin{equation}
\partial_t \bar m_\varphi^2 = \frac{\bar h^2}{6\pi^2}  \frac{k^5}{(k^2-\mu)^2},
\end{equation}
and the boson wave function renormalization
\begin{equation}
\partial_t \bar A_\varphi = -\frac{\bar h^2}{6\pi^2}\frac{k^5}{(k^2-\mu)^3}
\end{equation}
decouple from the other flow equations. These flow equations are very similar to the ones for the two-body sector of the BCS-BEC crossover model discussed in section \ref{sec:Twofermionspecies:Dimerformation}. The difference comes from the fact that we use a slightly different cutoff function (Eq.\ \eqref{eq:cutofftrions} to be compared with \eqref{eq7:DeltaSkfermions}). They can be solved analytically,
\begin{eqnarray}
\nonumber
\bar m_\varphi^2(k) &=& \bar m_\varphi^2(\Lambda)-\frac{\bar h^2}{6\pi^2}{\bigg [} (\Lambda-k)-\frac{\mu}{2}\left(\frac{\Lambda}{\Lambda^2-\mu}-\frac{k}{k^2-\mu}\right)\\
\nonumber
&+&\frac{3}{2}\sqrt{-\mu}\left(\text{arctan}\left(\frac{\sqrt{-\mu}}{\Lambda}\right)-\text{arctan}\left(\frac{\sqrt{-\mu}}{k}\right)\right){\bigg ]},\\
\nonumber
\bar A_\varphi(k) &=& 1
-\frac{\bar h^2}{6\pi^2}{\bigg [} \frac{\Lambda(5\Lambda^2-3\mu)}{8(\Lambda^2-\mu)^2}-\frac{k(5k^2-3\mu)}{8(k^2-\mu)^2}\\
&+&\frac{3}{8\sqrt{-\mu}}\left(\text{arctan}\left(\frac{\sqrt{-\mu}}{\Lambda}\right)-\text{arctan}\left(\frac{\sqrt{-\mu}}{k}\right)\right){\bigg ]}.
\label{eq:explicitsolutiontwobody}
\end{eqnarray}
Using this explicit solution \eqref{eq:explicitsolutiontwobody} we can relate the initial value of the boson gap parameter $\bar m_\varphi^2$ as well as the Yukawa coupling $\bar h$ to physical observables. The interaction between fermions $\psi$ is mediated by the exchange of the bound state $\varphi$. Again in terms of bare quantities, the scattering length between fermions is given by $a=-\bar h^2/(8\pi\bar m_\varphi^2)$ where the couplings $\bar h$ and $\bar m_\varphi^2$ are evaluated at the macroscopic scale $k=0$ and for vanishing chemical potential $\mu=0$. This fixes the initial value
\begin{equation}
\bar m_\varphi^2(\Lambda)=-\frac{\bar h^2}{8\pi}a^{-1}+\frac{\bar h^2}{6\pi^2}\Lambda-2\mu.
\end{equation}
In addition to the $\mu$-independent part we have added here the  chemical potential term $-2\mu$ where the factor $2$ accounts for the bosons consisting of two fermions. In vacuum, the gap parameter of the bosons $\bar{m}_\varphi^2$ is proportional to the detuning of the magnetic field $\bar m_\varphi^2(k=0,\mu=0)=\mu_M (B-B_0)$.
Here $\mu_M$ is the magnetic moment of the bosonic dimer and $B_0$ is the magnetic field at the resonance. From
\begin{equation}
a=-\frac{1}{8\pi} \frac{\bar h^2}{\mu_M (B-B_0)}
\end{equation}
one can read off that $\bar h^2$ is proportional to the width of the resonance. We have now fixed all initial values of the couplings at the scale $k=\Lambda$ or, in other words, the parameters of our microscopic model.

At vanishing density $n=0$, the chemical potential is negative or zero, $\mu\leq0$, and will be adjusted such that the lowest excitation of the vacuum is a gapless propagating degree of freedom in the infrared, i.~e. at $k=0$. Depending on the value of $a^{-1}$ this lowest energy level may be the original fermion $\psi$, the boson $\varphi$, or the composite fermion $\chi$. 

\subsubsection{Three-body problem}
Now that we have solved the equations for the two-body problem within our approximation, we can address the three-body sector. It is described by the flow equations for the trion gap parameter,
\begin{equation}
\partial_t m_\chi^2 = \frac{6 g^2}{\pi^2} \frac{k^5}{(3k^2-2\mu +2 m_\varphi^2)^2}+\eta_\chi m_\chi^2,
\label{eq:flowmchi}
\end{equation}
and the Yukawa-type coupling,
\begin{eqnarray}
\partial_t g &=& (\eta_\varphi+\eta_\chi)\frac{g}{2}+m_\chi^2\partial_t \alpha \notag \\ 
& & -\frac{2g h^2}{3\pi^2} \frac{k^5}{(k^2-\mu)^2}\frac{ 6k^2-5\mu + 2 m_\varphi^2}{(3k^2-2\mu+2m_\varphi^2)^2}.
\label{eq:flowg2}
\end{eqnarray}
These equations are supplemented by the anomalous dimension
\begin{equation}
\eta_\chi =-\frac{\partial_t \bar A_\chi}{\bar A_\chi} =\frac{24 g^2}{\pi^2} \frac{k^5}{ (3k^2-2\mu+2m_\varphi^2)^3}
\label{eq:etachi}
\end{equation}
and the variable
\begin{equation}
\partial_t \alpha = -\frac{h^4}{12\pi^2 g}\frac{ k^5}{(k^2-\mu)^3}\frac{9k^2-7\mu +4m_\varphi^2}{  (3k^2-2\mu+2m_\varphi^2)^2}.
\label{eq:alpha}
\end{equation}
Here $\alpha$ determines the scale dependence of the trion field $\chi$ 
\begin{equation}
\partial_t \chi =-(\partial_t \alpha)\varphi_i\psi_i-\eta_\chi \chi/2
\label{eq:scaledepfield}
\end{equation}
in the sense of the general coordinate transformation in Eq.\ \eqref{eq3:Wettericheqwithcoordtransf}. We neglect here and in the following correction terms coming from the connection in the space of fields since they are expected to be small. Instead of working with the coordinate transformation one might also work with the generalized flow equation \eqref{eq:flowequationGamma}. Since this is an exact equation and the structure is simpler compared to Eq.\ \eqref{eq3:Wettericheqwithcoordtransf} we propose to employ Eq.\ \eqref{eq:flowequationGamma} in future work. 
While the second term in Eq.\ \eqref{eq:scaledepfield} is the usual wave-function renormalization, the first term describes a nonlinear change of variables. It is chosen such that the flow of $\lambda_{\varphi\psi}$ vanishes on all scales, $\lambda_{\varphi\psi}(k)=0$. The scattering between bosons and fermions is then described by the exchange of the trion bound state $\chi$. This reparametrization, which is analogous to rebosonization \cite{Gies:2001nw, Pawlowski2007a}, is crucial for our description of the system in terms of the composite trion field $\chi$. 
Note that the flow equations \eqref{eq:flowmchi}, \eqref{eq:flowg2}, \eqref{eq:etachi}, \eqref{eq:alpha} that describe the three-body sector are independent from the flow of the boson-boson interaction $\lambda_\varphi$ that belongs to the four-body sector. 

\begin{figure}
\centering
\includegraphics[width=0.5\textwidth]{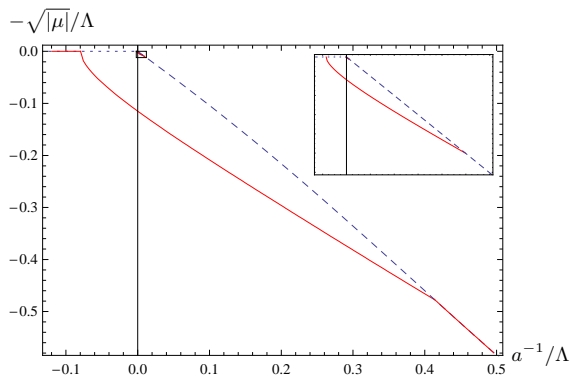}
\caption{Dimensionless chemical potential $-\sqrt{|\mu|}/\Lambda$ in vacuum as a function of the dimensionless scattering length $a^{-1}/\Lambda$. For comparison, we also plot the result for two fermion species where the original fermions are the propagating particles for $a^{-1}<0$ (dotted) and composite bosons for $a^{-1}>0$ (dashed). The inset is a magnification of the little box and shows the energy of the first excited Efimov state.}
\label{fig:Efimov}
\end{figure}

Since the three-body sector is driven by fermionic and bosonic fluctuations, it is not possible to find simple analytic solutions to the flow equations in the general case. However, it is no problem to solve them numerically. This is most conveniently done using again ``bare'' couplings. As a general feature we note that a negative chemical potential $\mu$ acts as an infrared cutoff for the fermionic fluctuations while a positive value of the bosonic gap $m_\varphi^2$ suppresses bosonic fluctuations in the infrared. We can find numerical solutions for different scattering length between the fermions $a$ and varying chemical potential $\mu$. We use an iteration process to determine the chemical potential $\mu\leq0$ where the lowest excitation of the vacuum is gapless. On the far BCS side for small and negative scattering length $a\to0_-$, where this lowest excitation is the fermion $\psi$, this implies simply $\mu=0$. On the far BEC side for small and positive scattering length $a\to0_+$ the lowest excitation is the boson $\varphi$, implying $\bar m_\varphi^2 =0$. We can then use our analytic solution of the two-body sector Eq.\ \eqref{eq:explicitsolutiontwobody} to obtain the chemical potential $\mu$ from the condition $\bar m_\varphi^2=0$ at the macroscopic scale $k=0$. In the limit $\Lambda/|\mu|\to\infty$ we find 
\begin{equation}
\mu =-\left(\sqrt{\frac{\bar h^4}{(32 \pi)^2}+\frac{a^{-1}\bar h^2}{16 \pi}}-\frac{\bar h^2}{32 \pi}\right)^2.
\end{equation}
As it should be, this is equivalent to the corresponding relation found for the two-fermion case Eq.\ \eqref{ScattLength}. In the broad resonance limit $\bar h^2\to \infty$ this reduces to the well-known result $\mu=-1/a^2$.
For scattering lengths close to the Feshbach resonance $a^{-1}_{c1}<a^{-1}<a^{-1}_{c2}$ we find that the lowest vacuum excitation is the trion $\chi$. Numerically, we determine $\mu$ from the implicit equation $m_\chi^2=0$ at $k=0$. Our result for $\mu$ obtained for $\bar h^2=100\Lambda$ as a function of the inverse scattering length $a^{-1}$ is shown in Fig. \ref{fig:Efimov}. With the choice $\Lambda=1/a_0$, this value of $\bar h^2$ corresponds to the width of the Feshbach resonance of $^{6}\textrm{Li}$ atoms in the $(m_F=1/2,\ m_F=-1/2)$-channel \cite{PhysRevLett.94.103201}. 

\begin{figure}
\centering
\includegraphics[width=0.5\textwidth]{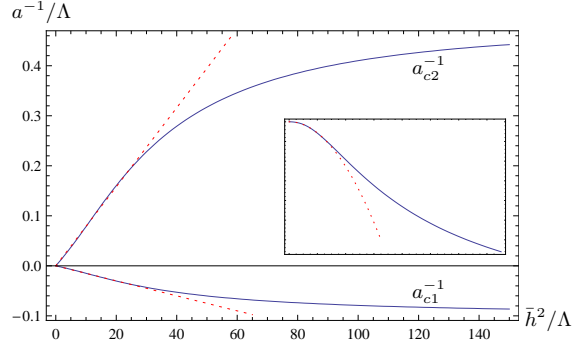}
\caption{Interval of scattering length $a^{-1}_{c1}<a^{-1}<a^{-1}_{c2}$ where the lowest vacuum excitation is the trimer fermion $\chi$ (solid lines). For comparison we also plot the linear fits $a^{-1}_{c1}=-0.0015\,\bar h^2$ and $a^{-1}_{c2}=0.0079\,\bar h^2$ (dotted). The inset shows the chemical potential $\mu_U/\Lambda^2$ at the resonance $a^{-1}=0$ as a function of $\bar h^2/\Lambda$ in the same range of $\bar h^2/\Lambda$. Here we also plot the curve $\mu_U=-3.5\times10^{-6}\bar h^4$ for comparison (dotted).}
\label{fig:ascaling}
\end{figure}

For small Yukawa couplings $\bar h^2/\Lambda\ll 1$, or narrow resonances we find that the range of scattering length where the trimer is the lowest excitation of the vacuum increases linear with $\bar h^2$ \cite{PhysRevLett.93.143201, gogolin:140404}. More explicit, we find $a^{-1}_{c1}=-0.0015\,\bar h^2$, $a^{-1}_{c2}=0.0079\,\bar h^2$. However, for very broad Feshbach resonances $\bar h^2/\Lambda\gg1$ the range depends on the ultraviolet scale $a^{-1}_{c1},a^{-1}_{c2}\sim \Lambda$. We show this behavior in Fig. \ref{fig:ascaling}.

The chemical potential $\mu_U$ at the unitarity point $a^{-1}=0$ is plotted in the inset of Fig. \ref{fig:ascaling}. It increases with the width of the resonance similar to $\mu_U=-3.5\times 10^{-6}\bar h^4$ for small $\bar h^2/\Lambda$. It also approaches a constant value which depends on the cutoff scale $\mu_U\sim\Lambda^2$ in the broad resonance limit $\bar h^2\to \infty$. 

The dispersion relation for the atoms, dimers and trimers can be computed by analytical continuation, $\tau=it$, of the inverse propagators to ``real frequencies'' $\omega$. In our truncation, they read (in terms of renormalized fields)
\begin{eqnarray}
P_\psi&=&\frac{\vec{p}^2}{2M}-\omega_\psi-\mu,\notag\\
P_\varphi&=&\frac{\vec{p}^2}{4M}-\omega_\varphi-2\mu+\nu_\varphi(\mu+\omega_\varphi/2-\vec{p}^2/(8M)),\notag\\
P_\chi&=&\frac{\vec{p}^2}{6M}-\omega_\chi-3\mu+\nu_\chi(\mu+\omega_\chi/3-\vec{p}^{2}/(18M)).\notag\\
\label{eq:propagators}
\end{eqnarray}
We note that the functions $\nu_\varphi$ and $\nu_\chi$ depend only on the particular combinations of $\mu$, $\omega$, and $\vec{p}^2$ given above. This is a result of the symmetries of our problem. The real-time microscopic action $S=\Gamma_{k=\Lambda}$ with $t=-i\tau$ is invariant under the time-dependent U(1) symmetry transformation of the fields $\psi\to e^{iEt}\psi$, $\varphi\to e^{i2Et}\varphi$, and $\chi\to e^{i3Et}\chi$ if we also change the chemical potential according to $\mu\to\tilde \mu=\mu+E$. Since the microscopic action has this symmetry and since we do not expect any anomalies, the quantum effective action $\Gamma=\Gamma_{k=0}$ is also invariant under these transformations. In consequence, any dependence on $\mu$ must be accompanied by a corresponding frequency dependence, such that only the invariant combinations $\omega_\psi+\mu$, $\omega_\varphi+2 \mu$, $\omega_\chi+3\mu$ appear in the effective action. The relation between the dependence on $\omega$ and on $\vec{p}^2$ is a consequence of Galilean symmetry. Even though we compute directly with the flow equations only $\nu_\varphi(\mu,\omega_\varphi=0,\vec{p}^2=0)$ and $\nu_\chi(\mu,\omega_\chi=0,\vec{p}^2=0)$, we can use the symmetry information (``Ward identities'') for an extrapolation to arbitrary $\omega$ and $\vec{p}^2$. The dispersion relation for the bosons $\mathcal{E}_\varphi(\vec{p}^2)$ follows from Eq.\ \eqref{eq:propagators} by solving $P_\varphi(\omega_\varphi=\mathcal{E}_\varphi)=0$ and similarly the dispersion relation of the fermions $\mathcal{E}_\psi(\vec{p}^2)$ and the trions $\mathcal{E}_\chi(\vec{p}^2)$.

We are interested in the dimer and trimer energy levels relative to the energy of the atoms. For this purpose we consider their dispersion relations at rest ($\vec{p}^2=0$) and substract twice or three times the atom energy $\mathcal{E}_\psi=-\mu$,
\begin{eqnarray}
E_\varphi&=&\mathcal{E}_\varphi(\vec{p}^2=0)+2 \mu=\nu_\varphi(E_\varphi/2),\notag\\
E_\chi&=&\mathcal{E}_\chi(\vec p^2=0)+3\mu=\nu_\chi(E_\chi/3).
\label{eq:energylevels}
\end{eqnarray}
When the dimers are the lowest states one has $m^2_\varphi=0$ and therefore $\nu_\varphi=2 \mu=E_\varphi$, while for lowest trimers $m_\chi^2=0$ implies $\nu_\chi=3\mu=E_\chi$. We have shown $E_\varphi$ and $E_\chi$ in Fig. \ref{fig:Energies}. A typical value for $\Lambda$ is of the order of the inverse Bohr radius $a_0^{-1}$. We also mention that the energy levels have been computed in the absence of a microscopic three-body interaction, $\lambda_3(\Lambda)=0$. The same computation may be repeated with nonzero $\lambda_3(\Lambda)$, and thus microscopic parameters may be fixed by a comparison to the experimentally measured spectrum.

So far we were only concerned with the lowest energy excitation of the vacuum. We found that near a Feshbach resonance this lowest excitation is given by a trimer. However, it is known from the work of Efimov \cite{Efimov1970, Efimov1973} that one can expect not only one trimer state close to resonance but a whole spectrum. In fact, the solution $\nu_\chi=3\mu$ may not be the only solution for the equation fixing the trimer energy levels, i. e.
\begin{equation}
\mathcal{E}_\chi=\nu_\chi\left(\mu+\frac{1}{3}\mathcal{E}_\chi\right)-3\mu.
\label{eq:TrionGroundEnergy}
\end{equation}
For an investigation of the energy dependence of $\nu_\chi$ we use the symmetry transformation and shift the chemical potential by $\mathcal{E}_\psi$, $\tilde{\mu}=\mu+\mathcal{E}_\psi=0$. We can therefore follow the flow at vanishing $\tilde{\mu}$ and Eq.\ \eqref{eq:TrionGroundEnergy} turns into a simple implicit equation for the trion energy levels
\begin{equation}
E_\chi=\nu_\chi\left(\frac{E_\chi}{3}\right)=m^2_\chi\left(\frac{E_\chi}{3}\right)
\label{eq:Echi}
\end{equation}

For $\tilde{\mu}=0$ the flow equations simplify considerably. For example, Eq.\ \eqref{eq:explicitsolutiontwobody} becomes
\begin{eqnarray}
\bar m^2_\varphi(k)&=&\bar m^2_\varphi(\Lambda)-\frac{\bar h^2}{6 \pi^2} (\Lambda-k)=\mu_M (B-B_0)+\frac{\bar h^2 k}{6 \pi^2},\notag \\
\bar A_\varphi(k)&=&1+\frac{\bar h^2}{6 \pi^2}\left(\frac{1}{k}-\frac{1}{\Lambda}\right).
\end{eqnarray}
We observe that the two-body sector is evaluated here for $\tilde{\omega}_\varphi=\omega_\varphi+2\mu=0$. The physical chemical potential $\mu$ remains, of course, negative in the vicinity of the Feshbach resonance, and our evaluation therefore corresponds to a positive energy $\omega_\varphi$ as compared to the lowest energy level of the trimer for which $\omega_\chi=0$.

An especially interesting point in the spectrum is the unitarity limit, $B=B_0$, where the scattering length diverges, $a^{-1}=0$. At that point all length scales drop out of the problem and we expect a sort of scaling solution for the flow equations. In the limit $k\to 0$ the solution for $\bar A_\varphi(k)$ is dominated completely by the term with $1/k$, $\bar A_\varphi(k)=\bar h^2/(6\pi^2 k)$, and we find $m_\varphi^2=\bar m_\varphi^2/\bar A_\varphi=k^2$. For $E_\chi\to 0$ the flow equations for the three-body sector simplify and we find 
\begin{eqnarray}
\nonumber
\partial_t \bar m_\chi^2 &=& \frac{36}{25} \bar g^2\frac{k^2}{\bar h^2}\\
\partial_t \bar g^2 &=& -\frac{64}{25}\bar g^2-\frac{13}{25}\bar m_\chi^2 \frac{\bar h^2}{k^2}.
\label{eq:mgsystem}
\end{eqnarray}
In contrast, for $E_\chi\neq 0$, as needed for Eq.\ \eqref{eq:Echi}, we will have to solve the flow with a nonvanishing ``effective chemical potential'' $\hat{\mu}=E_\chi/3$, which will cause a departure from the scaling flow.

\subsubsection{Limit cycle scaling} 
First, let us consider the scaling solution obeying Eq.\ \eqref{eq:mgsystem}. It is convenient to rescale the variables according to $\tilde m^2=\bar m_\chi^2 (k/\bar h)^\theta$, $\tilde g^2= \bar g^2(k/\bar h)^{2+\theta}$ which gives the linear differential equation
\begin{equation}
\partial_t \begin{pmatrix}\tilde m^2 \\ \tilde g^2\end{pmatrix}=\begin{pmatrix}\theta, & \frac{36}{25}\\ -\frac{13}{25}, & 2+\theta-\frac{64}{25}\end{pmatrix} \begin{pmatrix}\tilde m^2 \\ \tilde g^2\end{pmatrix}.
\label{eq:matrixdifferentialequation}
\end{equation}
The matrix on the right hand side of Eq.\ \eqref{eq:matrixdifferentialequation} has the eigenvalues
$\beta_{1/2}=\theta-(7/25)\pm i \sqrt{419}/25$.
The flow equation \eqref{eq:matrixdifferentialequation} leads therefore to an oscillating behavior. It is straightforward to solve Eq.\ \eqref{eq:matrixdifferentialequation} explicitly. Restricting to real solutions and using initially $\bar g^2(\Lambda)=0$ we find the following for the trimer gap parameter and coupling
\begin{eqnarray}
\bar m_\chi^2(k)&=&\left(\frac{k}{\Lambda}\right)^{-\frac{7}{25}}\bar m_\chi^2(\Lambda)\Big[\cos\left(s_0 \ln\frac{k}{\Lambda}\right)\notag \\ 
& &  +\frac{7}{\sqrt{419}}\sin\left( s_0 \ln \frac{k}{\Lambda}\right)\Big],\notag \\
\bar g^2(k)&=&-\frac{13 \,\bar h^2}{\sqrt{419}\,k^2}\left(\frac{k}{\Lambda}\right)^{-\frac{7}{25}} \bar m_\chi^2(\Lambda) \sin\left(s_0 \ln \frac{k}{\Lambda} \right).\notag\\
\label{eq:mgsolution}
\end{eqnarray}
As it should be, the initial value $\bar m_\chi^2(\Lambda)$ drops out of the ratio $\bar g^2/\bar m^2_\chi$ and we find for the three-body coupling
\begin{equation}
\lambda_3(k)=\frac{468 \pi^4}{\sqrt{419}}\frac{\sin\left(s_0 \ln \frac{k}{\Lambda}\right)}{\cos\left( s_0\ln \frac{k}{\Lambda} \right)+\frac{7}{\sqrt{419}}\sin\left(s_0 \ln \frac{k}{\Lambda}\right)}k^{-4}.
\label{eq:lambda3}
\end{equation}
We obtain for the ``frequency'' $s_0=\sqrt{419}/25\approx 0.82$. Since we use a truncation in the space of functionals $\Gamma_k$ this result is only a rough estimate. It has to be compared with the result of other methods which find $s_0\approx1.00624$ \cite{Efimov1970, Efimov1973, PhysRevLett.82.463, Bedaque1999444, PhysRevLett.85.908, BraatenHammer}. Considered the simplicity of our approximation, the agreement is quite reasonable. A more elaborate truncation that includes the full momentum dependence of vertices, confirms Efimovs value, indeed \cite{Moroz2008}.

For a determination of the trion energy levels we have to solve the flow with an effective negative chemical potential $\tilde{\mu}=E_\chi/3$. This acts as an infrared cutoff, such that the flow deviates from the limit cycle once $k^2\approx-\tilde{\mu}$. Qualitatively, the flow eventually stops once $k$ becomes smaller than $\sqrt{-\tilde{\mu}}$. For an evaluation of Eq.\ \eqref{eq:Echi} we may therefore use Eq.\ \eqref{eq:mgsolution} with a specific value for $k$, namely $k^2=-E_\chi/3$. The possible energy levels therefore obey \begin{equation}
3k^2+\bar A^{-1}_\chi(k)\,\bar m_\chi^2(k)=0.
\label{eq:mchicond}
\end{equation}
With
\begin{equation}
\partial_k\bar A_\chi(k)=\frac{24}{125\pi^2} \frac{\bar g^2(k)}{k^2}
\end{equation}
one finds, up to oscillatory behavior an increase of $\bar A_\chi \sim k^{-\frac{32}{25}}$. We can write Eq.\ \eqref{eq:mchicond} in the form
\begin{eqnarray}
F(k)&=&\cos\left(s_0 \ln\frac{k}{\Lambda}\right)+\frac{7}{\sqrt{419}}\sin\left(s_0 \ln\frac{k}{\Lambda}\right)\notag\\
&=&-3 \frac{k^2}{\bar m_\chi^2(\Lambda)}\left(\frac{k}{\Lambda}\right)^{\frac{7}{25}}\bar A_\chi(k).
\end{eqnarray}
For small $k\ll\Lambda$ we infer that $F(k)$ has to vanish $\sim -k/ \bar h^2$. Since $F(k)$ is periodic, solutions will occur for roughly equidistant values in $\ln\frac{k}{\Lambda}$.

For $k\ll\bar h^2$ the possible solutions simply correspond to $F(k)=0$. The first solution with the largest $k$ corresponds to the ground state level with $E_0<0$. The subsequent solutions obey 
\begin{equation}
s_0\left(\ln\frac{k_{n+1}}{\Lambda}-\ln\frac{k_{n}}{\Lambda} \right)=-\pi
\end{equation}
or
\begin{equation}
\frac{E_{n+1}}{E_n}=\exp\left( -\frac{2 \pi}{s_0} \right),\quad E_n=\exp\left( -\frac{2 \pi n}{s_0} \right)E_0.
\label{eq:energyscaling}
\end{equation}
This corresponds to the tower of trimer bound states at the unitarity limit, with $E_n$ approaching zero exponentially for $n\to\infty$. For Eq.\ \eqref{eq:energyscaling} we have actually taken into account all zeros of $F(k)$. Since $\bar g^2(k)$ oscillates periodically, only half of these zeros correspond to $\bar g^2(k)>0$, while the other half has formally $\bar g^2(k)<0$. We may use the mapping discussed after Eq.\ \eqref{eq:subst} to obtain an equivalent picture with positive $\bar g^2$.

\begin{figure}
\centering
\includegraphics[width=0.5\linewidth]{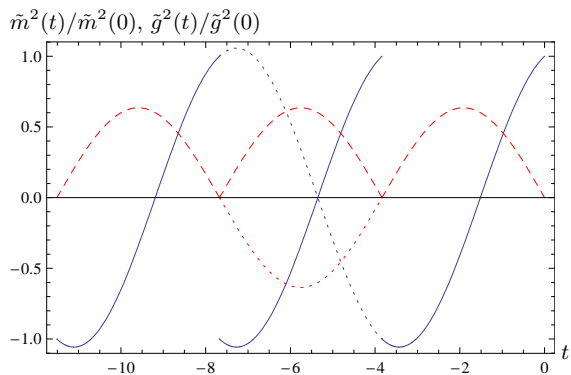}
\caption{Limit cycle in the renormalization group flow at the unitarity point $a^{-1}=0$, and for energy at the fermion threshold $\tilde{\mu}=\mu+E=0$. We plot the rescaled gap parameter of the trimer $\tilde m^2(t)$ (solid) and the rescaled Yukawa coupling $\tilde g^2(t)$ (dashed). The dotted curves would be obtained from naive continuation of the flow after the point where $\tilde g^2=0$.}
\label{fig:limitcycla}
\end{figure}

We may understand the repetition of states by the following qualitative picture. The coupling $\tilde m^2$, that is proportional to the energy gap of the trimer, starts on the ultraviolet scale with some positive value. The precise initial value is not important. The Yukawa-type coupling $\bar g$ vanishes initially so that the trion field $\chi$ is simply an auxiliary field which decouples from the other fields and is not propagating. However, quantum fluctuations lead to the emergence of a scattering amplitude between the original fermions $\psi$ and the bosons $\varphi$. We describe this by the exchange of a composite fermion $\chi$. 
This leads to an increase of the coupling $\tilde g^2$ and a decrease of the trion gap $\tilde m^2$. At some scale $t_1=\text{ln}(k_1/\Lambda)$ with $k_1^2\approx -\mu$ the coupling $\tilde m^2$ crosses zero which indicates that a trion state $\chi$ becomes the lowest energy excitation of the vacuum. Indeed, would we consider the flow without modifying the chemical potential, this would set an infrared cutoff that stops the flow at the scale $k_1\approx\sqrt{|\mu|}$ and the trion $\chi$ would be the gapless propagating particle while the original fermions $\psi$ and the bosons $\varphi$ are gapped since they have higher energy. 

Following the flow further to the infrared, we find that the Yukawa coupling $\tilde g^2$ decreases again until it reaches the point $\tilde g^2=0$ at $t=t_1^\prime$ (see Fig. \ref{fig:limitcycla}). Naive continuation of the flow below that scale would lead to $\tilde g^2<0$ and therefore imaginary Yukawa coupling $\tilde g$. However, since the trion field $\chi$ decouples from the other fields for $\tilde g=0$, we are not forced to use the same field $\chi$ as before. We can simply use another auxiliary field $\chi_2$ with very large gap $m^2_{\chi_2}=\bar m_{\chi_2}^2/\bar A_{\chi_2}$ to describe the scattering between fermions and bosons on scales $t<t_1^\prime$. We are then in the same position as on the scale $k=\Lambda$ and the process repeats. Starting from a positive value, the rescaled gap parameter $\tilde m_{\chi2}^2$ decreases as the infrared cutoff $k$ is lowered. At the scale $k_2$ it crosses zero which indicates that there is a second trimer bound state in the spectrum with energy per original fermion $E_2=-\mu-k_2^2$. Would we use the modified chemical potential $\tilde \mu =\mu+E_2=-k_2^2$, the flow would be stopped at the scale $k_2$ and the second trimer $\chi_2$ would be the propagating degree of freedom. This cycle repeats and corresponds precisely to the limit cycle scaling of Eq.\ \eqref{eq:matrixdifferentialequation}.

At the unitarity point with $a^{-1}=0$ and at the threshold energy $E=-\mu$ with $\tilde \mu=\mu+E=0$, this limit cycle scaling is not stopped and leads to an infinite tower of trimer bound states. The energy of this states comes closer and closer to the fermion-boson threshold energy $E=-\mu$.  In our language we recover the effect first predicted by Efimov \cite{Efimov1970}. 
A similar limit cycle description of the Efimov effect for identical bosons was given in the context of effective field theory in \cite{PhysRevLett.82.463, Bedaque1999444, PhysRevLett.85.908}.

The infinite limit cycle scaling occurs only directly at the Feshbach resonance with $a^{-1}=0$. Similar to a nonzero (modified) chemical potential $\tilde \mu$, also a nonzero inverse scattering length $a^{-1}$ provides an infrared cutoff that stops the flow. For example for negative scattering length $a$ and energy $E$ with $\tilde \mu=\mu+E=0$, the solution of the two body sector Eq.\ \eqref{eq:explicitsolutiontwobody} implies for the boson gap in the infrared $m_\varphi^2=-3\pi a^{-1}k/4 + k^2$ so that bosonic fluctuations are suppressed in comparison to the unitarity point with $a^{-1}=0$. This leads to a stop of the limit cycle scaling at the scale $k\approx3\pi |a^{-1}|/4$ so that the spectrum consists only of a finite number of trimer states. In the inset of Fig. \ref{fig:Efimov} we show the numerical result for the modified chemical potential $\tilde \mu$ of the first excited Efimov trimer.

\subsection{Experiments with Lithium}
From the work of Efimov it is known that the trion bound state (lowest energy Efimov state) is also expected to persist if the SU(3) symmetry is violated by a different location and strength for the Feshbach resonances between different pairs of atomic components \cite{Efimov1973, BraatenHammer}. In this subsection we generalize the SU(3) symmetric model discussed above to the case where the three resonances are at different positions. Recent measurements of the three-body loss coefficient in a three-component system of $^6$Li \cite{ottenstein:203202, Huckans2008} may find an interpretation in this way. 

\begin{figure}
\centering
\includegraphics[width=0.25\textwidth]{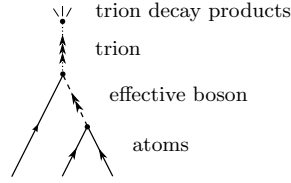}
\caption{Three-body loss process involving the trion.}
\label{fig:treeLossProcess}
\end{figure}

We investigate here a simple setting, where the loss arises from the formation of an intermediate trion bound state. The trion is not stable and subsequently decays into unspecified degrees of freedom -- possibly the ``molecule type'' dimers associated to the nearby Feshbach resonances. In turn, the trion formation from three atoms proceeds by the exchange of an effective bosonic field, as shown in Fig. \ref{fig:treeLossProcess}. 
Evaluating the matrix elements corresponding to the diagrams shown in Fig. \ref{fig:treeLossProcess}, we estimate the loss coefficient $K_3$ as being proportional to
\begin{equation}
p=\left| \sum_{i=1}^3 \frac{h_i g_i}{m_{\varphi i}^2}\frac{1}{\left(m_\chi^2-i \frac{\Gamma_\chi}{2}\right)}\right|^2.
\label{eq:decayprob}
\end{equation}
Here $m_\chi^2$ and $\Gamma_\chi$ are the trion gap parameter and decay width, while $m_{\varphi i}^2$ describes a type of gap parameter for the effective boson, such that its propagator can be approximated by $m_{\varphi i}^{-2}$. The Yukawa couplings $h_i$ couple the fermionic atoms to the effective boson, and the trion coupling $g_i$ accounts for the coupling between trion, atom and effective boson. We sum over the ``flavor'' indices $i=1,2,3$. We will estimate $m_\chi^2$, $m_{\varphi i}^2$, $h_i$ and $g_i$ using the flow equation method.

In the previous subsection we studied the SU(3) invariant Fermi gas close to a common Feshbach resonance and explored the manifestation of Efimov's effect. In contrast to this theoretical model, the system consisting of three-component $^{6}$Li atoms, which is of current experimental interest \cite{ottenstein:203202, Huckans2008}, does not possess this SU(3) symmetry. The main difference is that the resonances do not occur at the same magnetic field, and thus, for a given magnetic field $B$, the scattering lengths of different pairs of atoms, $(1,2)$, $(2,3)$, and $(3,1)$ differ from each other.

\begin{figure}
\centering
\includegraphics[width=0.5\textwidth]{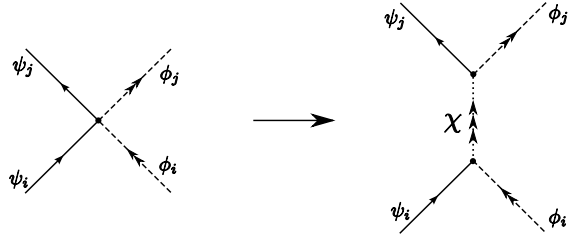}
\caption{Interaction between atoms $\psi$ and effective bosons $\varphi$ as mediated by the trion field $\chi$.}
\label{fig:tree1}
\end{figure}

In this section we adapt the model in Eq.\ \eqref{eq8:microscopicactiontrionmodel} to cope with this more general situation. Our truncation of the (euclidean) average action reads then
\begin{eqnarray}
\nonumber
\Gamma_k &=& \int_x {\bigg \{} \psi_i^*(\partial_\tau-\Delta-\mu)\psi_i\\
\nonumber
&& +\varphi_i^*\left[A_{\varphi i}(\partial_\tau-\Delta/2)+m_{\varphi i}^2\right]\varphi_i\\
\nonumber
&& + \chi^*\left[\partial_\tau-\Delta/3+m_\chi^2\right]\chi\\
\nonumber
&& +h_i \epsilon_{ijk}(\varphi_i^* \psi_j\psi_k-\varphi_i\psi_j^*\psi_k^*)\\
&&+ g_i(\varphi_i^* \psi_i^* \chi-\varphi_i \psi_i \chi^*) {\bigg \}},
\label{eq:action}
\end{eqnarray}
where we choose natural units $\hbar=2M=1$, with the atom mass $M$. We sum over the indices $i$, $j$, $k$ wherever they appear. Here $\psi=(\psi_1,\psi_2,\psi_3)^T$ denotes the fermionic atoms, $\varphi=(\varphi_1,\varphi_2,\varphi_3)^T$ a bosonic auxiliary field which mediates the four-fermion interaction and $\chi$ is a fermionic field representing the bound state of three atoms. Formally, this trion field is introduced via a generalized Hubbard-Stratonovich transformation on the microscopic scale and mediates the interaction between atoms $\psi$ and bosons $\varphi$. We show this schematically in Fig. \ref{fig:tree1}. 
In the limit $m_{\varphi i}^2\to\infty$, $h_i^2/m_{\varphi i}^2\to|\lambda_i|$, $m_\chi^2\to\infty$, $g_i^2/m_\chi^2\to |\lambda^{(3)}|$ the action describes pointlike two-body interactions between the atoms with strength $\lambda_i$ as well as a pointlike three-body interaction between atoms and dimers with strength $\lambda^{(3)}$. This corresponds to a zero-range approximation and describes the universal limit of dilute gases with large $s$-wave interactions. No detailed knowledge of the microscopic physics in necessary and physics depends only on the scattering length and a three-body parameter fixing the Efimov trimer energy spectrum \cite{BraatenHammer}.

We consider the ``vacuum limit'' where temperature and atom density go to zero. Then the chemical potential $\mu$ in Eq.\ \eqref{eq:action} satisfies $\mu\leq 0$. A negative chemical potential $\mu$ has the meaning of an energy gap for the fermions when some other particle (boson or trion) has a lower energy. 
The dominant difference to the SU(3) symmetric model arises from the different propagators of the bosonic fields $\varphi_1\widehat{=}\psi_2\psi_3$, $\varphi_2\widehat{=}\psi_3\psi_1$, and $\varphi_3\widehat{=}\psi_1\psi_2$. In addition, we allow in general for different Yukawa couplings $h_i$ corresponding to different widths of the three resonances. Also the Yukawa-like coupling $g_i$ that couples the different combinations of fermions $\psi_i$ and bosons $\varphi_i$ to the trion field $\chi\widehat{=}\psi_1\psi_2\psi_3$ is permitted to vary with the species involved. Although the SU(3) symmetry is explicitly broken, the system exhibits three global U(1) symmetries corresponding to the three conserved numbers of species of atoms.

The renormalization flow of the various couplings from the microscopic (UV), $k=\Lambda$, to the physical, macroscopic (IR) scale, $k=0$, is obtained by inserting the ``truncation'' \eqref{eq:action} into the exact flow equation \eqref{eq4:Wettericheqn}. The flow equations for the two-body sector, i.~e. for the boson propagator parameterized by $A_{\varphi i}$ and $m_{\varphi_i}^2$, are very similar to the SU(3) symmetric case ($t=\ln(k/\Lambda)$)
\begin{eqnarray}
\nonumber
\partial_t A_{\varphi i} &=& -\frac{h_i^2 k^5}{6\pi^2 (k^2-\mu)^2},\\
\partial_t m_{\varphi i}^2 &=& \frac{h_i^2 k^5}{6\pi^2 (k^2-\mu)^3}.
\label{eq:flowbosonprop}
\end{eqnarray}
Since the Yukawa couplings $h_i$ are not renormalized,
\begin{equation}
\partial_t h_i=0,
\label{eq:flowofh}
\end{equation}
we can immediately integrate the equations \eqref{eq:flowbosonprop}. The solutions can be found in Eq.\ \eqref{eq:explicitsolutiontwobody}. The microscopic values $m_{\varphi i}^2(\Lambda)$ (bare couplings) have to be chosen such that the physical scattering lengths (at $k=0$) between two fermions (renormalized couplings) are reproduced correctly. They are given by the exchange of the boson field $\varphi$. For example, the scattering length between the fermions 1 and 2 obeys
\begin{equation}
a_{12}=-\frac{h_3^2}{8\pi m_{\varphi 3}^2},
\label{mphifixing}
\end{equation} 
where all ``flowing parameters'' are evaluated at the macroscopic scale $k=0$ and for atoms at threshold energy and therefore $\mu=0$. We use this description for the scattering between fermions $\psi$ in terms of a composite boson field $\varphi$ also away from the resonance. We emphasize that the field $\varphi$ is not related to the closed channel Feshbach molecules of the nearby resonance. It rather describes an additional ``effective boson'', which may be seen as an auxiliary or Hubbard-Stratonovich field, allowing for a simple but effective description of the interaction between two fermions. For the numeric calculations in this note we will use large values of $h_i^2$ on the initial scale $\Lambda$. This corresponds to pointlike atom-atom interactions in the microscopic regime. 

Quite similar to the scattering between fermions $\psi$ in terms of the bosonic composite state $\varphi$ we use a description of the scattering between fermions $\psi$ and bosons $\varphi$ in terms of the trion field $\chi$. As an example, a process where the fermion $\psi_i$ and the boson $\varphi_i$ scatter to a fermion $\psi_j$ and a boson $\varphi_j$ is given by a tree level diagram as in Fig. \ref{fig:tree1}. For vanishing center-of-mass momentum the effective atom-boson coupling reads
\begin{equation}
\lambda_{i,j}^{(3)}=-\frac{g_i g_j}{m_\chi^2}.
\end{equation}
The flow equations for the three-body sector within our approximation are given by the flow of the ``mass term'' for the trion field
\begin{equation}
\partial_t m_\chi^2 = \sum_{i=1}^3 \frac{2 g_i^2 k^5}{\pi^2 A_{\varphi i}(3k^2-2\mu+2m_{\varphi i}^2/A_{\varphi i})^2}
\label{eq:flowofm}
\end{equation}
and the Yukawa-like coupling $g_1$ with flow equation
\begin{eqnarray}
\nonumber
\partial_t g_1  &=& -\frac{g_2 h_2 h_1 k^5\left(6k^2-5\mu+\frac{2m_{\varphi2}^2}{A_{\varphi 2}}\right)}{3\pi^2 A_{\varphi 2}(k^2-\mu)^2\left(3k^2-2\mu+\frac{2m_{\varphi2}^2}{A_{\varphi2}}\right)^2}\\
&&-\frac{g_3 h_3 h_1 k^5\left(6k^2-5\mu+\frac{2m_{\varphi3}^2}{A_{\varphi 3}}\right)}{3\pi^2 A_{\varphi 3}(k^2-\mu)^2\left(3k^2-2\mu+\frac{2m_{\varphi3}^2}{A_{\varphi3}}\right)^2}.
\label{eq:flowofg}
\end{eqnarray}
The flow equations for $g_2$ and $g_3$ can be obtained from Eq.\ \eqref{eq:flowofg} by permuting the indices $1$, $2$, $3$. 
For simplicity, we neglected in the flow equations \eqref{eq:flowofm} and \eqref{eq:flowofg} a contribution that arises from box-diagrams contributing to the atom-boson interaction. As described in the previous subsection this term can be incorporated into our formalism using scale-dependent fields, a procedure referred to as refermionization. Also terms of the form $\psi_i^*\psi_i \varphi^*_j\varphi_j$ with $i\neq j$, that are in principle allowed by the symmetries are neglected by our approximation in Eq.\ \eqref{eq:action}. We expect that their quantitative influence is sub dominant as it is the case for the SU(3) symmetric case \cite{Moroz2008}.

\begin{figure}
\centering
\includegraphics[width=0.6\textwidth]{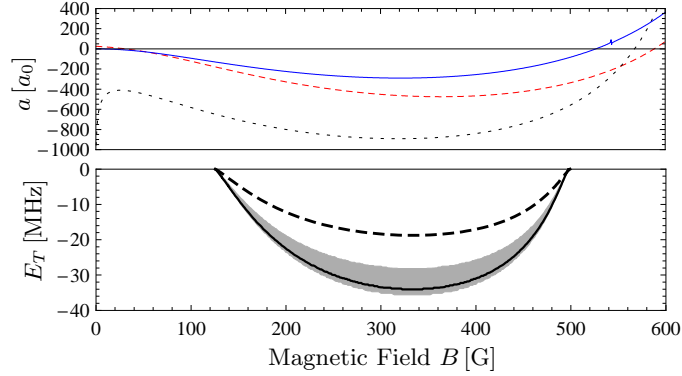}
\caption{{\itshape Upper panel:} Scattering length $a_{12}$ (solid), $a_{23}$ (dashed) and $a_{31}$ (dotted) as a function of the magnetic field $B$ for $^6$Li. These curves were calculated by P.~S.~Julienne 
and taken from Ref. 
{\itshape Lower panel:} Binding energy $E_T$ of the three-body bound state $\chi\widehat{=}\psi_1\psi_2\psi_3$. The solid line shows the result without the inclusion of box diagrams contributing to the atom-boson interaction and it corresponds to the initial value $h^2=100\, a_0^{-1}$, while the shaded region gives the result in the range $h^2=20\, a_0^{-1}$ (upper border) to $h^2=300\, a_0^{-1}$ (lower border). The dashed line corresponds to the calculated binding energy $E_T$ when the refermionization of the atom-boson interaction is taken into account.}
\label{fig:EnergiesLi}
\end{figure}

We apply our formalism to $^6$Li by choosing the initial values of $m_{\varphi i}^2$ at the scale $\Lambda$ such that the experimentally measured scattering lengths (see Fig. \ref{fig:EnergiesLi}) are reproduced. For $A_{\varphi i}(\Lambda)=1$, the value of $h_i$ parameterizes the momentum dependence of the interaction between atoms on the microscopic scale.  Close to the Feshbach resonance it is also connected to the width of the resonance $h_i^2\sim\Delta B$.  We choose here equal and large values for all three species $h_1=h_2=h_3=h$. This corresponds to pointlike interactions at the microscopic scale $\Lambda$. The zero-range limit is valid if the absolute values of the scattering lengths are much larger than the range of the interaction potentials which is typically given by the van der Waals length $l_\textrm{vdW}$. For $^6$Li one has $l_\textrm{vdW}\approx 62.5\, a_0$ and the loss resonances appear for values $a_{ij}>2\, l_\textrm{vdW}$. Therefore the zero-range approximation might be questionable. Since the precise value of $h$ is not known, we use the dependence of our results on $h$ as an estimate of their uncertainty within our truncation \eqref{eq:action}.
The initial values of the couplings $m_\chi^2$ and $g_i$ are parameters in addition to the scattering lengths which have to be fixed from experimental observation. For equal interaction between atoms $\psi$ and bosons $\varphi$ in the UV, the parameter to be fixed is
\begin{equation}
\lambda^{(3)}=-\frac{g^2(\Lambda)}{m_\chi^2(\Lambda)}
\end{equation}
with $g=g_1=g_2=g_3$. Pointlike interactions at the microscopic scale may be realized by $m_\chi^2(\Lambda)\to \infty$.

We solve the flow equations \eqref{eq:flowbosonprop}, \eqref{eq:flowofh}, \eqref{eq:flowofm} and \eqref{eq:flowofg} numerically. We find $m_\chi^2=0$ at $k=0$ 
for some range of $\lambda^{(3)}$ and $\mu\leq 0$ for large enough values of the scattering lengths $a_{12}$, $a_{23}$ and $a_{31}$. This indicates the presence of a bound state of three atoms $\chi\widehat{=}\psi_1\psi_2\psi_3$. The binding energy $E_T$ of this bound state is given by the chemical potential, $E_T=3|\mu|$ with $\mu$ fixed such that $m_\chi^2=0$. To compare with the recently performed experimental investigations of $^6$Li \cite{ottenstein:203202, Huckans2008}, we adapt the initial value $\lambda^{(3)}$ such that the appearance of this bound state corresponds to a magnetic field $B=125\, \text{G}$, the point where strong three-body losses have been observed. Using the same initial value of $\lambda^{(3)}$ also for other values of the magnetic field, all microscopic parameters are fixed. We can now proceed to the predictions of our model.

First we find that the bound state of three atoms exists in the magnetic field region from $B=125\, \text{G}$ to $B=498\, \text{G}$. The binding energy $E_T$ is plotted as the solid line in the lower panel of Fig. \ref{fig:EnergiesLi}. We choose here $h^2=100\, a_0^{-1}$, as appropriate for $^6$Li in the (1,2)-channel close to the resonance, while the shaded region corresponds to $h^2\in(20\, a_0^{-1},300\, a_0^{-1})$. If one includes the contribution to the flow of the atom-dimer interaction arising from box-diagrams by means of a refermionization procedure (see section \ref{ssect:SU3symmetricmodel}), the flow equations for the Yukawa couplings $g_i$, as in Eq.\ \eqref{eq:flowofg}, receive an additionial contribution
\begin{eqnarray}
m_\chi^2\left(-\frac{\partial_t \lambda_{ij}^{(3)}}{2 g_j}-\frac{\partial_t \lambda_{il}^{(3)}}{2 g_l}+\frac{g_i\partial_t \lambda_{jl}^{(3)}}{2 g_j g_l}\right).
\end{eqnarray}
Here we define $(i,j,l)=(1,2,3)$ and permutations thereof and we find
\begin{eqnarray}
\partial_t\lambda_{ij}^{(3)}=\frac{k^5 h_1 h_2 h_3 h_l(9k^2-7\mu+\frac{4 m_{\varphi l}^2}{A_{\varphi l}})}{6\pi^2 A_{\varphi l}(k^2-\mu)^3(3k^2-2\mu+\frac{2 m_{\varphi l}^2}{A_{\varphi l}})^2}.
\end{eqnarray}
This leads to a reduction of the trion binding energy $E_T$ and the result is shown as the dashed line in Fig. \ref{fig:EnergiesLi}.

As a second prediction, we present an estimate of the three-body loss coefficient $K_3$ that has been measured in the experiments by Jochim {\itshape et al.} \cite{ottenstein:203202} and O'Hara {\itshape et al.} \cite{Huckans2008}. For this purpose it is important to note that the fermionic bound state particle $\chi$ might decay into states with lower energies. These may be some deeply bound molecules not included in our calculation here. In order to include these decay processes we introduce a decay width $\Gamma_\chi$ of the trion. We first assume that such a loss process does not depend strongly on the magnetic field $B$ and therefore work with a constant decay width $\Gamma_\chi$. 
The decay width $\Gamma_\chi$ appears as an imaginary part of the trion propagator when continued to real time
\begin{equation}
G_\chi^{-1}=\omega-\frac{\vec p^2}{3}-m_\chi^2+i\frac{\Gamma_\chi}{2}.
\end{equation}
We now perform the calculation of the loss for the fermionic energy gap $\mu=0$, which corresponds to the open channel energy level. In the region from $B=125\, \text{G}$ to $B=498\, \text{G}$ the energy gap of the trion is then negative $m_\chi^2<0$.
The three-body loss coefficient $K_3$ for arbitrary $\Gamma_\chi$ is obtained as follows. The amplitude to form a trion out of three fermions with vanishing momentum and energy is given by $\sum_{i=1}^3 h_i g_i/m_{\varphi i}^2$.
The amplitude for the transition from an initial state of three atoms to a final state of the trion decay products (cf. Fig \ref{fig:treeLossProcess}) further involves the trion propagator that we evaluate in the limit of small momentum $\vec p^2=(\sum_i \vec p_i)^2\to0$, and small on-shell atom energies $\omega_i=\vec p_i^2$, $\omega=\sum_i \omega_i\to 0$. A thermal distribution of the initial momenta will induce some corrections. Finally, the loss coefficient involves the unknown vertices and phase space factors of the trion decay -- for this reason our computation contains an unknown multiplicative factor $c_K$. In terms of $p$ given by Eq.\ \eqref{eq:decayprob} we obtain the three-body loss coefficient
\begin{equation}
K_3=c_K \, p.
\end{equation}

\begin{figure}
\centering
\includegraphics[width=0.7\textwidth]{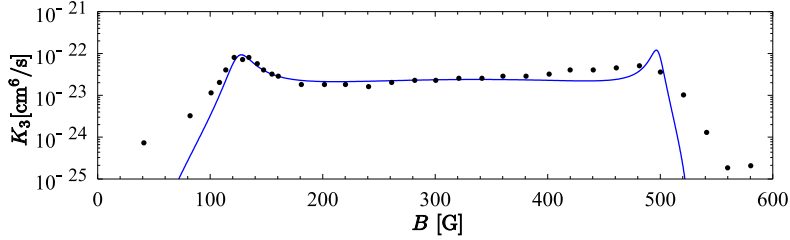}
\caption{Loss coefficient $K_3$ in dependence on the magnetic field $B$ as measured in \cite{ottenstein:203202} (dots). The solid line is the fit of our model to the experimental curve. We use here a decay width $\Gamma_\chi$ that is independent of the magnetic field $B$.}
\label{fig:Losscoefficient}
\end{figure}

Our result as well as the experimental data points \cite{ottenstein:203202} are shown in Fig. \ref{fig:Losscoefficient}. The agreement between the form of the two curves is already quite remarkable.

We have used three parameters, the location of the resonance at $B_0=125\,\text{G}$ which we translate into $\lambda^{(3)}$, the overall amplitude $c_K$ and the decay width $\Gamma_\chi$. They are essentially fixed by the peak at $B_0=125\,\text{G}$. The extension of the loss rate away from the peak involves then no further parameter. Our estimated decay width $\Gamma_\chi$ corresponds to a short lifetime of the trion of the order of $10^{-8}\,\textrm{s}$.

Our simple prediction involves a rather narrow second peak around $B_1\approx500\,\text{G}$, where the trion energy becomes again degenerate with the open channel, cf. Fig. \ref{fig:EnergiesLi}. The width of this peak is fixed so far by the assumption that the decay width $\Gamma_\chi$ is independent of the magnetic field. This may be questionable in view of the close-by Feshbach resonance and the fact that the trion may actually decay into the associated molecule-like bound states which have lower energy. We have tested several reasonable approximations, which indeed lead to a broadening or even disappearance of the second peak, without much effect on the intermediate range of fields $150\,\text{G}<B<400\,\text{G}$, see \cite{SchmidtDiplom2009}.

Shortly before finishing the calculations discussed here similar work by E.~Braaten {\itshape et al.} \cite{Braaten:2008wd} as well as P.\ Naidon and M.\ Ueda \cite{Naidon} was published. In ref.\ \cite{Braaten:2008wd} the scattering amplitudes of the loss process are calculated in the zero-range approximation with the help of generalized STM equations and a two parameter fit is given for the measured loss coefficient. Naidon and Ueda \cite{Naidon} perform a quantum mechanical calculation using the hyperspherical formalism as in \cite{Efimov1970}. In addition to a three parameter fit for the loss coefficient the authors calculate the Efimov trimer binding energy. The complex three-body parameters introduced in \cite{Braaten:2008wd, Naidon} correspond to the parameters $\lambda^{(3)}$ and $\Gamma_\chi$ in our setting. The authors of both papers conclude that the three-body loss is due to an Efimov state near the three-atom threshold which is consistent with the scenario depicted here. In the few-body regime our predictions should be equivalent to those obtained in ref.\ \cite{Braaten:2008wd, Naidon} and, indeed, the results for the loss coefficient are consistent with our calculation shown in Fig. \ref{fig:Losscoefficient}. Without refermionization we find a difference in the prediction of the trion binding energy compared to ref.\ \cite{Naidon}. However, with the improved truncation (dashed line in Fig. \ref{fig:EnergiesLi}) the agreement becomes better and the inclusion of an interaction of the type $\psi_i^*\psi_i \varphi^*_j\varphi_j$ with $i\neq j$ will induce further quantitative corrections.

In conclusion, a rather simple bound state exchange picture describes rather well the observed enhancement of the three-body loss coefficient in a range of magnetic fields between $100\,\text{G}$ and $520\,\text{G}$. A similar trion dominated three-body loss is possible for large $B$ ($B\gtrsim850\,\text{G}$), where also a trion bound state with energy below the open channel exists. However, the dimer bound states are now above the open channel level, such that the trion decay may be strongly altered. The role of trion bound states in the resonance region is an interesting subject by its own, that can be explored by our functional renormalization group methods with an extended truncation. The calculated energy spectrum may be experimentally tested by radio frequency spectroscopy although this might be difficult due to the short lifetimes of the trion. An advantage of the method we use is that the flow equations can also be extended to the case of nonvanishing temperature and density similar as for the BCS-BEC crossover in the two-component Fermi gas, see section \ref{sec:Particle-holefluctuationsandtheBCS-BECCrossover}. The few-body calculation we presented here is the necessary premise for finite temperature and density calculations and fixes the microscopic parameters for the three-component $^6$Li Fermi gas. Therefore this work provides a good starting point for the investigation of the interesting phase diagram of this system, a challenge for both theory and experiment.

\chapter{Many-body physics}
\label{ch:Many-bodyphysics}

	\section{Bose-Einstein Condensation in three dimensions}
	\label{sec:Bose-EinsteinCondensationinthreedimensions}
	In this section we discuss the many-body results obtained for the Bose gas model in Eq.\ \eqref{microscopicaction} using the approximation scheme described in section \ref{sec:TruncationBosegas}. We start with the phase diagram at zero temperature (quantum phase diagram) and move then to nonzero temperature. After describing the general features of the thermal phase transition we discuss the critical temperature for an interacting Bose gas and several thermodynamic observables ranging from pressure and energy density to different sound velocities.

\subsection{Different methods to determine the density}
The density sets a crucial scale for our problem. Its precise determination is mandatory for quantitative precision. We will discuss two different methods for its determination and show that the results agree within our precision. For $T=0$, we also find agreement with the Ward identity $n=\rho_0$.

The first method is to derive flow equations for the density. This has the advantage that the occupation numbers for a given momentum $\vec{p}$ are mainly sensitive to running couplings with $k^2=\vec{p}^2$. In the grand canonical formalism, the density is defined by
\begin{equation}
n=-\frac{\partial}{\partial \mu}\frac{1}{\Omega}\Gamma[\varphi]{\Big |}_{\varphi=\varphi_0,\mu=\mu_0}
\end{equation}
We can formally define a $k$-dependent density $n_k$ by
\begin{equation}
n_k=-\frac{\partial}{\partial \mu}\frac{1}{\Omega}\Gamma_k[\varphi]{\Big |}_{\varphi=\varphi_0,\mu=\mu_0}=-(\partial_\mu U)(\rho_0,\mu_0).
\end{equation}
The flow equation for $n_k$ is given in Eq.\ \eqref{eqFlowprescriptionnk} and the physical density obtains for $k=0$. The term $\partial_\mu \zeta \big{|}_{\rho_0,\mu_0}$ that enters Eq.\ \eqref{eqFlowprescriptionnk} is the derivative of the flow equation \eqref{eqFlowpotentialMatrix} for $U$ with respect to $\mu$. To compute it, we need the $\mu$-dependence of the propagator $G_k$ in the vicinity of $\mu_0$. Within a systematic derivative expansion, we use the expansion of $U(\rho,\mu)$ and the kinetic coefficients $Z_1$ and $Z_2$ to linear order in $(\mu-\mu_0)$, as described in section \ref{sec:Derivativeexpansionandwardidentities}. Here, $Z_1(\rho,\mu)$ and $Z_2(\rho,\mu)$ are the coefficient functions of the terms linear in the $\tau$-derivative and linear in $\Delta$, respectively. 
No reasonable qualitative behavior is found, if the linear dependence of $Z_1$ and $Z_2$ on $(\mu-\mu_0)$ is neglected. Also, the scale dependence of $\alpha$ and $\beta$ are quite important. The flow equations for $\alpha$ and $\beta$ can be obtained directly by differentiating the flow equation of the effective potential with respect to $\mu$ and $\rho$, cf. Eq.\ \eqref{eqflowalphabeta}. The situation is more complicated for the kinetic coefficients $Z_1^{(\mu)}=\partial_\mu Z_1(\rho_0,\mu_0)$ and $Z_2^{(\mu)}=\partial_\mu Z_2(\rho_0,\mu_0)$. Their flow equations have to be determined by taking the $\mu$-derivative of the flow equation for $Z_1(\rho, \mu)$ and $Z_2(\rho,\mu)$. As discussed in section \ref{sec:Derivativeexpansionandwardidentities}, we use in this paper the approximation $Z_1^{(\mu)}=Z_2^{(\mu)}=2V=2V_1(\rho_0,\mu_0)$.

As a check of both our method and our numerics, we also use another way to determine the particle density. This second method is more robust with respect to shortcomings of the truncation, but less adequate for high precision calculations as needed e.g. to determine the condensate depletion. The second method determines the pressure $p=-U(\rho_0,\mu_0)$ as a function of the chemical potential $\mu_0$. Here, the effective potential is normalized by $U(\rho_0=0,\mu_0)=0$ at $T=0$, $n=0$. The flow of the pressure can be read of directly from the flow equation of the effective potential and is independent of the couplings $\alpha$ and $\beta$. We calculate the pressure as a function of $\mu$ and determine the density $n=\frac{\partial}{\partial \mu}p$ by taking the $\mu$-derivative numerically. It turns out that $p$ is in very good approximation given by $p=c\,\mu^2$, where the constant $c$ can be determined from a numerical fit. The density is thus linear in $\mu$.  

\begin{figure}
\centering
\includegraphics[width=0.5\textwidth]{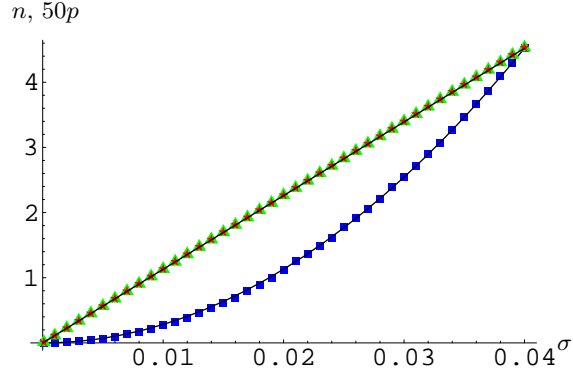}
\caption{Pressure and density as a function of the chemical potential at $T=0$. We use three different methods: $n=-\partial_\mu U_{\text{min}}$ from the flow equation (triangles), $n=\rho_0$ as implied by Galilean symmetry (stars) and $n=\partial_\mu p$, where the pressure $p=-U$ (boxes) was obtained from the flow equation and phenomenologically fitted by $p=56.5 \mu^2$ (solid lines). Units are arbitrary and we use $a=3.4\cdot 10^{-4}$, $\Lambda=10^3$.}
\label{densitycompared}
\end{figure}

At zero temperature and for $\tilde{v}=0$, we can additionally use the Ward identities connected to Galilean symmetry, which yield $n=\rho_0$. We compare our methods in figure \ref{densitycompared} and find that they give numerically the same result. We stress again the importance of a reliable method to determine the density, since we often rescale variables by powers of the density to obtain dimensionless variables.

\subsection{Quantum depletion of condensate}
We want to split the density into a condensate part $n_c$ and a density for uncondensed particles or "depletion density" $n_d=n-n_c$. For our model the condensate density is given by the "bare" order parameter
\begin{equation}
n_c=\bar{\rho}_0=\bar{\rho}_0(k=0).
\end{equation}
In order to show this, we introduce occupation numbers $n(\vec{p})$ for the modes with momentum $\vec{p}$ with normalization
\begin{equation}
\int_{\vec{p}}n(\vec{p})=n.
\end{equation}
One formally introduces a $\vec{p}$ dependent chemical potential $\mu(\vec{p})$ in the grand canonical partition function
\begin{equation}
e^{-\Gamma_{\text{min}}[\mu]}=\text{Tr}e^{-\beta(H-\Omega_3\int_{\vec{p}}\mu(\vec{p})n(\vec{p}))},
\end{equation}
with three dimensional volume $\Omega_3=\int_{\vec{x}}$. Then one can define the occupation numbers by
\begin{equation}
n(\vec{p})=-\frac{\delta}{\delta \mu(\vec{p})}\frac{1}{\beta \Omega_3}\Gamma[\varphi,\mu(\vec{p})]{\Big |}_{\varphi=\varphi_0,\mu(\vec{p})=\mu_0}.
\end{equation}
This construction allows us to use $k$-dependent occupation numbers by the definition
\begin{equation}
n_k(\vec{p})=-\frac{\delta}{\delta \mu(\vec{p})}\frac{1}{\beta \Omega_3}\Gamma_k[\varphi,\mu]{\Big |}_{\varphi=\varphi_0,\mu(\vec{p})=\mu_0}.
\end{equation}
One can derive a flow equation for this occupation number $n_k(\vec{p})$ \cite{WetterichOccupationNumbers}:
\begin{eqnarray}
\nonumber
\partial_k n_k(\vec{p}) &=& -\frac{1}{2}\frac{\delta}{\delta \mu(\vec{p})}\frac{1}{\beta \Omega_3}\text{Tr}\left\{(\Gamma^{(2)}+R_k)^{-1}\partial_k R_k\right\}\\
& & +\frac{\partial}{\partial \bar \rho}\frac{\delta}{\delta \mu(\vec{p})}\frac{1}{\beta \Omega_3}\Gamma[\varphi,\mu](\partial_k \bar\rho_0).
\label{flowofnp}
\end{eqnarray}

We split the density occupation number into a $\delta$-distribution like part and a depletion part, which is regular in the limit $\vec{p}\rightarrow0$
\begin{equation}
n_k(\vec{p})=n_{c,k}\,\delta(\vec{p})+n_{d,k}(\vec{p}).
\end{equation}
One can see from the flow equation for $n_k(\vec{p})$ that the only contribution to $\partial_k n_{c,k}$ comes from the second term in equation \eqref{flowofnp}. Within a more detailed analysis \cite{WetterichOccupationNumbers} one finds
\begin{equation}
\partial_k n_{c,k}=\partial_k\bar{\rho}_{0,k}.
\end{equation}
We therefore identify the condensate density with the bare order parameter
\begin{equation}
n_c=\bar{\rho}_0=\frac{\rho_0}{\bar{A}}=\bar{\varphi}_0^2.
\end{equation}
Correspondingly, we define the $k$-dependent quantities
\begin{equation}
n_{c,k}=\bar{\rho}_{0,k},\quad n_k=n_{c,k}+n_{d,k}
\end{equation}
and compute $n_d=n_d(k=0)$ by a solution of its flow equation. 

Even at zero temperature, the repulsive interaction connected with a positive scattering length $a$ causes a portion of the particle density to be outside the condensate. From dimensional reasons, it is clear, that $n_d/n=(n-n_c)/n$ should be a function of $an^{1/3}$. The prediction of Bogoliubov theory or, equivalently, mean field theory, is $n_d/n=\frac{8}{3\sqrt{\pi}}(an^{1/3})^{3/2}$. We may determine the condensate depletion from the solution to the flow equation for the particle density, $n=n_{k=0}$, and $n_c=\bar{\rho}_0=\bar{\rho}_0(k=0)$. 

\begin{figure}
\centering
\includegraphics[width=0.5\textwidth]{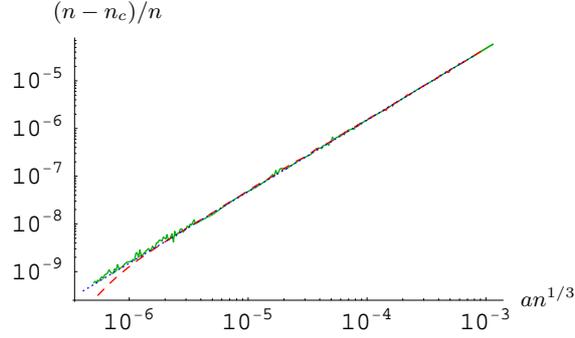}
\caption{Condensate depletion $(n-n_c)/n$ as a function of the dimensionless scattering length $a n^{1/3}$. For the solid curve, we vary $a$ with fixed $n=1$, for the dashed curve we vary the density at fixed $a=10^{-4}$. The dotted line is the Bogoliubov-Result $(n-n_c)/n=\frac{8}{3\sqrt{\pi}}(a n^{1/3})^{3/2}$ for reference. We find perfect agreement of the three determinations. The fluctuations in the solid curve for small $a n^{1/3}$ are due to numerical uncertainties. Their size demonstrates our numerical precision.}
\label{figDepletiond3}
\end{figure}

From Galilean invariance for $T=0$ and $\tilde{v}=0$, it follows that
\begin{equation}
\frac{n_d}{n}=\frac{\rho_0-\bar{\rho}_0}{\rho_0}=1-\frac{1}{\bar{A}},
\end{equation}
with $\bar{A}=\bar{A}(k=0)$. This gives an independent determination of $n_c$. 
In figure \ref{figDepletiond3} we plot the depletion density obtained from the flow of $n$ and $\bar{\rho}_0$ over several orders of magnitude. Apart from some numerical fluctuations for small $an^{1/3}$, we find that our result is in full agreement with the Bogoliubov prediction.

\subsection{Quantum phase transition}
\label{ssec:Quantumphasdiagram}
For $T=0$ a quantum phase transition separates the phases with $\rho_0=0$ and $\rho_0>0$.
In this section, we investigate the phase diagram at zero temperature in the cube spanned by the dimensionless parameters $\tilde{\mu}=\frac{\mu}{\Lambda^2}$, $\tilde{a}=a\Lambda$ and $\tilde{v}=\frac{V_\Lambda}{S_\Lambda^2}\Lambda^2$. This goes beyond the usual phase transition for nonrelativistic bosons, since we also include a microscopic second $\tau$-derivative $\sim\tilde{v}$, and therefore models with a generalized microscopic dispersion relation.
For non-vanishing $\tilde{v}$ (i.e. for a nonzero initial value of $V_1$ with $V_2=V_3=0$ in section \ref{sec:Derivativeexpansionandwardidentities}), the Galilean invariance at zero temperature is broken explicitly. For large $\tilde{v}$, we expect a crossover to the "relativistic" $O(2)$ model. If we send the initial value of the coefficient of the linear $\tau$-derivative $S_\Lambda$ to zero, we obtain the limiting case $\tilde{v}\rightarrow\infty$. The symmetries of the model are now the same as those of the relativistic O(2) model in four dimensions. The space-time-rotations or Lorentz symmetry replace Galilean symmetry. 

\begin{figure}
\centering
\includegraphics[width=0.5\textwidth]{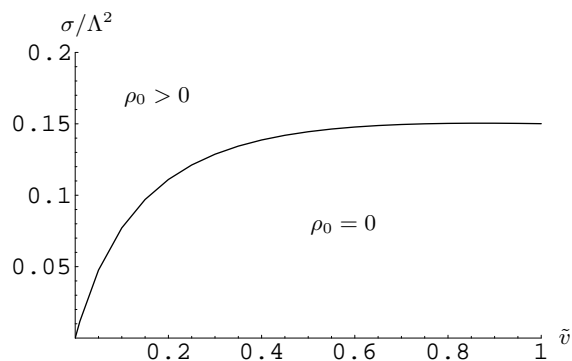}
\caption{Quantum phase diagram in the $\tilde{\mu}$-$\tilde{v}$ plane for $\tilde{a}=1$.}
\label{figQPTsigmava1}
\end{figure}

It is interesting to study the crossover between the two cases. Since our cutoff explicitly breaks Lorentz symmetry, we investigate in this paper only the regime $\tilde{v}\lesssim1$. Detailed investigations of the flow equations for $\tilde{v}\rightarrow\infty$ can be found in the literature \cite{Papenbrock:1994kf, Berges2000ew, PhysRevA.60.1442, PhysRevB.68.064421, PhysRevD.67.065004, Bervillier2007}. The phase diagram in the $\tilde{\mu}-\tilde{v}$ plane with $\tilde{a}=1$ is shown in figure \ref{figQPTsigmava1}. The critical chemical potential first increases linearly with $\tilde{v}$ and then saturates to a constant. The slope in the linear regime as well as the saturation value depend linearly on $\tilde{a}$ for $\tilde{a}<1$. 

At $T=0$, the critical exponents are everywhere the mean field ones ($\eta=0$, $\nu=1/2$). This is expected: It is the case for $\tilde{v}=0$ \cite{Wetterich:2007ba, Uzunov1981, SachdevQPT}, and for $\tilde{v}=\infty$ the theory is equivalent to a relativistic $O(2)$ model in $d=3+1$ dimensions. This is just the upper critical dimension of the Wilson-Fisher fixed point \cite{PhysRevLett.28.240}. 

\begin{figure}
\centering
\includegraphics[width=0.5\textwidth]{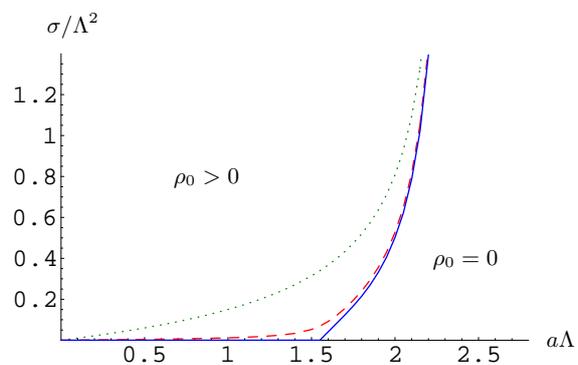}
\caption{Quantum phase diagram in the $\tilde{\mu}$-$\tilde{a}$ plane for $\tilde{v}=1$ (dotted), $\tilde{v}=0.01$ (dashed) and $\tilde{v}=0$ (solid).}
\label{figQPTsigmaofa}
\end{figure}

From section 3 we know that for $\tilde{v}=0$ the parameter $\tilde{a}$ is limited to $\tilde{a}<\frac{3\pi}{4}\approx 2.356$. For $\tilde{v}=0$ and a small scattering length $a\rightarrow0$, a second order quantum phase transition divides the phases without spontaneous symmetry breaking  for $\mu<0$ from the phase with a finite order parameter $\rho_0>0$ for $\mu>0$. It is an interesting question, whether this quantum phase transition at $\mu=0, \tilde{v}=0$ also occurs for larger scattering length $a$. We find in our truncation that this is indeed the case for a large range of $a$, but  not for $\tilde{a}>1.55$. Here, the critical chemical potential suddenly increases to large positive values as shown in Fig. \ref{figQPTsigmaofa}. For $\tilde{v}>0$ this increase happens even earlier. (For a truncation with $V_1\equiv0$, the phase transition would always occur at $\mu=0$.) We plot the $\tilde{\mu}-\tilde{a}$ plane of the phase diagram for different values of $\tilde{v}$ in figure \ref{figQPTsigmaofa}. The form of the critical line can be understood by considering the limits $\tilde{v}\rightarrow0$ as well as $\tilde{a}\rightarrow0$.  

\begin{figure}
\centering
\includegraphics[width=0.5\textwidth]{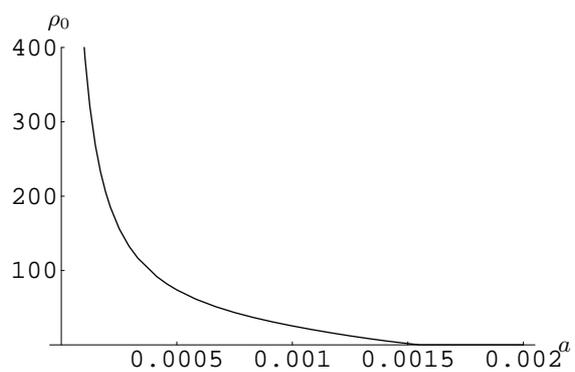}
\caption{Quantum phase transition for fixed chemical potential $\mu=1$, with $\Lambda=10^3$. The density $\rho_0=n$ as a function of the scattering length $a$ goes to zero at a critical $a_c \Lambda=1.55$, indicating a second order quantum phase transition at that point.}
\label{figQPTrhoofa}
\end{figure}

For a fixed chemical potential, the order parameter $\rho_0$ as a function of $a$ goes to zero at a critical value $a_c$ as shown in Fig. \ref{figQPTrhoofa}. This happens in a continuous way and the phase transition is therefore of second order.  For $\mu\rightarrow0$, we find $a_c=1.55 \Lambda^{-1}$. We emphasize, however, that $a_c$ is of the order of the microscopic distance $\Lambda^{-1}$. Universality may not be realized for such values, and the true phase transition may depend on the microphysics. For example, beyond a critical value for the repulsive interaction, the system may form a solid. Ultracold atom gases correspond to metastable states which may lose their relevance for $a\rightarrow\Lambda^{-1}$. For $\tilde{v}>0$ and $\mu\ll\Lambda^2$ the phase transition occurs for $a_c \Lambda \ll 1$ such that universal behavior is expected.

\subsection{Thermal depletion of condensate}
\label{ssec:Phasediagramatnonzerotemperature}
So far, we have only discussed the vacuum and the dense system at zero temperature. A non vanishing temperature $T$ will introduce an additional scale in our problem. For small $T\ll n^{2/3}$ we expect only small corrections. However, as $T$ increases the superfluid order will be destroyed. Near the phase transition for $T \approx T_c$ and for the disordered phase for $T>T_c$, the characteristic behavior of the boson gas will be very different from the $T\rightarrow0$ limit.

\begin{figure}
\centering
\includegraphics[width=0.5\textwidth]{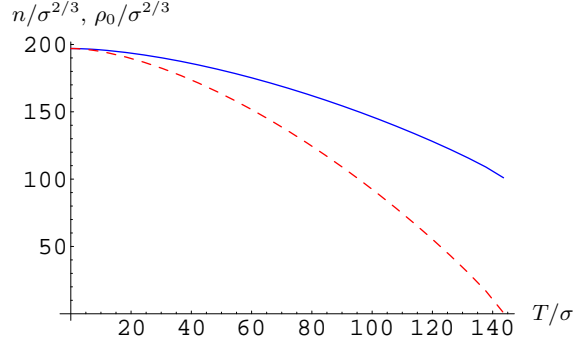}
\caption{Density $n/{\mu^{2/3}}$ (solid) and order parameter $\rho_0/{\mu^{2/3}}$ (dashed) as a function of the temperature $T/\mu$. The units are arbitrary with $a=2\cdot 10^{-4}$ and $\Lambda=10^3$. The plot covers only the superfluid phase. For higher temperatures, the density is given by the thermal contribution $n=n_T$ only.}
\label{fignoftemperature}
\end{figure}

For $T>0$ the particle density $n$ receives a contribution from a thermal gas of bosonic (quasi-) particles. It is no longer uniquely determined by the superfluid density $\rho_0$. We may write 
\begin{equation}
n=\rho_0+n_T
\label{eqDensityatT}
\end{equation}
and observe, that $n_T=0$ is enforced by Galilean symmetry only for $T=0, V_\Lambda=0$. The heat bath singles out a reference frame, such that for $T>0$ Galilean symmetry no longer holds. In our formalism, the thermal contribution $n_T$ appears due to modifications of the flow equations for $T\neq0$. We start for high $k$ with the same initial values as for $T=0$. As long as $k\gg \pi T$ the flow equations receive only minor modifications. For $k \approx \pi T$ or smaller, however, the discreteness of the Matsubara sum has important effects. We plot in Fig. \ref{fignoftemperature} the density as a function of $T$ for fixed $\mu=1$.

\begin{figure}
\centering
\includegraphics[width=0.5\textwidth]{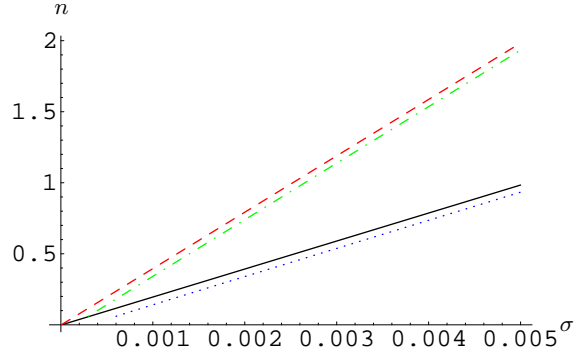}
\caption{Density $n$ for different temperatures and scattering length. We plot $n(\mu)$ in arbitrary units, with $\Lambda=10^3$, and for a scattering length $a=2\cdot10^{-4}$ (solid and dotted), $a=10^{-4}$ (dashed and dashed-dotted). The temperature is $T=0$ (solid and dashed) and $T=1$ (dotted and dashed-dotted).}
\label{fignofsigma}
\end{figure}

In Fig. \ref{fignofsigma} we show $n(\mu)$, similar to Fig. \ref{densitycompared}, but now for different $a$ and $T$. For $T=0$ the scattering length sets the only scale besides $n$ and $\mu$, such that by dimensional arguments $a^2\mu=f(a^3 n)$. Bogoliubov theory predicts 
\begin{equation}
f(x)=8\pi x(1+\frac{32}{3\sqrt{\pi}}x^{1/2}).
\end{equation}
The first term on the r.h.s. gives the contribution of the ground state, while the second term is added by fluctuation effects. For small scattering length $a$, the ground state contribution dominates. We have then $\mu\sim a$ for $n=1$ and $\mu/n$ can be treated as a small quantity. For $T\neq 0$ and small $a$ one expects $\mu=g(T/n^{2/3})an$. The curves in Fig. \ref{fignofsigma} for $T=1$ show that the density, as a function of $\mu$, is below the curve obtained at $T=0$. This is reasonable, since the statistical fluctuations now drive the order parameter $\rho_0$ to zero. At very small $\mu$, the flow enters the symmetric phase. The density is always positive, but for simplicity, we show the density as a function of $\mu$ in figure \ref{fignofsigma} only in those cases, where the flow remains in the phase with spontaneous $U(1)$ symmetry breaking. 

For temperatures above the critical temperature, the order parameter $\rho_0$ vanishes at the macroscopic scale and so does the condensate density $n_c=\bar{\rho_0}=\frac{1}{\bar{A}}\rho_0$. The density is now given by a thermal distribution of particles with nonzero momenta. 
Up to small corrections from the interaction $\sim aT$, it is described by a free Bose gas, 
\begin{equation}
n= \frac{T^{3/2}}{(4\pi)^{3/2}}g_{3/2}(e^{\beta \mu}),
\end{equation}
with the "Bose function"
\begin{equation}
g_p(z)=\frac{1}{\Gamma(p)}\int_0^\infty dx\,x^{p-1}\frac{1}{z^{-1}e^x-1}.
\end{equation}

\begin{figure}
\centering
\includegraphics[width=0.5\textwidth]{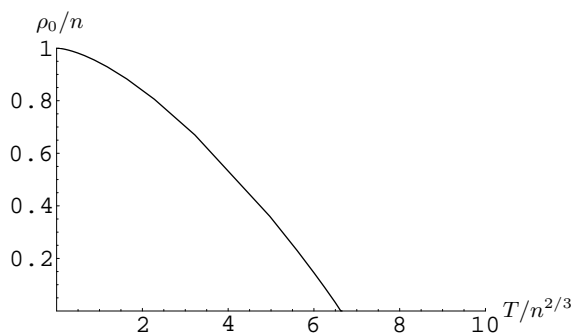}
\caption{Order parameter $\rho_0/n$ as a function of the dimensionless temperature $T/(n^{2/3})$ for scattering length $a=10^{-4}$. Here, we varied $T$ keeping $\mu$ fixed. Numerically, this is equivalent to varying $\mu$ with fixed $T$.\label{figrhooftemperature}}
\end{figure}

\begin{figure}
\centering
\includegraphics[width=0.5\textwidth]{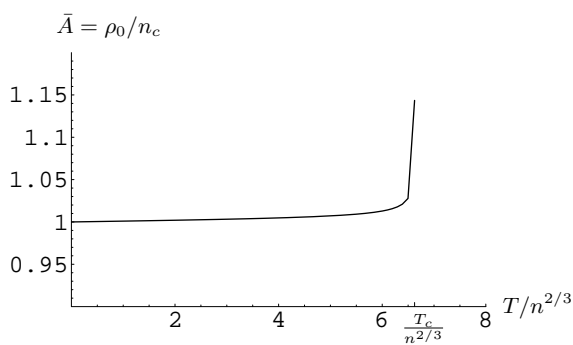}
\caption{Order parameter divided by the condensate density $\bar{A}=\rho_0/n_c$, as a function of the dimensionless temperature $T/(n^{2/3})$, and for scattering length $a=10^{-4}$. Here, we varied $T$ keeping $\mu$ fixed. Numerically, this is equivalent to varying $\mu$ with fixed $T$. The plot covers only the phase with spontaneous symmetry breaking. For higher temperatures, the symmetric phase has $\rho_0=n_c=0$. The divergence of $\bar{A}$ for $T\rightarrow T_c$ reflects the anomalous dimension $\eta$ of the Wilson-Fisher fixed point.}
\label{Abaroftemperature}
\end{figure}

In figure \ref{figrhooftemperature} we show the dimensionless order parameter $\rho_0/n$ as a function of the dimensionless temperature $T/n^{2/3}$. 
This plot shows the second order phase transition from the phase with spontaneous $U(1)$ symmetry breaking at small temperatures to the symmetric phase at higher temperatures. The critical temperature $T_c$ is determined as the temperature, where the order parameter just vanishes - it is investigated in the next section. Since we find $(\bar{A}-1)\ll1$, the condensate fraction $n_c/n=\bar{\rho_0}/n=\rho_0/(\bar{A}n)$ as a function of $T/n^{2/3}$ resembles the order parameter $\rho_0/n$. We plot $\bar{A}$ as a function of $T/n^{2/3}$ in Fig. \ref{Abaroftemperature}. Except for a narrow region around $T_c$, the deviations from one remain indeed small. Near $T_c$ the gradient coefficient $\bar{A}$ diverges according to the anomalous dimension, $\bar{A}\sim \xi^\eta$, with $\eta$ the anomalous dimension. The correlation length $\xi$ diverges with the critical exponent $\nu$, $\xi\sim |T-T_c|^{-\nu}$, such that
\begin{equation}
\bar{A}\sim |T-T_c|^{-\eta\nu}.
\end{equation}
Here, $\eta$ and $\nu$ are the critical exponents for the Wilson Fisher fixed point of the classical three-dimensional $O(2)$ model, $\eta=0.0380(4)$, $\nu=0.67155(27)$ \cite{Pelissetto2002549, Berges2000ew, PhysRevA.60.1442, PhysRevB.68.064421, PhysRevD.67.065004, Bervillier2007}.

\begin{figure}
\centering
\includegraphics[width=0.5\textwidth]{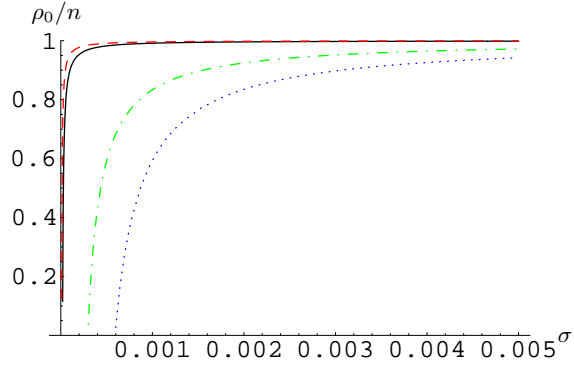}
\caption{Order parameter divided by the density, $\rho_0/n$, as a function of the chemical potential. We use arbitrary units with $\Lambda=10^3$. The curves are given for a scattering length $a=2\cdot10^{-4}$ (solid and dotted), $a=10^{-4}$ (dashed and dashed-dotted) and temperature $T=0.1$ (solid and dashed) and $T=1$ (dotted and dashed-dotted). At zero temperature, Galilean invariance implies $\rho_0=n$, which we find within our numerical resolution.}
\label{figrhoofsigma}
\end{figure}

In figure \ref{figrhoofsigma} we plot $\rho_0/n$ as a function of the chemical potential $\mu$ for different temperatures and scattering lengths. We find, that $\rho_0/n=1$ is indeed approached in the limit $T\rightarrow0$, as required by Galilean invariance. All figures of this section are for $\tilde{v}=0$. The modifications for $\tilde{v}\neq0$ are mainly quantitative, not qualitative. 

\subsection{Critical temperature}
The critical temperature $T_c$ for the phase transition between the superfluid phase at low temperature and the disordered or symmetric phase at high temperature depends on the scattering length $a$. By dimensional reasoning, the temperature shift $\Delta T_c=T_c(a)-T_c(a=0)$ obeys $\Delta T_c/T_c\sim an^{1/3}$. The proportionality coefficient cannot be computed in perturbation theory \cite{Andersen:2003qj}. It depends on $\tilde{v}$ and we concentrate here on $\tilde{v}=0$. Monte-Carlo simulations in the high temperature limit, where only the lowest Matsubara frequency is included, yield $\Delta T_c/T_c=1.3 \,a n^{1/3}$ \cite{PhysRevLett.87.120401, PhysRevLett.87.120402}. Within the same setting, renormalization group studies \cite{PhysRevLett.83.1703, blaizot:051116, blaizot:051117, PhysRevA.69.061601, PhysRevA.70.063621} yield a similar result, for  composite bosons see \cite{Diehl:2007th}. Near $T_c$, the long wavelength modes with momenta $\vec{p}^2\ll(\pi T)^2$ dominate the "long distance quantities". Then a description in terms of a classical three dimensional system becomes valid. This "dimensional reduction" is achieved by "integrating out" the nonzero Matsubara frequencies. 
However, both $\Delta T_c/T_c$ and $n$ are dominated by modes with momenta $\vec{p}^2\approx(\pi T_c)^2$ such that corrections to the classical result may be expected.

\begin{figure}
\centering
\includegraphics[width=0.5\textwidth]{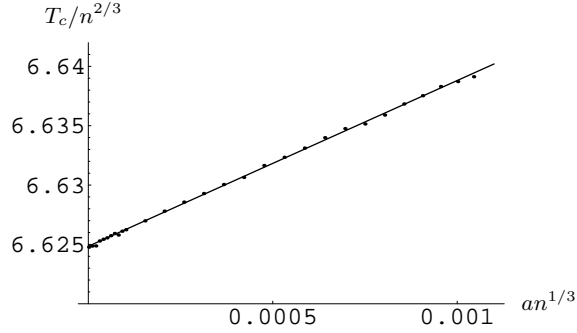}
\caption{Dimensionless critical temperature $T_c/(n^{2/3})$ as a function of the dimensionless scattering length $an^{1/3}$ (points). We also plot the linear fit $\Delta T_c/T_c=2.1\, a n^{1/3}$ (solid line).}
\label{Tcofa}
\end{figure}

We have computed $T_c$ numerically by monitoring the zero of $\rho_0$, as shown in Fig. \ref{figrhooftemperature}, $\rho_0(T\rightarrow T_c)\rightarrow0$. Our result is plotted in Fig. \ref{Tcofa}. In the limit $a\rightarrow0$ we find for the dimensionless critical temperature $T_c/(n^{2/3})=6.6248$, which is in good agreement with the expected result for the free theory $T_c/(n^{2/3})=\frac{4\pi}{\zeta(3/2)^{2/3}}=6.6250$. For the shift in $T_c$ due to the finite interaction strength, we obtain
\begin{equation}
\frac{\Delta T_c}{T_c}=\kappa \,a n^{1/3}, \quad \kappa=2.1.
\end{equation}
We expect that the result for $\kappa$ depends on the truncation and may change somewhat if additional higher order couplings are included.

\subsection{Zero temperature sound velocity}
The macroscopic sound velocity $v_S$ is a crucial quantity for the hydrodynamics of the gas or liquid. It is accessible to experiment. As a thermodynamic observable, the adiabatic sound velocity is defined as
\begin{equation}
v_S^2=\frac{1}{M}\frac{\partial p}{\partial n}\bigg{|}_s
\end{equation}
where $M$ is the particle mass (in our units $1/M=2$), $p$ is the pressure, $n$ is the particle density, and $s$ is the entropy per particle. It is related to the isothermal sound velocity $v_T$ by
\begin{equation}
v_S^2=\frac{1}{M}\left(\frac{\partial p}{\partial n}\bigg{|}_T+\frac{\partial p}{\partial T}\bigg{|}_n\frac{\partial T}{\partial n}\bigg{|}_s\right)=v_T^2+2\frac{\partial p}{\partial T}\bigg{|}_n \frac{\partial T}{\partial n}\bigg{|}_s
\label{eqSoundAdiabaticIsothermal}
\end{equation}
where we use our units $2M=1$. One needs the "equation of state" $p(T,n)$ and
\begin{equation}
s(T,n)=\frac{S}{N}=\frac{1}{n}\frac{\partial p}{\partial T}\bigg{|}_\mu.
\end{equation}
By dimensional analysis, one has
\begin{equation}
p=n^{5/3}{\cal F}(t,c), \quad t=\frac{T}{n^{2/3}}, \quad c=a n^{1/3},
\end{equation}
with ${\cal F}(0,c)=4\pi c$ (in Bogoliubov theory), and ${\cal F}(t,0)=\frac{\zeta(5/2)}{(4\pi)^{3/2}}t^{5/2}$ (free theory), such that for small $c$
\begin{equation}
{\cal F}=\frac{\zeta(5/2)}{(4\pi)^{3/2}}t^{5/2}+g(t)c.
\end{equation}

At zero temperature the second term in Eq.\ \eqref{eqSoundAdiabaticIsothermal} vanishes, such that $v_S=v_T$. For the isothermal sound velocity one has 
\begin{equation}
v_T^2=2\frac{\partial p}{\partial n}\bigg{|}_{T}=2 \frac{\partial p}{\partial \mu}\bigg{|}_T\left(\frac{\partial n}{\partial \mu}\bigg{|}_T\right)^{-1}.
\end{equation}
We can now use 
\begin{equation}
\frac{\partial p}{\partial \mu}\big{|}_T=-\frac{dU_{\text{min}}}{d\mu}=-\partial_\mu U(\rho_0)=n
\end{equation}
and infer
\begin{equation}
v_T^2=2\left(\frac{\partial\, \text{ln}\,n}{\partial \mu}\right)^{-1}.
\end{equation}

One may also define a microscopic sound velocity $c_S$, which characterizes the propagation of (quasi-) particles. At zero temperature, where we can perform the analytic continuation to real time, we can calculate the microscopic sound velocity from the dispersion relation $\omega(p)$ (with $p=|\vec{p}|$). In turn, the dispersion relation is obtained from the effective action by setting $\text{det}(G^{-1})=0$, where $G^{-1}$ is the full inverse propagator. We perform the calculation explicitly at the end of section \ref{sec:Derivativeexpansionandwardidentities} and find
\begin{equation}
c_S^{-2}=\frac{S^2}{2\lambda\rho_0}+V.
\end{equation}

\begin{figure}
\centering
\includegraphics[width=0.5\textwidth]{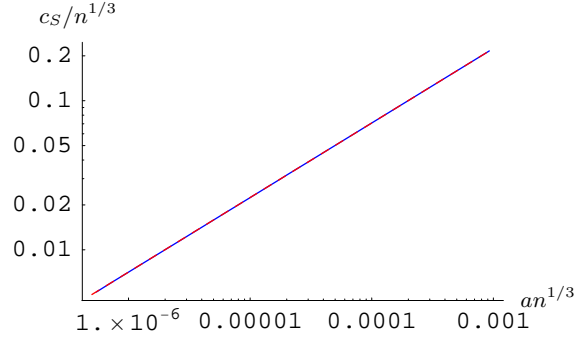}
\caption{Dimensionless sound velocity $c_s/(n^{1/3})$ at zero temperature, as a function of the scattering length $an^{1/3}$. Within the plot resolution the curves obtained by varying $a$ with fixed $n$, by varying $n$ with fixed $a$, and the Bogoliubov result, $c_s=\sqrt{16\pi}(an)^{1/2}$, coincide.}
\label{figSound}
\end{figure}

The Bogoliubov result for the sound velocity is in our units
\begin{equation}
c_S^2=2\lambda\rho_0=16\pi an.
\end{equation}
In three dimensions, the decrease of $S$ is very slow and the coupling $V$ remains comparatively small even on macroscopic scales, cf. Fig. \ref{figFlowKinetic}. We thus do not expect measurable deviations from the Bogoliubov result for the sound velocity at $T=0$. In figure \ref{figSound}, we plot our result over several orders of magnitude of the dimensionless scattering length and, indeed, find no deviations from Bogoliubovs result.

We finally show that for $T=0$ the macroscopic and microscopic sound velocities are equal, $v_S=v_T=c_S$. For this purpose, we use 
\begin{equation}
\frac{\partial n}{\partial \mu}\big{|}_T=-\frac{d}{d\mu}\left(\partial_\mu U(\rho_0)\right)=-\partial_\mu^2U(\rho_0)-\partial_\rho\partial_\mu U(\rho_0)\frac{d\rho_0}{d\mu}.
\end{equation}
From the minimum condition $\partial_\rho U=0$, it follows
\begin{equation}
\frac{d\rho_0}{d\mu}=-\frac{\partial_\rho\partial_\mu U}{\partial_\rho^2 U}=-\frac{\alpha}{\lambda}.
\end{equation}
Combining this with the Ward identities from section \ref{sec:Derivativeexpansionandwardidentities}, namely $\partial_\mu^2 U=-2V\rho_0$ and $\alpha=\partial_\rho\partial_\mu U=-S$, valid at $T=0$, it follows that the macroscopic sound velocity equals the microscopic sound velocity
\begin{equation}
v_S^2(T=0)=c_S^2.
\end{equation}
	\subsection{Thermodynamic observables}

We now come to the discussion of some thermodynamic properties at nonzero temperature. With our method we can determine many thermodynamic observables from the effective potential $U$. It is related to the pressure by
\begin{equation}
p(T,\mu)=-U_\text{min}(T,\mu){\big |}_{k=k_\text{ph}}
\end{equation}
which has the differential
\begin{equation}
dp=s\,dT+n\,d\mu.
\label{eq:differentialp}
\end{equation}
Here we use $s=S/V$ for the entropy density and $n=N/V$ for the particle density. The formal infinite volume limit corresponds to $k_\text{ph}=0$. Derivatives of $U$ with respect to $T$ and and $\mu$ are taken numerically by solving the flow equation for close enough values of $T$ and $\mu$. The numerical effort is reduced and the accuracy increased by using an additional flow equation for
\begin{equation}
n_k=-\frac{\partial}{\partial \mu} U_k {\big |}_{\bar \rho=\bar \rho_0(k)},
\end{equation}
with $n=n_{k_\text{ph}}$. The approximation scheme we use in the following is basically the same as the one described in \ref{sec:TruncationBosegas}. Since we use an infrared cutoff only for momenta but not for frequencies, the correct ultraviolet convergence for the sum of the Matsubara frequencies is not automatically obeyed for the flow equations. We have checked that all thermodynamic quantities discussed in the following show a satisfactory convergence of the Matsubara sum, except for the pressure. In the flow equation for $p_k$ we set the frequency coefficients to their microscopic values $\bar S=1$, $\bar V=0$ for very large Matsubara frequencies $|q_0|>\Lambda_\text{UV}^2$. 

For bosons with a pointlike repulsive interaction we found in section \ref{sec:Repulsiveinteractingbosons} that the scattering length is bounded by the ultraviolet scale $a<3\pi/(4\Lambda)$. This is an effect due to quantum fluctuations similar to the ``triviality bound'' for the Higgs scalar in the standard model of elementary particle physics. For a given value of the dimensionless combination $an^{1/3}$ we cannot choose $\Lambda/n^{1/3}$ larger then $3\pi/(4 an^{1/3})$. For our numerical calculations we use $\Lambda/n^{1/3}\approx 10$. Other momentum scales are set by the temperature and the chemical potential. The lowest nonzero Matsubara frequency gives the momentum scale $\Lambda_T^2=2\pi T$. For a Bose gas with $a=0$ one has $T_c/n^{2/3}\approx 6.625$ such that $\Lambda_{T_c}/n^{1/3}\approx 6.45$. The momentum scale associated to the chemical potential is $\Lambda_\mu^2=\mu$. For small temperatures and scattering length one finds $\mu\approx 8\pi a n$ and thus $\Lambda_\mu/n^{1/3}\approx \sqrt{8\pi a n^{1/3}}$.

We finally note that the thermodynamic relations for intensive quantities can only involve dimensionless ratios. We may set the unit of momentum by $n^{1/3}$. The thermodynamic variables are then $T/n^{2/3}$ and $\mu/n^{2/3}$. The thermodynamic relations will depend on the strength of the repulsive interaction $\lambda$ or the scattering length $a$, and therefore on a ``concentration'' type parameter $a n^{1/3}$.

\subsubsection{Density, superfluid density, condensate and correlation length}
\begin{figure}
\centering
\includegraphics[width=0.5\textwidth]{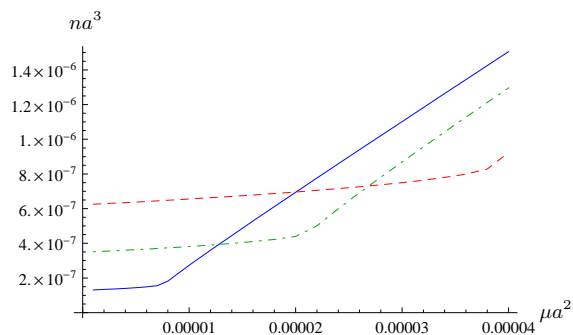}
\caption{Density in units of the scattering length $n a^3$ as a function of the (rescaled) chemical potential $\mu a^2$. We choose for the temperatures $T a^2=2\cdot 10^{-4}$ (solid curve), $T a^2=4 \cdot 10^{-4}$ (dashed-dotted curve) and $T a^2= 6 \cdot 10^{-4}$ (dashed curve). For all three curves we use $a\Lambda=0.1$.}
\label{fig:densityofmu}
\end{figure}

Let us start our discussion with the density. In the grand canonical formalism it is obtained by taking the derivative of the thermodynamic potential with respect to $\mu$
\begin{equation}
n=-\frac{1}{V}\frac{\partial}{\partial \mu}\Omega_G = \frac{\partial p}{\partial \mu}{\big |}_T.
\end{equation}
We could compute the $\mu$-derivative of $p$ numerically by solving the flow equation for U with neighboring values of $\mu$. As desribed above, we use also another method which employs a flow equation directly for $n$. Since we often express dimensionful quantities in units of the interparticle distance $n^{-1/3}$, it is crucial to have an accurate value for the density $n$. Comparison of the numerical evaluation and the solution of a separate flow equation for $n$ shows higher precision for the latter method and we will therefore employ the flow equation. We plot in Fig. \ref{fig:densityofmu} the density in units of the scattering length, $na^3$, as a function of the dimensionless combination $\mu a^2$. 

\begin{figure}
\centering
\includegraphics[width=0.5\textwidth]{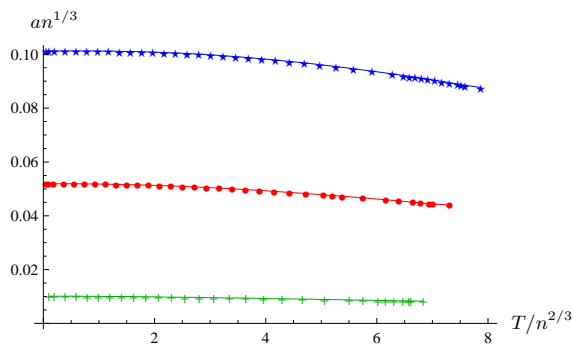}
\caption{Concentration $c=an^{1/3}$ as a function of temperature $T/(n^{2/3})$ for the three cases investigated in this paper. Case I corresponds to $an^{1/3}\approx 0.01$ (crosses), case II to $an^{1/3}\approx 0.05$ (dots) and case III has $an^{1/3}\approx 0.01$ (stars).}
\label{fig:an13}
\end{figure}

For a comparison with experimentally accessible quantities we have to replace the interaction parameter $\lambda$ in the microscopic action \eqref{microscopicaction} by a scattering length $a$ which is a macroscopic quantity. For this purpose we start the flow at the UV-scale $\Lambda_\text{UV}$ with a given $\lambda$, and then compute the scattering length in vacuum ($T=n=0$) by following the flow to $k=0$, see section \ref{sec:Repulsiveinteractingbosons}. This is a standard procedure in quantum field theory, where a ``bare coupling'' ($\lambda$) is replaced by a renormalized coupling ($a$). For an investigation of the role of the strength of the interaction we may consider different values of the ``concentration'' $c=an^{1/3}$ or of the product $\mu a^2$. While the concentration is easier to access for observation, it is also numerically more demanding since for every value of the parameters one has to tune $\mu$ in order to obtain the appropriate density. For this reason we rather present results for three values of $\mu a^2$, i.~e. $\mu a^2= 2.6\times 10^{-5}$ (case I), $\mu a^2=0.0040$ (case II) and $\mu a^2=0.044$ (case III). The prize for the numerical simplicity is a week temperature dependence of the concentration $c=a n^{1/3}$ for the three different cases, as shown in Fig. \ref{fig:an13}.
Here and in the following figures case I, which corresponds to $an^{1/3}\approx 0.01$, is represented by the little crosses, case II with $an^{1/3}\approx 0.05$ by the dots and case III with $an^{1/3}\approx 0.1$ by the stars. It is well known that the critical temperature depends on the concentration $c=an^{1/3}$. From our calculation we find $T_c/(n^{2/3})=6.74$ with $c=0.0083$ at $T=T_c$ in case I, $T_c/(n^{2/3})=7.16$ with $c=0.044$ at $T=T_c$ in case II and finally $T_c/(n^{2/3})=7.75$ with $c=0.088$ at $T=T_c$ in case III.

\begin{figure}
\centering
\includegraphics[width=0.5\textwidth]{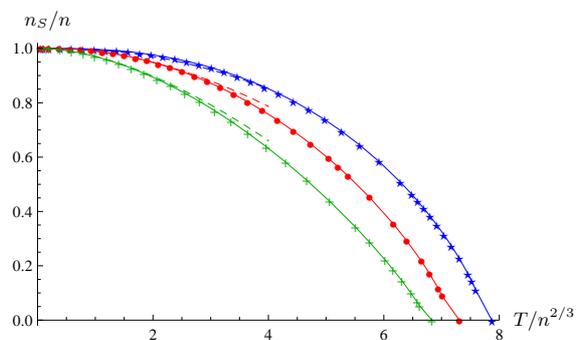}
\caption{Superfluid fraction of the density $n_S/n$ as a function of the temperature $T/n^{2/3}$ for the cases I, II, and III. For small $T/n^{2/3}$ we also show the corresponding curves obtained in the Bogoliubov approximation (dashed lines).}
\label{fig:superfluidfraction}
\end{figure}

\begin{figure}
\centering
\includegraphics[width=0.5\textwidth]{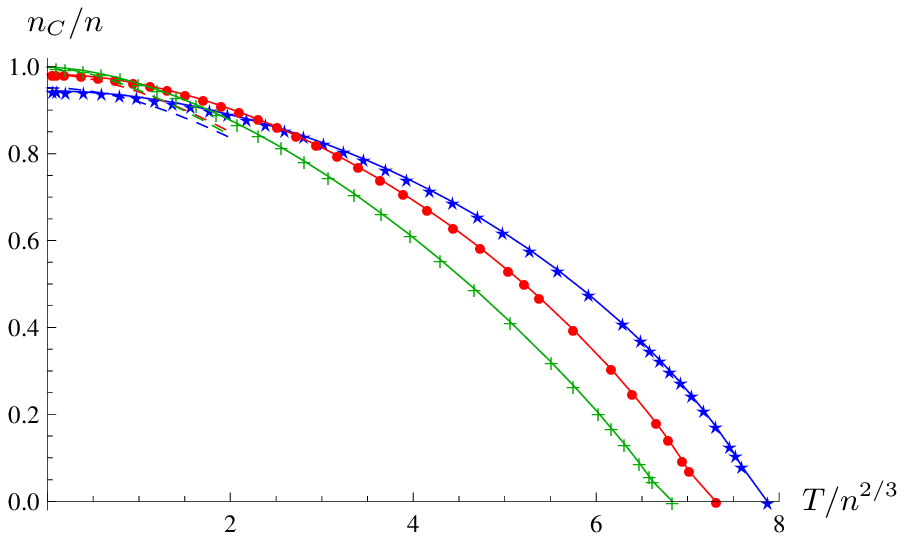}
\caption{Condensate fraction of the density $n_C/n$ as a function of the temperature $T/n^{2/3}$ for the cases I, II, and III. For small $T/n^{2/3}$ we also show the corresponding curves obtained in the Bogoliubov approximation (dashed lines).}
\label{fig:condensatefraction}
\end{figure}

This values can are obtained by following the superfluid fraction of the density $n_S/n$, or equivalently the condensate part of the density $n_C/n$ as a function of temperature. For small temperatures $T\to0$ all of the density is superfluid, which is a consequence of Galilean symmetry. However, in contrast to the ideal gas, not all particles are in the condensate. For $T=0$ this condensate depletion is completely due to quantum fluctuations. With increasing temperature both the superfluid density and the condensate decrease and vanish eventually at the critical temperature $T=T_c$. That the melting of the condensate is continuous shows that the phase transition is of second order. We plot our results for the superfluid fraction in Fig. \ref{fig:superfluidfraction} and for the condensate in Fig. \ref{fig:condensatefraction}. For small temperatures, we also show the corresponding result obtained in the framework of Bogoliubov theory \cite{Bogoliubov} (dashed lines). This approximation assumes a gas of non-interacting quasiparticles (phonons) with dispersion relation
\begin{equation}
\epsilon(p)=\sqrt{2\lambda n \vec p^2+\vec p^4}.
\end{equation} 
It is is valid in the regime with small temperatures $T\ll T_c$ and small interaction strength $an^{1/3}\ll 1$. For a detailed discussion of Bogoliubov theory and the calculation of thermodynamic observables in this framework we refer to ref. \cite{PitaevsikiiStringari2003}. Our curves for the superfluid fraction match the Bogoliubov result for temperatures $T/n^{2/3}\lesssim 1$ in all three cases I, II, and III. For larger temperatures there are deviations as expected. For the condensate density, there is already notable a deviation at small temperatures for case III with $an^{1/3}\approx 0.1$. This is also expected, since Bogoliubov theory gives only the first order contribution to the condensate depletion in a perturbative expansion for small $an^{1/3}$.  
For temperatures slightly smaller than the critical temperature $T_c$ one expects that the condensate density behaves like 
\begin{equation}
n_c(T)=B^2 \left(\frac{T_c-T}{T_c}\right)^{2\beta}
\label{eq:scalingnc}
\end{equation}
with $\beta=0.3485$ the critical exponent of the three-dimensional XY-universality class \cite{Pelissetto2002549}. Indeed, the condensate density is given by $n_C=\bar\varphi_0^*\bar\varphi_0$
where $\bar \varphi_0$ is the expectation value of the boson field which serves as an order parameter in close analogy to e.~g. the magnetization $\vec M$ in a ferromagnet. Eq.\ \eqref{eq:scalingnc} is compatible with our findings, although our numerical resolution does not allow for a precise determination of the exponent $\beta$. 

\begin{figure}
\centering
\includegraphics[width=0.5\textwidth]{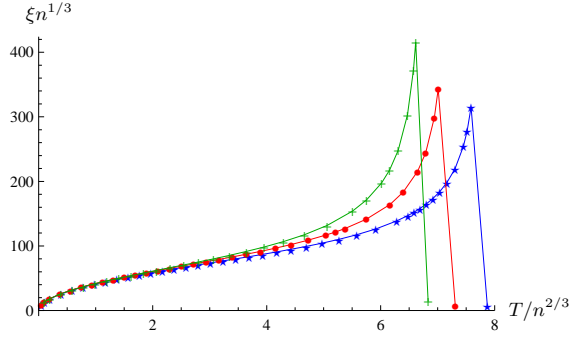}
\caption{Correlation length $\xi_R n^{1/3}$ for $T<T_c$ and $\xi n^{1/3}$ for $T>T_c$ as a function of the temperature $T/n^{2/3}$ for the cases I, II, and III.}
\label{fig:correlationlength}
\end{figure}

With our method we can also calculate the correlation length $\xi$. For temperatures $T<T_c$ one distinguishes between the Goldstone correlation length $\xi_G$ and the radial correlation length $\xi_R$. While the former is infinite, $\xi_G^{-1}=0$, the latter is finite for $T<T_c$. It is also known as the ``healing length'', given by 
\begin{equation}
\xi_R^{-2}=2\lambda \rho_0=2\frac{1}{\bar A}\frac{\partial^2 U}{\partial \bar \rho}\,\bar \rho_0
\end{equation} 
and diverges only close to the phase transition. In the symmetric regime for $T>T_c$ there is only one correlation length $\xi^{-1}=m=\frac{1}{\bar A}\frac{\partial U}{\partial \bar \rho}$, which also diverges for $T\to T_c$. From the theory of critical phenomena one expects close to $T_c$ the behavior
\begin{eqnarray}
\nonumber
\xi_R &=& f_R^- \left(\frac{T_c-T}{T_c}\right)^{-\nu} \quad \text{for}\quad T<T_c\\
\xi &=& f^+ \left(\frac{T-T_c}{T_c}\right)^{-\nu} \quad \text{for}\quad T<T_c.
\end{eqnarray}
The critical exponent $\nu=0.6716$ \cite{Pelissetto2002549} is again the one of the three-dimensional XY- or O(2) universality class. We plot our result for the correlation length in units of the interparticle distance $\xi_R n^{1/3}$ for $T<T_c$ and $\xi n^{1/3}$ for $T>T_c$ as a function of the temperature $T/n^{2/3}$ in Fig. \ref{fig:correlationlength}.

\subsubsection{Entropy density, energy density, and specific heat}
\begin{figure}
\centering
\includegraphics[width=0.5\textwidth]{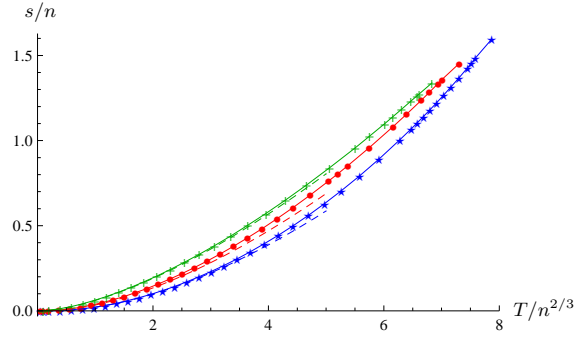}
\caption{Entropy per particle $s/n$ as a function of the dimensionless temperature $T/n^{2/3}$ for the cases I, II, and III. For $T/n^{2/3}<5$ we also plot the results obtained within the Bogoliubov approximation (dashed lines).}
\label{fig:entropypp}
\end{figure}

The next thermodynamic quantity we investigate is the entropy density $s$ and the entropy per particle $s/n$. We can obtain the entropy as
\begin{equation}
s=\frac{\partial p}{\partial T}{\big |}_\mu.
\end{equation}
We compute the temperature derivative by numerical differentiation, using flows with neighboring values of $T$ and show the result in Fig. \ref{fig:entropypp}. For small temperatures our result coincides with the entropy of free quasiparticles in the Bogoliubov approximation (dashed lines in Fig. \ref{fig:entropypp}). 
As it should be, the entropy per particle increases with the temperature. For small temperatures, the slope of this increase is smaller for larger concentration $c$.

\begin{figure}
\centering
\includegraphics[width=0.5\textwidth]{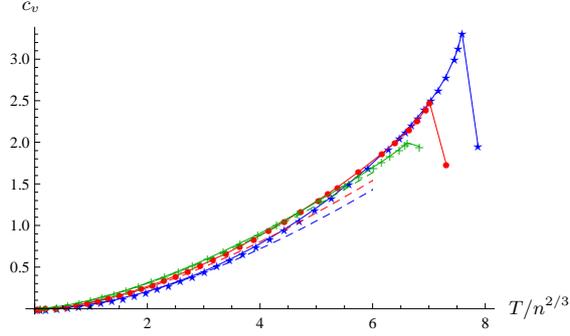}
\caption{Specific heat per particle $c_v$ as a function of the dimensionless temperature $T/n^{2/3}$. The dashed lines show the Bogoliubov result for $c_v$ which coincides with our findings for small temperature. However, the characteristic cusp behavior cannot be seen in a mean-field theory.}
\label{fig:specificheatpp}
\end{figure}

From the entropy density $s$ we infer the specific heat per particle,
\begin{equation}
c_v=\frac{T}{n}\frac{\partial s}{\partial T}{\bigg |}_n,
\end{equation}
as the temperature derivative of the entropy density at constant particle density.
Using the Jacobian, we can write
\begin{equation}
\frac{\partial s}{\partial T}{\big |}_n = \frac{\partial(s,n)}{\partial(T,n)}=\frac{\partial(s,n)}{\partial(T,\mu)}\frac{\partial(T,\mu)}{\partial(T,n)}.
\end{equation}
For the specific heat this gives
\begin{equation}
c_v=\frac{T}{n}\left(\frac{\partial s}{\partial T}{\big |}_\mu -\frac{\partial s}{\partial \mu}{\big |}_T\frac{\partial n}{\partial T}{\big |}_\mu \left(\frac{\partial n}{\partial \mu}{\big |}_T\right)^{-1} \right).
\end{equation}
Our result for the specific heat per particle is shown for different scattering lengths in Fig. \ref{fig:specificheatpp}.
While this quantity is positive in the whole range of investigated temperatures, it is interesting to observe the cusp at the critical temperature $T_c$ which is characteristic for a second order phase transition. This behavior cannot be seen in a mean-field approximation, where fluctuations are taken into account only to second order in the fields. Only for small temperatures, our curve is close to the Bogoliubov approximation, shown by the dashed lines in Fig. \ref{fig:specificheatpp}. 

In fact, close to $T_c$ the specific heat is expected to behave like
\begin{eqnarray}
\nonumber
c_v &\approx&  b_1-b_2^- \left(\frac{T_c-T}{T_c}\right)^{-\alpha} \quad \text{for} \quad T<T_c,\\
c_v &\approx&  b_1-b_2^+ \left(\frac{T-T_c}{T_c}\right)^{-\alpha} \quad \text{for} \quad T>T_c,
\end{eqnarray}
with the universal critical exponent $\alpha$ of the $3$-dimensional $XY$ universality class, $\alpha=-0.0146(8)$ \cite{Pelissetto2002549}. The critical region, where the law $c_v~\sim |T-T_c|^{-\alpha}$ holds, may be quite small.
Our numerical differentiation procedure cannot resolve the details of the cusp.  

\begin{figure}
\centering
\includegraphics[width=0.5\textwidth]{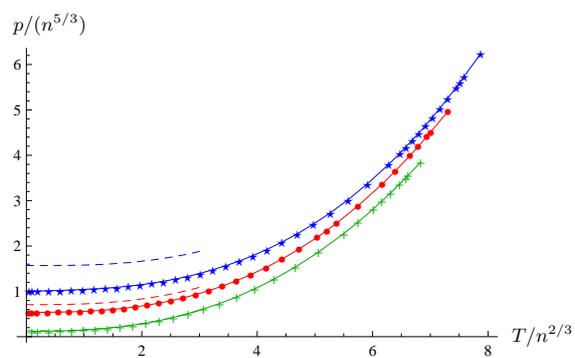}
\caption{Pressure in units of the density $p/n^{5/3}$ as a function of temperature $T/n^{2/3}$ for the cases I (crosses), II (dots), and III (stars). We also show the curves obtained in the Bogoliubov approximation for small temperatures (dashed lines).}
\label{fig:pressure}
\end{figure}

\begin{figure}
\centering
\includegraphics[width=0.5\textwidth]{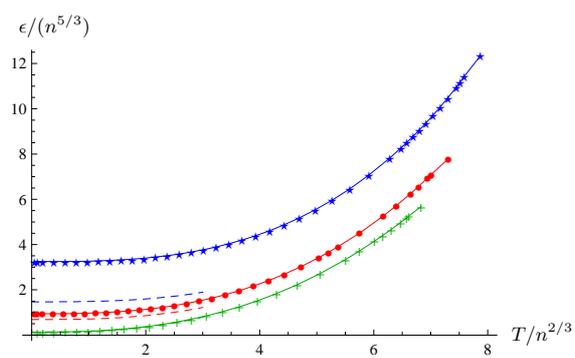}
\caption{Energy per particle $\epsilon/n^{5/3}$ as a function of temperature $T/n^{2/3}$ for the cases I (crosses), II (dots), and III (stars). We also show the curves obtained in the Bogoliubov approximation for small temperatures (dashed lines).}
\label{fig:energy}
\end{figure}

In the grand canonical formalism, the energy density $\epsilon$ is obtained as
\begin{equation}
\epsilon = -p +Ts+\mu n.
\end{equation}
We plot $p/(n^{5/3})$ as a function of temperature in Fig. \ref{fig:pressure} and the energy density $\epsilon/(n^{5/3})$ is plotted in Fig. \ref{fig:energy}.
We have normalized the pressure such that it vanishes for $T=\mu=0$. Technically we subtract from the flow equation of the pressure the corresponding expression in the limit $T=\mu=0$. This procedure has to be handled with care and leads to an uncertainty in the offset of the pressure, i.~e. the part that is independent of $T/n^{2/3}$ and $\mu/n^{2/3}$. 

For zero temperature, the pressure is completely due to the repulsive interaction between the particles. For nonzero temperature, the pressure is increased by the thermal kinetic energy, of course.
For the energy and the pressure we find some deviations from the Bogoliubov result already for small temperatures in cases II and III. These deviations may be partly due to the uncertainty in the normalization process described above. For weak interactions $an^{1/3}=0.01$ as in case I, the Bogoliubov prediction coincides with our result.

\subsubsection{Compressibility}
The isothermal compressibility is defined as the relative volume change at fixed temperature $T$ and particle number $N$ when some pressure is applied
\begin{equation}
\kappa_T=-\frac{1}{V}\frac{\partial V}{\partial p}{\big |}_{T,N}=\frac{1}{n}\frac{\partial n}{\partial p}{\big |}_T.
\label{eq:isothermalkomp}
\end{equation}
Very similar, the adiabatic compressibility is
\begin{equation}
\kappa_S=-\frac{1}{V}\frac{\partial V}{\partial p}{\big |}_{S,N}=\frac{1}{n}\frac{\partial n}{\partial p}{\big |}_{s/n}
\end{equation}
where now the entropy $S$ and the particle number $N$ are fixed. Let us first concentrate on the isothermal compressibility $\kappa_T$. To evaluate it in the grand canonical formalism, we have to change variables to $T$ and $\mu$. 
With $\partial p/\partial \mu{\big |}_{n,T}=n$ and $\partial p/\partial n{\big |}_T=n \partial \mu/\partial n{\big |}_T$ one obtains
\begin{equation}
\kappa_T=\frac{1}{n^2}\frac{\partial n}{\partial \mu}{\big |}_{T}.
\end{equation}
This expression can be directly evaluated in our formalism by numerical differentiation with respect to $\mu$.  

\begin{figure}
\centering
\includegraphics[width=0.5\textwidth]{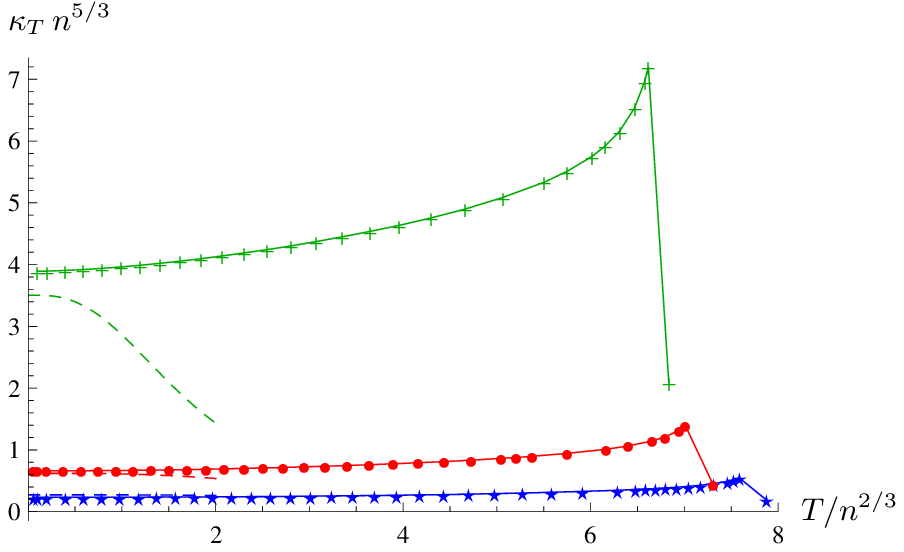}
\caption{Isothermal compressibility $\kappa_T\, n^{5/3}$ as a function of temperature $T/n^{2/3}$ for the cases I (crosses), II (dots), and III (stars). We also show the Bogoliubov result for small temperatures (dashed lines).}
\label{fig:Isothermalcompressibility}
\end{figure}

\begin{figure}
\centering
\includegraphics[width=0.5\textwidth]{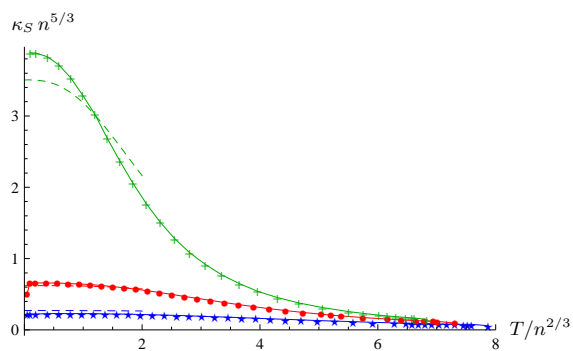}
\caption{Adiabatic compressibility $\kappa_S\, n^{5/3}$ as a function of temperature $T/n^{2/3}$ for the cases I (crosses), II (dots), and III (stars). We also show the Bogoliubov result for small temperatures (dashed lines).}
\label{fig:Adiabaticcompressibility}
\end{figure}

The approach to the adiabatic compressibility is similar. Using again the Jacobian we have
\begin{eqnarray}
\nonumber
\kappa_S &=& \frac{1}{n}\frac{\partial n}{\partial p}{\big |}_{s/n}=\frac{1}{n}\frac{\partial(n,s/n)}{\partial(p,s/n)}\\
&=& \frac{1}{n}\frac{\partial (n,s/n)}{\partial (\mu,T)}\frac{\partial(\mu,T)}{\partial(p,s/n)}.
\end{eqnarray}
We need therefore
\begin{equation}
\frac{\partial(n,s/n)}{\partial(\mu,T)}=\frac{1}{n}\left(\frac{\partial n}{\partial \mu}{\big |}_{T}\frac{\partial s}{\partial T}{\big |}_{\mu}-\frac{\partial n}{\partial T}{\big |}_{\mu}\frac{\partial s}{\partial \mu}{\big |}_{T}\right)
\end{equation}
and also
\begin{eqnarray}
\nonumber
\frac{\partial(p,s/n)}{\partial(\mu,T)} &=& \frac{1}{n}{\bigg (}\frac{\partial p}{\partial \mu}{\big |}_{T}\frac{\partial s}{\partial T}{\big |}_\mu - \frac{\partial p}{\partial \mu}{\big |}_{T} \frac{s}{n}\frac{\partial n}{\partial T}{\big |}_\mu\\
&& -\frac{\partial p}{\partial T}{\big |}_{\mu} \frac{\partial s}{\partial \mu}{\big |}_T+\frac{\partial p}{\partial T}{\big |}_\mu \frac{s}{n} \frac{\partial n}{\partial \mu}{\big |}_T {\bigg )}\\
\nonumber
&=& \left(\frac{\partial s}{\partial T}{\big |}_\mu-2\frac{s}{n}\frac{\partial n}{\partial T}{\big |}_\mu+\frac{s^2}{n^2}\frac{\partial n}{\partial \mu}{\big |}_T\right).
\end{eqnarray}
In the last equations we used the Maxwell identity $\frac{\partial n}{\partial T}{\big |}_\mu=\frac{\partial s}{\partial \mu}{\big |}_T$. Combining this we find
\begin{equation}
\kappa_S=\frac{\left(\frac{\partial n}{\partial \mu}{\big |}_T \frac{\partial s}{\partial T}{\big |}_\mu-\left(\frac{\partial n}{\partial T}{\big |}_\mu\right)^2\right)}{\left(n^2 \frac{\partial s}{\partial T}{\big |}_\mu-2 s n \frac{\partial n}{\partial T}{\big |}_\mu+s^2 \frac{\partial n}{\partial \mu}{\big |}_T\right)}.
\end{equation}
Since $\partial s/\partial T{\big |}_\mu=(\partial^2 p/\partial T^2){\big |}_\mu$ we need to evaluate a second derivative numerically.
We plot the isothermal and the adiabatic compressibility in Figs. \ref{fig:Isothermalcompressibility} and \ref{fig:Adiabaticcompressibility}.

For the isothermal compressibility the temperature dependence is qualitatively different than in Bogoliubov theory already for small temperatures, while there seem to be only quantitative differences for the adiabatic compressibility. The perturbative calculation of the compressibility is difficult since it is diverging in the non-interacting limit $an^{1/3}\to 0$.

\subsubsection{Isothermal and adiabatic sound velocity}
\begin{figure}
\centering
\includegraphics[width=0.5\textwidth]{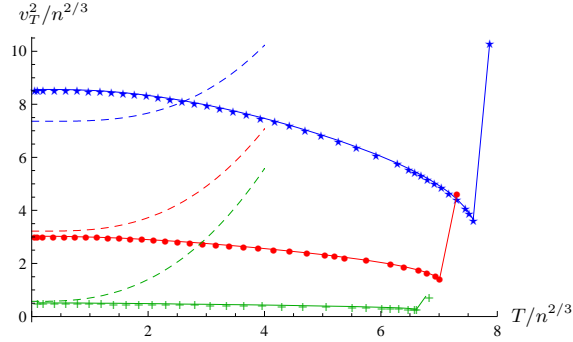}
\caption{Isothermal velocity of sound as appropriate for single fluid $v_T^2/n^{2/3}=1/(\kappa_T\, n^{5/3})$ as a function of the dimensionless temperature $T/n^{2/3}$ for the cases I (crosses), II (dots), and III (stars). We also show the Bogoliubov result for small temperatures (dashed lines).}
\label{fig:SingleFluidisothermalsound}
\end{figure}

The sound velocity of a normal fluid under isothermal conditions, i.~e. for constant temperature $T$ is given by
\begin{equation}
v_T^2=\frac{1}{M}\frac{\partial p}{\partial n}{\big |}_T.
\label{eq:singlefluidisothermalsound}
\end{equation} 
We can obtain this directly from the isothermal compressibility
\begin{equation}
M v_T^2=(n \kappa_T)^{-1}
\end{equation}
as follows from Eq.\ \eqref{eq:isothermalkomp}. We plot our result for $v_T^2$ in Fig. \ref{fig:SingleFluidisothermalsound}, recalling our units $2M=1$ such that $v_T^2$ stands for $2Mv_T^2$.
This plot also covers the superfluid phase where the physical meaning of $v_T^2$ is partly lost. This comes since the sound propagation there has to be described by more complicated two-fluid hydrodynamics. In addition to the normal gas there is now also a superfluid fraction allowing for an additional oscillation mode. We will describe the consequences of this in the next section.

\begin{figure}
\centering
\includegraphics[width=0.5\textwidth]{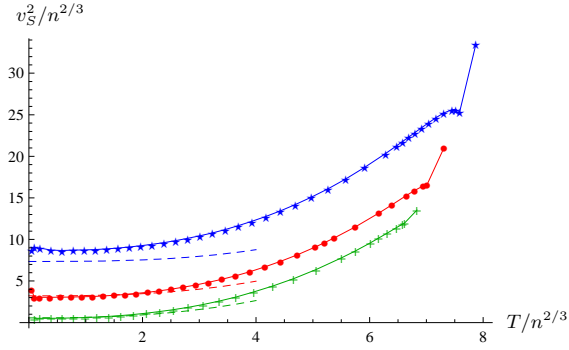}
\caption{Adiabatic velocity of sound as appropriate for single fluid $v_S^2/n^{2/3}=1/(\kappa_S \,n^{5/3})$ as a function of the dimensionless temperature $T/n^{2/3}$ for the cases I (crosses), II (dots), and III (stars). We also show the Bogoliubov result for small temperatures (dashed lines).}
\label{fig:SingleFluidadiabaticsound}
\end{figure}

For most applications the adiabatic sound velocity is more important then the isothermal sound velocity. Keeping the entropy per particle fixed, we obtain
\begin{equation}
v_S^2=\frac{1}{M}\frac{\partial p}{\partial n}{\big |}_{s/n}
\label{eq:singlefluidadiabaticsound}
\end{equation}
and therefore
\begin{equation}
M v_S^2 = (n\kappa_S)^{-1}.
\end{equation}
Our numerical result is plotted in Fig. \ref{fig:SingleFluidadiabaticsound}.
Again the plot covers both the superfluid and the normal part, but only in the normal phase the object $v_S^2$ has its physical meaning as a sound velocity.

\subsubsection{First and second velocity of sound}
For temperatures $0<T<T_c$ there are two components of the gas: the superfluid and the normal part. It was shown by Landau \cite{Landau41} that this leads to two-fluid hydrodynamics with two distinct velocities of sound $c_{1/2}$ corresponding to different kinds of excitations. 

The main reason for the existence of two sound velocities is that the entropy flow is carried only be the normal component while the particle flow (or equivalently mass-flow) is carried by both the normal and the superfluid part. The continuity equation for the conserved particle number reads
\begin{equation}
\partial_t n+\vec \nabla \cdot \vec j =0,
\label{eq:particlecontinuity}
\end{equation}
where $\vec j=n_N \vec v_N+n_S \vec v_S$ is the (complete) particle number current and $\vec v_N$, $\vec v_S$ are the velocities of the normal ($n_N$) and superfluid ($n_S$) parts of the density, $n=n_N+n_S$. The conservation equation for the entropy reads 
\begin{equation}
\partial_t s + \vec \nabla \cdot (s \vec v_N)=0.
\label{eq:entropycontinuity}
\end{equation}
We work in linear order in an expansion in the velocities $\vec v_N$ and $\vec v_S$.
To close the set of hydrodynamic equations for small $\vec v_N$, $\vec v_S$ we need the equations for momentum conservation
\begin{equation}
M \partial_t \vec j+\vec \nabla p=0,
\label{eq:momentumconservation}
\end{equation}
and for the change in the superfluid velocity
\begin{equation}
M\partial_t \vec v_S + \vec \nabla \mu=0.
\label{eq:changesuperfluidvelocity}
\end{equation}
The last equation guarantees that the superfluid flow remains irrotational, $\vec \nabla \times \vec v_S=0$.

From the combination of Eq.\ \eqref{eq:particlecontinuity} and \eqref{eq:momentumconservation} one obtains
\begin{equation}
M \partial_t^2 n = \Delta p.
\label{eq:wave1}
\end{equation}
To linear order in $\vec v_S$ and $\vec v_N$ one infers from the combination of Eq.\ \eqref{eq:particlecontinuity} and \eqref{eq:entropycontinuity}
\begin{equation}
n_S \vec \nabla\cdot(\vec v_N-\vec v_S)=-\frac{n^2}{s}\partial_t(s/n).
\label{eq:vnvs1}
\end{equation}
We recover $s/n=\text{const.}$ for $n_S=0$ as appropriate for the disordered phase. 
Similarly, the combination of Eq.\ \eqref{eq:momentumconservation} and \eqref{eq:changesuperfluidvelocity} gives
\begin{eqnarray}
\nonumber
M n_N \partial_t (\vec v_N-\vec v_S) &=& n\vec \nabla \mu -\vec \nabla p\\
&=& -s \vec \nabla T.
\label{eq:vnvs2}
\end{eqnarray}
The last equation uses the relation
\begin{equation}
\vec \nabla p=s\vec \nabla T+n \vec \nabla \mu
\end{equation}
which follows directly from the differential of $p$, 
\begin{equation}
dp = s\, dT+n\,d\mu.
\end{equation}
Combining now Eqs. \eqref{eq:vnvs1} and \eqref{eq:vnvs2} yields the analogue of Eq.\ \eqref{eq:wave1}. 
\begin{equation}
M\partial_t^2 (s/n)=\frac{s^2}{n^2}\frac{n_S}{n_N}\Delta T.
\label{eq:wave2}
\end{equation}

One next makes an ansatz for the thermodynamic variables in the form
\begin{eqnarray}
\nonumber
p=p_0+\delta p,\quad T=T_0+\delta T\\
n=n_0+\delta n, \quad s/n=s_0/n_0+\delta(s/n),
\end{eqnarray}
where $p_0$, $T_0$, $n_0$ and $s_0$ are constant in space and time whereas $\delta p$, $\delta T$, $\delta n$, and $\delta(s/n)$ are small and vary like $\text{sin}[p(x-ct)]$. We use $\delta T$ and $\delta n$ as independent variables, with
\begin{eqnarray}
\nonumber
\delta p &=& \frac{\partial p}{\partial T}{\big |}_n \,\delta T + \frac{\partial p}{\partial n}{\big |}_T \,\delta n,\\
\delta (s/n) &=& \frac{\partial (s/n)}{\partial T}{\big |}_n \,\delta T + \frac{\partial (s/n)}{\partial n}{\big |}_T \,\delta n,
\end{eqnarray}
in order to obtain from Eqs. \eqref{eq:wave1} and \eqref{eq:wave2} the wave equation
\begin{equation}
\begin{pmatrix}Mc^2\frac{\partial(s/n)}{\partial T}{\big |}_n-\frac{s^2 n_S}{n^2 n_N} &&, && Mc^2 \frac{\partial(s/n)}{\partial n}{\big |}_T \\ -\frac{\partial p}{\partial T}{\big |}_n &&, && Mc^2-\frac{\partial p}{\partial n}{\big |}_T\end{pmatrix}\begin{pmatrix}\delta T \\ \delta n\end{pmatrix}=0.
\end{equation}
As a condition for possible sound velocities $c$ one obtains
\begin{eqnarray}
\nonumber
(M c^2)^2-(Mc^2)\left[\frac{\partial p}{\partial n}{\bigg|}_{s/n}+\frac{s^2 n_S T}{n^2 n_N c_v}\right]\\
+\frac{s^2 n_S T}{n^2 n_N c_v}\frac{\partial p}{\partial n}{\bigg |}_T=0.
\label{eq:firstandsecondsound}
\end{eqnarray}
This relation uses
\begin{equation}
c_v=T\frac{\partial(s/n)}{\partial T}{\big |}_n
\end{equation}
as well as
\begin{equation}
\frac{\partial(s/n)}{\partial n}{\big |}_T=\frac{c_v}{T}\frac{\partial T}{\partial p}{\big |}_n\left[\frac{\partial p}{\partial n}{\big |}_T-\frac{\partial p}{\partial n}{\big |}_{s/n}\right].
\end{equation}
The latter relation follows from
\begin{equation}
\frac{\partial p}{\partial n}{\big |}_{s/n} = \frac{\partial p}{\partial n}{\big |}_T + \frac{\partial p}{\partial T}{\big |}_n \frac{\partial T}{\partial n}{\big |}_{s/n}
\end{equation}
together with 
\begin{eqnarray}
\nonumber
\frac{\partial T}{\partial n}{\big |}_{s/n} &=& \frac{\partial(T,s/n)}{\partial(n,s/n)} = -\frac{\partial(T,s/n)}{\partial(T,n)}\frac{\partial(T,n)}{\partial(s/n,n)}\\
&=& - \frac{\partial(s/n)}{\partial n}{\big |}_T \frac{T}{c_v}.
\end{eqnarray}

\begin{figure}
\centering
\includegraphics[width=0.5\textwidth]{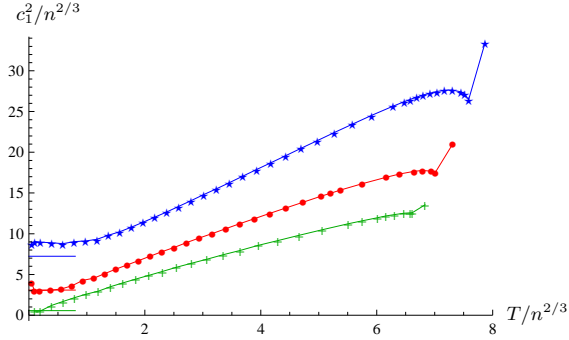}
\caption{First velocity of sound $c_1^2/n^{2/3}$ as a function of the dimensionless temperature $T/n^{2/3}$ for the cases I (crosses), II (dots), and III (stars). We also show the prediction from Bogoliubov theory for $T\to 0$ (short solid lines).}
\label{fig:Firstvelocityofsound}
\end{figure}

\begin{figure}
\centering
\includegraphics[width=0.5\textwidth]{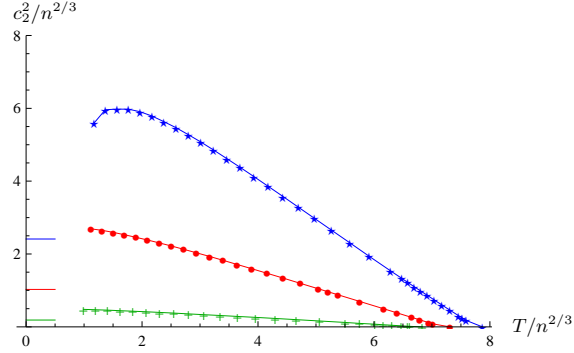}
\caption{Second velocity of sound $c_2^2/n^{2/3}$ as a function of the dimensionless temperature $T/n^{2/3}$ for the cases I (crosses), II (dots), and III (stars). We also show the prediction from Bogoliubov theory for $T\to 0$ (short solid lines). For $T/n^{2/3}<1$ our numerical determination becomes unreliable, since $c_2^2$ is dominated by a ratio of terms that vanish for $T\to 0$.}
\label{fig:Secondvelocityofsound}
\end{figure}

With these ingredients one can now solve Eq.\ \eqref{eq:firstandsecondsound} for the first and second velocity of sound. The numerical results as a function of temperature are shown in Fig. \ref{fig:Firstvelocityofsound} and \ref{fig:Secondvelocityofsound}.
We also show there the prediction from Bogoliubov theory for $T\to 0$ (short solid lines). For $c_1^2$ the agreement with our findings is rather good, although there are some deviations for strong interactions as in case III. For $c_2^2$ our numerical determination becomes unreliable for $T/n^{2/3}<1$ since $c_2^2$ is dominated by the term $(s^2 n_S T)/(n^2 n_N c_v)$ in Eq.\ \eqref{eq:firstandsecondsound}. In the limit $T\to 0$ the quantities $s$, $n_N$, and $c_v$ also go to zero so that the numerical value for $c_2^2$ is sensitive to the precise way how this limit is approached.

We observe that Eq.\ \eqref{eq:firstandsecondsound} can be written as
\begin{eqnarray}
\nonumber
(M c^2)^2-\left[M v_S^2+\frac{n_S T s^2}{(n-n_S) c_v n^2}\right](Mc^2)\\
+\frac{n_S T s^2}{(n-n_S)c_v n^2}M v_T^2=0,
\end{eqnarray}
with the single fluid isothermal and adiabatic sound velocities $v_T$ and $v_S$ given by Eqs. \eqref{eq:singlefluidisothermalsound} and \eqref{eq:singlefluidadiabaticsound}. This shows that $c$ coincides with $v_S$ in the disordered phase where $n_S=0$. 
An intuitive form of the wave equation can be written as
\begin{eqnarray}
\nonumber
\partial_t^2 \delta n &=& v_T^2\Delta \delta n+ (v_S^2-v_T^2)\Delta \delta \tilde T,\\
\partial_t^2\delta \tilde T &=& (v_S^2-v_T^2+\bar v^2)\Delta\delta \tilde T + v_T^2 \Delta \delta n,
\end{eqnarray}
with
\begin{equation}
M\bar v^2 = \frac{s^2 n_S T}{n^2(n-n_S)c_v}, \quad \delta\tilde T=\frac{\delta T}{\eta}
\end{equation}
and
\begin{eqnarray}
\eta &=& -\frac{T}{c_v}\frac{\partial(s/n)}{\partial n}{\big |}_T = \frac{\partial T}{\partial n}{\big |}_{s/n}.
\end{eqnarray}
For fluctuations of $\delta n$ and $\delta \tilde T$ only $v_T$, $v_S$ and $\bar v$ matter. In the limit $T\to0$ one observes $v_S^2\to v_T^2$ such that the fluctuations $\delta n$ are governed by the isothermal sound velocity $v_T$. On the other hand, the the velocity $\bar v$ characterizes the dynamics of a linear combination of $\delta \tilde T$ and $\delta n$.
		
	\section{Superfluid Bose gas in two dimensions}
	\label{sec:SuperfluidBosegasintwodimensions}
	We now come to the many-body properties of the Bose gas model (Eq.\ \eqref{microscopicaction}) in the case of two spatial dimensions. Although the formalism is the same as in three dimensions (only one number is changed!) the physical properties are quite different. Many quantities have a logarithmic scale-dependence in two dimensions and the effect of fluctuations is more important. As discussed in section \ref{sec:Bosegasintwodimensions} it is important to realize that experimental observables are associated with a characteristic momentum scale $k_\text{ph}$ and probe the flowing action $\Gamma_{k_\text{ph}}$. We start with the investigation of the flow equations at zero temperature and come then to the phase diagram at nonzero temperature which is governed by the Kosterlitz-Thouless phase transition.

\subsection{Flow equations at zero temperature}

In this section we investigate the many body problem for a nonvanishing density $n$ at zero temperature $T=0$. A crucial new ingredient as compared to the vacuum discussed in section \ref{sec:Repulsiveinteractingbosons} is the nonzero superfluid density 
\begin{equation}
n_S=\rho_0(k_\text{ph}).
\end{equation}
For interacting bosons at zero temperature the density $n$ and the superfluid density $n_S$ are equal. In contrast, the condensate density is given by the unrenormalized order parameter 
\begin{equation}
n_C=\bar{\rho}_0(k_\text{ph})=\bar{A}^{-1}(k_\text{ph})\rho_0(k_\text{ph}).
\end{equation}
Due to the repulsive interaction, $n_C$ may be smaller than $n$, the difference $n-n_C$ being the condensate depletion. (In the limit $\lambda\rightarrow 0$ we have $n=n_S=n_C$.) To obtain a nonzero density at temperature $T=0$ we have to go to positive chemical potential $\mu>0$. At the microscopic scale $k=\Lambda$ the minimum of the effective potential $U$ is then at $\rho_{0,\Lambda}=\mu/\lambda>0$. 

The superfluid density $\rho_0$ is connected to a nonvanishing ``renormalized order parameter'' $\varphi_0$, with $\rho_0=\varphi_0^*\varphi_0$. It is responsible for an effective spontaneous breaking of the U(1)-symmetry. Indeed, the expectation value $\varphi_0$ points out a direction in the complex plane so that the global U(1)-symmetry of phase rotations is broken by the ground state of the system. Goldstone's theorem implies the presence of a gapless Goldstone mode, and the associated linear dispersion relation $\omega\sim|\vec{q}|$ accounts for superfluidity. The Goldstone physics is best described by using a real basis in field space by decomposing the complex field $\varphi=\varphi_0+\frac{1}{\sqrt{2}}(\varphi_1+i\varphi_2)$. Without loss of generality we can choose the expectation value $\varphi_0$ to be real. The real fields $\varphi_1$ and $\varphi_2$ describe then the radial and Goldstone mode. respectively. For $\mu=\mu_0$ the inverse propagator reads in our truncation
\begin{equation}
G^{-1}=\bar{A}\begin{pmatrix} \vec{p}^2+V p_0^2+U^\prime+2\rho U^{\prime\prime}, & -S p_0 \\ S p_0, & \vec{p}^2+V q_0^2+U^\prime \end{pmatrix}.
\label{eqprop}
\end{equation}
Here $\vec{p}$ is the momentum of the collective excitation, and for $T=0$ the frequency obeys $\omega=-ip_0$.
In the regime with spontaneous symmetry breaking, $\rho_0(k)\neq 0$, the propagator for $\rho=\rho_0$ has $U^\prime=0$, $2\rho U^{\prime\prime}=2\lambda \rho_0\neq 0$, giving rise to the linear dispersion relation characteristic for superfluidity. This strongly modifies the flow equations as compared to the vacuum flow equations once $k^2 \ll 2\lambda \rho_0$. For $n\neq0$ the flow is typically in the regime with $\rho_0(k)\neq0$. In practice, we have to adapt the initial value $\rho_{0,\Lambda}$ such that the flow ends at a given density $\rho_0(k_\text{ph})=n$. For $k_\text{ph}\ll n^{1/2}$ one finds that $\rho_0(k_\text{ph})$ depends only very little on $k_\text{ph}$. As mentioned above, we will often choose the density to be unity such that effectively all length scales are measured in units of the interparticle distance $n^{-1/2}$. 

\begin{figure}
\centering
\includegraphics[width=0.5\textwidth]{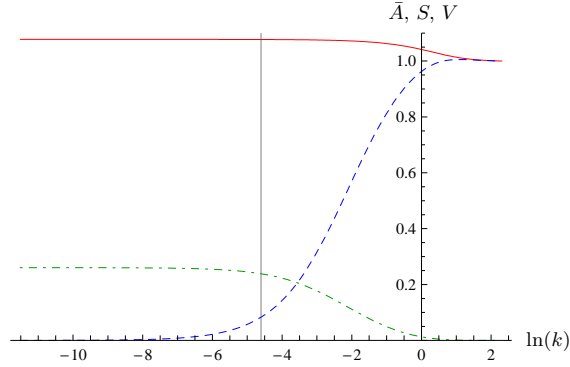}
\caption{Flow of the kinetic coefficients $\bar{A}$ (solid), $S$ (dashed), and $V$ (dashed-dotted) at zero temperature $T=0$, density $n=1$, and vacuum interaction strength $\lambda=1$.}
\label{figFlowKinetic}
\end{figure}

In contrast to the vacuum with $T=\rho_0=0$, the flow of the propagator is nontrivial in the phase with $\rho_0>0$ and spontaneous U(1) symmetry breaking. In Fig. \ref{figFlowKinetic} we show the flow of the kinetic coefficients $\bar{A}$, $V$, $S$ for a renormalized or macroscopic interaction strength $\lambda=1$. 
The wavefunction renormalization $\bar{A}$ increases only a little at scales where $k\approx n^{1/2}$ and saturates then to a value $\bar{A}>1$. As will be explained below, we can directly infer the condensate depletion from the value of $\bar{A}$ at macroscopic scales. The coefficient of the linear $\tau$-derivative $S$ goes to zero for $k\rightarrow 0$. The frequency dependence is then governed by the quadratic $\tau$-derivative with coefficient $V$, which is generated by the flow and saturates to a finite value for $k\rightarrow 0$. 

\begin{figure}
\centering
\includegraphics[width=0.5\textwidth]{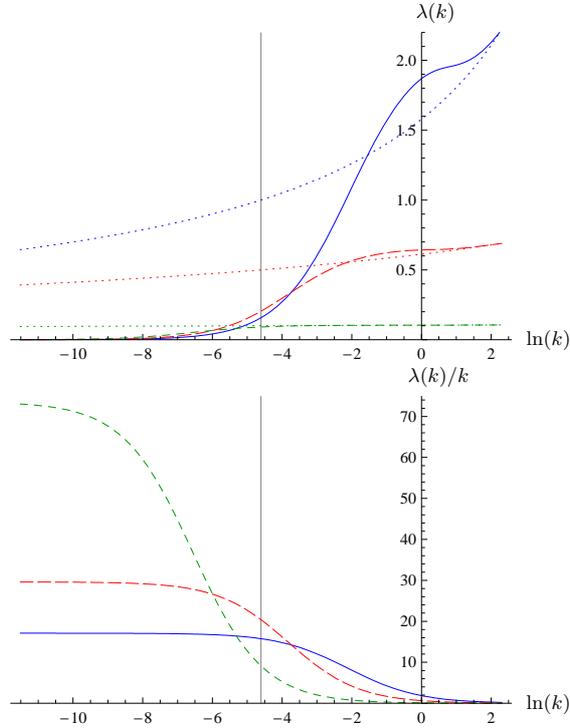}
\caption{Flow of the interaction strength $\lambda(k)$ at zero temperature $T=0$, density $n=1$, for different initial values $\lambda_\Lambda$. The dotted lines are the corresponding graphs in the vacuum $n=0$. The vertical line labels our choice of $k_\text{ph}$. The lower plot shows $\lambda(k)/k$ for the same parameters, demonstrating that $\lambda(k)\sim k$ for small $k$.}
\label{figFlowLambda}
\end{figure}

The flow of the interaction strength $\lambda(k)$ for different values of $\lambda=\lambda(k_\text{ph})$ is shown in Fig. \ref{figFlowLambda}.
While the decrease with the scale $k$ is only logarithmic in vacuum, it becomes now linear $\lambda(k)\sim k$ for $k\ll n^{1/2}$. It is interesting that the ratio $\lambda(k)/k$ reaches larger values for smaller values of $\lambda_\Lambda$. 

\subsection{Quantum depletion of condensate}

\begin{figure}
\centering
\includegraphics[width=0.5\textwidth]{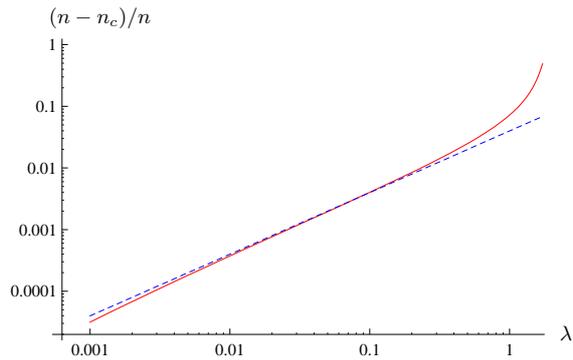}
\caption{Condensate depletion $(n-n_c)/n$ as a function of the vacuum interaction strength $\lambda$. The dashed line is the Bogoliubov result $(n-n_c)/n=\frac{\lambda}{8\pi}$ for reference.}
\label{figDepletion}
\end{figure}

As $k$ is lowered from $\Lambda$ to $k_\text{ph}$, the renormalized order parameter or the superfluid density $\rho_0$ increases first and then saturates to $\rho_0=n=1$. This is expected since the superfluid density equals the total density at zero temperature. In contrast, the bare order parameter $\bar{\rho}_0=\rho_0/\bar{A}$ flows to a smaller value $\bar{\rho}_0<\rho_0$. As argued in section \ref{sec:Bose-EinsteinCondensationinthreedimensions}, the bare order parameter is just the condensate density, such that
\begin{equation}
n-n_C=\rho_0-\bar{\rho}_0=\rho_0(1-\frac{1}{\bar{A}})
\end{equation}
is the condensate depletion. Its dependence on the interaction strength $\lambda$ is shown in Fig. \ref{figDepletion}.
For small interaction strength $\lambda$ the condensate depletion follows roughly the Bogoliubov form
\begin{equation}
\frac{n-n_c}{n}=\frac{\lambda}{8\pi}.
\end{equation}
However, we find small deviations due to the running of $\lambda$ which is absent in Bogoliubov theory. For large interaction strength $\lambda\approx 1$ the deviation from the Bogoliubov result is quite substantial, since the running of $\lambda$ with the scale $k$ is more important. 

\subsection{Dispersion relation and sound velocity}

We also investigate the dispersion relation at zero temperature. The dispersion relation $\omega(p)$ follows from the condition
\begin{equation}
\text{det}\, G^{-1}(\omega(p),p)=0
\label{eqdispersionfromprop}
\end{equation}
where $G^{-1}$ is the inverse propagator after analytic continuation to real time $p_0\rightarrow i\omega$. As was shown at the end of section\ref{sec:Derivativeexpansionandwardidentities} the generation of the kinetic coefficient $V$ by the flow leads to the emergence of a second branch of solutions of Eq.\ \eqref{eqdispersionfromprop}. In our truncation the dispersion relation for the two branches $\omega_+(\vec{p})$ and $\omega_-(\vec{p})$ are
\begin{eqnarray}
\nonumber
\omega_\pm(\vec{p})&=&{\Bigg (}\frac{1}{V}(\vec{p}^2+\lambda \rho_0)+\frac{S^2}{2V^2}\\
&&\pm{\Bigg (}\left(\frac{1}{V}(\vec{p}^2+\lambda\rho_0)+\frac{S^2}{2V^2}\right)^2-\frac{1}{V^2}\vec{p}^2(\vec{p}^2+2\lambda \rho_0){\Bigg )}^{1/2}{\Bigg )}^{1/2}.
\label{eqdispersionrelation}
\end{eqnarray}
In the limit $V\rightarrow 0$, $S\rightarrow 1$ we find that the lower branch approaches the Bogoliubov result $\omega_-\rightarrow \sqrt{\vec{p}^2(\vec{p}^2+2\lambda \rho_0)}$ while the upper branch diverges $\omega_+\rightarrow \infty$ and thus disappears from the spectrum. The lower branch is dominated by phase changes (Goldstone mode), while the upper branch reflects waves in the size of $\rho_0$ (radial mode). 

\begin{figure}
\centering
\includegraphics[width=0.5\textwidth]{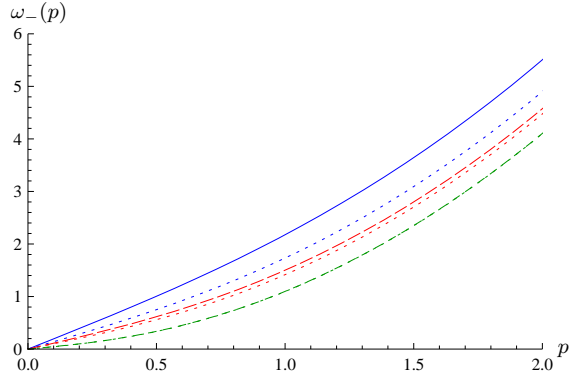}
\caption{Lower branch of the dispersion relation $\omega_{-}(p)$ at temperature $T=0$ and for the vacuum interaction strength $\lambda=1$ (solid curve), $\lambda=0.5$ (upper dashed curve), and $\lambda=0.1$ (lower dashed curve). The units are set by the density $n=1$. We also show the Bogoliubov result for $\lambda=1$ (upper dotted curve) and $\lambda=0.5$ (lower dotted curve). For $\lambda=0.1$ the Bogoliubov result is identical to our result within the plot resolution.}
\label{figDispersionlinear}
\end{figure}

\begin{figure}
\centering
\includegraphics[width=0.5\textwidth]{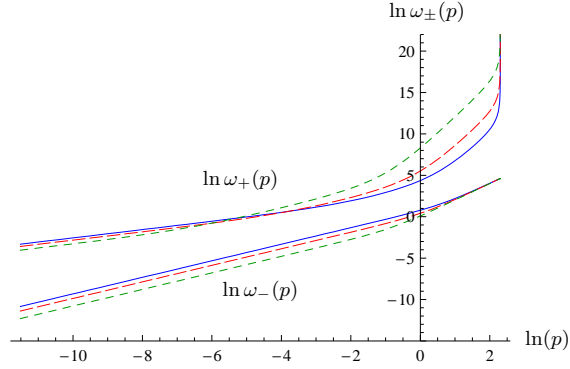}
\caption{Dispersion relation $\omega_{-}(p)$, $\omega_{+}(p)$ at temperature $T=0$ and for vacuum interaction strength $\lambda=1$ (solid), $\lambda=0.5$ (long dashed), and $\lambda=0.1$ (short dashed). The units are set by the density $n=1$.}
\label{figDispersion}
\end{figure}

In principle, the coupling constants on the right-hand side of Eq.\ \eqref{eqdispersionrelation} also depend on the momentum $p=|\vec{p}|$. Since an external momentum provides an infrared cutoff of order $k\approx |\vec{p}|$ we can approximate the $|\vec{p}|$-dependence by using on the right-hand side of Eq.\ \eqref{eqdispersionrelation} the $k$-dependent couplings with the identification $k=|\vec{p}|$. Our result for the lower branch of the dispersion relation is shown in Fig. \ref{figDispersionlinear}.
We also plot the Bogoliubov result $\omega=\sqrt{\vec{p}^2(\vec{p}^2+2\lambda \rho_0)}$ for comparison. For small $\lambda$ our result is in agreement with the Bogoliubov result, while we find substantial deviations for large $\lambda$. Both branches $\omega_+$ and $\omega_-$ are shown in Fig. \ref{figDispersion} on a logarithmic scale. 
Since we start with $V=0$ at the microscopic scale $\Lambda$ we find $\omega_+(\vec{p})\rightarrow\infty$ for $|\vec{p}|\rightarrow \Lambda$.

\begin{figure}
\centering
\includegraphics[width=0.5\textwidth]{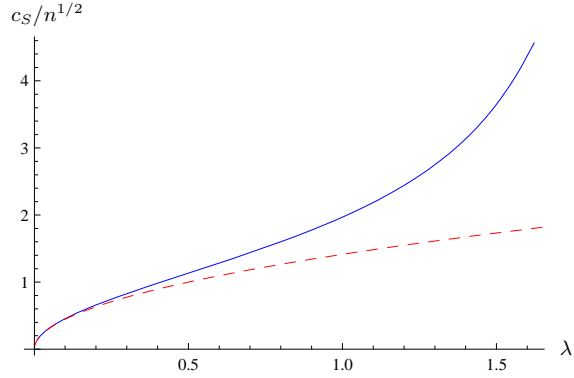}
\caption{Dimensionless sound velocity $c_S/n^{1/2}$ as a function of the vacuum interaction strength (solid). We also show the Bogoliubov result $c_S=\sqrt{2\lambda \rho_0}$ for reference (dashed).}
\label{figsoundvelocity}
\end{figure}

The sound velocity $c_S$ can be extracted from the dispersion relation. More precisely, we compute the microscopic sound velocity for the lower branch $\omega_-(\vec{p})$ as $c_S=\frac{\partial \omega}{\partial p}$ at $p=0$. In our truncation we find
\begin{equation}
c_S^2=\frac{2\lambda\rho_0}{S^2+2\lambda\rho_0 V}.
\end{equation}
Our result for $c_S$ at $T=0$ is shown in Fig. \ref{figsoundvelocity} as a function of the interaction strength $\lambda$. 
For a large range of small $\lambda$ we find good agreement with the Bogoliubov result $c_S^2=2\lambda \rho_0$. However, for large $\lambda$ or result for $c_S$ exceeds the Bogoliubov result by a factor up to 2.

\subsection{Kosterlitz-Thouless physics}

\subsubsection{Superfluidity and order parameter}
At nonzero temperature and for infinite volume, long range order is forbidden in two spatial dimensions by the Mermin-Wagner theorem. Because of that, no proper Bose-Einstein condensation is possible in a two-dimensional homogeneous Bose gas at nonvanishing temperature. However, even if the order parameter vanishes in the thermodynamic limit of infinite volume, one still finds a nonzero superfluid density for low enough temperature. The superfluid density can be considered as the square of a renormalized order parameter $\rho_0=|\varphi_0|^2$ and the particular features of the low-temperature phase can be well understood by the physics of the Goldstone boson for a phase with effective spontaneous symmetry breaking \cite{Wetterich:1991be}. The renormalized order parameter $\varphi_0$ is related to the expectation value of the bosonic field $\bar{\varphi}_0$ and therefore to the condensate density $\bar{\rho}_0=\bar{\varphi}_0^2$ by a wave function renormalization, defined by the behavior of the bare propagator $\bar{G}$ at zero frequency for vanishing momentum
\begin{equation}
\varphi_0=\bar{A}^{1/2}\bar{\varphi}_0,\quad \rho_0=\bar{A}\bar{\rho}_0,\quad \bar{G}^{-1}(\vec{p}\rightarrow 0)=\bar{A}\vec{p}^2.
\end{equation}
While the renormalized order parameter $\rho_0(k)$ remains nonzero for $k\rightarrow 0$ if $T<T_c$, the condensate density $\bar{\rho}_0=\rho_0/\bar{A}$ vanishes since $\bar{A}$
diverges with the anomalous dimension, $\bar{A}\sim k^{-\eta}$. After restoring dimensions the relation
\begin{equation}
\rho_0=\lim\limits_{\vec{p}\to 0}{\frac{\bar{\rho}_0}{\vec{p}^2\bar{G}(\vec{p})}}
\end{equation}
is the Josephson relation \cite{Josephson1966608}. 

The strict distinction between a zero Bose-Einstein condensate $\bar{\rho}_0=0$ and a nonzero superfluid density $\rho_0>0$ for nonzero temperature $0<T<T_c$ is valid only in the infinite volume limit of a homogeneous system. For a finite size of the system, as atoms in a trap, the running of $\bar{A}(k)$ is effectively stopped at some scale $k_\text{ph}$. There are simply no collective modes with wavelength larger than the size of the system, whose fluctuations would be responsible for a further increase of $\bar{A}$. With a finite $\bar{A}_\text{ph}$ both $\bar{\rho}_0$ and $\rho_0$ are nonzero for $T<T_c$, and the distinction between a Bose-Einstein condensate and superfluidity is no longer relevant in practice. For large systems $\bar{A}(k_\text{ph})$ can be large, however, such that the condensate density can be suppressed substantially as compared to the superfluid density. In any case, there is only one critical temperature $T_c$, defined by $\rho_0(T<T_c)>0$.

\subsubsection{Critical temperature}
\begin{figure}
\centering
\includegraphics[width=0.5\textwidth]{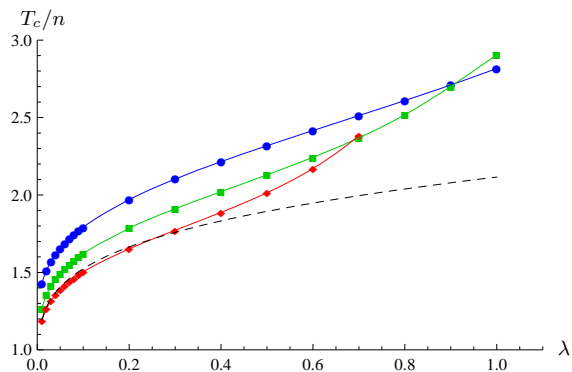}
\caption{Critical temperature $T_c/n$ as a function of the interaction strength $\lambda$. We choose here $k_\text{ph}=10^{-2}$ (circles), $k_\text{ph}=10^{-4}$ (boxes) and $k_\text{ph}=10^{-6}$ (diamonds). For the last case the bound on the scattering length is $\lambda<\frac{4\pi}{\text{ln}(\Lambda/k_\text{ph})}\approx 0.78$. We also show the curve $\frac{T_c}{n}=\frac{4\pi}{\text{ln}(\zeta/\lambda)}$ (dashed) with the Monte-Carlo result $\zeta=380$ \cite{PhysRevLett.87.270402, PhysRevA.66.043608, PhysRevB.69.144504, PhysRevB.59.14054} for reference.}
\label{figtcoflambda}
\end{figure}

The flow equations permit a straightforward computation of $\rho_0(T)$ for arbitrary $T$, once the interaction strength of the system has been fixed at zero temperature and density. We have extracted the critical temperature as a function of $\lambda=\lambda(k_\text{ph})$ for different values of $k_\text{ph}$. The behavior for small $\lambda$,
\begin{equation}
\frac{T_c}{n}=\frac{4\pi}{\text{ln}(\zeta/\lambda)}
\label{Tcperturbative}
\end{equation}
is compatible with the free theory where $T_c$ vanishes for $k_\text{ph}\rightarrow0$ and with the perturbative analysis in Ref. \cite{Popov1983, PhysRevB.37.4936, MarkusHolzmann01302007}. We find that the value of $\zeta$ depends on the choice of $k_\text{ph}$. For $k_\text{ph}=10^{-2}$ we find $\zeta=100$, while $k_\text{ph}=10^{-4}$ corresponds to $\zeta=225$ and $k_\text{ph}=10^{-6}$ to $\zeta=424$. In Fig. \ref{figtcoflambda} we show or result for $T_c/n$ as a function of $\lambda$ for these choices. We also plot the curve in Eq.\ \eqref{Tcperturbative} with the Monte-Carlo result $\zeta=380$ from Ref. \cite{PhysRevLett.87.270402, PhysRevA.66.043608, PhysRevB.69.144504, PhysRevB.59.14054}.
We find that $T_c$ vanishes for $k_\text{ph}\to0$ in the interacting theory as well. This is due to the increase of $\zeta$ and, for a fixed microscopical interaction, to the decrease of $\lambda(k_\text{ph})$. Since the vanishing of $T_c/n$ is only logarithmic in $k_\text{ph}$, a phase transition can be observed in practice. We find agreement with Monte-Carlo results \cite{PhysRevLett.87.270402} for small $\lambda$ if $k_\text{ph}/\Lambda\approx 10^{-7}$. The dependence of $T_c/n$ on the size of the system $k_\text{ph}^{-1}$ remains to be established for the Monte-Carlo computations.

The critical behavior of the system is governed by a Kosterlitz-Thouless phase transition. Usually this is described by considering the thermodynamics of vortices. In Refs. \cite{Grater:1994qx, VonGersdorff:2000kp} it was shown that functional renormalization group can account for this ``nonperturbative'' physics without explicitly taking vortices into account. The correlation length in the low-temperature phase is infinite. In our picture, this arises due to the presence of a Goldstone mode if $\rho_0>0$. The system is superfluid for $T<T_c$. The powerlike decay of the correlation function at zero frequency 
\begin{equation}
\bar{G}(\vec{p})\sim (\vec{p}^2)^{-1+\eta/2}
\label{eq:momentumdependenceofpropagator}
\end{equation}
is directly related to the running of $\bar{A}$. As long as $k^2\gg \vec{p}^2$ the bare propagator obeys approximately
\begin{equation}
\bar{G}=\frac{1}{\bar{A}(k)\vec{p}^2},\quad \bar{A}(k)\sim k^{-\eta}.
\label{eq:momentumdependenceofpropfinitek}
\end{equation}
Once $k^2\ll \vec{p}^2$, the effective infrared cutoff is given by $\vec{p}^2$ instead of $k^2$, and therefore $\bar{A}(k)$ gets replaced by $\bar{A}(\sqrt{\vec{p}^2})$, turning Eq.\ \eqref{eq:momentumdependenceofpropfinitek} into Eq.\ \eqref{eq:momentumdependenceofpropagator}. For large $\rho_0$ the anomalous dimension depends on $\rho_0$ and $T$, $\eta=T/(4\pi \rho_0)$.

\subsubsection{Superfluid fraction}	
Another characteristic feature of the Kosterlitz-Thouless phase transition is a jump in the superfluid density at the critical temperature. However, a true discontinuity arises only in the thermodynamic limit of infinite volume ($k_\text{ph}\rightarrow 0$), while for finite systems ($k_\text{ph}>0$) the transition is smoothened. In order to see the jump, as well as essential scaling for $T$ approaching $T_c$ from above, our truncation is insufficient. These features become visible only in extended truncations that we will briefly describe next. 

For very small scales $\frac{k^2}{T}\ll 1$, the contribution of Matsubara modes with frequency $q_0=2\pi T n$, $n\neq 0$, is suppressed since nonzero Matsubara frequencies act as an infrared cutoff. In this limit a dimensionally reduced theory becomes valid. The long distance physics is dominated by classical two-dimensional statistics, and the time dimension parametrized by $\tau$ no longer plays a role. 

The flow equations simplify considerably if only the zero Matsubara frequency is included, and one can use more involved truncations. Such an improved truncation is indeed needed to account for the jump in the superfluid density. In Ref. \cite{VonGersdorff:2000kp} the next to leading order in a systematic derivative expansion was investigated. It was found that for $k\ll T$ the flow equation for $\rho_0$ can be well approximated by
\begin{equation}
\partial_t \rho_0=2.54\,T^{-1/2}(0.248\, T-\rho_0)^{3/2}\,\theta(0.248 \,T-\rho_0).
\label{eq:improvedflowofrho}
\end{equation}
We switch from the flow equation in our more simple truncation to the improved flow equation \eqref{eq:improvedflowofrho} for scales $k$ with $k^2/T<10^{-3}$. We keep all other flow equations unchanged. A similar procedure was also used in Ref. \cite{DrHCK}.

\begin{figure}
\centering
\includegraphics[width=0.5\textwidth]{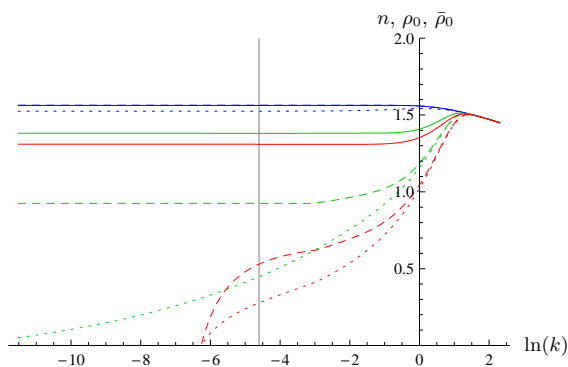}
\caption{Flow of the density $n$ (solid), the superfluid density $\rho_0$ (dashed), and the condensate density $\bar{\rho}_0$ (dotted) for chemical potential $\mu=1$, vacuum interaction strength $\lambda=0.5$ and temperatures $T=0$ (top), $T=2.4$ (middle) and $T=2.8$ (bottom). The vertical line marks our choice of $k_\text{ph}$. We recall $n=\rho_0$ for $T=0$ such that the upper dashed and solid lines coincide.}
\label{figFlowofnrho}
\end{figure}

\begin{figure}
\centering
\includegraphics[width=0.5\textwidth]{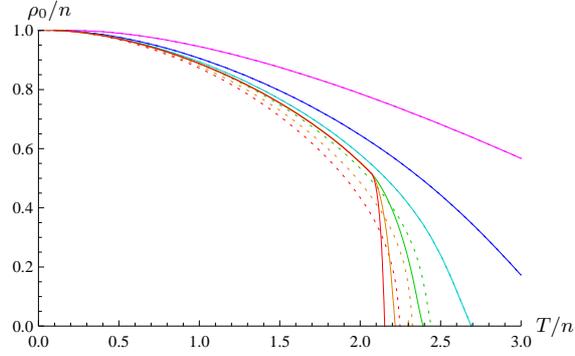}
\caption{Superfluid fraction of the density $\rho_0/n$ as a function of the dimensionless temperature $T/n$ for interaction strength $\lambda=0.5$ at different macroscopic scales $k_\text{ph}=1$ (upper curve), $k_\text{ph}=10^{-0.5}$, $k_\text{ph}=10^{-1}$, $k_\text{ph}=10^{-1.5}$, $k_\text{ph}=10^{-2}$, $k_\text{ph}=10^{-2.5}$ (bottom curve). We plot the result obtained with the improved truncation for small scales (solid) as well as the result obtained with our more simple truncation (dotted). (The truncations differ only for the three lowest lines.)}
\label{figrhodnoftemperature}
\end{figure}

\begin{figure}
\centering
\includegraphics[width=0.5\textwidth]{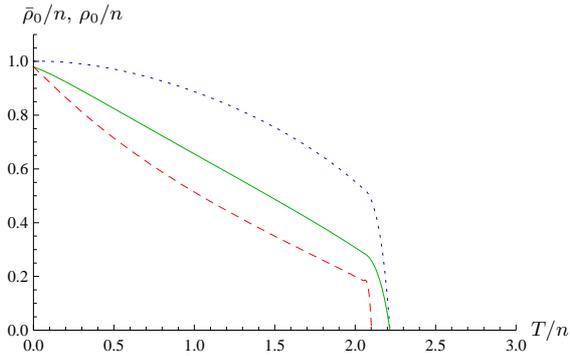}
\caption{Condensate fraction of the density $\bar{\rho}_0/n$ as a function of the dimensionless temperature $T/n$ for interaction strength $\lambda=0.5$ at macroscopic scale $k_\text{ph}=10^{-2}$ (solid curve) and $k_\text{ph}=10^{-4}$ (dashed curve). For comparison, we also plot the superfluid density $\rho_0/n$ at $k_\text{ph}=10^{-2}$ (dotted). These results are obtained with the improved truncation.}
\label{figcondensatedensity}
\end{figure}

In Fig. \ref{figFlowofnrho} we show the flow of the density $n$, the superfluid density $\rho_0$ and the condensate density $\bar{\rho}_0$ for different temperatures.
In Fig. \ref{figrhodnoftemperature} we plot our result for the superfluid fraction of the density as a function of the temperature for different scales $k_\text{ph}$. 
One can see that with the improved truncation the jump in the superfluid density is indeed found in the limit $k_\text{ph}\rightarrow 0$. Fig. \ref{figcondensatedensity} 
shows the condensate fraction $\bar{\rho}_0/n$ and the superfluid density fraction $\rho_0/n$ as a function of $T/n$. We observe the substantial $k_\text{ph}$ dependence of the condensate fraction, as well as an effective jump at $T_c$ for small $k_\text{ph}$. We recall that the infinite volume limit $k_\text{ph}=0$ amounts to $\bar{\rho}_0=0$ for $T>0$. 

The Kosterlitz-Thouless description is only valid if the zero Matsubara frequency mode ($n=0$) dominates. For a given nonzero $T$ this is always the case if the the characteristic length scale goes to infinity. In the infinite volume limit the characteristic length scale is given by the correlation length $\xi$. The description in terms of a classical two dimensional system with U(1) symmetry is the key ingredient of the Kosterlitz-Thouless description and holds for $\xi^2 T\gg 1$. In the infinite volume limit this always holds for $T<T_c$ or near the phase transition, where $\xi$ diverges or is very large. For a finite size system the relevant length scale becomes $k_\text{ph}^{-1}$ if this is smaller than $\xi$. Thus the Kosterlitz-Thouless picture holds only for $T>k_\text{ph}^2$. 

For very small temperatures $T<k_\text{ph}^2$ one expects a crossover to the characteristic behavior near a quantum critical phase transition, governed by the quantum critical fixed point. The crossover between the different characteristic behaviors for $T>k_\text{ph}^2$ and $T<k_\text{ph}^2$ can be observed in several quantities. As an example we may take Fig. \ref{figFlowofnrho} and compare the flow of $\rho_0$ and $\bar{\rho}_0$ for low $T$ (close to the $T=0$ curve) or large $T$ (other curves).

	\section{Particle-hole fluctuations and the BCS-BEC Crossover}
	\label{sec:Particle-holefluctuationsandtheBCS-BECCrossover}
	In this chapter we discuss the many-body properties of the BCS-BEC crossover model in Eq.\ \eqref{eqMicroscopicAction}. We treat with our method the whole crossover phase diagram but concentrate the discussion on the effect of particle-hole fluctuations. These fluctuations give rise to a the first nontrivial correction to BCS theory on the fermionic side of the crossover. For small negative scattering length their effect can be included in a perturbative setting as was shown by Gorkov and Melik-Barkhudarov \cite{Gorkov}.

\subsection{Particle-hole fluctuations}
\label{sec:ParticleHole}
The BCS theory of superfluidity in a Fermi gas of atoms is valid for a small attractive interaction between the fermions \cite{PhysRev.104.1189, Bardeen:1957kj, Bardeen:1957mv}. In a renormalization group setting, the features of BCS theory can be described in a purely fermionic language. The only scale dependent object is the fermion interaction vertex $\lambda_\psi$. The flow depends on the temperature and the chemical potential. 
For positive chemical potential ($\mu>0$) and small temperatures $T$, the appearance of pairing is indicated by the divergence of $\lambda_\psi$.

\begin{figure}
\centering
\includegraphics[width=0.5\textwidth]{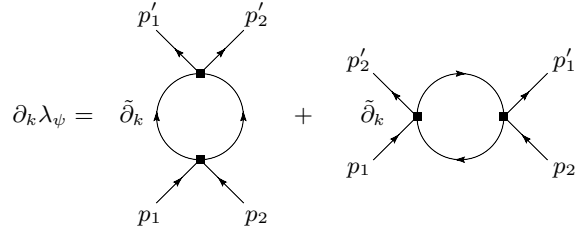}
\caption{Running of the momentum dependent vertex $\lambda_{\psi}$. Here $\tilde{\partial}_k$ indicates derivatives with respect to the cutoff terms in the propagators and does not act on the vertices in the depicted diagrams. We will refer to the first loop as the particle-particle loop (pp-loop) and to the second one as the particle-hole loop (ph-loop).}
\label{fig:lambdaflow}
\end{figure}

In general, the interaction vertex is momentum dependent and represented by a term
\begin{eqnarray}
\Gamma_{\lambda_\psi}&=&\int_{p_1,p_2,p_1^\prime,p_2^\prime}\lambda_{\psi}(p_1^\prime,p_1,p_2^\prime,p_2)\nonumber\\
& &\times\psi_1^{\ast}(p_1^\prime)\psi_1(p_1)\psi_2^{\ast}(p_2^\prime)\psi_2(p_2)
\label{eq:momentumdepvertex}
\end{eqnarray}
in the effective action. In a homogeneous situation, momentum conservation restricts the expression in Eq.\ \eqref{eq:momentumdepvertex} to three independent momenta, $\lambda_{\psi}\sim \delta(p_1^\prime+p_2^\prime-p_1-p_2)$. The flow of $\lambda_\psi$ has two contributions which are depicted in Fig. \ref{fig:lambdaflow}. 
The first diagram describes particle-particle fluctuations. For $\mu>0$ its effect increases as the temperature $T$ is lowered. For small temperatures $T\leq T_{c,\text{BCS}}$ the logarithmic divergence leads to the appearance of pairing, as $\lambda_\psi\to \infty$. 

In the purely fermionic formulation the flow equation for $\lambda_{\psi}$ has the general form \cite{Ellwanger1994137, Aoki2000, PTPS.160.58, PTP.105.1}

\begin{equation}\label{eq:lambdapsi2}
\partial_k \lambda_{\psi}^{\alpha}=A^{\alpha}_{\beta\gamma}\lambda_{\psi}^{\beta}\lambda_{\psi}^{\gamma}\,, 
\end{equation}
with $\alpha, \beta, \gamma$ denoting momentum as well as spin labels. A numerical solution of this equation is rather involved due to the rich momentum structure. The case of the attractive Hubbard model in two dimensions, which is close to our problem, has recently been discussed in \cite{strack:014522}.
The BCS approach concentrates on the pointlike coupling, evaluated by setting all momenta to zero. For $k \rightarrow 0,\ \mu_0 \rightarrow 0,\ T \rightarrow 0$ and $n \rightarrow 0$ this coupling is related the scattering length, $a= \frac{1}{8\pi} \lambda_{\psi}(p_i=0)$. In the BCS approximation only the first diagram in Fig. \ref{fig:lambdaflow} is kept, and the momentum dependence of the couplings on the right-hand side of Eq.\ \eqref{eq:lambdapsi2} is neglected, by replacing $\lambda_{\psi}^{\alpha}$ by the pointlike coupling evaluated at zero momentum. In terms of the scattering length $a$, Fermi momentum $k_F$ and Fermi temperature $T_F$, the critical temperature is found to be 
\begin{equation}
 \frac{T_c}{T_F}\approx 0.61 e^{\pi/(2 a k_F)}\,.
\end{equation}
This is the result of the original BCS theory. However, it is obtained by entirely neglecting the second loop in Fig. \ref{fig:lambdaflow}, which describes particle-hole fluctuations. At zero temperature the expression for this second diagram vanishes if it is evaluated for vanishing external momenta. Indeed, the two poles of the frequency integration are always either in the upper or lower half of the complex plane and the contour of the frequency integration can be closed in the half plane without poles. 

The dominant part of the scattering in a fermion gas occurs, however, for momenta on the Fermi surface  rather than for zero momentum. For non-zero momenta of the "external particles" the second diagram in Fig. \ref{fig:lambdaflow} - the particle-hole channel - makes an important contribution. 

Setting the external frequencies to zero, we find that the inverse propagators in the particle-hole loop are 
\begin{equation}\label{eq:loopmom1}
P_\psi(q)=i q_0 +(\vec{q}-\vec{p}_1)^2-\mu\,,
\end{equation}
and 
\begin{equation}\label{eq:loopmom2}
P_\psi(q)=i q_0 +(\vec{q}-\vec{p}_2^{\,\prime})^2-\mu.
\end{equation}
Depending on the value of the momenta $\vec{p}_1$ and $\vec p_2^{\,\prime}$, there are now values of the loop momentum $\vec q$ for which the poles of the frequency integration are in different half planes so that there is a nonzero contribution even for $T=0$.

To include the effect of particle-hole fluctuations one could try to take the full momentum dependence of the vertex $\lambda_\psi$ into account. However, this leads to complicated expressions which are hard to solve even numerically. 
One therefore often restricts the flow to the running of a single coupling $\lambda_\psi$ by choosing an appropriate projection prescription to determine the flow equation. In the purely fermionic description with a single running coupling $\lambda_\psi$, this flow equation has a simple structure. The solution for $\lambda_{\psi}^{-1}$ can be written as a contribution from the particle-particle (first diagram in Fig. \ref{fig:lambdaflow}, pp-loop) and the particle-hole (second diagram, ph-loop) channels 
\begin{equation}\label{PHComp}
 \frac{1}{\lambda_{\psi}(k=0)}=\frac{1}{\lambda_{\psi}(k=\Lambda)} + \mbox{pp-loop} + \mbox{ph-loop}\,.
\end{equation}
Since the ph-loop depends only weakly on the temperature, one can evaluate it at $T=0$ and add it to the initial value $\lambda_\psi(k=\Lambda)^{-1}$. Since $T_c$ depends exponentially on the "effective microscopic coupling"
\begin{equation}
 \left(\lambda_{\psi,\Lambda}^{\text{eff}}\right)^{-1}=\lambda_{\psi}(k=\Lambda)^{-1} + \text{ph-loop}\,,
\end{equation}
any shift in $\left(\lambda_{\psi,\Lambda}^{\text{eff}}\right)^{-1}$ results in a multiplicative factor for $T_c$. The numerical value of the ph-loop and therefore of the correction factor for $T_c/T_F$ depends on the precise projection description.

Let us now choose the appropriate momentum configuration. For the formation of Cooper pairs, the  relevant momenta lie on the Fermi surface, 

\begin{equation}\label{absmom}
\vec{p}^2_1=\vec{p}^2_2=\vec{p}^{{\,\prime}2}_1=\vec{p}^{{\,\prime}2}_2=\mu\,,
\end{equation}
and point in opposite directions

\begin{equation}\label{oppmom}
 \vec{p}_1=-\vec{p}_2,\ \vec{p}^{\,\prime}_1=-\vec{p}^{\,\prime}_2\,.
\end{equation}
This still leaves the angle between $\vec{p}_1$ and $\vec{p}^{\,\prime}_1$ unspecified. Gorkov's approximation uses Eqs. \eqref{absmom} and \eqref{oppmom} and projects on the $s$-wave by averaging over the angle between $\vec{p}_1$ and $\vec{p}^{\,\prime}_1$. One can shift the loop momentum such that the internal propagators depend on $\vec{q}^2$ and $(\vec{q}+\vec{p}_1-\vec{p}^{\,\prime}_1)^2$. In terms of spherical coordinates the first propagator depends only on the magnitude of the loop momentum $q^2=\vec{q}^2$, while the second depends additionally on the transfer momentum $\tilde{p}^2=\frac{1}{4}(\vec{p}_1-\vec{p}^{\,\prime}_1)^2$ and the angle $\alpha$ between $\vec{q}$ and $(\vec{p}_1-\vec{p}^{\,\prime}_1)$,

\begin{equation}
 (\vec{q}+\vec{p}_1-\vec{p}^{\,\prime}_1)^2=q^2+4\tilde{p}^2+4\,q\, \tilde{p}\,\text{cos}(\alpha)\,.
\end{equation}
Performing the loop integration involves the integration over $q^2$ and the angle $\alpha$. The averaging over the angle between $\vec{p}_1$ and $\vec{p}_1^{\,\prime}$ translates to an averaging over $\tilde{p}^2$. Both can be done analytically \cite{Heiselberg} for the fermionic particle-hole diagram and the result gives the well-known Gorkov correction to BCS theory, resulting in

\begin{equation}
T_c=\frac{1}{(4e)^{1/3}}T_{c,\text{BCS}}\approx \frac{1}{2.2} T_{c,\text{BCS}}\,.
\end{equation}

In our treatment we will use a numerically simpler projection by choosing $\vec{p}^{\,\prime}_1=\vec{p}_1$, and $\vec{p}_2=\vec{p}^{\,\prime}_2$, without an averaging over the angle between $\vec{p}^{\,\prime}_1$ and $\vec{p}_1$. The size of $\tilde p^2 = \vec{p}^2_1$ is chosen such that the one-loop result reproduces exactly the result of the Gorkov correction, namely $\tilde p = 0.7326 \sqrt{\mu}$. Choosing different values of $\tilde p$ demonstrates the dependence of $T_c$ on the projection procedure and may be taken as an estimate for the error that arises from the limitation to one single coupling $\lambda_{\psi}$ instead of a momentum dependent function.

\subsection{Bosonization}
In Sec. \ref{sec:BCS-BECCrossover} we describe an effective four-fermion interaction by the exchange of a boson. In this picture the phase transition to the superfluid phase is indicated by the vanishing of the bosonic  ``mass term'' $m^2 = 0$. Negative $m^2$ leads to the spontaneous breaking of U(1)-symmetry, since the minimum of the effective potential occurs for a nonvanishing superfluid density $\rho_0>0$.

\begin{figure}
\centering
\includegraphics[width=0.35\textwidth]{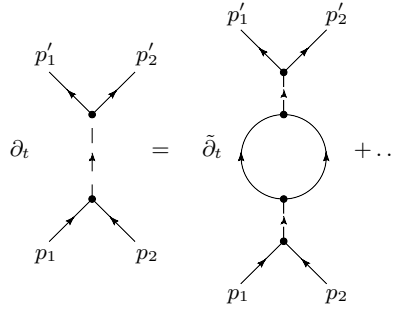}
\caption{Flow of the boson propagator.}
\label{fig:bosonexchangeloop}
\end{figure}
For $m^2 \geq 0$ we can solve the field equation for the boson $\varphi$ as a functional of $\psi$ and insert the solution into the effective action. This leads to an effective four-fermion vertex describing the scattering $\psi_1(p_1)\psi_2(p_2)\to \psi_1(p_1^{\,\prime})\psi_2(p_2^{\,\prime})$
\begin{equation}
\lambda_{\psi,\text{eff}}=\frac{-h^2}{i(p_1+p_2)_0+\frac{1}{2}(\vec p_1+\vec p_2)^2+m^2}.
\label{eq:lambdapsieff}
\end{equation}
To investigate the breaking of U(1) symmetry and the onset of superfluidity, we first consider the flow of the bosonic propagator, which is mainly driven by the fermionic loop diagram. For the effective four-fermion interaction this accounts for the particle-particle loop (see r.h.s. of Fig. \ref{fig:bosonexchangeloop}). In the BCS limit of a large microscopic $m_\Lambda^2$ the running of $m^2$ for $k\to0$ reproduces the BCS result \cite{PhysRev.104.1189, Bardeen:1957kj, Bardeen:1957mv}.

The particle-hole fluctuations are not accounted for by the renormalization of the boson propagator. Indeed, we have neglected so far that a term

\begin{equation}\label{eq:fourfermionvertex}
 \int_{\tau,\vec{x}}\lambda_{\psi}\psi_1^{\ast}\psi_1\psi_2^{\ast}\psi_2\,,
\end{equation}
in the effective action is generated by the flow. This holds even if the microscopic pointlike interaction is absorbed by a Hubbard-Stratonovich transformation into an effective boson exchange such that $\lambda_\psi(\Lambda)=0$. The strength of the total interaction between fermions

\begin{equation}\label{eq:lambdapsieff2}
\lambda_{\psi,\text{eff}}=\frac{-h^2}{i(p_1+p_2)_0+\frac{1}{2}(\vec p_1+\vec p_2)^2+m^2} + \lambda_{\psi}
\end{equation}
adds $\lambda_\psi$ to the piece generated by boson exchange. 
\begin{figure}
\centering
\includegraphics[width=0.35\textwidth]{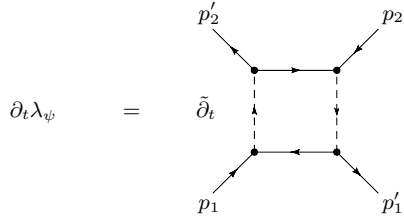}
\caption{Box diagram for the flow of the four-fermion interaction.}
\label{fig:boxes}
\end{figure}
In the partially bosonized formulation, the flow of $\lambda_\psi$ is generated by the box-diagrams depicted in Fig. \ref{fig:boxes}. We may interpret these diagrams and establish a direct connection to the particle-hole diagrams depicted in Fig. \ref{fig:lambdaflow} on the BCS side of the crossover and in the microscopic regime. There the boson gap $m^2$ is large. In this case, the effective fermion interaction in Eq.\ \eqref{eq:lambdapsieff2} becomes momentum independent. Diagrammatically, this is represented by contracting the bosonic propagator. One can see, that the box-diagram in Fig. \ref{fig:boxes} is then equivalent to the particle-hole loop investigated in Sec. \ref{sec:ParticleHole} with the pointlike approximation $\lambda_{\psi,\text{eff}}\to-\frac{h^2}{m^2}$ for the fermion interaction vertex. As mentioned above, these contributions vanish for $T=0$, $\mu<0$ for arbitrary $\vec p_i$. Indeed, at zero temperature, the summation over the Matsubara frequencies becomes an integral. All the poles of this integration are in the upper half of the complex plane and the integration contour can be closed in the lower half plane. We will evaluate $\partial_k \lambda_\psi$ for $\vec p_1=\vec p_1^{\,\prime}=-\vec p_2=-\vec p_2^{\,\prime}$, $|\vec p_1|=\tilde p = 0.7326 \sqrt{\mu}$, as discussed in the Sec. \ref{sec:ParticleHole}. For $\mu>0$ this yields a nonvanishing flow even for $T=0$.

Another simplification concerns the temperature dependence. While the contribution of particle-particle diagrams becomes very large for small temperatures, this is not the case for particle-hole diagrams. For nonvanishing density and small temperatures, the large effect of particle-particle fluctuations leads to the spontaneous breaking of the U(1) symmetry and the associated superfluidity. In contrast, the particle-hole fluctuations lead only to quantitative corrections and depend only weakly on temperature. This can be checked explicitly in the pointlike approximation, and holds not only in the BCS regime where $T/\mu \ll 1$, but also for moderate $T/\mu$ as realized at the critical temperature in the unitary regime. We can therefore evaluate the box-diagrams in Fig. \ref{fig:lambdaflow} for zero temperature. We note that an implicit temperature dependence, resulting from the couplings parameterizing the boson propagator, is taken into account.

After these preliminaries, we can now incorporate the effect of particle-hole fluctuations in the renormalization group flow. A first idea might be to include the additional term \eqref{eq:fourfermionvertex} in the truncation and to study the effects of $\lambda_{\psi}$ on the remaining flow equations. On the initial or microscopic scale one would have $\lambda_{\psi}=0$, but it would then be generated by the flow. This procedure, however, has several shortcomings. First, the appearance of a local condensate would now be indicated by the divergence of the effective four-fermion interaction

\begin{equation}
 \lambda_{\psi,\text{eff}}=-\frac{h^2}{m^2}+\lambda_{\psi}\,.
\end{equation}
This might lead to numerical instabilities for large or diverging $\lambda_{\psi}$. The simple picture that the divergence of $\lambda_{\psi,\text{eff}}$ is connected to the onset of a nonvanishing expectation value for the bosonic field $\varphi_0$, at least on intermediate scales, would not hold anymore. Furthermore, the dependence of the box-diagrams on the center of mass momentum would be neglected completely by this procedure. Close to the resonance the momentum dependence of the effective four-fermion interaction in the bosonized language as in Eq.\ (\ref{eq:lambdapsieff2}) is crucial, and this might also be the case for the particle-hole contribution.

Another, much more elegant way to incorporate the effect of particle-hole fluctuations is provided by the method of bosonization \cite{Gies:2001nw, Gies:2002kd, Pawlowski2007a}, see also chapter \ref{ch:Generalizedflowequation}. For this purpose, we use scale dependent fields in the average action. The scale dependence of $\Gamma_k[\chi_k]$ is modified by a term reflecting the $k$-dependence of the argument $\chi_k$ \cite{Gies:2001nw, Gies:2002kd}

\begin{equation}\label{eq:scalefield}
 \partial_k \Gamma_k[\chi_k]=\int\frac{\delta \Gamma_k}{\delta \chi_k}\partial_k\chi_k+\frac{1}{2}\mathrm{STr}\left[ \left(\Gamma_k^{(2)}+R_k \right)^{-1} \partial_k R_k\right] \,.
\end{equation}

For our purpose it is sufficient to work with scale dependent bosonic fields $\bar\varphi$ and keep the fermionic field $\psi$ scale independent. In practice, we employ bosonic fields $\bar\varphi_k^*$, and $\bar\varphi_k$ with an explicit 
scale dependence which reads in momentum space
\begin{eqnarray}
\nonumber
\partial_k \bar \varphi_k(q) & = & (\psi_1\psi_2)(q) \partial_k \upsilon\,,\\
\partial_k \bar \varphi_k^*(q) & = & (\psi_2^\dagger\psi_1^\dagger)(q) \partial_k \upsilon.
\label{eq:scaledependenceoffields}
\end{eqnarray}
In consequence, the flow equations in the symmetric regime get modified

\begin{eqnarray}
 \partial_k \bar{h} &=& \partial_k \bar{h}{\big |}_{\bar \varphi_k}-\bar{P}_{\varphi}(q)\partial_k \upsilon\,,\\
 \partial_k \lambda_{\psi} &=& \partial_k \lambda_{\psi}{\big |}_{\bar \varphi_k}-2\bar{h}\partial_k \upsilon.
\label{eq:modifiedflowequations}
\end{eqnarray}
Here $q$ is the center of mass momentum of the scattering fermions. In the notation of Eq.\ \eqref{eq:lambdapsieff} we have $q=p_1+p_2$ and we will take $\vec q=0$, and $q_0=0$. The first term on the right hand side in Eq.\ \eqref{eq:modifiedflowequations} gives the contribution of the flow equation which is valid for fixed field $\bar \varphi_k$. The second term comes from the explicit scale dependence of $\bar \varphi_k$. The inverse propagator of the complex boson field $\bar{\varphi}$ is denoted by $\bar{P}_{\varphi}(q)=\bar A_\varphi P_\varphi(q)=\bar A_\varphi (m^2+i Z_\varphi q_0+\vec q^2/2)$, cf. Eq.\ \eqref{eq:Bosonpropagator}. 

We can choose $\partial_k \upsilon$ such that the flow of the coupling $\lambda_{\psi}$ vanishes, i.e. that we have $\lambda_{\psi}=0$ on all scales. This modifies the flow equation for the renormalized Yukawa coupling according to

\begin{equation}
 \partial_k h = \partial_k h{\big |}_{\bar \varphi_k}-\frac{m^2}{2h}\partial_k \lambda_{\psi}{\big |}_{\bar \varphi_k}\,,
 \label{eq:modfiedflowofh}
\end{equation}
with $\partial_k h{\big |}_{\bar \varphi_k}$ the contribution without bosonization and $\partial_k \lambda_\psi{\big |}_{\bar \varphi_k}$ given by the box diagram in Fig. \ref{fig:boxes}. Since $\lambda_\psi$ remains zero during the flow, the effective four-fermion interaction $\lambda_{\psi,\text{eff}}$ is now purely given by the boson exchange. However, the contribution of the particle-hole exchange diagrams is incorporated via the second term in Eq.\ \eqref{eq:modfiedflowofh}. 

In the regime with spontaneously broken symmetry we use a real basis for the bosonic field
\begin{equation}
 \bar{\varphi}=\bar{\varphi}_0+\frac{1}{\sqrt{2}}(\bar{\varphi}_1+i\bar{\varphi}_2),
\end{equation}
where the expectation value $\bar{\varphi}_0$ is chosen to be real without loss of generality. The real fields $\bar \varphi_1$ and $\bar \varphi_2$ then describe the radial and the Goldstone mode, respectively. To determine the flow equation of $\bar{h}$, we use the projection description

\begin{equation}\label{eq:projectiononh}
 \partial_k \bar{h}=i\sqrt{2}\Omega^{-1}\frac{\delta}{\delta\varphi_2(0)}\frac{\delta}{\delta\psi_1(0)}\frac{\delta}{\delta\psi_2(0)}\partial_k \Gamma_k\,,
\end{equation}
with the four volume $\Omega=\frac{1}{T}\int_{\vec{x}}$. Since the Goldstone mode has vanishing ``mass'', the flow of the Yukawa coupling is not modified by the box diagram (Fig. \ref{fig:boxes}) in the regime with spontaneous symmetry breaking.

We emphasize that the non-perturbative nature of the flow equations for the various couplings provides for a resummation similar to the one in Eq.\ \eqref{PHComp}, and thus goes beyond the treatment by Gorkov and Melik-Barkhudarov \cite{Gorkov} which includes the particle-hole diagrams only in a perturbative way. Furthermore, the inner bosonic lines $h^2/P_\varphi(q)$ in the box-diagrams represent the center of mass momentum dependence of the four-fermion vertex. This center of mass momentum dependence is neglected in Gorkov's pointlike treatment, and thus represents a further improvement of the classic calculation. Actually, this momentum dependence becomes substantial -- and should not be neglected in a consistent treatment -- away from the BCS regime where the physics of the bosonic bound state sets in. Finally, we note that the truncation \eqref{eq:truncation} supplemented with \eqref{eq:fourfermionvertex} closes the truncation to fourth order in the fields except for a fermion-boson vertex $\lambda_{\psi\varphi}\psi^\dagger\psi\varphi^*\varphi$ which plays a role for the scattering physics deep in the BEC regime \cite{DKS} but is not expected to have a very important impact on the critical temperature in the unitarity and BCS regimes.

\subsection{Critical temperature}
To obtain the flow equations for the running couplings of our truncation Eq.\ \eqref{eq:truncation} we use projection prescriptions similar to Eq.\ \eqref{eq:projectiononh}. The resulting system of ordinary coupled differential equations is then solved numerically for different chemical potentials $\mu$ and temperatures $T$. For temperatures sufficiently small compared to the Fermi temperature $T_F=(3\pi^2n)^{2/3}$, $T/T_F\ll 1$ we find that the effective potential $U$ at the macroscopic scale $k=0$ develops a minimum at a nonzero field value $\rho_0>0$, $\partial_\rho U(\rho_0)=0$. The system is then in the superfluid phase. For larger temperatures we find that the minimum is at $\rho_0=0$ and that the ``mass parameter'' $m^2$ is positive, $m^2=\partial_\rho U(0)>0$. The critical temperature $T_c$ of this phase transition between the superfluid and the normal phase is then defined as the temperature where one has
\begin{equation}
\rho_0=0,\quad \partial_\rho U(0)=0\quad \text{at} \quad k=0.
\end{equation}
Throughout the whole crossover the transition $\rho_0\to0$ is continuous as a function of $T$ demonstrating that the phase transition is of second order.

\begin{figure}
 \centering
	\includegraphics[width=0.45\textwidth]{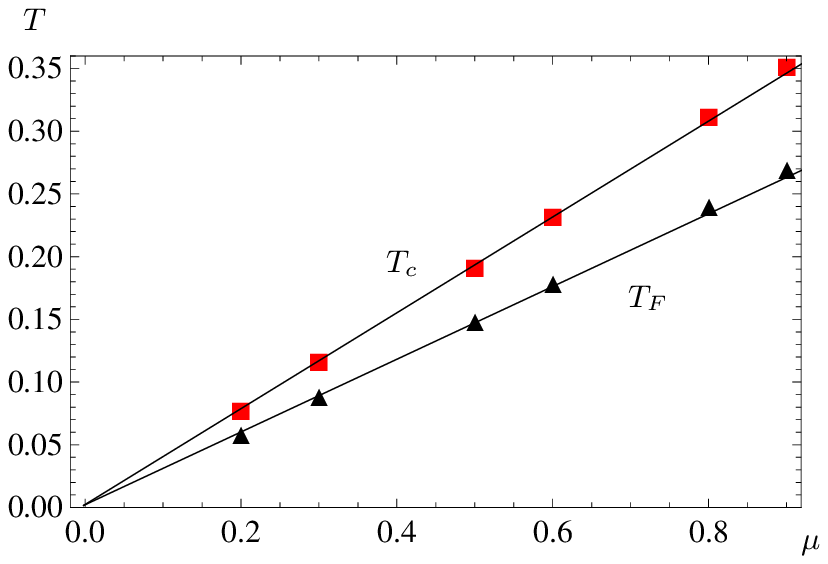}
	\caption{Critical temperature $T_c$  (boxes) and Fermi temperature $T_F=(3\pi^2 n)^{2/3}$ (triangles) as a function of the chemical potential $\mu$. For convenience the Fermi temperature is scaled by a factor 1/5. We also plot the linear fits $T_c=0.39\mu$ and $T_F=1.48\mu$. The units are arbitrary and we use $\Lambda=e^7$.}
	\label{fig:tcrit2}
\end{figure}

In Fig. \ref{fig:tcrit2} we plot our result obtained for the critical temperature $T_c$ and the Fermi temperature $T_F$ as a function of the chemical potential $\mu$ at the unitarity point with $a^{-1}=0$. From dimensional analysis it is clear that both dependencies are linear, $T_c, T_F\sim \mu$, provided that non-universal effects involving the ultraviolet cutoff scale $\Lambda$ can be neglected. That this is indeed found numerically can be seen as a nontrivial test of our approximation scheme and the numerical procedures as well as the universality of the system. Dividing the slope of both lines gives $T_c/T_F=0.264$, a result that will be discussed in more detail below. 
We emphasize that part of the potential error in this estimates is due to uncertainties in the precise quantitative determination of the density or $T_F$.

\subsection{Phase diagram}
\begin{figure}
\centering
\includegraphics[width=0.45\textwidth]{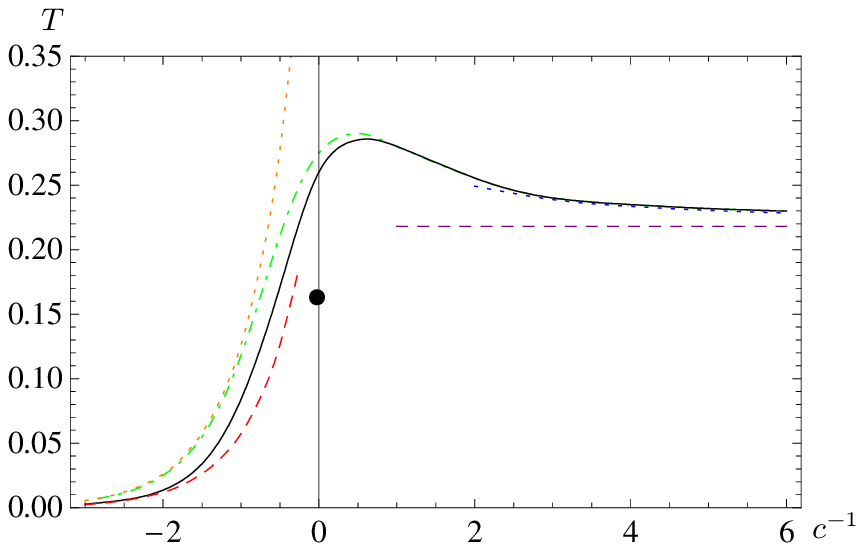}
\caption{Dimensionless critical temperature $T_c/T_F$ as a function of the inverse concentration $c^{-1}=(a k_F)^{-1}$. The black solid line includes the effect of particle-hole fluctuations. We also show the result obtained when particle-hole fluctuations are neglected (dot-dashed line). For comparison, we plot the BCS result without (left dotted line) and with Gorkov's correction (left dashed). On the BEC side with $c^{-1}>1$ we show the critical temperature for a gas of free bosonic molecules (horizontal dashed line) and a fit to the shift in $T_c$ for interacting bosons, $\Delta T_c\sim c$ (dotted line on the right). The black solid dot gives the QMC results \cite{bulgac:090404, bulgac:023625, burovski:160402}.}
\label{fig:tcrit}
\end{figure}

The effect of the particle-hole fluctuations shows most prominently in the result for the critical temperature. With our approach we can compute the critical temperature for the phase transition to superfluidity throughout the crossover. The results are shown in Fig. \ref{fig:tcrit}. We plot the critical temperature in units of the Fermi temperature $T_c/T_F$ as a function of the scattering length measured in units of the inverse Fermi momentum, i.~e. the concentration $c=a k_F$.
We can roughly distinguish three different regimes. On the left side, where $c^{-1}\lesssim-1$, the interaction is weakly attractive. Mean field or BCS theory is qualitatively valid here. In Fig. \ref{fig:tcrit} we denote the BCS result by the dotted line on the left ($c^{-1}<0$). However, the BCS approximation has to be corrected by the effect of particle-hole fluctuations, which lower the value for the critical temperature by a factor of $2.2$. This is the Gorkov correction (dashed line on the left side in Fig. \ref{fig:tcrit}). The second regime is found on the far right side, where the interaction again is weak, but now we find a bound state of two atoms. In this regime the system exhibits Bose-Einstein condensation of molecules as the temperature is decreased. The dashed horizontal line on the right side shows the critical temperature of a free Bose-Einstein condensate of molecules. In-between there is the unitarity regime, where the two-atom scattering length diverges ($c^{-1} \rightarrow 0$) and we deal with a system of strongly interacting fermions.

Our result including the particle-hole fluctuations is given by the solid line. This may be compared with a functional renormalization flow investigation without including particle-hole fluctuations (dot-dashed line) \cite{Diehl:2007th}. For $c\to 0_-$ the solid line of our result matches the BCS theory including the correction by Gorkov and Melik-Barkhudarov \cite{Gorkov},
\begin{equation}
\frac{T_c}{T_F}=\frac{e^C}{\pi}\left(\frac{2}{e}\right)^{7/3} e^{\pi/(2c)}\approx 0.28 e^{\pi/(2c)}.
\end{equation}
In the regime $c^{-1}>-2$ we see that the non-perturbative result given by our RG analysis deviates from Gorkov's result, which is derived in a perturbative setting. 

On the BEC-side for very large and positive $c^{-1}$  our result approaches the critical temperature of a free Bose gas where the bosons have twice the mass of the fermions $M_B=2M$. In our units the critical temperature is then
\begin{equation}
\frac{T_{c,\text{BEC}}}{T_F}=\frac{2\pi}{\left[6\pi^2 \zeta(3/2)\right]^{2/3}}\approx 0.218.
\end{equation} 
For $c\to 0_+$ this value is approached in the form
\begin{equation}
\frac{T_c-T_{c,\text{BEC}}}{T_{c,\text{BEC}}}=\kappa a_M n_M^{1/3}=\kappa \frac{a_M}{a}\frac{c}{(6\pi^2)^{1/3}}.
\end{equation}
Here, $n_M=n/2$ is the density of molecules and $a_M$ is the scattering length between them. For the ratio $a_M/a$ we use our result $a_M/a=0.718$ obtained from solving the flow equations in vacuum, i.~e. at $T=n=0$, see section \ref{DimerDimer}. For the coefficients determining the shift in $T_c$ compared to the free Bose gas we find $\kappa=1.55$. 

For $c^{-1}\gtrsim0.5$ the effect of the particle-hole fluctuations vanishes. This is expected since the chemical potential is now negative $\mu<0$ and there is no Fermi surface any more. Because of that there is no difference between the new curve with particle-hole fluctuations (solid in Fig. \ref{fig:tcrit}) and the one obtained when particle-hole contributions are neglected (dot-dashed in Fig. \ref{fig:tcrit}). Due to the use of an optimized cutoff scheme and a different computation of the density our results differ slightly from the ones obtained in \cite{Diehl:2007th}.

In the unitary regime ($c^{-1}\approx 0$) the particle-hole fluctuations still have a quantitative effect. We can give an improved estimate for the critical temperature at the resonance ($c^{-1}=0$) where we find $T_c/T_F=0.264$. Results from quantum Monte Carlo simulations are $T_c/T_F = 0.15$ \cite{bulgac:090404, bulgac:023625, burovski:160402} and $T_c/T_F = 0.245$ \cite{akkineni:165116}. The measurement by Luo \textit{et al.} \cite{Luo2007} in an optical trap gives $T_c/T_F = 0.29 (+0.03/-0.02)$, which is a result based on the study of the specific heat of the system.

\subsection{Crossover to narrow resonances}
Since we use a two channel model (Eq.\ \eqref{eqMicroscopicAction}) we can not only describe broad resonances with $h_\Lambda^2\to \infty$ but also narrow ones with $h_\Lambda^2\to0$. This corresponds to a nontrivial limit of the theory which can be treated exactly \cite{Diehl:2005an, Gurarie2007}. In the limit $h_\Lambda\to 0$ the microscopic action Eq.\ \eqref{eqMicroscopicAction} describes free fermions and bosons. The essential feature is, that they are in thermodynamic equilibrium so that they have equal chemical potential. (There is a factor 2 for the bosons since they consist of two fermions.) For vanishing Yukawa coupling $h_\Lambda$ the theory is Gaussian and the macroscopic propagator equals the microscopic propagator. There is no normalization of the ``mass''-term $m^2$ so that the detuning parameter in Eq.\ \eqref{eqMicroscopicAction} is $\nu=\mu_M(B-B_0)$ and
\begin{equation}
m^2=\mu_M(B-B_0)-2\mu.
\end{equation}

To determine the critical density for fixed temperature, we have to adjust the chemical potential $\mu$ such that the bosons are just at the border to the superfluid phase. For free bosons this implies $m^2=0$ and thus
\begin{equation}
\mu=\frac{1}{2}\mu_M (B-B_0)=-\frac{1}{16\pi}h^2 a^{-1}.
\label{eq:ChemicalpotetialatTc}
\end{equation}
In the last equation we use the relation between the detuning and the scattering length
\begin{equation}\label{eq:detuningandscatlenth}
a=-\frac{h^2}{8\pi \mu_M(B-B_0)}.
\end{equation}
The critical temperature $T_c$ is now determined from the implicit equation
\begin{equation}
\int \frac{d^3 p}{(2\pi)^3}\left\{\frac{2}{e^{\frac{1}{T_c}(\vec p^2-\mu)}+1}+\frac{2}{e^{\frac{1}{2T_c}\vec p^2}-1}\right\}=n.
\label{eq:TcNarrowresonanceimplicit}
\end{equation}

\begin{figure}
\centering
\includegraphics[width=0.45\textwidth]{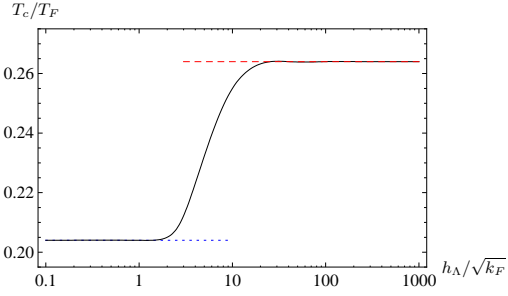}
\caption{The critical temperature divided by the Fermi temperature $T_c/T_F$ as a function of the dimensionless Yukawa coupling $h_\Lambda/\sqrt{k_F}$ for $c^{-1}=0$ (solid line). One can clearly see the plateaus in the narrow resonance limit ($T_c/T_F \approx 0.204$, dotted line) and in the broad resonance limit ($T_c/T_F \approx 0.264$, dashed line).}
\label{fig:narrowbroad}
\end{figure}

While the BCS-BEC crossover can be studied as a function of $B-B_0$ or $\mu$, Eq.\ \eqref{eq:detuningandscatlenth} implies that for $h_\Lambda^2\to 0$ a finite scattering length $a$ requires $B\to B_0$. For all $c\neq 0$ the narrow resonance limit implies for the phase transition $B=B_0$ and therefore $\mu=0$. (A different concentration variable $c_\text{med}$ was used in \cite{Diehl:2005an, Diehl:2005ae}, such that the crossover could be studied as a function of $c_\text{med}$ in the narrow resonance limit, see the discussion at the end of this section.) For $\mu=0$ Eq.\ \eqref{eq:TcNarrowresonanceimplicit} can be solved analytically and gives
\begin{equation}
\frac{T_c}{T_F}=\left(\frac{4\sqrt{2}}{3(3+\sqrt{2})\pi^{1/2}\zeta(3/2)}\right)^{2/3}\approx 0.204.
\end{equation} 
This result is confirmed numerically by solving the flow equations for different microscopic Yukawa couplings $h_\Lambda$ and taking the limit $h_\Lambda\to 0$. In Fig. \ref{fig:narrowbroad}, we show the critical temperature $T_c/T_F$ as a function of the dimensionless Yukawa coupling $h_\Lambda/\sqrt{k_F}$ in the ``unitarity limit'' $c^{-1}=0$ (solid line). For small values of the Yukawa coupling, $h_\Lambda/\sqrt{k_F} \lesssim 2$ we enter the regime of the narrow resonance limit and the critical temperature is independent of the precise value of $h_\Lambda$. The numerical value matches the analytical result $T_c/T_F\approx 0.204$ (dotted line in Fig. \ref{fig:narrowbroad}). For large Yukawa couplings, $h_\Lambda/\sqrt{k_F} \gtrsim 40$, we recover the result of the broad resonance limit as expected. In between there is a smooth crossover of the critical temperature from narrow to broad resonances.

We use here a definition of the concentration $c=a k_F$ in terms of the vacuum scattering length $a$. This has the advantage of a straightforward comparison with experiment since $a^{-1}$ is directly related to the detuning of the magnetic field $B-B_0$, and the ``unitarity limit'' $c^{-1}=0$ precisely corresponds to the peak of the resonance $B=B_0$. However, for a nonvanishing density other definitions of the concentration parameter are possible, since the effective fermion interaction $\lambda_{\psi,\text{eff}}$ depends on the density. For example, one could define for $n\neq 0$ a ``in medium scattering length'' $\bar a=\lambda_{\psi,\text{eff}}/(8\pi)$, with $\lambda_{\psi,\text{eff}}=-h^2/m^2$ evaluated for $T=0$ but $n\neq 0$ \cite{Diehl:2005an}. The corresponding ``in medium concentration'' $c_\text{med}=\bar a k_F$ would differ from our definition by a term involving the chemical potential, resulting in a shift of the location of the unitarity limit if the latter is defined as $c_\text{med}^{-1}=0$. While for broad resonances both definitions effectively coincide, for narrow resonances a precise statement how the concentration is defined is mandatory when aiming for a precision comparison with experiment and numerical simulations for quantities as $T_c/T_F$ at the unitarity limit. For example, defining the unitarity limit by $c_\text{med}^{-1}=0$ would shift the critical temperature in the narrow resonance limit to $T_c/T_F=0.185$ \cite{Diehl:2005an}.
			
	\section{BCS-Trion-BEC Transition}
	\label{sec:BCS-Trion-BECTransitionlong}
	In this section we discuss briefly the many-body physics for the SU(3) symmetric model in Eq.\ \eqref{eq8:microscopicactiontrionmodel} that describes three fermion species close to a common Feshbach resonance. No proper many-body calculation for this model has been performed so far. However, based on the knowledge of the BCS-BEC crossover physics and the vacuum calculation presented in section \ref{ssect:SU3symmetricmodel} we can already infer some qualitative features of the quantum (zero temperature) phase diagram.

\begin{figure}
\centering
\includegraphics[width=0.5\textwidth]{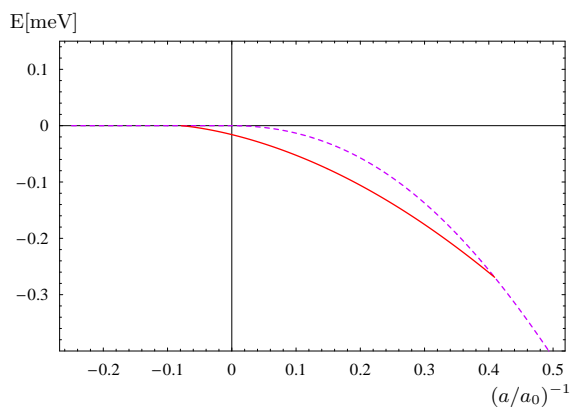}
\caption{Chemical potential or binding energy per atom $\mu=E$ in vacuum ($n=T=0$) as a function of the inverse scattering length $(a/a_0)^{-1}$ for three fermion species at a common Feshbach resonance. The solid line gives the energy per atom of the trion with respect to the fundamental fermion (zero energy line). The dashed line is the energy of the dimer per atom. Here $a_0$ is the Bohr radius which we also use as the ultraviolet scale $\Lambda=1/a_0$. Parameters are chosen to correspond to $^6\text{Li}$ in the ($m_F=1/2$, $m_F=-1/2$)-channel \cite{PhysRevLett.94.103201}.}
\label{fig:Energies}
\end{figure}

For increasing density the chemical potential increases compared to the vacuum chemical potential shown in Fig. \ref{fig:Energies}. 
The BEC phase occurs for small density for $a^{-1}>(a_{c2})^{-1}$. Due to the symmetries of the microscopic action, the effective potential for the bosonic field depends only on the $\text{SU(3)}\times\text{U(1)}$ invariant combination $\rho=\varphi^\dagger\varphi=\varphi_1^*\varphi_1+\varphi_2^*\varphi_2+\varphi_3^*\varphi_3$. It reads for small $\mu-\mu_0$, with $\mu_0$ the vacuum chemical potential,
\begin{equation}
U(\rho)=\frac{\lambda_\varphi}{2}\rho^2-2(\mu-\mu_0)\rho,
\label{eq::effpot}
\end{equation}
where we use the fact that the term linear in $\rho$ vanishes for $\mu=\mu_0$, i.e. $m_\varphi^2(\mu_0)=0$. The minimum of the potential shows a nonzero superfluid density
\begin{equation}
\rho_0=\frac{2(\mu-\mu_0)}{\lambda_\varphi},\ U(\rho_0)=-\frac{2}{\lambda_\varphi}(\mu-\mu_0)^2,
\end{equation}
with density
\begin{equation}
n=-\frac{\partial U(\rho_0)}{\partial \mu}=\frac{4}{\lambda_\varphi}(\mu-\mu_0)=2\rho_0.
\end{equation}
The factor two is expected since every boson contributes two fermions to the density. The superfluid density is twice the renormalized order parameter, $n_S=2\rho_0$. At zero temperature the density equals the superfluid density $n=n_S$ in the BEC regime.

For the BCS phase for $a<a_{c1}$ ($a_{c1}<0$), one has $\mu_0=0$. Nonzero density corresponds to positive $\mu-\mu_0$. In this region the contribution of fermionic atom fluctuations to the renormalization flow drives $m_\varphi^2$ always to zero at some finite $k_c$, with BCS spontaneous symmetry breaking ($\rho_0>0$) induced by the flow for $k<k_c$. Both the BEC and BCS phases are therefore characterized by superfluidity with a nonzero expectation value of the boson field $\varphi_0\neq0$ with $\rho_0=\varphi_0^*\varphi_0$. 

As an additional feature to the BCS-BEC crossover for a Fermi gas with two components, the expectation value for the bosonic field $\varphi_0$ in the three component case also breaks the spin symmetry of the fermions (SU(3)). Due to the analogous in QCD this was called ``color superfluidity'' \cite{PhysRevB.70.094521}. For any particular direction of $\varphi_0$ a continuous symmetry $\text{SU(2)}\times\text{U(1)}$ remains. According to the symmetry breaking $\text{SU(3)}\times\text{U(1)}\to \text{SU(2)}\times\text{U(1)}$, the effective potential has five flat directions.

For two identical fermions the BEC and BCS phases are not separated, since in the vacuum either $\mu_0=0$ or $m_\varphi^2=0$. There is no phase transition, but rather a continuous crossover. For three identical fermions, however, we find a new \textit{trion phase} for $(a^{(1)}_{c1})^{-1}<a^{-1}<(a^{(1)}_{c2})^{-1}$. In this region the vacuum has $\mu_0<0$ and $m_\varphi^2>0$. The atom fluctuations are cut off by the negative chemical potential and do not drive $m_\varphi^2$ to zero, such that for small density $m_\varphi^2$ remains positive. Adding a term $m_\varphi^2 \rho$ to the effective potential \eqref{eq::effpot} we see that the minimum remains at $\rho_0=0$ as long as $m_\varphi^2>2(\mu-\mu_0)$. No condensate of bosons occurs. The BEC and BCS phases that show both extended superfluidity through a spontaneous breaking of the $\text{SU(3)}\times\text{U(1)}$ symmetry, are now separated by a phase where $\varphi_0=0$, such that the $\text{SU(3)}\times\text{U(1)}$ symmetry remains unbroken (or will be only partially broken).

Deep in the trion phase, e.g. for very small $|a^{-1}|$, the atoms and dimers can be neglected at low density since they both have a gap. The thermodynamics at low density and temperature is determined by a single species of fermions, the trions. In our approximation it is simply given by a noninteracting Fermi gas, with fermion mass $3M$ and chemical potential $3(\mu-\mu_0)$. Beyond our approximation, we expect that trion interactions are induced by the fluctuations. While local trion interactions $\sim (\chi^*\chi)^2$ are forbidden by Fermi statistics, momentum dependent interactions are allowed. These may, however, be ``irrelevant interactions'' at low density, since also the relevant momenta are small such that momentum dependent interactions will be suppressed. Even if attractive interactions would induce a di-trion condensate, this has atom number six and would therefore leave a $\textrm{Z}_6$ subgroup of the U(1) transformations unbroken, in contrast to the BEC and BCS phase where only $\textrm{Z}_2$ remains. Furthermore, the trions are $\text{SU(3)}$-singlets such that the $\text{SU(3)}$ symmetry remains unbroken in the trion phase. The different symmetry properties between the possible condensates guarantee true quantum phase transitions in the vicinity of $a_{c1}$ and $a_{c2}$ for small density and $T=0$. We expect that this phase transition also extends to small nonzero temperature.

While deep in the trion phase the only relevant scales are given by the density and temperature, and possibly the trion interaction, the situation becomes more complex close to quantum phase transition points. For $a\approx a_{c1}$ we have to deal with a system of trions and atoms, while for $a\approx a_{c2}$ a system of trions and dimers becomes relevant. The physics of these phase transitions may be complex and rather interesting.

We finally comment on the precision of our computation. At nonzero density, our truncation may be improved by including in the two-body sector a four fermion coupling $\sim (\psi^{\dagger}\psi)^2$, which may be partially bosonized in favor of a running $\bar h$, see section \ref{sec:Particle-holefluctuationsandtheBCS-BECCrossover}. Also the momentum dependence of the interactions $\sim \lambda_{\varphi\psi}$ or $\lambda_\psi$ may be resolved beyond the pointlike approximation. Furthermore we expect an improvement from including a second atom-dimer scattering channel $\sim \lambda_{AD} \varphi^\dagger\varphi \psi^\dagger\psi$. These improvements are expected to change the quantitative values of $a_{c1}$, $a_{c2}$, $\mu_0$, and $s_0$, but not the qualitative situation. We believe that already the present truncation will yield a reliable picture of the qualitative properties of the phase diagram, once it is extended to nonzero density and temperature.

\chapter{Conclusions}
\label{ch:Conclusions}
In this thesis we applied the method of functional renormalization to the theoretical description of ultracold quantum gases. We derived flow equations for various physical systems, namely for a Bose gas with approximately pointlike, repulsive interaction, for a Fermi gas with two hyperfine components in the BCS-BEC crossover and for a Fermi gas with three components. The implications of these flow equations are investigated both in the few-body and the many-body regime.

For repulsive bosons with approximately pointlike interaction we find an upper bound for the scattering length $a$. Quantum fluctuations lead to a screening of the interaction and imply that the scattering length is of the order of the inverse microscopical momentum scale $a\lesssim \Lambda^{-1}$ in three dimensions. In two dimensions, the scale dependence of the interaction strength is logarithmic and the bound is correspondingly weaker. At low temperatures we find that Bogoliubov theory gives a surprisingly accurate account of many properties of an interacting Bose gas although it treats fluctuations only to quadratic order and neglects for example the scale dependence of the interaction strength. On the other hand, perturbative extensions beyond Bogoliubov theory are plagued with infrared divergences. Our renormalization group treatment solves these problems and is free of infrared problems. It is also possible to apply our method to the case of stronger interactions. However, one should keep in mind that very large ratios of the scattering length to the interparticle distance $n^{-1/3}$ are only possible if the microscopic scale $\Lambda$ is of the same order as $n^{1/3}$. In this case, the details of the interaction become important and universality might be lost. 

We also calculate different thermodynamic observables at nonvanishing temperatures. This includes for example the critical temperature which is larger for an interacting gas compared to a free Bose gas. The shift in the critical temperature is linear in the concentration parameter, $\Delta T_c / T_c \sim a n^{1/3}$. The region around the phase transition is mainly governed by the critical exponents of the three-dimensional XY universality class. In principle, these critical exponents can also be calculated from our flow equations. However, they are obtained also from the simpler flow equations of the classical O(N) model which are studied in much detail in the literature. Instead, we calculated the thermodynamic quantities in the full range from zero temperature to the phase transition. Besides the superfluid and the condensate fraction this includes the correlation length (or healing length), the entropy and energy density, the pressure, the specific heat, the isothermal and the adiabatic compressibility. All these quantities are well described by Bogoliubov theory for small temperatures and interaction strength while deviations are found away from this limit. This is also the case for the first and second velocity of sound.

The calculations concerning the Bose gas in three dimensions could be improved in different directions. One might use a regulator function that acts as an ultraviolet cutoff also in the space of Matsubara frequencies. This should improve the flow equation for the pressure and its derivatives such as density and entropy density. One could also improve the approximation by going to the next order in a systematic derivative expansion. It would also be interesting to study the flow equation for the effective potential using a method different from an expansion around the (scale-dependent) minimum. With a numerical solution of the partial differential equation for $U$ it might be possible to derive also some predictions for attractive interactions between two atoms corresponding to a negative scattering length. Finally, an interesting extension would be to consider gases trapped by some confining potential. Much of the formalism developed in this thesis can be transferred from the homogeneous space to this more complicated case using a local density approximation. One assumes that the relevant length scale for the fluctuations is small compared to the extension of the trap and works with a position dependent chemical potential according to $\mu(\vec x)=\mu_0-V(\vec x)$ where $V(\vec x)$ is the potential. Besides the thermodynamic observables, it would be interesting to study hydrodynamic excitations and their dependence on temperature and density. 

We studied the flow equations for the homogeneous Bose gas also in two spatial dimensions. The formalism is pretty similar -- one has to adapt a single number, only. The physical properties, on the other side, are very different. The effect of fluctuations is more important than in three dimensions and many quantities have a logarithmic scale dependence. In the low temperature regime, the emergence of a quadratic (Matsubara-) time derivative from quantum fluctuations plays a more important role. Nevertheless, we find at small temperatures that some physical observables such as the quantum depletion of the condensate and the velocity of sound are described by Bogoliubov theory to good approximation. There are sizable deviations for large temperatures, however. Our calculations show, that a second branch in the dispersion relation becomes relevant at small frequencies. In contrast to the sound mode which describes fluctuations in the phase of the oder parameter, the second branch describes fluctuations in its absolute value. 

The thermal phase diagram of the two-dimensional Bose gas is governed by the Kosterlitz-Thouless phase transition. The flow equation method is especially valuable here, since the dependence of various quantities on the infrared scale is crucial. For a homogeneous system of infinite extension, long-range order and Bose-Einstein condensation in the strict sense are forbidden by the Mermin-Wagner theorem. Nevertheless, one observes superfluidity at small temperatures. For a finite system the size of the probe sets an infrared cutoff scale. No fluctuations with momenta smaller then this scale contribute and the Mermin-Wagner theorem does not apply any more. Our investigations show, that the difference between superfluidity and long-range order looses its importance for systems with finite extension or for observables that are connected with a momentum scale that acts as an effective infrared cutoff. The superfluid density $n_S=\rho_0=\bar A \bar \rho_0$ and the condensate density $n_C=\bar \rho_0$ are connected by the wavefunction renormalization $\bar A$. When the infrared cutoff scale $k_\text{ph}$ is lowered, the condensate density as the square of the order parameter, decreases. At the same time, the wavefunction renormalization factor $\bar A$ grows (and goes to infinity for $k_\text{ph}\to0$). The superfluid density $n_S$ remains positive for small enough temperatures. 

For the dependence of the critical temperature on the interaction strength $\lambda$ we find for small $\lambda$ the functional form $T_c/n = 4\pi/\ln(\xi/\lambda)$. This is in agreement with an perturbative analysis. In contrast to quantum Monte-Carlo simulations, we find that the coefficient $\xi$ depends on the infrared cutoff scale $k_\text{ph}$. In the infinite size limit $k_\text{ph}\to0$ we find that the critical temperature vanishes, $T_c/n\to 0$. At least in parts this feature is attributed to the scale dependence of the interaction strength, a feature that was not taken into account for the numerical simulations.

In conclusion, a rather detailed picture of the superfluid Bose gas in two dimensions has been obtained from the flow equation method. Nevertheless, there is still room for some extensions of the theory. Most of the points discussed for the Bose gas in three dimensions apply here as well. In addition, an investigation of occupation numbers as a function of temperature and interaction strength would be of interest. 

Besides the bosonic systems we also investigated models for fermions. For two fermion species close to a Feshbach resonance, the flow equations in the vacuum limit show the expected dimer formation. The binding energy in the broad resonance limit matches the expectation from quantum mechanical calculations. In the many-body regime the discussion presented here concentrates on the effect of particle-hole fluctuations. On the BCS side of the crossover we recover Gorkovs correction to the original BCS theory, i.e. a suppression of the critical temperature by a factor $\sim 2.2$. We extend the calculation by Gorkov and Melik-Barkhudarov to the whole crossover regime including the unitarity point where the scattering length diverges. At this point we can also compare our estimate of the critical temperature in units of the Fermi temperature, $T_c/T_F\approx 0.264$, to the result from Monte-Carlo simulations $T_c/T_F\approx 0.15$. The deviations are rather large and may have different reasons. To calculate the ratio $T_c/T_F$ one has to determine for given chemical potential both the critical temperature and the density. (In our units the latter is connected to the Fermi temperature by $T_F=(3\pi^2 n)^{2/3}$.) Both steps are equally important -- and nontrivial. For example, would we estimate the density by the corresponding formula in the non-interacting case, the resulting ratio would be $T_c/T_F\approx 0.165$. This shows that a reliable method to calculate the density is crucial -- both for the flow equation method and the Monte-Carlo simulations. 

Apart from the density, other improvements of our calculation are conceivable. The inclusion of a vertex $\lambda_{\varphi\psi}$, which describes the scattering of a fermionic atom with a dimer, is work in progress. On the BEC side of the crossover and in the vacuum limit it leads to a reduction in the ratio $a_M/a$ \cite{Scherer2007}. One expects that the inclusion of this term in the many-body calculation slightly reduces the critical temperature on the BEC side and in the crossover regime. Another improvement concerns the rebosonization method. The calculations performed so far employ the approximate flow equation proposed in ref.\ \cite{Gies:2001nw}. Using the exact flow equation derived in chapter \ref{ch:Generalizedflowequation} might improve the result. Once the truncation has reached a satisfactory quantitative precision, it would be interesting to study different thermodynamic observables. For a comparison to experimental investigations it is also necessary to transfer the formalism to the trapped Fermi gas. With increasing experimental precision, the goal of a accurate test for non-perturbative methods seems to be in reach. 

The knowledge of the phase diagram is much less developed for the Fermi gas with three components than for the two-component case. Before quantitative questions can be answered for this system, one has to settle the qualitative features. In this thesis we investigated a simple SU(3) symmetric model for fermions with equal properties (except for hyperfine-spin, of course) close to a common Feshbach resonance. As expected, we find that the two-body problem is equivalent to the one for two species. In contrast, the three-body problem is now governed by Efimov physics. In our model and within our approximation scheme we recovered the infinite tower of three-body bound states (trimers) at the resonance and for energies close to the free atom threshold. In the renormalization group picture the Efimov effect shows as a limit cycle scaling. With our simple truncation we find a scaling parameter $s_0\approx0.82$ while a calculation that takes the full momentum dependence of vertices into account confirms Efimovs value $s_0\approx1.00624$.

More important, for the phase diagram then the tower of bound states is the state with lowest energy and the corresponding chemical potential. In a region $a^{-1}_{c1}<a^{-1}<a^{-1}_{c2}$ with $a_{c1}<0$ and $a_{c2}>0$ we find that the many-body ground state is dominated by trions (the trimers with lowest energy). The chemical potential in this region is lower than in the two-component case. The mechanisms that lead to BCS superfluidity for $a\to0_-$ or BEC-like superfluidity for $a\to 0_+$ do not work in this region. While SU(3) symmetry is broken spontaneously in the BCS and BEC phases, this is not the case in the trion phase. We therefore expect true and rather interesting quantum phase transitions around the points $a_{c1}^{-1}$ and $a_{c2}^{-1}$ (``BCS-Trion-BEC transition''). 

In addition to the SU(3) symmetric model we also investigate a model for three fermion species where this symmetry is broken explicitly by different positions and widths of the resonances. Although the Efimov tower of bound states is not expected for three distinct resonances, one still finds three-body bound states provided the scattering lengths are large enough. We apply our method to the case of $^6$Li which is currently of experimental interest. The trion is not stable and can decay into lower lying bound states (possibly the two-body bound states of the nearby Feshbach resonance). Introducing a decay width $\Gamma_\chi$ as an additional parameter to be fixed from experiment, we can explain the form of the observed three-body loss coefficient as a function of magnetic field.

The calculations performed for the three-component Fermi gas so far are a first step towards a more detailed investigation of the interesting phase diagram. A proper many-body calculation at zero temperature could shed more light on the quantum phase transitions while the inclusion of temperature effects will allow for an investigation of the complete phase diagram. Functional renormalization provides a very good method for this purposes. Note that since three-body effects are important, it is not possible to work with Mean-Field theory. Although one might first investigate the SU(3) symmetric model, it would also be interesting to consider $^6$Li where comparison to future experiments seems possible. 

Besides the application to concrete models, we also considered some more formal aspects of the renormalization group method. We derived a generalization of the flow equation to scale-dependent composite operators. In the new equation, the usual renormalization flow with one-loop form is supplemented by an intuitive tree-level term. One application of this generalized flow equation is a scale-dependent translation from ultraviolet to infrared degrees of freedom as discussed in \cite{Gies:2001nw, Gies:2002kd}. It may also provide a method to treat momentum dependence of vertices. Indeed, one could introduce scale-dependent composite operators for different ``scattering channels''. The difficult problem of treating momentum-dependent vertices would be reduced to the momentum dependence of propagators for composite operator fields. This might be simpler in some respects and also allows to trace the renormalization group flow in the regime with spontaneous symmetry breaking.

\begin{appendix}
\chapter{Some ideas on functional integration and probability}
\label{ch:Foundationsofquantumtheory:someideas}
The wave-particle duality is discussed since the early days of quantum mechanics. Schr\"{o}\-ding\-ers equation \cite{Schroedinger} is a non-relativistic equation for a wave function $\psi(\vec x,t)$. On the other side this wave function, or more precisely its modulus square $|\psi(\vec x,t)|^2$, can be interpreted as the probability density to detect a particle at the point $\vec x$ at time $t$. This was first pointed out by Born \cite{Born}. In this sense an electron for example has properties of both a particle and a field.
In the modern understanding of quantum field theory the situation is basically still the same. The formulation of the theory is mainly in terms of wave functions or fields, but one still needs the interpretation in terms of probabilities to ``find a particle in a certain state'' to bring the formalism into contact with experiments. 

The formulation of quantum field theory using the functional integral \cite{Feynman1948, Feynman1950} is quite close to formulations of statistical field theory \cite{ZinnJustin, Weinberg}, \cite{Wetterich:2002fy}. The far reaching parallels in both formalisms were important for many achievements, for example in the development of the renormalization group theory \cite{PhysRevB.4.3174}, \cite{PhysRevA.8.401}. One might ask whether this duality between quantum and statistical field theory has a deeper physical origin and whether it can help us for a better understanding of the wave-particle duality. One might hope that some difficult points in the foundations of quantum theory related to the collapse of the wave function could be clarified.

What prevents us from using our good knowledge and intuition about statistical field theory to investigate these questions is at the same time the most important difference between the two formalisms. In statistical field theory one calculates expectation values, correlation functions etc.\ by taking the sum over all possible field configurations $\varphi$ weighted by the real and positive semi-definite probability measure
\begin{equation}
p[\varphi]=e^{-S[\varphi]}.
\end{equation}
For a local theory in $d$ dimensions the action $S[\varphi]$ is given as an integral over the Lagrange density
\begin{equation}
S[\varphi]=\int d^d x {\cal L}
\end{equation}
where ${\cal L}$ is a function of $\varphi$ and its derivatives. In quantum field theory, the corresponding measure under the functional integral
\begin{equation}
e^{iS[\varphi]}
\end{equation}
is complex and can therefore not be interpreted as a probability. In the following we show, that the formalism of quantum field theory can be reformulated such that the weighting factor under the functional integral is real and discuss possible consequences for our understanding of what ``particles'' are.

\subsubsection{Functional integral with probability measure}
In this section we reconsider the functional integral formulation of quantum field theory and formulate an representation with a (quasi-) probability distribution. Let us start with a simple Gaussian theory
\begin{equation}
S= \sum_{\alpha,\beta} \varphi_\alpha^* \left(P_{\alpha\beta}+i \epsilon \delta_{\alpha\beta} \right)  \varphi_\beta.
\label{eq:Gauusianaction}
\end{equation}
We use here an abstract index notation where e.g. $\alpha$ stands for both continuous variables such as position or momentum and internal degrees of freedom. In principle, the field $\varphi$ may have both bosonic and fermionic components, the latter are Grassmann-valued. For simplicity we assume in the following that $\varphi$ is a bosonic field. The formalism can be extended to cover also the case of fermions with minor modifications. 

We included in Eq.\ \eqref{eq:Gauusianaction} a small imaginary part $\sim i\epsilon$ to make the functional integral convergent and to enforce the correct frequency integration contour (Feynman prescription). Although $\epsilon$ is usually taken to be infinitesimal, we work with an arbitrary positive value here and take the limit $\epsilon \to 0_+$ only at a later point in our investigation. 
For simplicity, we will often drop the abstract index and use a short notation with e.~g.
\begin{equation}
S=\varphi^* (P+ i \epsilon ) \varphi.
\end{equation}
The operator $P$ is the real part of the inverse microscopic propagator. As an example we consider a relativistic theory for a scalar field where $P$ reads in position space
\begin{equation}
P(x,y)=\delta^{(4)}(x-y)\left(-g^{\mu\nu}\frac{\partial}{\partial y^\mu}\frac{\partial}{\partial y^\nu}-m^2\right).
\end{equation}
Another example is the nonrelativistic case with 
\begin{equation}
P(x,y)=\delta^{(4)}(x-y)\left(i\frac{\partial}{\partial y_0} + \frac{1}{2M}\vec \nabla_y^2\right).
\end{equation}

For a Gaussian theory the microscopic propagator coincides with the full propagator. The latter is obtained for general actions $S$ from
\begin{eqnarray}
\nonumber
i G_{\alpha\beta} &=& \langle \varphi_\alpha \varphi^*_\beta\rangle_c\\
&=& (-i)^3 \frac{\delta}{\delta J^*_\alpha}\frac{\delta}{\delta J_\beta} W[J]
\end{eqnarray}
with
\begin{equation}
e^{-i W[J]} = Z[J] = \int D \varphi e^{i S[\varphi]+i\int \{J^*\varphi+\varphi^* J\}}.
\label{eq:Schwingerfunctional}
\end{equation}
For $\alpha=(x_0,\vec x)$ and $\beta=(y_0,\vec y)$ the object $G_{\alpha\beta}$ can be interpreted as the {\itshape probability amplitude} for a particle to propagate from the point $\vec y$ at time $y_0$ to the point $\vec x$ at time $x_0$. More general, one might label by $\alpha$ some single-particle state $|\varphi_\alpha\rangle$ at time $t_\alpha$ and with $\beta$ the state $|\varphi_\beta\rangle$ at time $t_\beta$. The propagator $G_{\alpha\beta}$ describes then the probability amplitude for the transition between the two states.
However, the description of an actual physical experiment involves the transition probability given by the modulus square of the probability amplitude (no summation over $\alpha$ and $\beta$)
\begin{equation}
q_{\alpha\beta} = |G_{\alpha\beta}|^2= G^*_{\alpha\beta} G_{\alpha\beta}.
\end{equation}
This transition probability can also directly be obtained from functional derivatives of
\begin{eqnarray}
\nonumber
\tilde Z[J_1,J_2]= \int D\varphi_1 \int D \varphi_2 e^{i S[\varphi_1]} e^{-i S^*[\varphi_2]} \\
e^{i \int \{J_1^* \varphi_1+\varphi_1^* J_1\}} e^{-i \int \{J_2^* \varphi_2+\varphi_2^* J_2\}}.
\label{eq:tildedpartfunction}
\end{eqnarray}
We note that Eq.\ \eqref{eq:tildedpartfunction} contains twice the functional integral over the field configuration $\varphi$. One may also write this as a single functional integral over fields that depend in addition to the position variable $\vec x$ on the contour time $t_c$ which is integrated along the Keldysh contour \cite{Keldysh}. For our purpose it will be more convenient to work directly with the expression in Eq.\ \eqref{eq:tildedpartfunction}.

For $\langle \varphi \rangle = \langle \varphi^* \rangle = 0$ we can write
\begin{equation}
q_{\alpha\beta} =\frac{1}{\tilde Z} \frac{\delta}{\delta (J_1^*)_\alpha} \frac{\delta}{\delta (J_1)_\beta} \frac{\delta}{\delta (J_2)_\alpha} \frac{\delta}{\delta (J_2^*)_\beta} \tilde Z[J_1,J_2].
\end{equation}
This is immediately clear since $\tilde Z[J_1,J_2] = Z[J_1] Z^*[J_2]$ and
\begin{eqnarray}
\nonumber
i G_{\alpha\beta} &=& (-i)^2 \frac{1}{Z[J_1]} \frac{\delta}{\delta (J_1^*)_\alpha} \frac{\delta}{\delta (J_1)_\beta} Z[J_1]\\
-i G^*_{\alpha\beta} &=& i^2 \frac{1}{Z^*[J_2]} \frac{\delta}{\delta (J_2)_\alpha} \frac{\delta}{\delta (J_2^*)_\beta} Z^*[J_2].
\end{eqnarray}
for
\begin{equation}
\langle \varphi \rangle = \frac{-i}{Z} \frac{\delta}{\delta J^*} Z[J] = \langle \varphi^* \rangle = \frac{-i}{Z} \frac{\delta}{\delta J} Z[J]=0.
\end{equation}
Usually one obtains $q_{\alpha\beta}$ by first calculating $G_{\alpha\beta}$ and then taking the modulus square thereof. The way we go here seems to be more complicated from a technical point of view, but has the advantage that it will allow for an intuitive physical interpretation. 

We first concentrate on $\tilde Z[J_1,J_2]$. This object plays a similar role as the partition function in statistical field theory. In some sense it is a sum over microscopic states weighted with some ``probability''. However, in contrast to statistical physics, the summation does not go over states of a system at some fixed time $t$ but over field configurations that depend both on the position variable $\vec x$ and the time variable $t$. The summation seems to go over a even larger space since the functional integral appears twice
\begin{equation}
\int D\varphi_1 \int D \varphi_2
\end{equation}
so that the configuration space seems to be the tensor product of twice the space that contains the field configurations in space-time $\varphi(\vec x,t)$. In addition the ``probability weight''
\begin{equation}
e^{i S[\varphi_1]} e^{-i S^*[\varphi_2]}
\end{equation}
is not positive semi-definite and has even complex values in general. This last two features (``doubled'' configuration space and missing positivity) prevent us from interpreting quantum field theory in a similar way as statistical field theory.ting quantum field theory in a similar way as statistical field theory.

An idea to overcome these difficulties is to partially perform the functional integral in Eq.\ \eqref{eq:tildedpartfunction}. For this purpose we make a change of variables of the form
\begin{eqnarray}
\nonumber
\varphi_1 = \frac{1}{\sqrt{2}}\phi + \frac{1}{\sqrt{2}}\chi, &\quad& J_1 = \frac{1}{\sqrt{2}}J_\phi +\frac{1}{\sqrt{2}} J_\chi, \\
\varphi_2 = \frac{1}{\sqrt{2}}\phi - \frac{1}{\sqrt{2}}\chi, &\quad& J_2 = -\frac{1}{\sqrt{2}}J_\phi + \frac{1}{\sqrt{2}} J_\chi.
\label{eq:12phichitranslation}
\end{eqnarray}
For $\tilde Z$ this gives then
\begin{equation}
\tilde Z = \int D\phi\,\, v[\phi,J_\chi]\,\,e^{i\int \{J_\phi^* \phi+\phi^* J_\phi\}}
\label{eq:ZwithJ1J2}
\end{equation}
with
\begin{eqnarray}
\nonumber
v[\phi, J_\chi] &=& \int D \chi\, e^{iS[(\phi+\chi)/\sqrt{2}]}\, e^{-i S^*[(\phi-\chi)/\sqrt{2}]}\\
&& \times\, e^{i \int \{J_\chi^* \chi+\chi^* J_\chi\}}.
\label{eq:vphiJchigeneralcase}
\end{eqnarray}
We note that $v[\phi,J_\chi]$ as a functional of $\phi$ and $J_\chi$ is real. This follows from comparison with the complex conjugate together with the change of variables $\chi\to-\chi$. If it is also positive, we can interpret this object as a probability for the field configurations $\phi(x)$. We call $v$ the {\itshape functional probability} for the field configuration $\phi$. 

Before we discuss the general properties of $v[\phi,J_\chi]$ in more detail, we consider it explicitly for a Gaussian action $S[\varphi]$ as in Eq.\ \eqref{eq:Gauusianaction}. In that case we can perform the functional integral
\begin{eqnarray}
\nonumber
v[\phi,J_\chi] &=& \int D \chi e^{-\epsilon \{\phi^*\phi+\chi^*\chi\}} e^{i\{\phi^* P \chi +\chi^* P \phi\}}\\
\nonumber
&& \times e^{i\{J_\chi^* \chi + \chi^* J_\chi\}}\\
&=& e^{-\epsilon \phi^*\phi}\, e^{-\frac{1}{\epsilon}(J_\chi^*+\phi^* P)(J_\chi+P\phi)}.
\label{eq:vphijGaussian}
\end{eqnarray}
The last line holds up to a multiplicative factor that is irrelevant for us here. For $\tilde Z$ we are left with
\begin{eqnarray}
\nonumber
\tilde Z[J_\phi,J_\chi] &=& \int D\phi \,\,v[\phi, J_\chi]\,  \,e^{i\int\{J_\phi^*\phi+\phi^*J_\phi\}}\\
\nonumber
&=& \int D \phi \, e^{-\frac{1}{\epsilon}\phi^*(P^2+\epsilon^2)\phi} \, e^{i\{J_\phi^* \phi +\phi^* J_\phi\}}\\
&&\times \, e^{-\frac{1}{\epsilon}\{J_\chi^* P \phi+\phi^* P J_\chi\}}\, e^{-\frac{1}{\epsilon}J_\chi^*J_\chi}.
\label{eq:partitionfunctionphiJphiJchi}
\end{eqnarray}
For $J_\phi=J_\chi=0$ the integrand in Eq.\ \eqref{eq:partitionfunctionphiJphiJchi} is strictly positive. Eq.\ \eqref{eq:partitionfunctionphiJphiJchi} can therefore be interpreted in a similar way as the partition function in statistical field theory. The probability measure is
\begin{equation}
v=e^{-\frac{1}{\epsilon}\phi^*(P^2+\epsilon^2)\phi}.
\label{eq:probmeasuregaussian}
\end{equation}

We can distinguish three different classes of field configurations $\phi$. In the simplest case the norm vanishes,
\begin{equation}
|\phi|^2=\phi^*\phi=\sum_\alpha \phi^*_\alpha\phi_\alpha\to 0.
\end{equation} 
The functional probability for this case is of order $1$. The second class contains field configurations where the norm is nonzero, $\phi^*\phi\neq 0$, but where $\phi$ satisfies the on-shell condition, i.~e. $\phi^*P^2\phi=0$. The functional probability for this case is of order $e^{-\epsilon}$ (for $\phi^*\phi\sim1$). Finally, in the third class the norm is nonzero and the field configuration is off-shell, i.~e.
\begin{equation}
\phi^*\phi\sim1,\quad \phi^*P^2\phi \sim 1.
\end{equation}
The functional probability in Eq.\ \eqref{eq:probmeasuregaussian} for this case is only of the order $e^{-1/\epsilon}$. This shows that off-shell configurations are strongly suppressed in the limit $\epsilon\to0$ compared to the trivial case $|\phi|=0$ and the on-shell fields with
\begin{equation}
\phi^* P^2\phi = 0.
\label{eq:onshellcond}
\end{equation}
However, for $\epsilon>0$ the probability for off-shell configurations is not strictly zero and they give also contributions to $\tilde Z$. This is in contrast to classical statistics where only states that fulfill the equation of motion are included. (At nonzero temperature states with different energies are weighted according to a thermal distribution.) 

There are more differences between the partition function in classical statistics and the quantum partition function in Eq.\ \eqref{eq:partitionfunctionphiJphiJchi}. In classical statistics the averaging over some phase space is directly linked to time averaging by the ergodic hypothesis. Indeed, this hypothesis says that a mean value calculated by taking the average of some quantity over the accessible phase space is equal to the average of that quantity over a -- sufficiently long -- time interval. 
In classical statistics time plays an outstanding role. The formalism breaks space-time symmetries such as Lorentz- or Galilean symmetry explicitly. For the case of quantum field theory this point is different. The theory in the vacuum (zero temperature and density, $T=n=0$) is symmetric under Lorentz symmetry (or Galilean symmetry in the nonrelativistic case). 

The summation over possible field configurations in Eq.\ \eqref{eq:partitionfunctionphiJphiJchi} is not related to time averaging. We directly interpret it in the following way. Every physical experiment or ``measurement-like situation'' corresponds to a different microscopic field configuration $\phi$. This field configuration does not necessarily have to fulfill the on-shell condition Eq.\ \eqref{eq:onshellcond} but has a probability $v$ that strongly favors on-shell fields. There is, however, a subtlety in this interpretation. When the action $S$ is given as an integral over the $3+1$ dimensional spacetime $\Omega$, then $v$ describes the probability for a field configuration $\phi(x_0,\vec x)$ on this manifold $\Omega$. Since we experience only one universe with one configuration one might ask why we should take the sum over different configurations weighted with some probability. To answer that question it is important to realize that our information about the field configuration $\phi(x_0,\vec x)$ is limited. 

First we can investigate only limited regions in space-time (around our own ``world-line''). Regions that are too far away either in the spatial or the temporal sense are not accessible. However, in the framework of a local field theory, the experiments in some region of space-time depend on the other (not accessible) regions only via the boundary conditions. Second, and more important, we have only access to the field configuration in some ``momentum range''. No experiment has an arbitrary large resolution and can resolve infinitely small wavelength. Therefore the true microscopic field configuration is inevitably hidden from our observation. In a Gaussian or non-interacting theory this issue seems to be not so important since different momentum modes decouple from each other. In a theory with interactions this is different, however. Modes with different momenta $p^\mu$ (and different values of $p^\mu p_\mu$) are coupled via the interaction. The microscopic regime does influence the macroscopic states. 

Our interpretation of Eq.\ \eqref{eq:partitionfunctionphiJphiJchi} is therefore that the functional integral sums over possible microscopic configurations $\phi(x)$ with probability (up to a factor) given by
\begin{equation}
e^{-\frac{1}{\epsilon}\phi^*(P^2+\epsilon^2)\phi}
\end{equation}
for the Gaussian theory considered above and more general (for $J_\chi=0$) by
\begin{equation}
v[\phi]=v[\phi,J_\chi=0]=\int D\chi e^{iS[(\phi+\chi)/\sqrt{2}]}\,e^{-i S^*[(\phi-\chi)/\sqrt{2}]}.
\end{equation}
In this general case, the functional probability for nonzero source terms is given by Eq.\ \eqref{eq:vphiJchigeneralcase}. As argued there, it is always real. This is made more explicit in the expression
\begin{eqnarray}
\nonumber
v[\phi,J_\chi] &=& \int D \chi \,\text{cos}\left(S_1[\phi+\chi]-S_1[\phi-\chi]\right)\\
&& \times \, \text{exp}\left(-S_2[\phi+\chi]-S_2[\phi-\chi]\right)
\label{eq:vphiJchiexplrealform}
\end{eqnarray}
with
\begin{eqnarray}
S_1[\varphi] &=& \text{Re}\, S[\varphi/\sqrt{2}]+ \frac{1}{\sqrt{2}}\left\{J_\chi^*\varphi+\varphi^*J_\chi\right\}
\end{eqnarray}
and
\begin{equation}
S_2[\varphi]=\text{Im}\, S[\varphi/\sqrt{2}].
\end{equation}
We note that the functional integral in Eq.\ \eqref{eq:vphiJchiexplrealform} converges when $S_2[\varphi]$ increases with $\varphi^*\varphi$ fast enough. For $S_2[\varphi]\sim \varphi^*\varphi$ as in our Gaussian example, the convergence is quite good. Although for arbitrary actions $S[\varphi]$ the ``probability'' $v[\phi,J_\chi]$ does not have to be positive, this is expected to be the case for many choices of $S[\varphi]$. 

When $v[\phi,J_\chi]$ is negative for some choices of $\phi$, this indicates that different values for $\phi$ do not directly correspond to independent physical configurations. One might come to positive definite probabilities when the space of possible fields is restricted to a physically subspace. However, $v[\phi,J_\chi]$ as defined above can in any case be seen an quasi-probability for $\phi$. This is in some respect similar to Wigner's representation of density matrices \cite{Wigner}. 

Let us make another comment on the case of non-Gaussian actions $S[\varphi]$. When $S[\varphi]$ contains terms of higher then quadratic order in the fields $\varphi$ the form of the action is subject to renormalization group modifications. Usually the true microscopic action $S[\varphi]$ is not known. Measurements have only access to the effective action $\Gamma[\varphi]$ which already includes the effect of quantum fluctuations. (Measurements at some momentum scale $k^2=|p_\mu p^\mu|$ might probe the average action or flowing action $\Gamma_k[\varphi]$ \cite{Wetterich1993b}.) The microscopic action $S$ is connected to $\Gamma$ by a renormalization group flow equation \cite{Wetterich1993b}, however it is in most cases not possible to construct $S$ from the knowledge of $\Gamma$. Typically many different microscopic actions $S$ lead to the same effective action $\Gamma$. It may therefore often be possible that a microscopic action $S$ exists that is consistent with experiments and allows for a positive probability $v[\phi,J_\chi]$. 

Finally we comment on the general properties of $v[\phi,J_\chi]$. Since it is defined as a functional integral over a (local) complex action one expects that $v[\phi,J_\chi]$ is local to a similar degree as the the effective action $\Gamma[\varphi]$ or the Schwinger functional $W[J]$ defined in Eq.\ \eqref{eq:Schwingerfunctional}. For general non-Gaussian microscopic actions $S[\varphi]$ the functional $v[\phi,J_\chi]$ may be quite complicated and not necessarily local in the sense that it can be written in the form
\begin{equation}
v[\phi,J_\chi] = e^{-\int_x{\cal L}_v}
\end{equation}
where ${\cal L}_v$ is a local ``Lagrange density'' that depends only on $\phi(x)$, $J_\chi(x)$ and derivatives thereof at the space-time point $x$. 

Since $v[\phi,J_\chi]$ is similarly defined as the effective action $\Gamma[\varphi]$ or the Schwinger functional $W[J]$ we expect that it respects the same symmetries as the microscopic action $S[\varphi]$ when no anomalies are present. For example, when $S[\varphi]$ is invariant under some $U(1)$ symmetry transformation $\varphi \to e^{i\alpha} \varphi$, we expect that $v[\phi,J_\chi]$ has a corresponding symmetry under the transformation
\begin{equation}
\phi \to e^{i\alpha} \phi, \quad J_\chi \to e^{i\alpha} J_\chi.
\end{equation} 

\subsubsection{Correlation functions from functional probabilities}
In this subsection we use the expression for $\tilde Z$ in Eq.\ \eqref{eq:partitionfunctionphiJphiJchi} to derive functional integral representations of some correlation functions. In the following we denote by $\langle \cdot \rangle$ the ``expectation value'' in the quantum field theoretic sense, e.~g. for an operator $A[\varphi]$
\begin{equation}
\langle A[\varphi] \rangle = \frac{1}{Z} \int D \varphi e^{i S[\varphi]} \,A[\varphi].
\end{equation}
In contrast, we use $\langle\langle\cdot\rangle\rangle$ to denote the expectation value with respect to the functional integral over $\phi$, i.~e.
\begin{equation}
\langle\langle A[\phi]\rangle\rangle = \frac{1}{\tilde Z} \int D \phi \,\,e^{-\frac{1}{\epsilon}\phi^*(P^2+\epsilon^2)\phi} \,A[\phi],
\label{eq:gaussiandistroperator}
\end{equation}
or more general
\begin{equation}
\langle\langle A[\phi]\rangle\rangle = \frac{1}{\tilde Z} \int D \phi\,\, v[\phi]\, A[\phi]. 
\end{equation}
For the discussion of the correlation functions it is useful to express $\tilde Z$ in Eq.\ \eqref{eq:partitionfunctionphiJphiJchi} again in terms of $J_1$ and $J_2$. Using Eq.\ \eqref{eq:12phichitranslation} we find
\begin{eqnarray}
\nonumber
\tilde Z[J_1,J_2] &=& \int D\phi \,\,e^{-\frac{1}{\epsilon}\phi^*(P^2+\epsilon^2)\phi}\\
\nonumber
&& \times\, e^{-\frac{1}{\sqrt{2}\epsilon}\left\{J_1^*(P-i\epsilon)\phi+\phi^*(P-i\epsilon)J_1\right\}}\\
\nonumber
&& \times \,e^{-\frac{1}{\sqrt{2}\epsilon}\left\{J_2^*(P+i\epsilon)\phi+\phi^*(P+i\epsilon)J_2\right\}}\\
&& \times \,e^{-\frac{1}{2\epsilon}\left\{J_1^*J_1+J_1^*J_2+J_2^*J_1+J_2^*J_2\right\}}.
\end{eqnarray}
We start with the modulus square of the quantum field theoretic one-point function (no summation convention used in the following)
\begin{eqnarray}
\nonumber
|\langle\phi_\alpha\rangle|^2 &=& \frac{1}{\tilde Z} \frac{\delta}{\delta (J_1^*)_\alpha} \frac{\delta}{\delta (J_2)_\alpha} \tilde Z \\
\nonumber
&=& \frac{1}{\tilde Z} \int D\phi \,\, e^{-\frac{1}{\epsilon}\phi^*(P^2+\epsilon^2)\phi}\\
\nonumber
&& \times \frac{1}{2\epsilon}\left[\sum_{\beta,\gamma}\frac{1}{\epsilon}(P-i\epsilon)_{\alpha\beta}\phi_\beta \phi^*_\gamma(P+i\epsilon)_{\gamma\alpha}-\delta_{\alpha\alpha}\right]\\
\nonumber
&=& \frac{1}{2\epsilon} \left[\sum_{\beta,\gamma}\frac{1}{\epsilon}(P-i\epsilon)_{\alpha\beta}\langle\langle\phi_\beta \phi^*_\gamma\rangle\rangle(P+i\epsilon)_{\gamma\alpha}-\delta_{\alpha\alpha}\right]\\
&=& 0.
\label{eq:vanishingexpectationvalue}
\end{eqnarray}
In the last line of Eq.\ \eqref{eq:vanishingexpectationvalue} we used the standard property of the Gaussian distribution Eq.\ \eqref{eq:gaussiandistroperator}
\begin{equation}
\langle\langle\phi_\beta\phi^*_\gamma\rangle\rangle=\epsilon (P^2+\epsilon^2)_{\beta\gamma}^{-1}.
\end{equation}
Next we turn to the two-point function or ``transition probability''
\begin{eqnarray}
\nonumber
q_{\alpha\beta} &=& \frac{1}{\tilde Z} \frac{\delta}{\delta (J_1^*)_\alpha}\frac{\delta}{\delta (J_2)_\alpha}\frac{\delta}{\delta (J_1)_\beta}\frac{\delta}{\delta (J_2^*)_\beta} \tilde Z[J_1,J_2]\\
\nonumber
&=& \frac{1}{\tilde Z}\int D\phi \,\, e^{\frac{1}{\epsilon}\phi^*(P^2+\epsilon^2)\phi}\\
\nonumber
&&\times \frac{1}{2\epsilon}\left[\sum_{\eta,\gamma}\frac{1}{\epsilon}(P-i\epsilon)_{\alpha\eta}\phi_\eta \phi^*_\gamma(P+i\epsilon)_{\gamma\alpha}-\delta_{\alpha\alpha}\right]\\
\nonumber
&&\times \frac{1}{2\epsilon}\left[\sum_{\kappa,\lambda}\frac{1}{\epsilon}(P+i\epsilon)_{\beta\kappa}\phi_\kappa\phi^*_\lambda(P-i\epsilon)_{\lambda\beta}-\delta_{\beta\beta}\right]\\
&=& \left\langle\langle\rho_\alpha\rho_\beta\right\rangle\rangle-\langle\langle\rho_\alpha\rangle\rangle\langle\langle\rho_\beta\rangle\rangle.
\label{eq:qxyconnexptvalue}
\end{eqnarray}
This expression is the (connected) two-point correlation function of the operator
\begin{equation}
\rho_\alpha=\frac{1}{2\epsilon^2}\sum_{\gamma,\eta}(P-i\epsilon)_{\alpha\eta}\phi_\eta \phi^*_\gamma(P+i\epsilon)_{\gamma\alpha}
\end{equation}
with respect to averaging over the possible field configurations $\phi(x)$. Note that $\rho_\alpha$ is real and positive for all field configurations $\phi$. Indeed, we can write with $P^\dagger =P$
\begin{equation}
\rho_\alpha=\frac{1}{2\epsilon^2} \left|\sum_\eta (P-i\epsilon)_{\alpha\eta}\phi_\eta\right|^2
\end{equation}
showing this more explicit. The multiplicative normalization of $\rho$ is somewhat arbitrary and could be changed by rescaling the fields according to $\phi\to\phi^\prime=c\phi$. 
Note that for on-shell modes with $\phi^* P^2\phi=0$ the operator $\rho$ reads
\begin{equation}
\rho_\alpha=\frac{1}{2} \phi^*_\alpha\phi_\alpha.
\end{equation}

Although the description of $q_{\alpha\beta}$ as a connected correlation function of the operators $\rho_\alpha$ and $\rho_\beta$ is appealing, its meaning as a transition probability is not yet completely clear. In a typical experiment one asks for the probability to find a particle both at the space-time point $y=(y_0,\vec y)$ and at the space-time point $x=(x_0,\vec x)$. We denote the probability for this by $p(x \cap y)$. Quite generally, one would calculate this quantity as a sum over all field configurations $\phi$ weighted by the product
\begin{equation}
p[\phi]\,p(x|\phi] \,p(y|\phi].
\end{equation}
Here $p(x|\phi]$ gives the probability for the event ``particle measured at $x$'' under the condition that the field configuration $\phi$ is realized. The expression $p[\phi]$ is the probability for the field configuration $\phi$. In combination, we find
\begin{eqnarray}
p(x \cap y) &=& \int D\phi\, p[\phi]\,p(x|\phi] \,p(y|\phi].
\end{eqnarray}
Let us now compare this to our expression for $q_{\alpha\beta}$ in Eq.\ \eqref{eq:qxyconnexptvalue}. If we identify $\alpha=x=(x_0,\vec x)$ and $\beta=y=(y_0,\vec y)$ and neglect for the moment the second term in the last line of Eq.\ \eqref{eq:qxyconnexptvalue}, we can write
\begin{equation}
q(x,y)=\int D\phi\, v[\phi]\, \rho(x)\, \rho(y).
\end{equation}
The expressions for $p(x\cap y)$ and $q(x,y)$ are proportional when the probability for the field configuration $\phi$ is
\begin{equation}
p[\phi] \sim v[\phi]
\end{equation}
and the probability to find a particle at $x=(x_0,\vec x)$ for the field configuration $\phi$ is given by
\begin{equation}
p(x|\phi] \sim \rho(x). 
\end{equation}
The subtraction of the term $\langle\langle\rho_\alpha\rangle\rangle\langle\langle\rho_\beta\rangle\rangle$ in Eq.\ \eqref{eq:qxyconnexptvalue} provides for the two events ``particle measured at $y$'' and ``particle measured at $x$'' not to be in a coincidence. Instead, there has to be a ``causal connection'' between them. Only in that case would we speak of ``two measurements on the same particle''. Moreover, fluctuations at different space-time points that are uncorrelated would not show the characteristics of particles at all. Let us assume for definiteness that we use a cloud chamber as a particle detector. The vapor would only condense if neighboring points in space are stimulated during a small but nonzero period of time. Stimulations at random points in space-time would not lead to the detection of a particle. The disconnected part of the two point function $\langle\langle\rho_\alpha\rangle\rangle\langle\langle\rho_\beta\rangle\rangle$ should therefore be seen as part of the nontrivial vacuum structure in quantum field theory. 

To end this subsection let us comment of the general, not necessary Gaussian case. We can obtain the quantum field theoretic one-point function from
\begin{eqnarray}
\nonumber
|\langle\phi_\alpha\rangle|^2 &=& \frac{1}{\tilde Z} \frac{\delta}{\delta (J_1^*)_\alpha} \frac{\delta}{\delta (J_2)_\alpha} \tilde Z\\
\nonumber
&=& \frac{1}{2 \tilde Z} \left(\frac{\delta}{\delta (J_\phi^*)_\alpha}+\frac{\delta}{\delta (J_\chi^*)_\alpha}\right)\\
&&\times \left(-\frac{\delta}{\delta (J_\phi)_\alpha}+\frac{\delta}{\delta (J_\chi)_\alpha}\right) \tilde Z[J_\phi,J_\chi]. 
\end{eqnarray}
With Eq.\ \eqref{eq:ZwithJ1J2} this gives
\begin{eqnarray}
\nonumber
|\langle\phi_\alpha\rangle|^2 &=& \frac{1}{2\tilde Z} \int D\phi\\
&& \times\, \left[\phi^*_\alpha\phi_\alpha+\frac{\delta}{\delta (J_\chi^*)_\alpha}\frac{\delta}{\delta (J_\chi)_\alpha}\right] \, v[\phi,J_\chi]. 
\end{eqnarray}
Here we used that $v[\phi,J_\chi]$ and $|\langle \phi_\alpha\rangle|^2$ have to be real. The general expression for the two-point function $q_{\alpha\beta}$ is somewhat more complicated, but straightforward to obtain in an analogous way as the calculations above. 

\subsubsection{Conservation laws for on-shell excitations}
Although particles are created and annihilated in quantum field theory, these processes are constraint by several conservation laws. For example, in quantum electrodynamics, the electric charge is a conserved quantum number. Electrons and positrons can only be created in pairs such that the total charge remains constant. In a formalism where particles are described as excitations of fields, one must show that these excitations fulfill the usual conservation constraints. 

In quantum field theory, conserved quantities such as charge or also energy are associated to a continuous symmetry via Noether's theorem. However, only the combination of a symmetry together with some field equation leads to a conservation law. For example, for a field that satisfies the on-shell condition
\begin{equation}
P\phi = (-\partial_\mu\partial^\mu-m^2)\phi=0
\end{equation}
one can easily show that the current
\begin{equation}
j^\mu=i(\partial^\mu\phi^*)\phi-i\phi^*(\partial^\mu\phi)
\label{eq:currentrelativistic}
\end{equation}
is conserved, i.~e. $\partial_\mu j^\mu=0$. This current is directly linked to the symmetry of the action
\begin{equation}
S[\varphi]=\int_x \varphi^*(-\partial_\mu\partial^\mu-m^2)\varphi
\end{equation}
under global U(1) transformations $\varphi\to e^{i\alpha}\varphi$, $\varphi^*\to e^{-i\alpha}\varphi^*$. As discussed in the last section, the functional probability $v[\phi]$ is invariant under the same symmetries as the microscopic action $S[\varphi]$ if no anomalies are present. This implies that there should be conservation laws associated with these symmetries for on-shell excitations, that fulfill a field equation as Eq.\ \eqref{eq:onshellcond}. 
We emphasize again that e.~g. the current in Eq.\ \eqref{eq:currentrelativistic} is not conserved for general field configurations with $P\phi\neq0$. However, if particles correspond to on-shell field excitations, the usual conservation laws are indeed fulfilled. 

\subsubsection{Conclusions}
In this appendix we discussed a (quasi-) probability representation of quantum field theory based on the functional integral. We showed for a Gaussian theory of bosonic fields that the functional integral can be reordered such that an interpretation in terms of real and positive probabilities for field configurations (``functional probabilities'') is possible. Our formalism is also applicable to the more general case of non-Gaussian microscopic actions where it may be necessary to work also with negative (quasi-) probabilities. We believe that a description using only positive probabilities is possible in many cases. However, it is not excluded that for some physical theories negative probabilities are needed. This would be highly interesting and demonstrate -- once again -- the extraordinariness of quantum theory. In any case the (quasi-) probability representation developed here might be useful as a theoretical tool, for example in studies of non-equilibrium quantum field dynamics. The formalism can be extended with minor modifications to fermionic or Grassmann valued fields.

The concept of functional probabilities addresses both classical field configurations and particles. The former are described by a nonzero expectation value $\langle\langle\phi\rangle\rangle$ while particles correspond to on-shell excitations, described by the connected two-point function $\langle\langle\phi\phi\rangle\rangle-\langle\langle\phi\rangle\rangle\langle\langle\phi\rangle\rangle$. For quadratic microscopic actions as in Eq.\ \eqref{eq:Gauusianaction} the functional probability is local (Eq.\ \eqref{eq:probmeasuregaussian}). This does no longer have to be the case once interactions are included. For example in a perturbation theory for weak interactions it should be possible to derive explicit expressions beyond the Gaussian case. Higher order correlation functions can then be studied which might shed more light on the question of locality. Interesting features of quantum mechanics as entanglement and the implications of Bells inequalities \cite{Bell} can then be studied in this framework.

\chapter{Technical additions}
\label{ch:Technicaladditions}

\section{Flow of the effective potential for Bose gas}
\label{sec:FlowoftheeffectivepotentialforBosegas}
In this appendix we derive a flow equation for the effective potential for a Bose gas. We use the truncation presented at the beginning of Sect.\ \ref{sec:Derivativeexpansionandwardidentities} and specialize at a later stage to the more simple truncation in Eq.\ \eqref{eqSimpleTruncation}. We derive the flow equation for the effective potential by evaluating the flow equation for the average action \eqref{eq4:Wettericheqn} for constant fields. Inserting a real constant field $\phi(x)=\sqrt{\rho}$ one finds for $U=\Gamma_k/\Omega$ the flow at fixed $\rho$
\begin{eqnarray}
\nonumber
&& \partial_t U(\rho,\mu) \,=\, \eta \rho U^\prime\,+ \,\zeta(\rho,\mu),\\
\nonumber
&& \zeta(\rho,\mu)=T\sum_n 2 v_d \int_0^\infty dp\, p^{d-1}\theta(k^2-p^2-m^2)\\
\nonumber
&& \left(2k^2-\eta(k^2-p^2-m^2)+\partial_t m^2\right)\\
&& \frac{g_1+g_2+2(V_1+\rho V_1^\prime)\omega_n^2}{h^2 \omega_n^2+(g_1+(V_1+2\rho V_1^\prime)\omega_n^2)(g_2+V_1 \omega_n^2)}.
\end{eqnarray}
Here, $d$ is the number of spatial dimensions and we use the abbreviations
\begin{eqnarray}
\nonumber
g_1 &=& k^2-m^2+(Z_2-1+2\rho Z_2^\prime)p^2-(V_3+2\rho V_3^\prime)p^4\\
\nonumber
&&+U^\prime+2\rho U^{\prime\prime},\\
\nonumber
g_2 &=& k^2-m^2+(Z_2-1)p^2-V_3p^4+U^\prime,\\
\nonumber
h &=& Z_1+\rho Z_1^\prime -(V_2+\rho V_2^\prime)p^2,\\
\nonumber
\omega_n&=&2\pi T n,\\
v_d &=& (2^{d+1}\pi^{d/2}\Gamma(d/2))^{-1}.
\label{eqabbeffpot}
\end{eqnarray}
We dropped the arguments $(\rho,\mu)$ at several places on the right hand side. Primes denote derivatives with respect to $\rho$. In the phase with spontaneous symmetry breaking, we have $m^2=\partial_tm^2=0$. 

The Matsubara sums over $n$ can be carried out by virtue of the formulas
\begin{eqnarray}
\nonumber
&& \sum_{n=-\infty}^{\infty}\frac{1}{an^4+bn^2+c}=\frac{\pi}{d\sqrt{2c}} \\
\nonumber
&& \left(\sqrt{b+d}\,\text{coth}(\sqrt{\frac{b-d}{2a}}\pi)-\sqrt{b-d}\,\text{coth}(\sqrt{\frac{b+d}{2a}}\pi)\right), \\
\nonumber
&& \sum_{n=-\infty}^{\infty}\frac{n^2}{an^4+bn^2+c}=\frac{\pi}{d\sqrt{2a}} \\
\nonumber
&& \left(\sqrt{b+d}\,\text{coth}(\sqrt{\frac{b+d}{2a}}\pi)-\sqrt{b-d}\,\text{coth}(\sqrt{\frac{b-d}{2a}}\pi)\right),\\
\label{eqMatsubara}
\end{eqnarray}
with $d=\sqrt{b^2-4ac}$. This brings us to
\begin{eqnarray}
\nonumber
&& \zeta(\rho,\mu)= 2 v_d\int_0^{\sqrt{k^2-m^2}}dp\,p^{d-1}\\
\nonumber
&& (2k^2-\eta(k^2-p^2-m^2)+\partial_tm^2)\frac{1}{\sqrt{8}D}\\
\nonumber
&& \Bigg{(}\left(\sqrt{B+D}\frac{E}{\sqrt{C}}-2\sqrt{B-D}\right)\,\text{coth}(\frac{\sqrt{B-D}}{\sqrt{8A}\,T})\\
\nonumber
&& + \left(2\sqrt{B+D}-\sqrt{B-D}\frac{E}{\sqrt{C}}\right)\,\text{coth}(\frac{\sqrt{B+D}}{\sqrt{8 A}\,T})\Bigg{)},\\
\label{eqFlowEffectivPotentialSystematic}
\end{eqnarray}
where we introduced 
\begin{eqnarray}
\nonumber
A & = & V_1(V_1+2\rho V_1^\prime),\\
\nonumber
B &=& h^2+g_1V_1+g_2(V_1+2\rho V_1^\prime),\\
\nonumber
C &=& g_1 g_2,\\
\nonumber
D &=& \sqrt{B^2-4AC},\\
E &=& g_1+g_2.
\end{eqnarray}
In our simple truncation with $S=Z_1+Z_1^\prime\,\rho_0$, $V=V_1$, $Z_2^\prime=V_2=V_3=V_1^\prime=V_2^\prime=V_3^\prime=0$, and at $\mu=\mu_0$, the integrand in eq. \eqref{eqFlowEffectivPotentialSystematic} becomes mostly independent of the spatial momentum. The integral can than be carried out and we find
\begin{eqnarray}
\nonumber
&& \zeta(\rho,\mu_0)= (1-\frac{\eta}{d+2}) \frac{\sqrt{2}\,v_d}{d \,D}\\
\nonumber
&& \Bigg{(}\left(\sqrt{B+D}\frac{E}{\sqrt{C}}-2\sqrt{B-D}\right)\,\text{coth}(\frac{\sqrt{B-D}}{\sqrt{8A}\,T})\\
\nonumber
&& + \left(2\sqrt{B+D}-\sqrt{B-D}\frac{E}{\sqrt{C}}\right)\,\text{coth}(\frac{\sqrt{B+D}}{\sqrt{8 A}\,T})\Bigg{)},\\
\end{eqnarray}
with
\begin{eqnarray}
\nonumber
A & = & V^2,\\
\nonumber
B &=& S^2+2 V(k^2+U^\prime+\rho U^{\prime\prime}),\\
\nonumber
C &=& (k^2+U^\prime+2\rho U^{\prime\prime})(k^2+U^\prime),\\
\nonumber
D &=& \sqrt{B^2-4AC},\\
E &=& 2(k^2+U^\prime+\rho U^{\prime\prime}).
\end{eqnarray}
That the momentum integral can be performed analytically is a nice feature of the cutoff \eqref{eq7:DeltaSkbosons}. The limit $T\rightarrow0$ is obtained by substituting the $\text{coth}$ functions with unity. 

The flow of the effective potential contains a subtlety that can be seen in the limit $V_i\rightarrow 0$ ($i=1,2,3$), where we find
\begin{eqnarray}
\nonumber
\partial_t U(\rho,\mu)&=& \eta \rho U^\prime+2 v_d\int_0^{\sqrt{k^2-m^2}}dp\,p^{d-1}\\
\nonumber
&& (2k^2-\eta(k^2-p^2-m^2)+\partial_tm^2)\\
&& \left(\frac{g_1+g_2}{2h\sqrt{g_1 g_2}}\,\text{coth}(\frac{\sqrt{g_1 g_2}}{2\sqrt{h}\,T})+\frac{1}{h}\right).
\end{eqnarray}
The term $1/h$ in the last line is not present if $V_1$ is set to zero from the outset. If $Z_1$ is independent of $\rho$, this term is independent of $\rho$ and gives only an overall shift of the effective potential.

\section{Flow of the effective potential for BCS-BEC Crossover}
\label{sec:FlowoftheeffectivepotentialforBCSBECcrossover}
Now we come to the flow of the effective potential for the BCS-BEC crossover model. Again, we evaluate the flow equation \eqref{eq4:Wettericheqn} with a field $\varphi$ that is constant in space and in (Matsubara-) time $\varphi(\vec x,\tau)=\varphi$. The expectation value of the fermionic field $\psi$ vanishes due to its Grassmann property. Using the truncation in Eq. \eqref{eq:truncation} we find the flow equation for the potential
\begin{eqnarray}
\nonumber
\partial_k U_k{\big |}_{\bar\rho} &=& \frac{1}{2} \int\limits_{q_0,\vec q} {\bigg \{} \left[(G_{k\varphi})_{11}+(G_{k\varphi})_{22}\right]A_\varphi^{-1}\partial_k \left(A_\varphi k^2 r_{k\varphi}\right)\\
\nonumber
&-& \left[ (G_{k\psi})_{13}+(G_{k\psi})_{24}-(G_{k\psi})_{31}-(G_{k\psi})_{42}\right]\\
&& \times \partial_k\left(k^2 r_{k\psi}\right) {\bigg \} }
\label{eq:flowUk1}
\end{eqnarray}
The dimensionless function $r_{k\varphi}$ depends on $y=\vec q^2/(2k^2)$ while $r_{k\psi}$ depends on $z=(\vec q^2-\mu_0)/k^2$. The propagators $G_{k\varphi}$ and $G_{k\psi}$ that appear in Eq. \eqref{eq:flowUk1} are modified by the presence of the ultraviolett regulator
\begin{eqnarray}
\nonumber
G_{k\varphi}^{-1} &=& G_\varphi^{-1}+\begin{pmatrix}  k^2 r_{k\varphi} && 0 \\ 0 && k^2 r_{k\varphi} \end{pmatrix}\\
G_{k\psi}^{-1}  &=& G_\psi^{-1} + \begin{pmatrix} 0 && -k^2 r_{k\psi} \\ k^2 r_{k\psi} && 0 \end{pmatrix}.
\end{eqnarray}
We can now perform the summation over the Matsubara frequencies $q_0=2\pi T n$ for the bosons and $q_0=2\pi T (n+1/2)$ for the fermions. The integration over $\vec q$ is performed quite generally in $d$ spatial dimensions.
The result can be expressed in terms of the dimensionless variables
\begin{eqnarray}\label{DimlessVar}
\nonumber
w_1=\frac{U_k^\prime}{k^2},\quad w_2=\frac{U_k^\prime+2\rho U_k^{\prime\prime}}{k^2},\\
\nonumber 
w_3=\frac{h_\varphi^2 \rho}{k^4}, \quad \tilde \mu=\frac{\mu_0}{k^2},\quad \Delta\tilde\mu=\frac{\mu-\mu_0}{k^2},\\
\nonumber
\tilde T=\frac{T}{k^2},\quad S_\varphi=\frac{Z_\varphi}{A_\varphi},
\end{eqnarray}
and the anomalous dimension 
\begin{equation}\label{AnDim}
\eta_{A_\varphi}=-\frac{k\partial_k A_\varphi}{A_\varphi}.
\end{equation}
The flow of the effective potential reads then after the variable change $\bar \rho\to \rho=A_\varphi \bar \rho$
\begin{eqnarray}
\nonumber
k\partial_k U_k &=& \eta_{A_\varphi} \,\rho\, U_k^\prime+8\sqrt{2} \frac{k^{d+2} v_d }{d\,S_\varphi} \left(1-\frac{2}{d+2}\eta_{A_\varphi}\right) s_{\text{B}}^{(0)}\\
&& - \,8  \frac{k^{d+2} v_d}{d}\,l(\tilde \mu)\, s_{\text{F}}^{(0)}.
\label{eq:flowofUk2}
\end{eqnarray}
Here, the $v_d$ is proportional to the surface of the $d$-dimensional unit sphere, which is $(2\pi)^d 4 v_d$ with $v_d^{-1}=2^{d+1}\pi^{d/2} \Gamma(d/2)$. In particular one has $v_3=1/(8\pi^2)$.

The threshold functions $s_{\text{B}}^{(0)}=s_{\text{B}}^{(0)}(w_1,w_2,\tilde T,S_\varphi,\eta_{A_\varphi})$ and $s_{\text{F}}^{(0)}=s_{\text{F}}^{(0)}(w_3,\tilde \mu,\Delta\tilde\mu,\tilde T)$ as well as the function $l(\tilde \mu)$ used in Eq. \eqref{eq:flowofUk2} depend on the choice of the infrared  regulator functions $r_{k\varphi}$ and $r_{k\psi}$. They describe the decoupling of modes when the effective ``masses'' $w_j$ or $-\tilde \mu$ get large. The threshold functions for the bosonic fluctuations reads
\begin{eqnarray}
\nonumber
\left(1-\frac{2\,\eta_{A_\varphi}}{d+2}\right)s_{\text{B}}^{(0)} = d\int\limits_0^\infty dy\, y^{\frac{d}{2}-1} \left(r_{k\varphi}-y r_{k\varphi}^\prime-\eta_{A_\varphi}r_{k\varphi}\right)\\
\nonumber
\times \frac{\frac{1}{2}(w_1+w_2)+y+r_{k\varphi}}{\sqrt{w_1+y+r_{k\varphi}}\sqrt{w_2+y+r_{k\varphi}}}\\
\nonumber
\times \left[\frac{1}{2}+N_{\text{B}}(\sqrt{w_1+y+r_{r\varphi}}\sqrt{w_2+y+r_{k\varphi}}/S_\varphi)\right]\\
\end{eqnarray}
with the Bose function
\begin{equation}\label{BoseF}
N_{\text{B}}(\epsilon)=\frac{1}{e^{\epsilon/\tilde T}-1}.
\end{equation}
We take it as a condition for the cutoff function $r_{k\varphi}$ that $s_{\text{B}}^{(0)}=1$ for $w_1=w_2=\tilde T=0$. For the calculations below we choose the cutoff function (see chapter \ref{ch:Cutoffchoices})
\begin{equation}
r_{k\varphi}(y)=(1-y)\theta(1-y)
\end{equation}
which gives the particular simple expression
\begin{eqnarray}
\nonumber
s_{\text{B}}^{(0)}&=&\left[\sqrt{\frac{1+w_1}{1+w_2}}+\sqrt{\frac{1+w_2}{1+w_1}}\right]\\
&&\times\left[\frac{1}{2}+N_{\text{B}}(\sqrt{1+w_1}\sqrt{1+w_2}/S_\varphi)\right].
\label{eq:Bosonicth}
\end{eqnarray}
The threshold function for the fermionic fluctuations is obtained similar. For a generic cutoff that addresses the spatial momentum, it reads
\begin{eqnarray}
\nonumber
l(\tilde \mu) s_{\text{F}}^{(0)}=d\int_{-\tilde\mu}^\infty dz\,(z+\tilde\mu)^{\frac{d}{2}-1} \left(r_{k\psi}-z r_{k\psi}^\prime\right)\\
\nonumber
\times \frac{(z+r_{k\psi}-\Delta\tilde\mu)}{\sqrt{w_3+(z+r_{k\psi}-\Delta\tilde\mu)^2}}\\
\times \left[\frac{1}{2}-\,N_{\text{F}}\left(\sqrt{w_3+(z+r_{k\psi}-\Delta\tilde \mu)^2}\right)\right].
\label{eq:thresholdfermionsimplicitdef}
\end{eqnarray}
Here we employ the Fermi function
\begin{equation}\label{FermiF}
N_{\text{F}}(\epsilon)=\frac{1}{e^{\epsilon/\tilde T}+1}.
\end{equation}
Note that for a generic cutoff the right hand side of equation \eqref{eq:thresholdfermionsimplicitdef} does not necessarily factorize. In that case one migth work with a threshold function $s_{\text{F}}^{(0)}$ that also depends on $\tilde \mu$.

Again it is a condition for possible cutoff functions $r_{k\psi}$ to fulfill $S_{\text{F,Q}}^{(0)}=1$ for $w_3=\tilde\mu=\Delta\tilde\mu=0$. We choose the form
\begin{equation}
r_{k\psi}=(\text{sign}(z)-z)\theta(1-|z|).
\end{equation}
This implies for $\mu=\mu_0$ and therefore $\Delta\tilde\mu=0$ the simple form
\begin{equation}
\nonumber
s_{\text{F,Q}}^{(0)}=\frac{2}{\sqrt{1+w_3}}\left[\frac{1}{2}-\,N_{\text{F}}\left(\sqrt{1+w_3}\right)\right]
\label{eq:Fermionicth}
\end{equation}
and
\begin{equation}
l(\tilde \mu)=\theta(\tilde \mu+1)(\tilde \mu+1)^{d/2}-\theta(\tilde \mu-1)(\tilde \mu-1)^{d/2}.
\end{equation}
In the limit $\tilde T=T/k^2\to0$ the thermal contributions to the flow of $U_k$ vanish, $N_{\text{B}}=N_{\text{F}}=0$.

Taking a derivative with respect to $\rho$ on both sides of Eq. \eqref{eq:flowofUk2} we obtain
\begin{eqnarray}
\nonumber
k\partial_k U_k^\prime &=& \eta_{A_\varphi}(U_k^\prime+\rho U_k^{\prime\prime})-8\sqrt{2} \frac{k^d v_d}{dS_\varphi}\left(1-\frac{2}{d+2}\eta_{A_\varphi}\right)\\
\nonumber
&&\times \left[U_k^{\prime\prime}s_{\text{B}}^{(1,0)}+(3U_k^{\prime\prime}+2\rho U_k^{(3)})s_{\text{B}}^{(0,1)}\right]\\
&& +8\frac{k^{d-2}v_d}{d} h_\varphi^2\,l(\tilde \mu)\,s_{\text{F}}^{(1)}.
\end{eqnarray}
Here we introduced the derivatives of the threshold functions
\begin{eqnarray}
\nonumber
s_{\text{B}}^{(1,0)} &=& -\frac{\partial}{\partial w_1}s_{\text{B}}^{(0)}\\
\nonumber
s_{\text{B}}^{(0,1)} &=& -\frac{\partial}{\partial w_2}s_{\text{B}}^{(0)}\\
\nonumber
s_{\text{F}}^{(1)} &=& -\frac{\partial}{\partial w_3}s_{\text{F}}^{(0)}.\\
\end{eqnarray}
We may devide these into contributions from quantum and thermal fluctuations
\begin{eqnarray}
\nonumber
s_{\text{B}}^{(1,0)}&=&(w_2-w_1) s_{\text{B,Q}}^{(1,0)}+s_{\text{B},T}^{(1,0)},\\
\nonumber
s_{\text{B}}^{(0,1)}&=&(w_2-w_1) s_{\text{B,Q}}^{(0,1)}+s_{\text{B},T}^{(0,1)},\\
s_{\text{F}}^{(1)}&=& s_{\text{F,Q}}^{(1)}+s_{\text{F},T}^{(1)}.
\end{eqnarray}
For $\tilde T\to0$ the thermal contribution vanishes $s_{\text{B},T}^{(0,1)}=s_{\text{B},T}^{(1,0)}=s_{\text{F},T}^{(1)}=0$. 
We extracted a factor $(w_2-w_1)$ from the threshold functions $s_{\text{B,Q}}^{(1,0)}$ and $s_{\text{B,Q}}^{(0,1)}$ to make explicit that these contributions vanish for $w_1=w_2$ which holds for $\rho=0$.

For our choice of the regulator functions $r_{k,\psi}$ and $r_{k\varphi}$ we find the explicit expressions
\begin{eqnarray}
\nonumber
s_{\text{B,Q}}^{(1,0)} &=& \frac{1}{4(1+w_1)^{3/2}(1+w_2)^{1/2}},\\
\nonumber
s_{\text{B,Q}}^{(0,1)} &=& -\frac{1}{4(1+w_1)^{1/2}(1+w_2)^{3/2}},\\
\nonumber
s_{\text{F,Q}}^{(1)} &=& \frac{1}{2(1+w_3)^{3/2}}.
\end{eqnarray}
For $\tilde T>0$ the thermal fluctuations lead to the additional contributions from the bosons
\begin{eqnarray}
\nonumber
s_{\text{B},T}^{(1,0)}&=& 2(w_2-w_1)\, s_{\text{B,Q}}^{(1,0)}\, N_{\text{B}}\left(\sqrt{(1+w_1)(1+w_2)}/S_\varphi\right)\\
\nonumber
&& + s_{\text{B,Q}}^{(0)}\frac{\sqrt{1+w_2}}{\sqrt{1+w_1}S_\varphi}\, N_{\text{B}}^\prime\left(\sqrt{(1+w_1)(1+w_2)}/S_\varphi\right),\\
\nonumber
s_{\text{B},T}^{(0,1)}&=& 2(w_2-w_1)\, s_{\text{B,Q}}^{(1,0)}\, N_{\text{B}}\left(\sqrt{(1+w_1)(1+w_2)}/S_\varphi\right)\\
\nonumber
&& + s_{\text{B,Q}}^{(0)}\frac{\sqrt{1+w_1}}{\sqrt{1+w_2}S_\varphi}\, N_{\text{B}}^\prime\left(\sqrt{(1+w_1)(1+w_2)}/S_\varphi\right).\\
\end{eqnarray}
Here, we use the derivative of the Bose function
\begin{equation}
N_{\text{B}}^\prime(\epsilon)=\frac{\partial}{\partial \epsilon}N(\epsilon).
\end{equation}
Similarly, the fermionic part of the thermal contribution reads
\begin{eqnarray}
\nonumber
s_{\text{F},T}^{(1)} &=& -2 \,s_{\text{F,Q}}^{(1)}\,N_{\text{F}}\left(\sqrt{1+w_3}\right)\\
&&-s_{\text{F,Q}}^{(0)}\frac{1}{\sqrt{1+w_3}}\, N_{\text{F}}^\prime\left(\sqrt{1+w_3}\right),
\end{eqnarray}
with the derivative of the Fermi function
\begin{equation}
N_{\text{F}}^\prime(\epsilon)=\frac{\partial}{\partial \epsilon} N_{\text{F}}(\epsilon).
\end{equation}

\section{Hierarchy of flow equations in vacuum}
\label{sec:Hierarchyofflowequationsinvacuum}
In this appendix we sketch the proof of the theorem mentioned at the beginning of chapter \ref{ch:Few-bodyphysics}. The theorem is about a hierarchy of flow equations in the vacuum limit $n\to0$, $T\to0$. In this limit the effective action describes few-body physics which might also be described using the formalism of quantum mechanics. 

We use a notation where the field $\phi$ is a spinor that may contain in general both fermionic and bosonic degrees of freedom. We demand that the microscopic model is invariant under the global $U(1)$ transformation
\begin{equation}
\phi \to e^{i\alpha} \phi,
\label{eq9:GlobalU1}
\end{equation}
where the charge $\alpha$ is the same for all components of $\phi$. To fulfill the condition above it will sometimes be necessary to ``integrate out'' composite fields. 

The second premise is that the microscopic propagator for the field $\varphi$ is of the nonrelativistic form $i q_0+\vec q^2/(2M)+\nu-\mu$. Here we work with a imaginary-time (or Matsubara) frequency $q_0$. The mass $M$ and the gap parameter $\nu$ might be different for the different components of $\phi$. One can then proof the following  

\paragraph{Theorem:} The flow equation for the $n$-point function $G_n$ is independent of correlation functions $G_m$ with order $m>n$. Here we use a notation where $G_n = \Gamma_k^{(n)}$ for $n>2$ and $G_2=(\Gamma_k^{(2)}+R_k)^{-1}$ is the regularized propagator.

\begin{figure}
\centering
\includegraphics[width=0.14\textwidth]{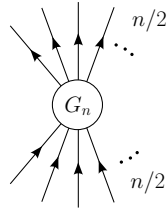}
\caption{Graphical representation of the $n$-point function $G_n$.}
\label{fig9:Gn}
\end{figure}

\begin{figure}
\centering
\includegraphics[width=0.63\textwidth]{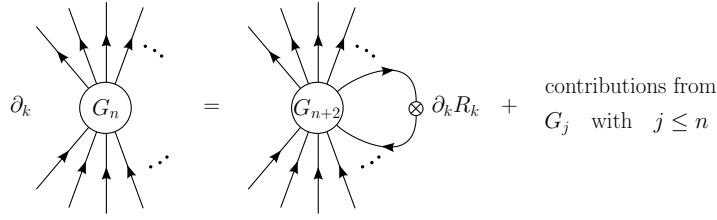}
\caption{Flow equation of the $n$-point function $G_n$. On the right hand side we show only the contribution involving the correlation function $G_{n+2}$. Additional contributions to the flow from lower order function $G_j$ with $j\leq n$ are not shown.}
\label{fig9:flowGn}
\end{figure}

\paragraph{}For the proof we use a basin with independent variables $\phi$ and $\phi^*$. We expand the flowing action in orders of field
\begin{eqnarray}
\nonumber
\Gamma_k[\phi] &=& \Gamma_k(0,0) + \Gamma_k(0,1) + \Gamma_k(0,2) + \ldots\\
\nonumber
&+& \Gamma_k(1,0) + \Gamma_k(1,1) + \Gamma_k(1,2) + \ldots\\
\nonumber
&+& \Gamma_k(2,0) + \Gamma_k(2,1) + \Gamma_k(2,2) +\ldots\\
&+& \ldots.
\end{eqnarray}
Here we denote by $\Gamma_k(i,j)$ a term that is of order $i$ in the conjugate field $\phi^*$ and  of order $j$ in the field $\phi$. We choose the cutoff function to be invariant under the global U(1) transformation in Eq.\ \eqref{eq9:GlobalU1}. Since we expect no anomalies, the flowing action $\Gamma_k[\phi]$ is also invariant. This implies that all contributions $\Gamma_k(i,j)$ vanish except for those with $i=j$,
\begin{equation}
\Gamma_k[\phi] = \Gamma_k(0,0)+\Gamma_k(1,1)+\Gamma_k(2,2)+\ldots.
\end{equation}
For simplicity we write this also as
\begin{equation}
\Gamma_k[\phi] = \Gamma_k(0)+\Gamma_k(2)+\Gamma_k(4)+\ldots.
\end{equation}
The term $\Gamma_k(n)$ contains the information of the $n$-point function $G_n$. Graphically, $G_n$ is represented by a vertex with $n$ external lines. Half of these lines represent incoming particles and the other $n/2$ outgoing particles. We show this schematically in Fig.\ \ref{fig9:Gn}.
Since the flow equation has a one-loop structure, the flow of a term $\Gamma_k(n)$ (or $G_n$) can only have contributions from $G_{n+2}$ if we close one line as shown in Fig.\ \ref{fig9:flowGn}.
There can be no contribution from $G_j$ with $j>n+2$ since we would have to contract a second line which would lead to a two-loop diagram. We have to show that the loop shown on the right hand side in Fig.\ \ref{fig9:flowGn}, i.\ e.\ the contribution of $G_{n+2}$, vanishes. 

The argument uses the dependence of the $n$-point function on the frequency $q_0$. With $q_0$ we denote the imaginary (or Matsubara) frequency of an incoming particle (line with arrow pointing inwards in Figs.\ \ref{fig9:Gn} and \ref{fig9:flowGn}). One of the outgoing particles has frequency $q_0+\Delta q_0$. For the case $n=2$ $q_0$ is just the frequency argument of the regularized propagator. We need for our proof that all poles of the $n$-point function $G_n$ as a function of $q_0$ are in the upper half of the complex plane. For the microscopic propagator (which equals the full propagator in vacuum according to our theorem) this is indeed the case. The two-point function $G_2$ has the form
\begin{equation}
G_2 = \frac{1}{i q_0 + \vec q^2+\nu -\mu+R_k(q_0,\vec q)}.
\label{eq:nonrelativisticpropa}
\end{equation}
With a regulator of the simple form $R_k=k^2$ the frequency pole is at
\begin{equation}
q_0=i(\vec q^2+\nu-\mu+k^2).
\end{equation}
For $\nu-\mu>0$ this is always in the upper half of the complex plane. For more general cutoff functions $R_k(q_0,\vec q)$ we take it as a condition that this feature is not affected. 

For $k\to 0$ and after analytic continuation $q_0\to i\omega$, the pole of $G_2$ determines the dispersion relation, e.\ g.\ 
\begin{equation}
\omega = \vec q^2+\nu-\mu.
\end{equation}
This describes the energy as a function of momentum. As usually in nonrelativistic quantum theory there is a certain ambiguity in the absolute scale for energy. Using the semilocal U(1) symmetry described in section \ref{sec:Derivativeexpansionandwardidentities}, we can (formally) shift the chemical potential corresponding to a shift in the energy scale. The origin of this energy shift ambiguity is in the transition from a relativistic (inverse) propagator
\begin{equation}
-\omega^2+\vec q^2 c^2 + M^2 c^4-i \epsilon
\label{eq:esnonrelpropagator}
\end{equation}
to a nonrelativistic one. In Eq.\ \eqref{eq:esnonrelpropagator} the small imaginary part is to enforce the correct frequency integration contour (Feynman prescription). We write Eq.\ \eqref{eq:esnonrelpropagator} as
\begin{equation}
(\omega+\sqrt{M^2 c^4+\vec q^2 c^2}-i \epsilon)(-\omega+\sqrt{M^2 c^4+\vec q^2 c^2}-i \epsilon).
\label{eq:relativproptwobrackets}
\end{equation}
The first bracket leads to a pole in the propagator that corresponds to antiparticles (negative frequency) while the second bracket describes particles (positive frequency). Close to the point where the second bracket vanishes, one can expand the square root and approximate the first bracket in Eq.\ \eqref{eq:relativproptwobrackets} by a constant. One obtains
\begin{equation}
2M c^2 (-\omega + M c^2+\vec q^2/(2M)-i\epsilon).
\label{eq:relativproptwobrackets2}
\end{equation}
Due to the $i\epsilon$ term the frequency pole of the propagator is always in the lower half of the complex plane. The terms $Mc^2$ in Eq.\ \eqref{eq:relativproptwobrackets2} can now be absorbed into a redefinition of the fields. This includes a simple rescaling and a shift in frequency, similar to the semilocal U(1) transformation described in Sect.\ \ref{sec:Derivativeexpansionandwardidentities}. However, this transformation is not unique and the origin of the energy scale remains undetermined. One can always perform the frequency shift such that all states of a number of particles have positive energies. This holds within the nonrelativistic theory where all energies are small compared to $Mc^2$. The transition to imaginary (or Matsubara-) frequencies $\omega+i\epsilon\to iq_0$ involves an additional relabeling $q_0\to-q_0$ such that we arrive at the form of the propagator in Eq.\ \eqref{eq:nonrelativisticpropa}. 

Now we consider the general $n$-point function $G_n$. A pole at frequency $q_0$ corresponds now to a state of $n/2$ particles with energy $\omega = -i q_0$ of the considered particle. The energy per particle of such a state might be above or below the energy of $n/2$ free particles. In the latter case one speaks of a bound state, in the former of a resonance. In any case we can use the freedom to choose the energy scale to obtain $q_0>0$. In other words we set the energy scale such that all relevant states have positive energy. 

We can now come back to the proof of our theorem. From the above discussion it follows for the loop integral corresponding to Fig.\ \ref{fig9:flowGn} that all poles are in the upper half of the complex plane. We can close the contour in the lower half, implying that the loop integral vanishes. This closes the proof.

\end{appendix}



\chapter*{Danken...}
m\"{o}chte ich ganz besonders Prof.\ Dr.\ Christof Wetterich f\"ur die hervorragende Betreuung und die ausgezeichnete Zusammenarbeit. Von den vielen Gespr\"achen und guten Diskussionen habe ich sehr profitiert.\\

F\"ur tolle Zusammenarbeit und viele interessante Diskussionen danke ich auch ganz herzlich Michael Scherer, Richard Schmidt, Sergej Moroz, Dr.\ Sebastian Diehl, Dr.\ Hans-Christian Krahl, Dr.\ Philipp Strack, Prof.\ Dr.\ Holger Gies, Prof.\ Dr.\ Jan Pawlowski, Prof.\ Dr.\ Markus Oberthaler, Prof.\ Dr.\ Selim Jochim, dessen Arbeitsgruppe sowie insbesondere auch allen Teilnehmern des Seminars ,,Kalter Quantenkaffee''. \\

Prof. Dr. Holger Gies danke ich auch f\"ur die bereitwillige \"{U}bernahme des Zweitgutachtens und der damit verbundenen M\"{u}hen.\\

Anne Doster danke ich f\"ur die sehr sch\"{o}ne, gemeinsam verbrachte Zeit, willkommene und notwendige Ablenkung sowie viel Verst\"{a}ndniss und Unterst\"{u}tzung. Sehr dankbar f\"{u}r sehr Vieles bin ich auch meinen Eltern, Geschwistern und Freunden.
\end{document}